\documentclass[preprint]{aastex}
\usepackage{graphicx}
\usepackage{color}
\newcommand{\myfigA}{\ref{myfigA}}
\newcommand{\myfigB}{\ref{myfigB}}
\newcommand{\myfigC}{\ref{myfigC}}
\newcommand{\myfigD}{\ref{myfigD}}
\newcommand{\myfigE}{\ref{myfigE}}
\newcommand{\myfigF}{\ref{myfigF}}
\newcommand{\myfigG}{\ref{myfigG}}
\newcommand{\myfigH}{\ref{myfigH}}
\newcommand{\myfigI}{\ref{myfigI}}
\newcommand{\myfigJ}{\ref{myfigJ}}
\newcommand{\myfigK}{\ref{myfigK}}
\newcommand{\mytabA}{\ref{mytabA}}
\newcommand{\mytabB}{\ref{mytabB}}
\newcommand{\mytabC}{\ref{mytabC}}
\newcommand{\mytabD}{\ref{mytabD}}

\begin{document}
%%%%%%%%%%%%%%%%%%%%%%%%%%%%%%%%%%%%%%%%%%%%%%%%%%%%%%%%%%%%%%%%%%%%%%%%%
%%%%%%%%%%%%%%%%%%%%%%%%%%%%%%%%%%%%%%%%%%%%%%%%%%%%%%%%%%%%%%%%%%%%%%%%%
\title{A comprehensive analysis of {\em Fermi} Gamma-Ray Burst Data: IV. Spectral lag and Its Relation to $E_p$ Evolution}
\author{Rui-Jing Lu\altaffilmark{1}, Yun-Feng Liang\altaffilmark{2}, Da-Bin Lin\altaffilmark{1}, Jing L\"{u}\altaffilmark{1}, Xiang-Gao Wang\altaffilmark{1}, Hou-Jun L\"{u}\altaffilmark{1}, Hong-Bang Liu\altaffilmark{1}, En-Wei Liang\altaffilmark{1}, Bing Zhang\altaffilmark{3,1}}
\altaffiltext{1}{Guangxi Key Laboratory for Relativistic Astrophysics, Department of Physics, Guangxi University, Nanning 530004, China; lindabin@gxu.edu.cn, luruijing@gxu.edu.cn, lew@gxu.edu.cn}
\altaffiltext{2}{Key Laboratory of Dark Matter and Space Astronomy, Purple Mountain Observatory, Chinese Academy of Sciences, Nanjing 210008, China}
\altaffiltext{3}{Department of Physics and Astronomy, University of Nevada Las Vegas, NV 89154, USA; zhang@physics.unlv.edu}
%%%%%%%%%%%%%%%%%%%%%%%%%%%%%%%%%%%%%%%%%%%%%%%%%%%%%%%%%%%%%%%%%%%%%%%%%
%%%%%%%%%%%%%%%%%%%%%%%%%%%%%%%%%%%%%%%%%%%%%%%%%%%%%%%%%%%%%%%%%%%%%%%%%\textcolor[rgb]{1.00,0.00,0.00} {}
\begin{abstract}
The spectral evolution and spectral lag behavior of 92 bright pulses from 84 gamma-ray bursts (GRBs) observed by the Fermi GBM telescope are studied. These pulses can be classified into hard-to-soft pulses (H2S, 64/92), H2S-dominated-tracking pulses (21/92), and other tracking pulses (7/92). We focus on the relationship between spectral evolution and spectral lags of H2S and H2S-dominated-tracking pulses.
%in hard-to-soft pulses (H2S, 64/92) and H2S-dominating-tracking (21/92) pulses.
The main trend of spectral evolution (lag behavior)
is estimated with $\log E_p\propto k_E\log(t+t_0)$ (${\hat{\tau}} \propto k_{\hat{\tau}}\log E$),
where $E_p$ is the peak photon energy in the radiation spectrum,
$t+t_0$ is the observer time relative to the beginning of pulse $-t_0$,
and ${\hat{\tau}}$ is the spectral lag of photons with energy $E$
with respect to the energy band $8$-$25$~keV.
For H2S and H2S-dominated-tracking pulses,
a weak correlation between $k_{{\hat{\tau}}}/W$ and $k_E$ is found, where $W$ is the pulse width.
We also study the spectral lag behavior with peak time $t_{\rm p_E}$ of pulses for 30 well-shaped pulses
and estimate the main trend of the spectral lag behavior with $\log t_{\rm p_E}\propto k_{t_p}\log E$.
It is found that $k_{t_p}$ is correlated with $k_E$.
We perform simulations under a phenomenological model of spectral evolution, and find that these correlations are reproduced.
We then conclude that spectral lags are closely related to spectral evolution within the pulse.
The most natural explanation of these observations is that the emission is from the electrons in the same fluid unit at an emission site moving away from the central engine, as expected in the models invoking magnetic dissipation in a moderately-high-$\sigma$ outflow.
\end{abstract}

%%%%%%%%%%%%%%%%%%%%%%%%%%%%%%%%%%%%%%%%%%%%%%%%%%%%%%%%%%%%%%%%%%%%%%%%%
%%%%%%%%%%%%%%%%%%%%%%%%%%%%%%%%%%%%%%%%%%%%%%%%%%%%%%%%%%%%%%%%%%%%%%%%%
\section{Introduction}\label{sec:intro}
It was theoretically speculated and observationally confirmed that
long gamma-ray bursts (GRBs) are from the collapse of massive stars and
short GRBs are from mergers of compact binaries
(e.g., \citealp{Colgate_SA-1974}; \citealp{Paczynski_B-1986}; \citealp{Eichler_D-1989-Livio_M}; \citealp{Narayan_R-1992-Paczynski_B}; \citealp{Woosley_SE-1993}; \citealp{Woosley_SE-2006-Bloom_JS}; \citealp{Kumar_P-2015-Zhang_B}).
A spectral lag, defined as the time delay of high-energy photons with
respect to low-energy photons, is commonly observed in long GRBs
(\citealp{Norris_JP-1986-Share_GH}; \citealp{Cheng_LX-1995-Ma_YQ}; \citealp{Band_DL-1997}; \citealp{Norris_JP-2000-Marani_GF}),
but is not in short ones (\citealp{Yi_T-2006-Liang_E}; \citealp{Norris_JP-2006-Bonnell_JT}).
An extensive analysis of the data from the Burst And Transient Source Experiment (BATSE)
also shows that long wide-pulse GRBs tend to have long spectral lags (\citealp{Norris_JP-2005-Bonnell_JT}).
The distinguished features of spectral lags in these two types of GRBs are
proposed to be a phenomenological indicator in GRB classification
(\citealp{Yi_T-2006-Liang_E}; \citealp{Norris_JP-2006-Bonnell_JT};
\citealp{gehrels06}; \citealp{McBreen_S-2008-Foley_S}; \citealp{zhangbb09}).
An anti-correlation between the spectral lag and peak luminosity
is found in a few redshift-known BATSE GRBs (\citealp{Norris_JP-2000-Marani_GF}).
The low-luminosity GRB~060218 detected with the Burst Alert Telescope (BAT)
onboard Swift satellite is consistent with this correlation (\citealp{Liang_EW-2006-Zhang_BB}).
The precursor and main prompt emission in GRB 061121 show different spectral lags,
but both are consistent with the above correlation (\citealp{Page_KL-2007-Willingale_R}).
The above correlation also holds in \emph{Swift}/BAT GRB sample and even in the
X-ray flares observed with the X-Ray Telescope onboard {\em Swift},
albeit with large scatter (\citealp{Ukwatta_TN-2010-Stamatikos_M}; \citealp{Margutti_R-2010-Guidorzi_C};
\citealp{Sultana_J-2012-Kazanas_D}; \citealp{Sonbas_E-2013-MacLachlan_GA}; \citealp{Bernardini_MG-2015-Ghirlanda_G}).
The relation between spectral lags and peak luminosity was
proposed to be a distance indicator of GRBs (\citealp{Norris_JP-2000-Marani_GF}; \citealp{Schaefer_BE-2007})
for the purpose of using GRBs
to constrain cosmological parameters.

GRB light curves are usually composed of overlapping pulses (\citealp{Norris_JP-1996-Nemiroff_RJ}; \citealp{Hu_YD-2014-Liang_EW}).
It is generally speculated that each pulse is related to one individual radiation episode,
which represents the fundamental unit of the GRB temporal profile (see, e.g., \citealp{Fishman_GJ-1994-Bhat_PN}).
The spectral lags measured in different emission episodes could be different (e.g., \citealp{Page_KL-2007-Willingale_R}).
However, the measured spectral lag is likely dominated by the wide pulses or the brightest pulse in a light curve.
As a result, \cite{Hakkila_J-2008-Giblin_TW} suggested that the spectral lag is better defined using individual pulses
rather than the whole burst light curve profile.
Significant spectral evolution is also a common feature of GRB pulses
(e.g., \citealp{Liang_E-1996-Kargatis_V}; \citealp{Preece_RD-2000-Briggs_MS}; \citealp{Lu_RJ-2010-Hou_SJ,Lu_RJ-2012-Wei_JJ}; \citealp{Racz_II-2018-Balazs_LG}).
Two evolutionary patterns have been observed.
One is the so-called hard-to-soft (hereafter H2S) pattern,
which can be defined as the peak photon energy ($E_{p}$) in $\nu$-$\nu f_\nu$ spectrum decreasing monotonically
within an individual pulse.
The second is the so-called intensity tracking pattern, which is defined by $E_{p}$ tracking
the flux (\citealp{Wheaton_WA-1973-Ulmer_MP}; \citealp{Norris_JP-1986-Share_GH}; \citealp{Golenetskii_SV-1983-Mazets_EP}; \citealp{Lu_RJ-2012-Wei_JJ}).
It was suggested that the spectral lag behavior of GRB prompt emission may be related to spectral evolution
(\citealp{Ukwatta_TN-2012-Dhuga_KS}; \citealp{uhm16}; \citealp{Preece2016}).
Some authors also pointed out that the spectral evolution is related to the pulse profile (Hakkila \& Preece 2011; Hakkila et al. 2015, 2018).
For example, the hard-to-soft spectral evolution may primarily occur in hard and/or asymmetric pulses,
and the intensity-tracking spectral evolution may be more prevalent in soft and/or symmetric pulses (Hakkila et al. 2015, 2018).
In addition, Hakkila \& Preece (2014) found that the residuals of single pulse fits to GRB lightcurves leave behind a mysterious triple-peaked structure. They argued that the interplay between such a triple-peaked structure and pulse fits may complicate spectral evolution in pulses.
%observed in GRB pulses may give rise to an asymmetric pulse
%(detailed information please refers to Hakkila et al. 2014).
In this paper, we explore the relation between the spectral lags and spectral evolution.

Gamma-Ray Burst Monitor (GBM) onboard the Fermi satellite has established a large GRB sample.
We present a comprehensive analysis of the \emph{Fermi} GRB data
and report our results in a series of papers.
We revealed the spectral components and their temporal evolution of the \emph{Fermi} GRBs in the first two papers
of this series (\citealp{zhangbb11}; \citealp{Lu_RJ-2012-Wei_JJ})
and studied energy dependence of the burst duration
in the third paper (\citealp{Qin_Y-2013-Liang_EW}). This paper is dedicated to investigating the spectral lag and its relation to spectral evolution within bright GRB pulses. We describe our sample selection and data reduction in Section~2.
The relations between the spectral evolution and the spectral lag behavior are studied in Section~3.
Based on the results in Section~2,
we explore the possible origin of spectral lags by performing simulations within the framework of a phenomenological model in Section~4.
We summarize our results in Section~5.
Throughout of this paper,
a flat $\Lambda$CDM cosmology with the parameters $H_{0}=71  \rm km \cdot s^{-1}\cdot Mpc^{-1}$, $\Omega_{\rm M}=0.3$,
and $\Omega_{\Lambda}=0.7$ is adopted.
The quoted uncertainties are at $1\sigma$ confidence level.

\section{Samples and Data Reduction}\label{sec:data}
We download the GBM data of GRBs by August, 2015 from Fermi Archive FTP websites\footnote{ftp://legacy.gsfc.nasa.gov/fermi/data/}.
The energy-dependent light curves are extracted with the python source package
$gtBurst$\footnote{http://sourceforge.net/projects/gtburst/.},
the standard \texttt{HEASOFT} tools (version 6.20), and the {\em Fermi} \texttt{Science Tool} (v10r0p5).
The TTE data from the brightest detectors
(with the smallest angle between this detector and the source object)
are used in our analysis.
We select bright and well-shaped pulses for our analysis.
To identify such a pulse, we employ an empirical pulse model (\citealp{Kocevski_D-2003-Ryde_F}), i.e.,
\begin{equation}\label{KRL}
I(t)=I_{\rm m}\left(\frac{t+t_0}{t_{\rm m}+t_0}\right)^r\left[\frac{d}{d+r}+\frac{r}{d+r}\left(\frac{t+t_0}{t_{\rm m}+t_0}\right)^{r+1}\right]^{-\frac{r+d}{r+1}},
\end{equation}
to fit the bright pulses,
where $t_{0}$ measures the offset of the pulse zero time relative to
the GRB trigger time, $t_{\rm m}$ is the time of the peak flux
($I_{\rm m}$), and $r$ and $d$ are the power-law rising and decaying indices,
respectively.
Note that the values of $r$ and $d$ are degenerated with $t_0$ and
thus one needs to set a $t_0$ value for estimating the value of $r$ and $d$ for a pulse.
In this analysis, we simply set $t_0$ at the time corresponding to $0.1I_m$ in the rising phase.
To obtain the value of $t_0$, we first give a trial fit to a pulse profile with Equation~(\ref{KRL}).
With the fitting result, one can easily identify the time of $0.1I_{\rm m}$,
which is set as the value of $t_0$.
Then, we fit the pulse profile with Equation~(\ref{KRL}).
We would like to select relatively smooth pulses and adopt $\chi^2_{\rm r}<1.5$ as our pulse selection criterion.
Owing to the mini-pulses or the triple-peaked structure (e.g., bn0901310901, Hakkila and Preece 2014), a few pulses have larger $\chi^2_{\rm r}$.
However, the spectral lags of these pulses are dominated by the bright pulse.
Similar to the work of \cite{Hakkila_J-2008-Giblin_TW},
we also include these pulses in our analysis.
We finally obtain 92 pulses from 84 GRBs for our analysis.
These pulses along with our fittings (red solid lines) are shown in Figures~{\myfigA} (H2S spectral evolution)
and {\myfigB} (tracking spectral evolution),
where the spectral evolution behavior is identified by eye (see Section~\ref{Sec:Spectral_Evolution} for detail discussion). The results of our light curve fittings are reported in Tables~{\mytabA}.
Note that there might be a triple-peaked structure in the residual of our defined pulses (\citealp{Hakkila2014}).
Equation~(\ref{KRL}) only provides a first-order approximation to the pulse profiles.

We then perform an analysis for the spectral evolution and the spectral lag.
The Gamma-Ray Spectral Fitting Package (RMFIT, Version 4.3.2)\footnote{http://fermi.gsfc.nasa.gov/ssc/data/analysis/rmfit/}
is used to extract the time-dependent spectra with a signal-to-noise ratio $\rm SNR>40$
and perform joint spectral fitting with the Band function (\citealp{Band_IM-1993-Trzhaskovskaya_MB}).
The spectral fitting results are reported in Tables~{\mytabB}.
The values of $E_p$, which are marked as violet ``$\star$'' and labeled with bottom $x$- and right $y$-axis in the left panel
and top $x$- and right $y$-axis in the right panel of each sub-figure, are plotted Figures~{\myfigA} and {\myfigB}.
In Figures~{\myfigA} and {\myfigB},
the two vertical dotted lines in the left panel of each sub-figure
mark the time period (see first column of Table~{\mytabC}) adopted to perform the cross correlation function (CCF) analysis in order to obtain the spectral lag.
The procedure to obtain the spectral lag ${\hat{\tau}}$ and its error is shown as follows.
(1) For each energy band, we produce $10^{3}$ sets of simulated light curves with the Monte Carlo method
by assuming that the observed flux error is normal distribution.
(2) We calculate the CCF(${\hat{\tau}}'$) of our simulated light curve for high-energy photons and that for lowest-energy photons (\citealp{Band_DL-1997}; \citealp{Norris_JP-2000-Marani_GF,Norris_JP-2005-Bonnell_JT}; \citealp{Yi_T-2006-Liang_E}; \citealp{Ackermann_M-2010-Asano_K}).
(3) We fit the ${\hat{\tau}}'-\rm CCF$ relation with a Gaussian function
and obtain the maximum value of CCF and its corresponding value of ${\hat{\tau}}'={\hat{\tau}}'_p$.
(4) The value of ${{\hat{\tau}}}$ and its $1\sigma$ uncertainty are obtained by fitting the distribution of
${\hat{\tau}}'_p$ with a Gaussian function for a sample of $10^3$ pairs of our simulated light curves.
In our analysis,
we estimate the spectral lag (${\hat{\tau}}$) with respect to 8-25 keV for four energy bands
(\textcircled{1}: 25-50 keV, \textcircled{2}: 50-100 keV, \textcircled{3}: 100-300 keV,
\textcircled{4}: 300-1000 keV) of NaI detectors
and two energy bands (\textcircled{5}: 300-1000 keV, \textcircled{6}: 1000-5000 keV) of the BGO detectors if it is available.
In some GRBs, especially soft pulses, we also estimate the spectral lag for other energy bands,
such as $\textcircled{7}$: 25-40 keV, $\textcircled{8}$: 40-60 keV,
$\textcircled{9}$: 60-100 keV, $\textcircled{10}$: 16-25 keV.
The same energy band (300-1000 keV) in the NaI (\textcircled{4}) and BGO (\textcircled{5}) detectors
is chosen in order to perform cross-check of the lag in different detectors.
Our obtained spectral lag ${\hat{\tau}}$ can be found in Table~{\mytabC}
and is plotted in the right panel of sub-figure of Figures~{\myfigA} and {\myfigB} with the black ``$\bullet$'' (bottom $x$- and left $y$-axis), where the value of $E$ is the median value of each energy band in the logarithmic space.
In this work, we also estimate the spectral lag (${{\hat{\tau}}}_{31}$) between 100-300 keV and 25-50 keV
in order to compare our results with those found in the previous works.
The value of ${{\hat{\tau}}}_{31}$ is reported in Table~{\mytabD}.

\section{Spectral Evolution and Spectral Lag}\label{Sec:Lag vs. spectral evolution}
\subsection{Spectral Evolution}\label{Sec:Spectral_Evolution}
In this work, we plan to investigate the possible relation between spectral evolution and spectral lags and
identity the main trend of spectral evolution. For this purpose,
we make a linear fit to the data sets $(\log (t+t_0),\log E_{p})$,
i.e., $\log E_{p}=k_{E}\log (t+t_0)+b$.
A steeper slope (i.e., larger $|k_E|$) indicates a faster softening.
The results of our fittings can be found in the right panel of each sub-figure of Figures~{\myfigA} (for H2S pulses) and {\myfigB} (for tracking pulses),
and the derived slope $k_{E}$ of each burst is reported in Table~{\mytabD}.
The left panel of Figure~{\myfigC} shows the distribution of $k_{E}$, where we have defined three types of pulses:
H2S pulses (the green ``$/$'' hatch), H2S-dominated-tracking pulses (blue ``$\setminus$'' hatch), and other tracking pulses
(``$o$'' hatch). The definitions of the latter two categories are as follows.
From Figure~{\myfigC}, we find that all the H2S pulses have a value of $k_{E}<-0.2$,
ranging from $-0.23$ to $-1.55$.
In addition, the tracking pulses in Figure~{\myfigB} with a value of $k_{E}<-0.2$ (21/28) have essentially the same distribution as the H2S pulses at the 5\% significance level
by performing the Anderson-Darling test.
In these pulses, the dominating feature of $E_{p}$
evolution is still H2S during their decay phase and the tracking
feature is illustrated by only one or two hard spectra in their rising phase.
Therefore, we define that these tracking pulses (21/28) with a value of $k_{E}<-0.2$
%dominate by a H2S evolution pattern, calling these pulses
as H2S-dominated-tracking pulses.
%which are shown with blue ``$\setminus$'' hatches in Figure~{\myfigC}.
The H2S and H2S-dominated-tracking pulses are included in our statistical analysis about the relationship
between spectral evolution and spectral lag.
For other pulses in Figure~{\myfigB},
four tracking pulses (4/28), i.e., GRBs 110622158, 120222021, 120402669 and 130612456
have a value of  $k_{E}\simeq0$, indicating no clear $E_{p}$ evolution.
The other three tracking pulses (3/28), i.e., GRBs 100122616, 131214705 and 120308588,
have a positive  $k_{E}$, indicating that those pulses are dominated by a soft-to-hard spectral evolution pattern.
These seven pulses are shown with the black ``$o$'' hatches in Figure~{\myfigC}, and are grouped into the category of other tracking pulses.

It is suggested that the spectral evolution may be related to the pulse profile
(\citealp{Hakkila2011,Hakkila2015,Hakkila2018}).
For example, the H2S spectral evolution may primarily occur in hard and/or asymmetric pulses,
and the intensity-tracking spectral evolution may be more prevalent in soft and/or symmetric pulses
(\citealp{Hakkila2015, Hakkila2018}).
We then study the relationship between pulse width ($W$) and pulse asymmetry ($\kappa_a$)
in Figure~{\myfigK}, where $\kappa_a=T_d/T_r$ with $T_d$ ($T_r$) being the time interval between $I_m$ and $0.1I_m$ in the decay (rising) phase and the pulse width $W=T_d+T_r$.
From this figure, we find that the tracking pulses are systematically more symmetrical compared with the H2S pulses, which is consistent with those found in \cite{Hakkila2015,Hakkila2018}.
This reveals that pulse asymmetry can be taken as an indicator of its spectral evolution (\citealp{Norris_JP-1996-Nemiroff_RJ}).
\cite{Preece2016} pointed out that pulse asymmetry is directly produced by $E_p$ evolution.
Here we only focus on the relationship between spectral evolution and spectral lags.

It should be also noted that the light curve of some GRB pulses may be non-monotonic,
i.e., there is an underlying the triple-peaked profile superposed on the simple pulse fits (\citealp{Hakkila2014}).
The non-monotonicity behavior of the pulses may affect spectral evolution.
The triple-peaked pulse found in the light curve fitting residuals
can be decomposed into a precursor peak, a central peak, and a decay peak (\citealp{Hakkila2014}).
The precursor peak is always harder than
the decay profile in which these two peaks can be measured.
As a result, the spectral evolution of the pulses would always take on a H2S behavior (\citealp{Hakkila2015}).
However, some pulses often exhibit re-hardening at or around the time of the central/decay peak.
These pulses may then take an intensity-tracking behavior,
depending on the hardness contribution made by each of these peaks.
The spectral evolution in these pulses would be complex
and thus the relation of spectral evolution and spectral lags may be difficult to estimate.
To avoid these complications, our analysis about the relation of spectral evolution and spectral lags
is performed based on the H2S and H2S-dominated pulses.

\subsection{Spectral Lag Behavior}\label{Sec:Spectral_lag_Behavior}
In Figure~{\myfigA}, a clear dependence of ${{\hat{\tau}}}$ on $E$ can be found.
Since ${{\hat{\tau}}}$ is almost a linear function of $\log(E)$ in this figure,
we estimate the relation of ${\hat{\tau}}-\log E$ with a linear fit, i.e.,
${{{\hat{\tau}}}}=k_{{{\hat{\tau}}}}\log E + b$. Our fitting results are shown in Figure~{\myfigA}
and the derived $k_{{\hat{\tau}}}$ is reported in Table~{\mytabD}.
A similar dependence of ${{\hat{\tau}}}$ on $E$ can be also found in the H2S-dominated-tracking pulses.
We then also estimate the relation of ${\hat{\tau}}-\log E$ by performing a linear fitting for these pulses.
The fitting results can also be found in Figure~{\myfigB} and the derived value of $k_{{\hat{\tau}}}$ is also reported in Table~{\mytabD}.
The distribution of $k_{\hat{\tau}}$ can be found in the right panel of Figure~{\myfigC}.

In Figure~{\myfigD},
we demonstrate the relations of $k_{\hat{\tau}}$-$W$ (left panel), $k_{E}$-$W$ (middle panel),
and $k_{\hat{\tau}}/W$-$k_{E}$ (right panel) for H2S pulses (black ``$\bullet$'') and H2S-dominated-tracking pulses (blue ``$\star$'').
Here, $W$ is the full width at half maximum of the pulse observed at 8keV-1MeV energy band
and can be found in Table~{\mytabD}.
We perform a Kendall's tau correlation test to the $k_{{\hat{\tau}}}$-$W$ sample
and obtain a tau statistic value of 0.578 with the probability $p\ll 10^{-4}$ by testing non-correlation.
We then employ a linear fit to the data from the H2S pulses and H2S-dominated-tracking pulses.
The relation of $\log k_{{\hat{\tau}}}=(-0.821\pm0.057)+(1.269\pm0.088)\log W$
is obtained.
This relation indicates that a wider pulse gives a steeper slope for the relation of ${{\hat{\tau}}}$ and $E$.
We also notice that the value of the $k_{{\hat{\tau}}}$ found in the H2S-dominated-tracking pulses is systematically smaller than
that found in the H2S pulses (see the left panel in Figure~{\myfigD}).
This is owing to the hardening behavior found in the rising phase of the H2S-dominated-tracking pulses.
In the middle panel of Figure~{\myfigD},
we do not find any relation between $k_{E}$ and $W$ in the both H2S pulses and H2S-dominated-tracking pulses.
In the right panel of Figure~{\myfigD}, a Kendall's tau correlation test indicates a weak correlation between $\log (k_{{\hat{\tau}}}/W)$ and $\log (-k_E)$ with a tau statistic value of 0.307 and the probability $p < 10^{-4}$.
We perform a linear fit to the data
and obtain the correction of
$\log (k_{{{\hat{\tau}}}}/W)=(-0.558\pm0.032)+(0.594\pm0.115)\log (-k_{E})$.
This relation indicates that a quick evolution of $E_{p}$ would form a strong evolution of ${{\hat{\tau}}}$ relative to the photon energy $E$.

To compare our analysis with that in the literature,
we also study the relations of ${{\hat{\tau}}}_{\rm 31}$ and $W$
for the H2S pulses (blue ``$\bullet$'') and the H2S-dominated-tracking pulses (violet ``$\star$'') in the left panel of Figure~{\myfigE}.
It is shown that the value of ${{\hat{\tau}}}_{31}$ is positively related to $W$.
We find a correction of $\log{{\hat{\tau}}}_{31}\propto \varepsilon \log(W/{\rm s})$ with $\varepsilon =0.88\pm0.03$,
which is consistent with that found in \cite{Arimoto_M-2010-Kawai_N}
($\varepsilon =0.86$ in the 6-25 keV band and $\varepsilon =1.06$ in the 50-400 keV band),
but slightly shallower than that obtained by \cite{Hakkila_J-2008-Giblin_TW}
($\varepsilon =1.17$) and that of \cite{Norris_JP-2005-Bonnell_JT} ($\varepsilon =1.42$).
This relation shows that a wide pulse tends to have a large spectral lag (\citealp{Norris_JP-2005-Bonnell_JT}), but it is not the case in a tracking pulse.
We also notice that the value of ${{\hat{\tau}}}_{31}$ in the H2S-dominated-tracking pulses is systematically smaller than
that found in the H2S pulses.
The reason for this behavior is the same as that found in the left panel of Figure~{\myfigD}.

\subsection{$t_{\rm p_E}$-$E$ and $W_{E}$-$E$ Relation}
To further study the spectral lag behavior,
we select the pulses which are well shaped in at least three energy bands.
Thirty pulses are obtained.
We then plot the $t_{\rm p_E}$ (left $y$-axis) and $W_{E}$ (right $y$-axis) as a function of $E$ in Figure~{\myfigF},
where $t_{\rm p_E}$ ($W_{E}$) is the peak time (full width at half maximum) of the pulse
and $t_0$ is from Table~{\mytabA}.
One can find that both $t_{\rm p_E}$ and $W_{E}$ are negatively related to the photon energy $E$
for the H2S pulses (19/30) and H2S-dominated-tracking pulses (4/30).
These behaviors suggest
that the pulse profile in a lower energy band tends to peak earlier and be wider than that in a higher energy band
for the H2S pulses and H2S-dominated-tracking pulses.
In 4/30 other tracking pulses (GRBs 100122616, 120222021, 120402669 and 130612456),
$W_{E}$ tends to become narrower with an increasing photon energy $E$
while $t_{\rm p_E}$ is positively correlated to $E$.
On the other hand, in the other 3/30 tracking pulses (GRBs 110622158, 120308588 and 131214705),
both $t_{\rm p_E}$ and $W_{E}$ are positively related to the energy $E$.
We estimate the $E$-dependent behavior of $t_{\rm p_E}$ and $W_{E}$
by performing a linear fit to the relations of $\log(t_{\rm p_E}+t_0)$-$\log E$ and $\log W_{E}$-$\log E$,
i.e., $\log (t_{\rm p_E}+t_0) =k_{t_p}\log E + b$ and $\log W_{E} =k_{_W}\log E + c$, respectively.
Our best fitting results are plotted in Figure~{\myfigF}.
One can find that for smooth H2S pulses (e.g., bn081224887 and bn090922539),
a linear model does fit the data well.
For tracking pulses (e.g., bn131214705),
the rising part is too short and thus the evolution trend of $E_p$ is difficult to estimate.
However, a linear fit to the data can roughly estimate the main trend of the spectral evolution.
In addition, \cite{uhm16} found from their pulse modeling that the peak time and the width of a pulse are
all linearly related to $E$ in logarithm space. As a result,
we perform linear fits to the data in Figure~{\myfigF}.
In Figure~{\myfigG}, we show the relations of $k_{W}$-$k_{t_p}$ and $k_{t_p}$-$k_{E}$
together with $k_{W}$, $k_{t_p}$, and $k_{E}$ distributions.
One can find from this figure that the values of $k_{W}$, $k_{t_p}$, and $k_{E}$
are distributed at the center of $-0.25$, $-0.08$, and $-0.8$, respectively.
In addition, the tentative correlations, i.e.,
$k_W=(-0.16\pm0.02)+(1.03\pm0.20)k_{t_p}$ with $(r,p)=(0.77,\;\ll{0.0001})$
and $k_{t_p}=(-0.46\pm0.21)+(0.89\pm0.29)k_E$ with $(r,p)=(0.40,6.09\times 10^{-2})$,
are obtained for H2S and H2S-dominated-tracking pulses, respectively.
The relation of $k_{tp}-k_E$ reveals that
the dependence of spectral lag on $E$ is likely related to the spectral evolution.

\section{Origin of Spectral Lags}
As shown in Section~\ref{Sec:Lag vs. spectral evolution},
the observed spectral lag behavior is likely related to spectral evolution. To illustrate this,
in this section we develop a phenomenological model and investigate the spectral lag behavior based
on simulations. We assume
the flux density $f(t, E)$ of a GRB observed at a given time $t$ and photon energy $E$ as
\begin{equation}\label{flux}
f(t, E)=I(t)\phi(E,t),
\end{equation}
where $I(t)$ and $\phi(E,t)$ are the intensity and normalized radiation spectrum, respectively.
Equation (\ref{KRL}) is adopted to describe the evolution of $I(t)$\footnote{
One caveat here is that the results obtained in this section is the first-order approximation. The
introduction of the underlying triple-peaked structure may somewhat complicate the situaltion.}.
For the radiation spectrum $\phi(E,t)$,
we take the Band function, which is characterized with three parameters,
i.e., low photon energy index $\alpha$, high energy photon index $\beta$,
and break energy $E_{b}$ (Band et al. 1993).
For the Band function, the value of $E_p$ can be straightforwardly derived using $E_{p}(t)=(2+\alpha)E_b$ if $\beta<-2$ is satisfied.
We take the observed spectral evolution pattern to model the $E_{p}(t)$ evolution in our simulations, i.e.,\\
$\bullet\;\;\;\;$Case (I): H2S spectral evolution
\begin{equation} \label{H2S}
\log E_{p}=a +k_E\log t\;\;{\rm with}\;\; k_E<0,
\end{equation}
$\bullet\;\;\;\;$Case (II): Tracking spectral evolution
\begin{equation} \label{tracking}
\log {E_p} = \left\{ {\begin{array}{*{20}{c}}
{b + {k_1}\log t,}&{t \le {t_{\rm{m}}} + \xi ,}\\
{b + ({k_1} - {k_2})\log ({t_{\rm{m}}} + \xi ) + {k_2}\log t,}&{{t>t_{\rm{m}}} + \xi,}
\end{array}} \right.
\end{equation}
where $k_1>0$ and $k_2<0$ are adopted,
and $\xi$ is adopted to describe the difference between the intensity peak and the peak time of $E_{p}(t)$ evolution pattern.

The simulated light curves for Case (I) are plotted in the left panel of Figure~{\myfigH},
where $F_{\rm m}=1$, $t_{\rm m}=5$, $r=1$, $d=2$, $t_{\rm 0}=0$, $\alpha=-1$, $\beta=-2.3$, and $k_{E}=-1$ are used in our simulations. Eight energy bands are adopted which are uniformly distributed in the logarithm space in the range of (10, 10$^{4}$)~keV.
The spectral lags for different energy bands are estimated based on our simulated light curves
and are plotted in the middle panel of Figure~{\myfigH}.
It can be found that the value of ${\hat{\tau}}$ increases with $E$ and saturates at around $10^{3}$ MeV,
which is the value of $E_{p}(t=0)$.
This behavior can be also easily found in the left panel of Figure~{\myfigH}.
The relation of ${{\hat{\tau}}}_{31}$ and $W$ is also studied in the right panel of Figure~{\myfigH},
where the value of $r$ or $d$ in Equation~(\ref{KRL}) is changed in order to produce different $W$ from our simulations.
We also estimate the values of $k_{\hat{\tau}}$ and $k_{t_p}$ for different values of $k_E$ and $E<1\rm MeV$.
The relations of $k_{\hat{\tau}}$-$k_E$ and $k_{t_p}$-$k_E$ are shown in Figure~{\myfigI},
which indicates strong correlation between $k_{\hat{\tau}}$ and $k_E$ or $k_{t_p}$ and $k_E$.

For Case (II), our simulated light curves and the $E$-dependent ${\hat{\tau}}$ are shown in the left and right panels of Figure~{\myfigJ},
respectively.
Here, the value of $F_{\rm m}=1$, $t_{\rm m}=5$, $r=1$, $d=2$, $\xi=-1$, $\beta=-2.3$, $k_1=5$, and $k_2=-2$ are adopted in the simulations,
and the sub-cases with $\xi= -1.5, 0,  -1.5$, are adopted in the upper, middle, and lower sub-figures, respectively.
One can find that there is no spectral lag for all energy bands in the sub-case with $\xi=0$~s.
However, the value of ${\hat{\tau}}$ increases with the observed energy band for the sub-case with $\xi=-1.5$~s
and decreases with photon energy for the sub-case with $\xi=1.5$~s.
It should be noted that the spectral evolution in the sub-case with $\xi=-1.5$~s
is similar to that of H2S-dominated-tracking pulses and that the sub-case with $\xi=1.5$~s
is similar to that of other tracking pulses.
The spectral lag behaviors found in these two sub-cases are similar to those found in Section~\ref{Sec:Lag vs. spectral evolution}.

As shown in the above,
our simulations with different spectral evolution patterns can reproduce the observational results.
This may indicate that spectral lag is the result of temporal evolution of $E_{p}$
across energy bands (e.g., \citealp{Ukwatta_TN-2012-Dhuga_KS}).
That is to say, spectral lag is associated with spectral evolution.
This is also consistent with the physically-motivated simulations within the framework of synchrotron radiation at a large
emission radius (\citealp{uhm16}).

%%%%%%%%%%%%%%%%%%%%%%%%%%%%%%%%%%%%%%%%%%%%%%%%%%%%%%%%%%%%%%%%%%%%%
\section{Conclusions and Discussion}
This work studies the spectral evolution and spectral lag behavior
in 92 bright pulses from 84 GRBs observed by the Fermi GBM telescope.
We focus on the relation of spectral evolution and the spectral lag behavior
in H2S (64/92) and H2S-dominated-tracking (21/92) pulses.
It is found that
the spectral lag (${\hat{\tau}}$) is usually photon-energy-dependent.
In the H2S and H2S-dominated-tracking pulses,
the spectral lag increases with increasing photon energy $E$ (given the same reference band in 8-25 keV).
In addition, the dependence of spectral lag on $E$
may be different from pulse to pulse, even for those in the same GRB.
We then adopt the slope $k_{\hat{\tau}}$ in the relation of ${\hat{\tau}} \propto k_{\hat{\tau}}\log E$
to describe the behavior of $E$-dependent ${\hat{\tau}}$ for different pulses.
The main trend of $E_p$ evolution is approximated by performing a log-linear
fit to the $E_p$-$t$ relation with $\log E_p(t)\propto k_E\log(t+t_0)$
and the $k_E$ is used to describe the behavior of spectral evolution.
For H2S and H2S-dominated-tracking pulses,
a weak relation of $k_{{\hat{\tau}}}/W$ and $k_E$ is found, where $W$ is the full width
at half maximum for the pulses observed at $8$-$10^3$~keV energy band.
For further studying the relation between spectral evolution and spectral lag,
we also investigate the evolution of peak time $t_{\rm p_E}$ in different energy bands
for 30 well shaped pulses.
A log-linear relation of $t_{\rm p_E}$-$E$ is found, i.e. $\log t_{\rm p_E}\propto k_{t_p}\log E$,
in H2S and H2S-dominated-tracking pulses.
In addition, a weak relation between $k_{t_p}$ and $k_E$ is also obtained.
The relations of $k_{{\hat{\tau}}}$-$k_E$ and $k_{t_p}$-$k_E$ together with the spectral evolution pattern
are reproduced in our simulations within the framework of a phenomenological model
that invokes the evolution of a Band-function spectrum with time.
Based on these results, we conclude that spectral lag is related to the spectral evolution in GRBs.

The discovery reported in this paper sheds light on the GRB prompt emission mechanism that is highly debated.
The fact that the spectrum evolves uniformly within each pulse (which has a typical duration of seconds)
suggests that pulses are fundamental units of GRB radiation. Within the framework of GRB models, there
are two scenarios to interpret these broad pulses. The first scenario is that the pulse shape is defined by the
central engine activity history, so that each emission epoch in the light curve corresponds to emission from
different groups of electrons when they reach a certain radius (e.g., photosphere or internal shocks).
The second scenario is that the broad pulse is emission from electrons within the same fluid unit, and different
emission episodes correspond to the same fluid unit emitting at different locations as it streams outwards. This second possibility
corresponds to an emission radius $R_{\rm GRB,pulse} \sim \Gamma^2 c t_{\rm pulse} \sim 10^{15} \ {\rm cm}
(\Gamma/100)^2 (t_{\rm pulse}/ 3 \ {\rm s})$. Any model that invokes an emission radius smaller than this
radius, e.g., the photosphere model (unless $\Gamma$ is extremely small) and the internal shock model
(unless the variability time $\Delta t$ is of the order of pulse duration), belong to the first scenario.
Our data and simulations suggest that spectral lags are closely related to the spectral evolution of
the emission spectrum. This is much easier to realize if emission comes from electrons within the same
fluid unit, so that emission properties can evolve continuously as the fluid unit moves in space (so that the
magnetic field strength, bulk Lorentz factor, and probably characteristic electron Lorentz factors) can
evolve continuously (see the generic physical model discussed by \citealp{uhm16} and \citealp{uhm18}).
Such a scenario naturally produces asymmetric pulse profiles and H2S evolution. Under certain conditions,
one can also reproduce tracking pulses (most are H2S-dominated)\footnote{ The mysterious triple-peaked
structure claimed by \cite{Hakkila2014} is, however, difficult to interpret within the framework of any physically
oriented models.}.
It is consistent with dissipation of magnetic energy at a radius of the order of $R_{\rm GRB,pulse}$,
as has been invoked in the ICMART model of GRB prompt emission (\citealp{zhangyan11}; \citealp{lazarian18}).
Within this picture, the smaller variability timescale overlapping on the broad pulses are produced by the
local Lorentz-boosted regions due to magnetic reconnection within a moderately-high-$\sigma$ bulk flow
(\citealp{zhangyan11}; \citealp{zhangzhang14}; \citealp{deng15}).
Within the framework of the first scenario, on the other hand, the
requirement to have emission spectrum evolves corporately as a function of time is much more demanding, since electrons
are from different fluid units. Indeed within the framework of the photosphere model, it has been shown that it is quite
challenging to produce H2S evolution pulses (\citealp{deng14}). The small-radii internal shock model
(which is needed to interpret the rapid variability in the light curves) also faces the similar problem, since it
requires that the electron Lorentz factors, magnetic fields, and bulk Lorentz factors in different internal
shocks would behave corporately with conspiracy to give rise to the observed spectral evolution.

%%%%%%%%%%%%%%%%%%%%%%%%%%%%%%%%%%%%%%%%%%%%%%%%%%%%%%%%%%%%%%%%%%%%%%%%%%
%%%%%%%%%%%%%%%%%%%%%%%%%%%%%%%%%%%%%%%%%%%%%%%%%%%%%%%%%%%%%%%%%%%%%%%%%%
\acknowledgments
We thank the referee for a constructive report. This work is supported by the National Basic Research Program of China
(973 Program, grant No. 2014CB845800), the National Natural
Science Foundation of China (grant Nos. 11573034, 11773007, 11533003, 11673006, 11603006),
the Guangxi Science Foundation (grant Nos. AD17129006, 2016GXNSFDA380027, 2016GXNSFFA380006, 2017GXNSFFA198008, 2016GXNSFCB380005),the Natural Science Key Foundation of Guangxi under Grant No. 2015GXNSFDA139002, the Special Funding for Guangxi Distinguished Professors (Bagui Yingcai \& Bagui
Xuezhe), and the Innovation Team and Outstanding Scholar Program in Guangxi Colleges.

%%%%%%%%%%%%%%%%%%%%%%%%%%%%%%%%%%%%%%%%%%%%%%%%%%%%%%%%%%%%%%%%%%%%%%%%%%%
%%%%%%%%%%%%%%%%%%%%%%%%%%%%%%%%%%%%%%%%%%%%%%%%%%%%%%%%%%%%%%%%%%%%%%%%%%%%
\clearpage
\begin{figure*}
\vspace{-4em}\hspace{-4em}
\centering
\begin{tabular}{cc}
\includegraphics[angle=0,scale=1,width=0.50\textwidth,height=0.22\textheight]{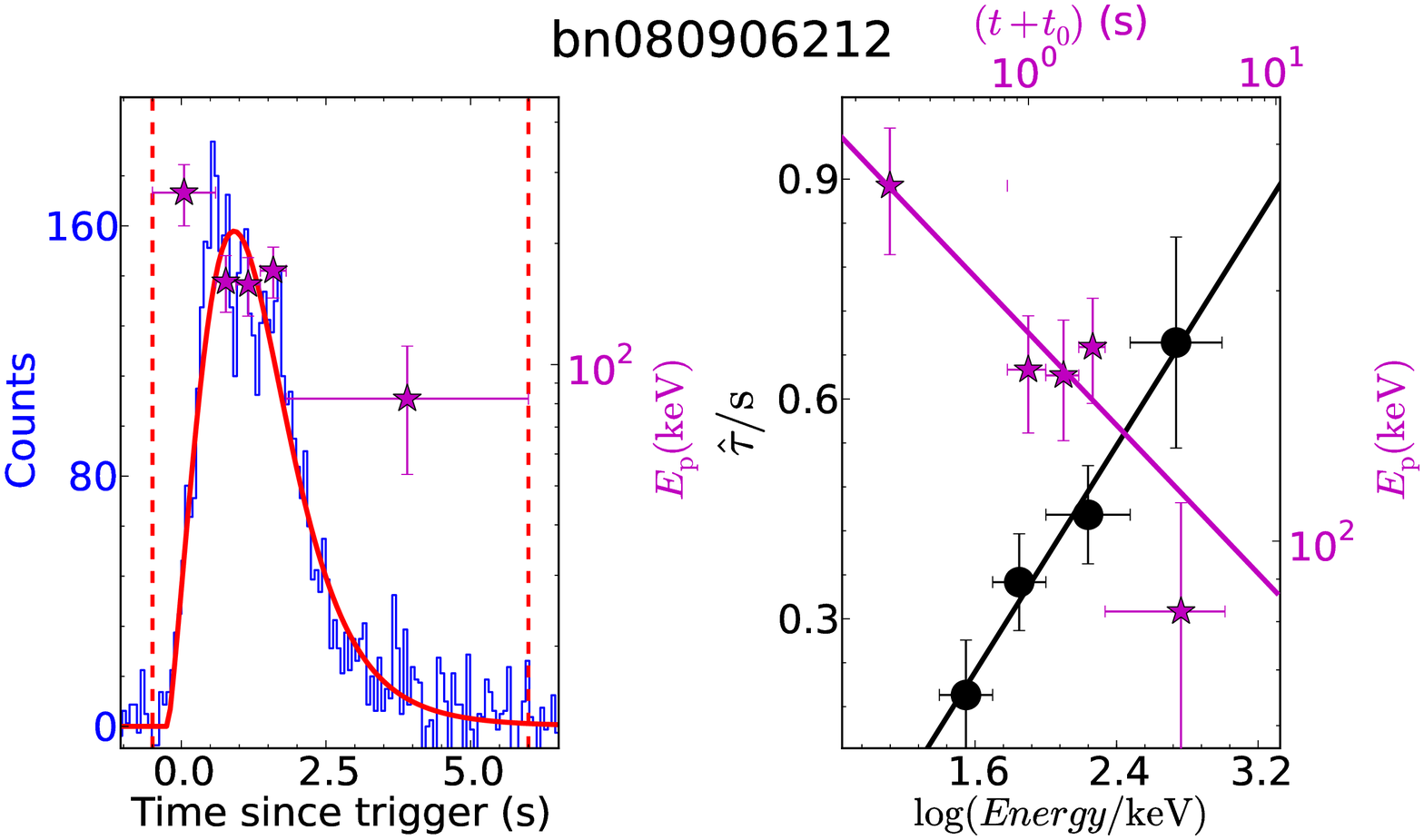} &
\includegraphics[angle=0,scale=1,width=0.50\textwidth,height=0.220\textheight]{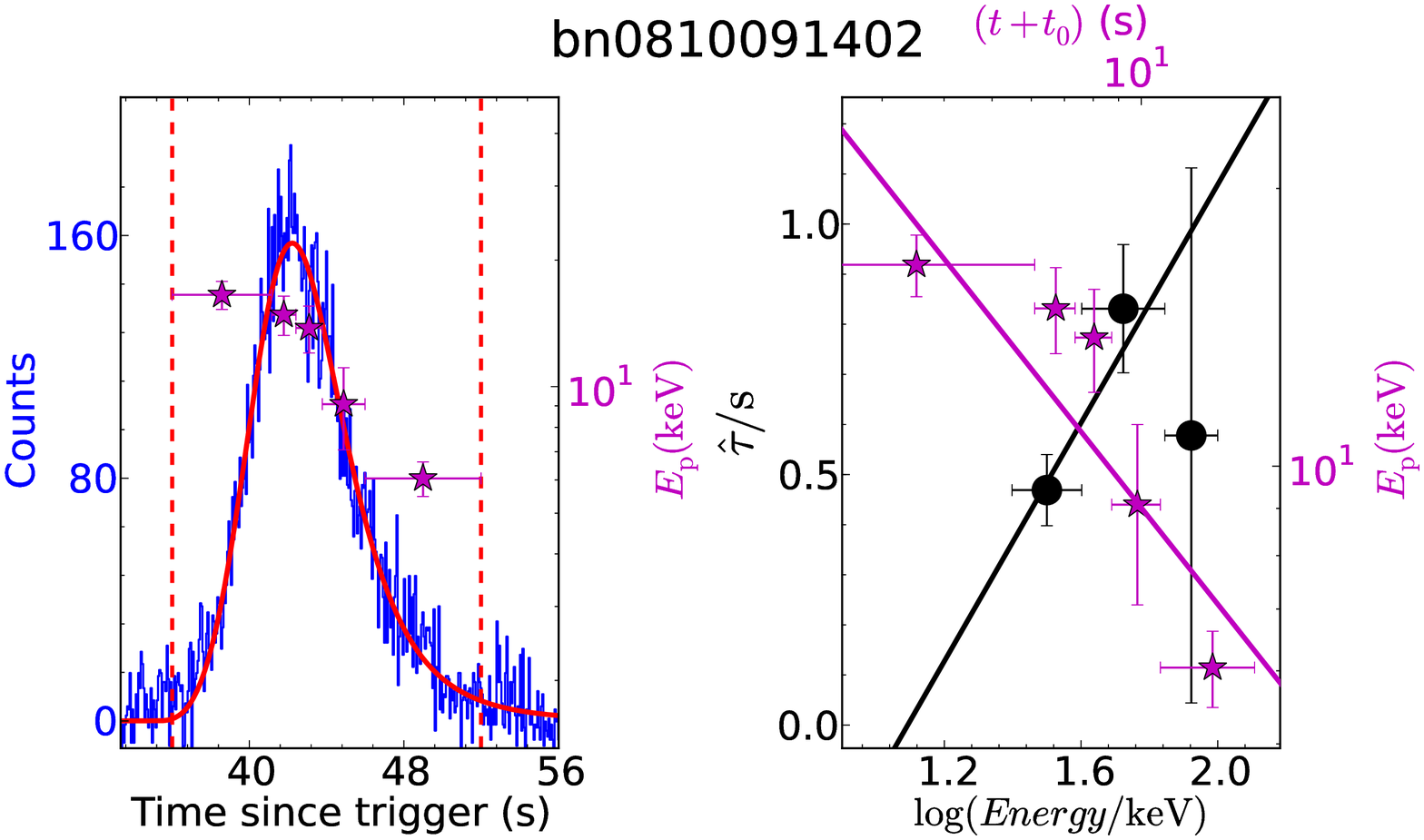} \\
\includegraphics[origin=c,angle=0,scale=1,width=0.50\textwidth,height=0.22\textheight]{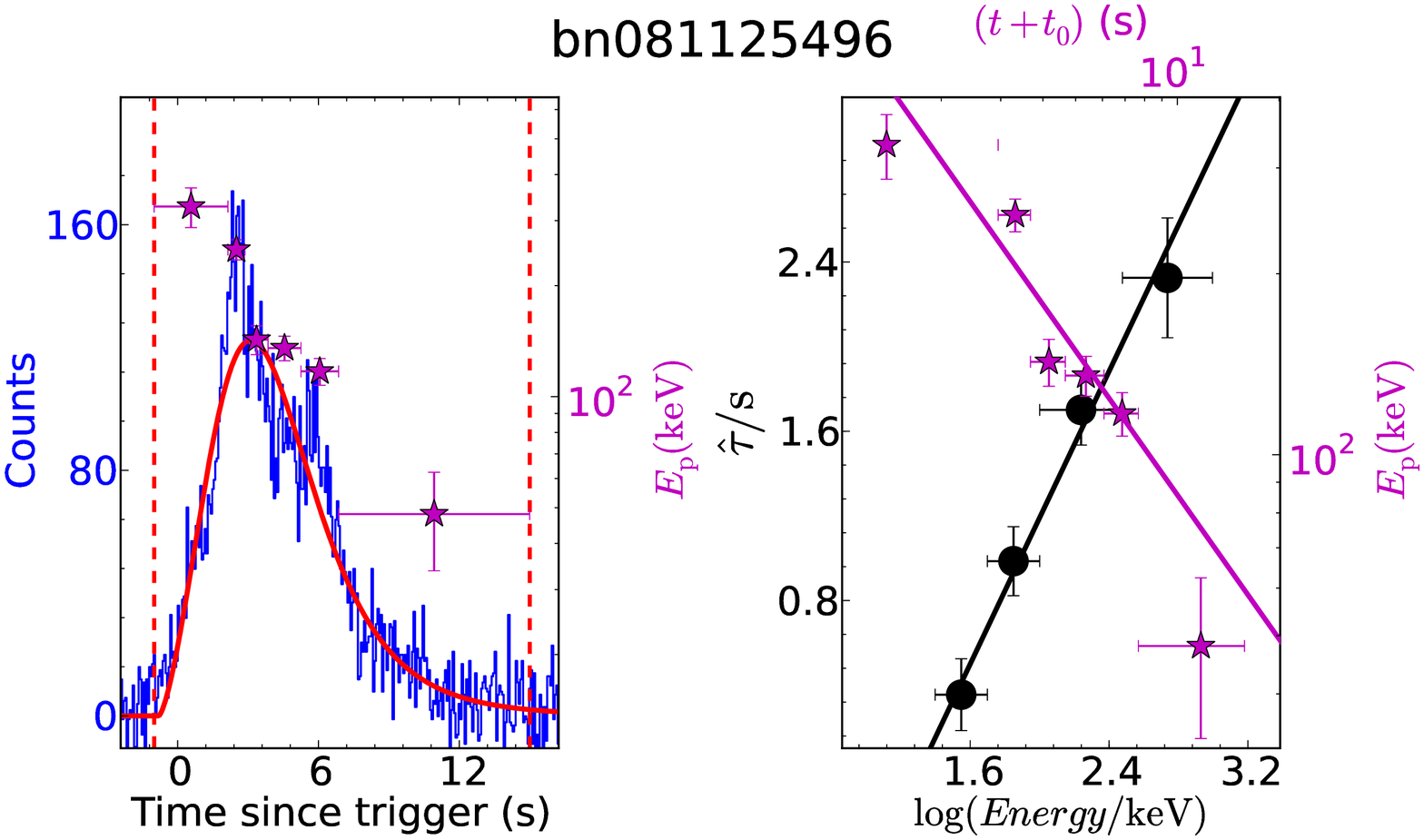} &
\includegraphics[origin=c,angle=0,scale=1,width=0.50\textwidth,height=0.22\textheight]{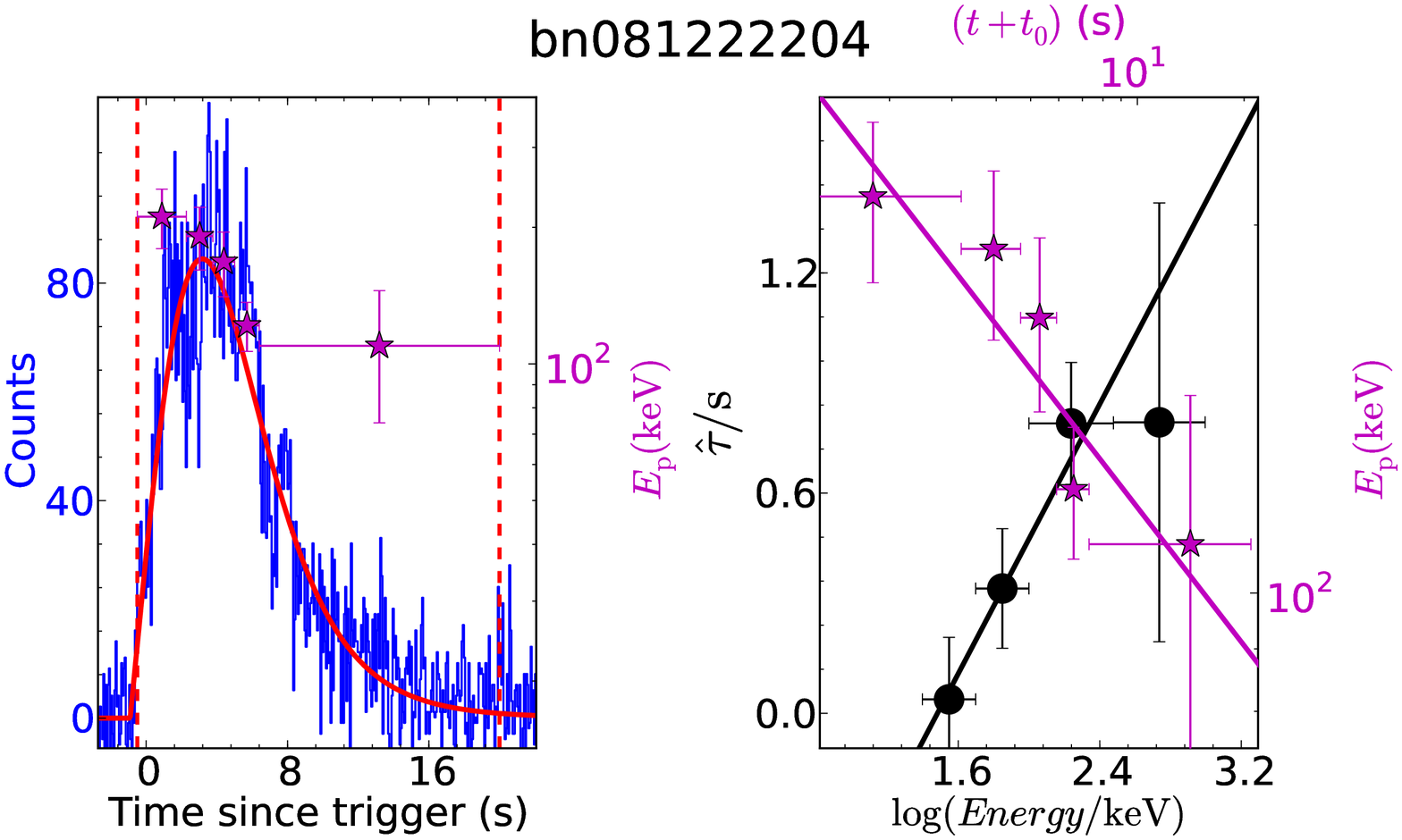} \\
\includegraphics[origin=c,angle=0,scale=1,width=0.500\textwidth,height=0.22\textheight]{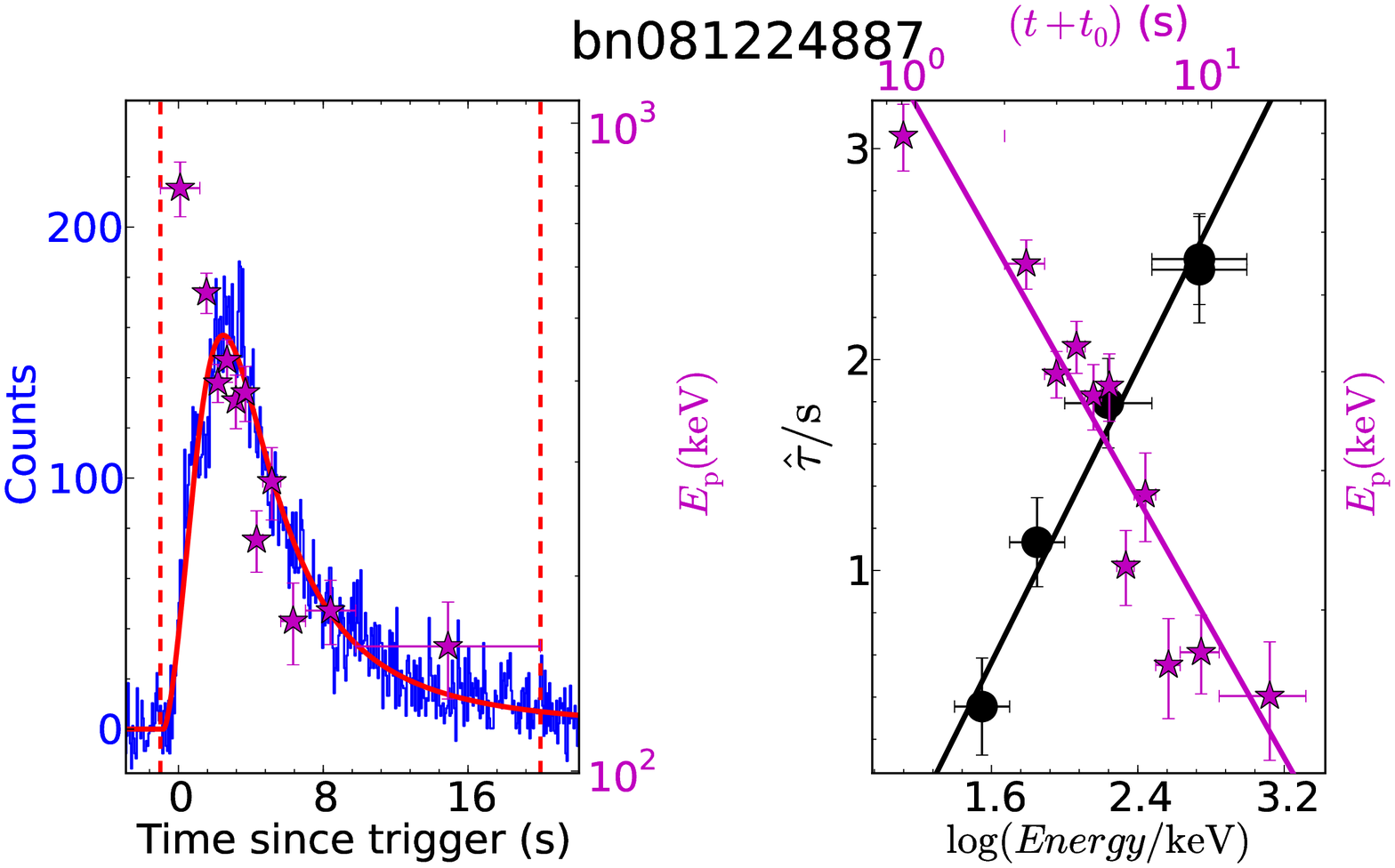} &
\includegraphics[origin=c,angle=0,scale=1,width=0.500\textwidth,height=0.22\textheight]{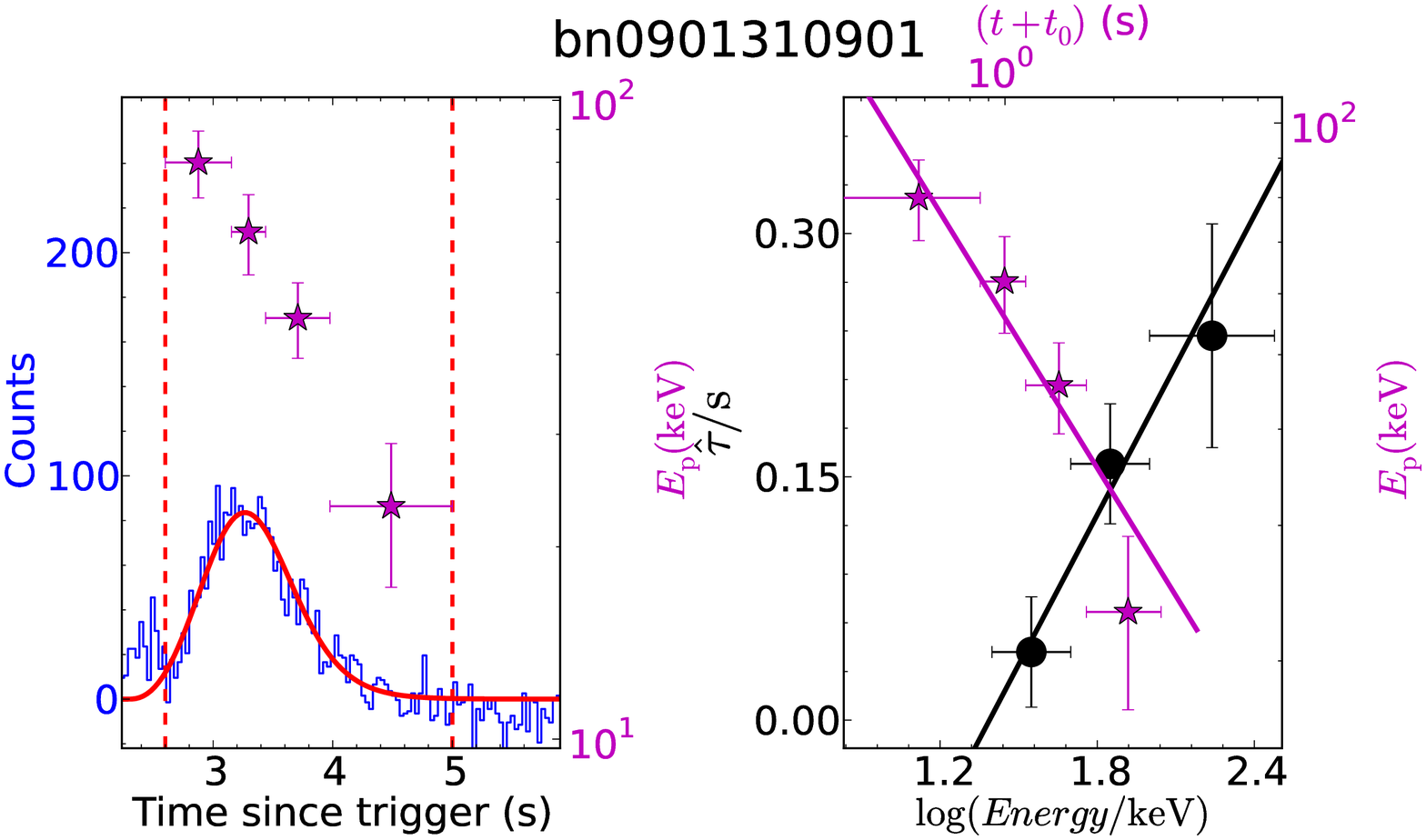} \\
\includegraphics[origin=c,angle=0,scale=1,width=0.500\textwidth,height=0.22\textheight]{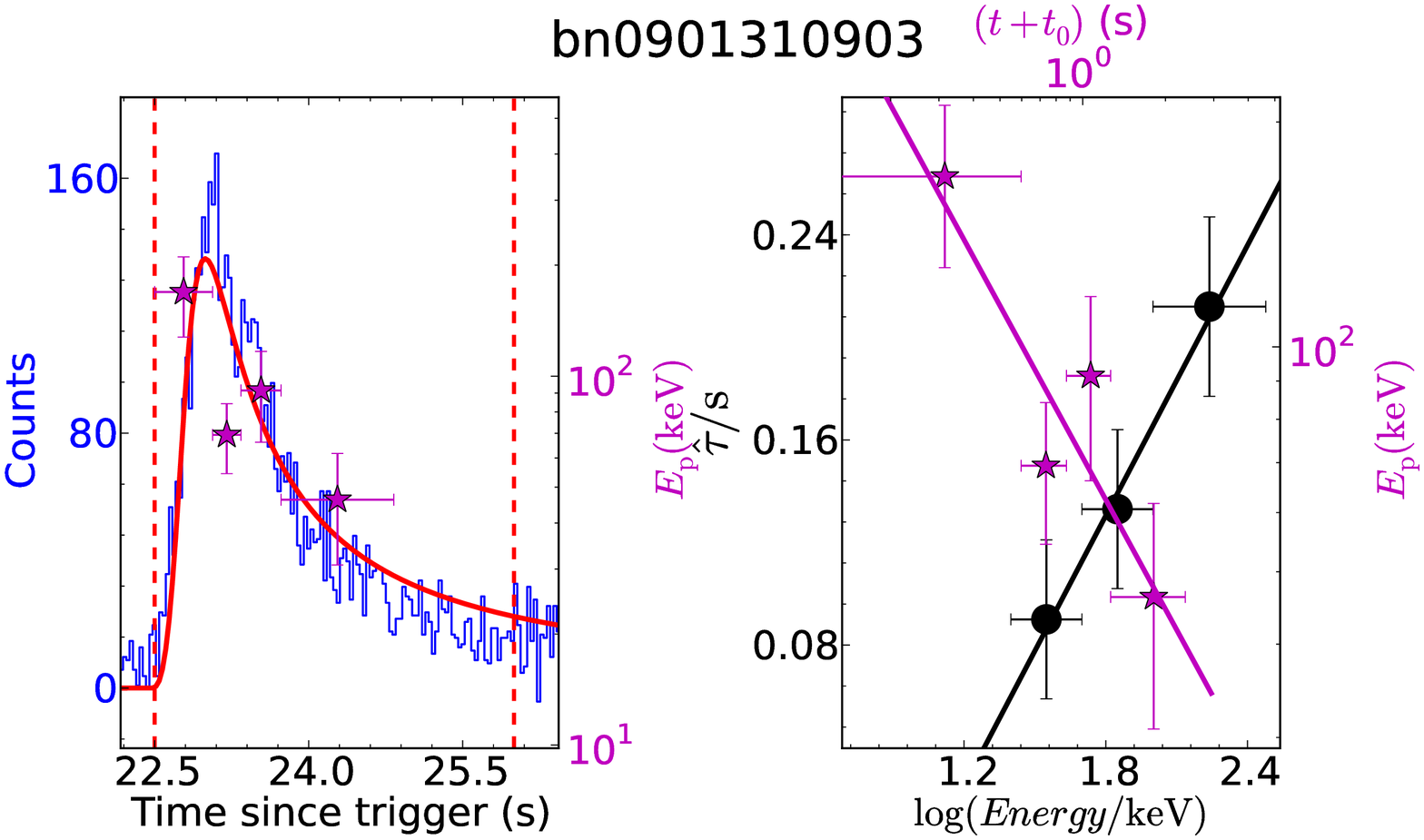} &
\includegraphics[origin=c,angle=0,scale=1,width=0.500\textwidth,height=0.22\textheight]{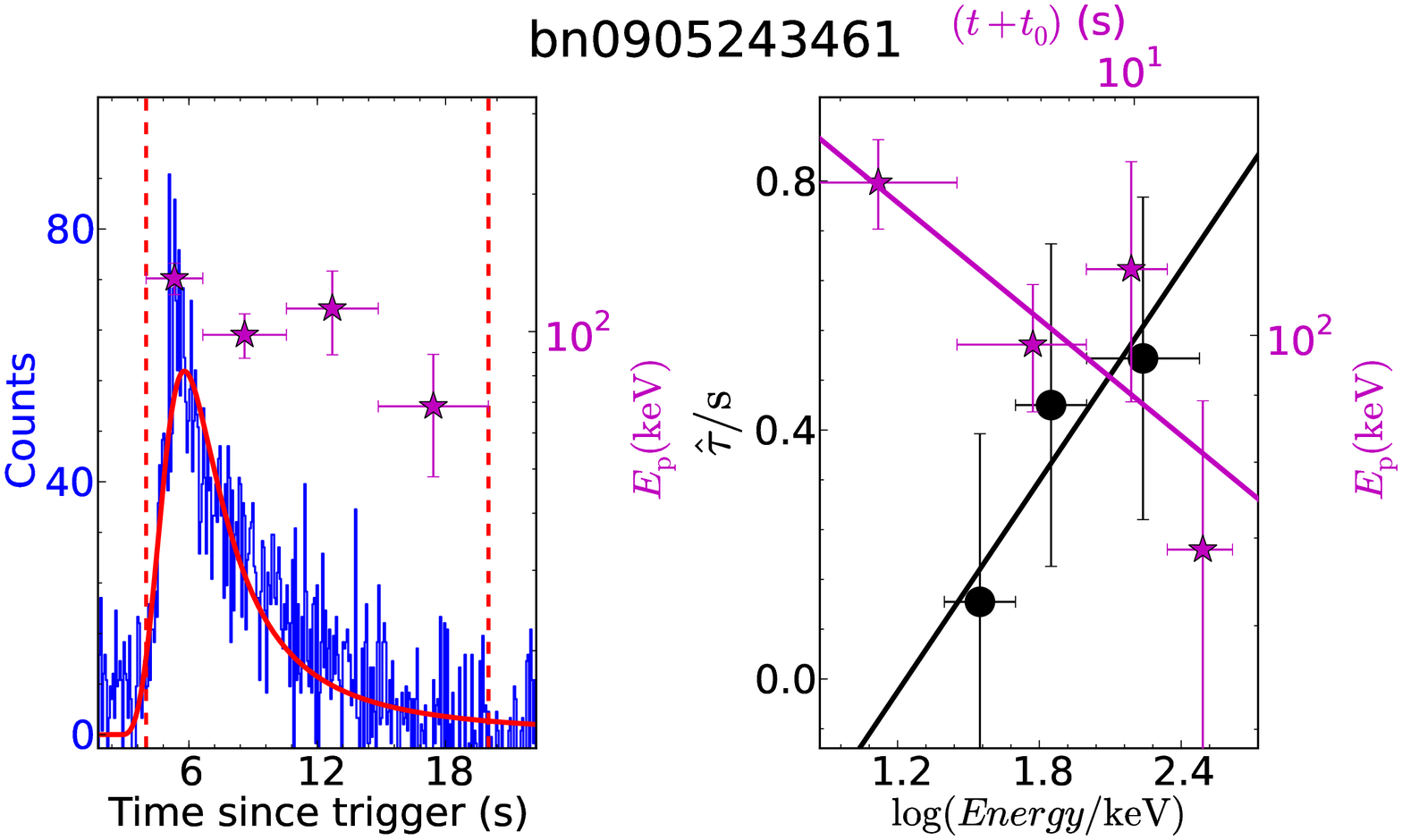} \\
\end{tabular}
\hfill
\vspace{4em}
\caption{Our selected pulses with H2S spectral evolution.
The light curve of pulse (blue curves), best light curve fitting result (red solid line) with Equation \ref{KRL}, and $E_{\rm p}$ (violet ``$\star$'') are shown in the left panel of each sub-figure.
Here, the two vertical dotted lines mark the time period for performing CCF analysis.
The energy dependent ${\hat{\tau}}$ (black ``$\bullet$'', bottom $x$- and left $y$-axis)
and time dependent $E_{\rm p}$ (violet ``$\star$'', top $x$- and right $y$- axis)
are shown in the right panel of each sub-figure,
where the solid lines are the best linear fitting result for the spectral lag behavior (black solid line) and
spectral evolution (violet solid line).
}\label{myfigA}
\end{figure*}

\clearpage % p2
\begin{figure*}
\vspace{-4em}\hspace{-4em}
\centering
\begin{tabular}{cc}
\includegraphics[origin=c,angle=0,scale=1,width=0.500\textwidth,height=0.22\textheight]{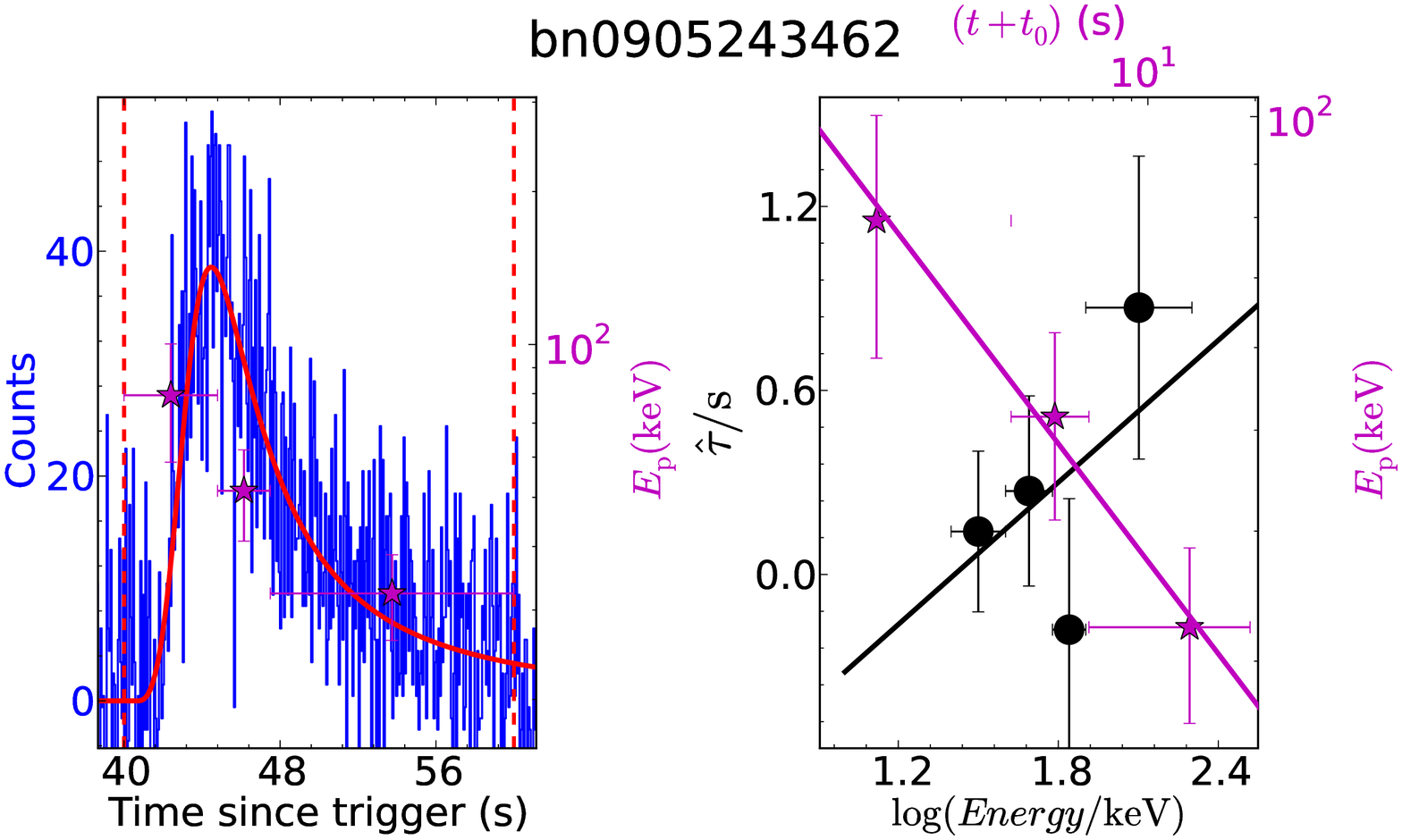} &
\includegraphics[origin=c,angle=0,scale=1,width=0.500\textwidth,height=0.22\textheight]{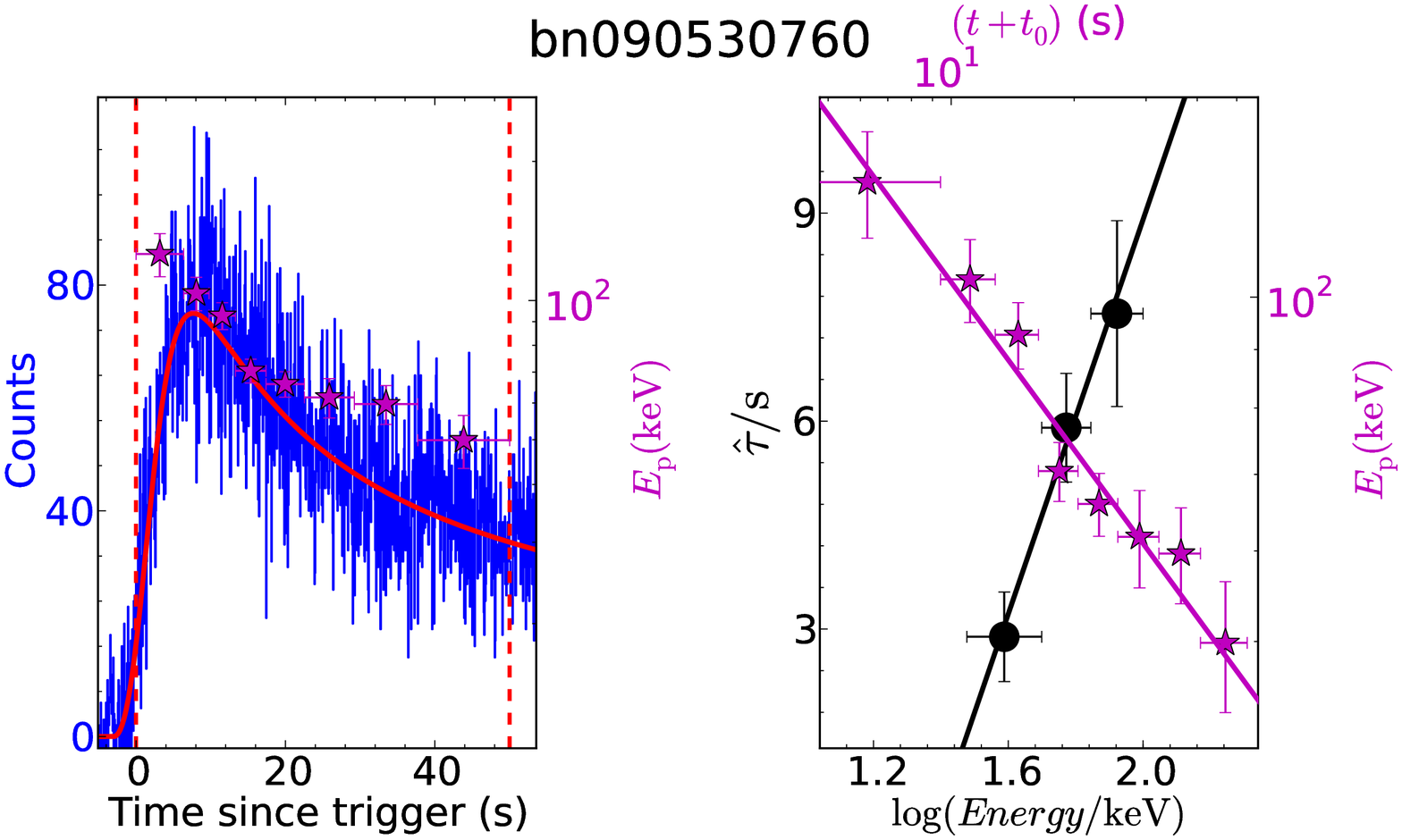} \\
\includegraphics[origin=c,angle=0,scale=1,width=0.500\textwidth,height=0.22\textheight]{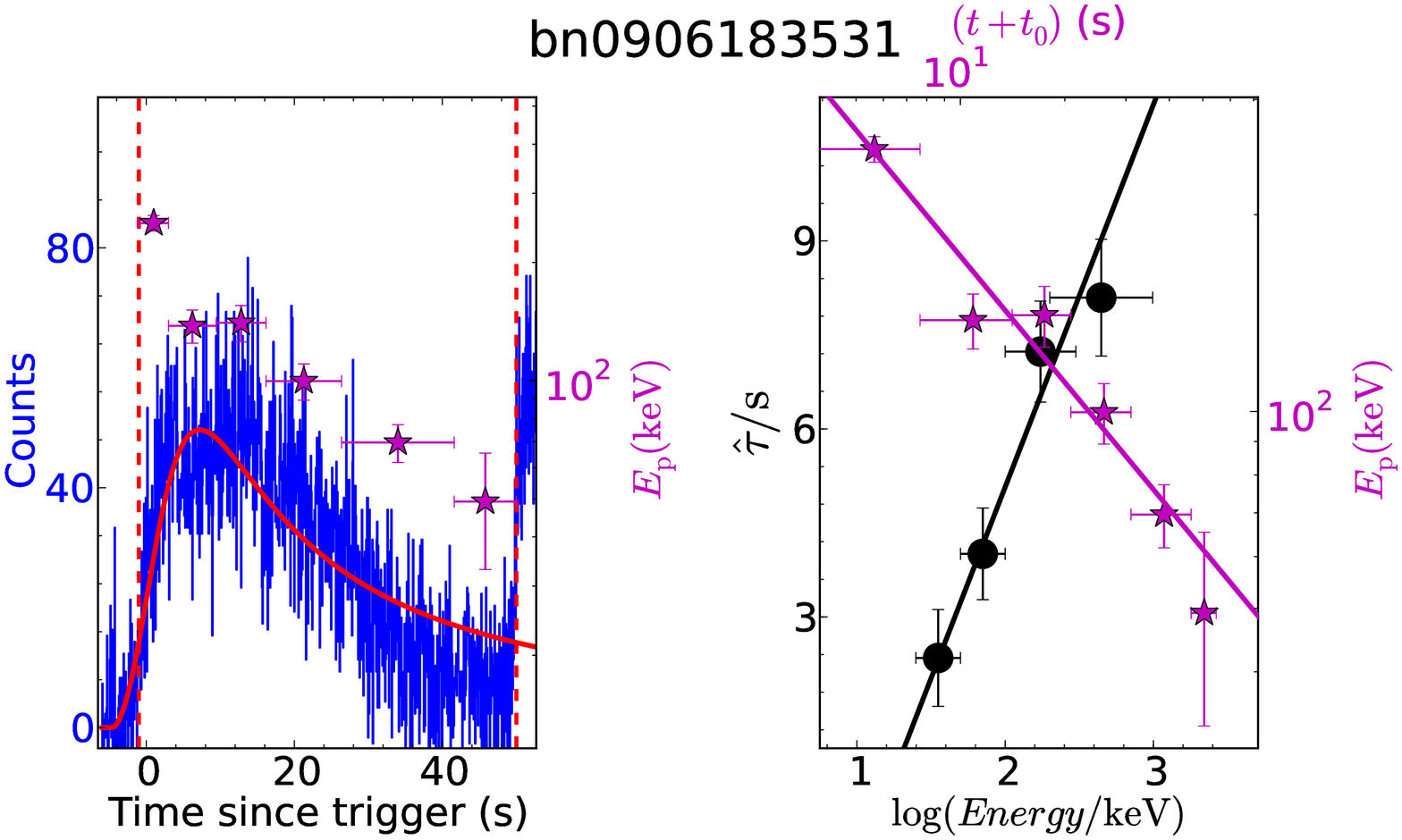} &
\includegraphics[origin=c,angle=0,scale=1,width=0.500\textwidth,height=0.22\textheight]{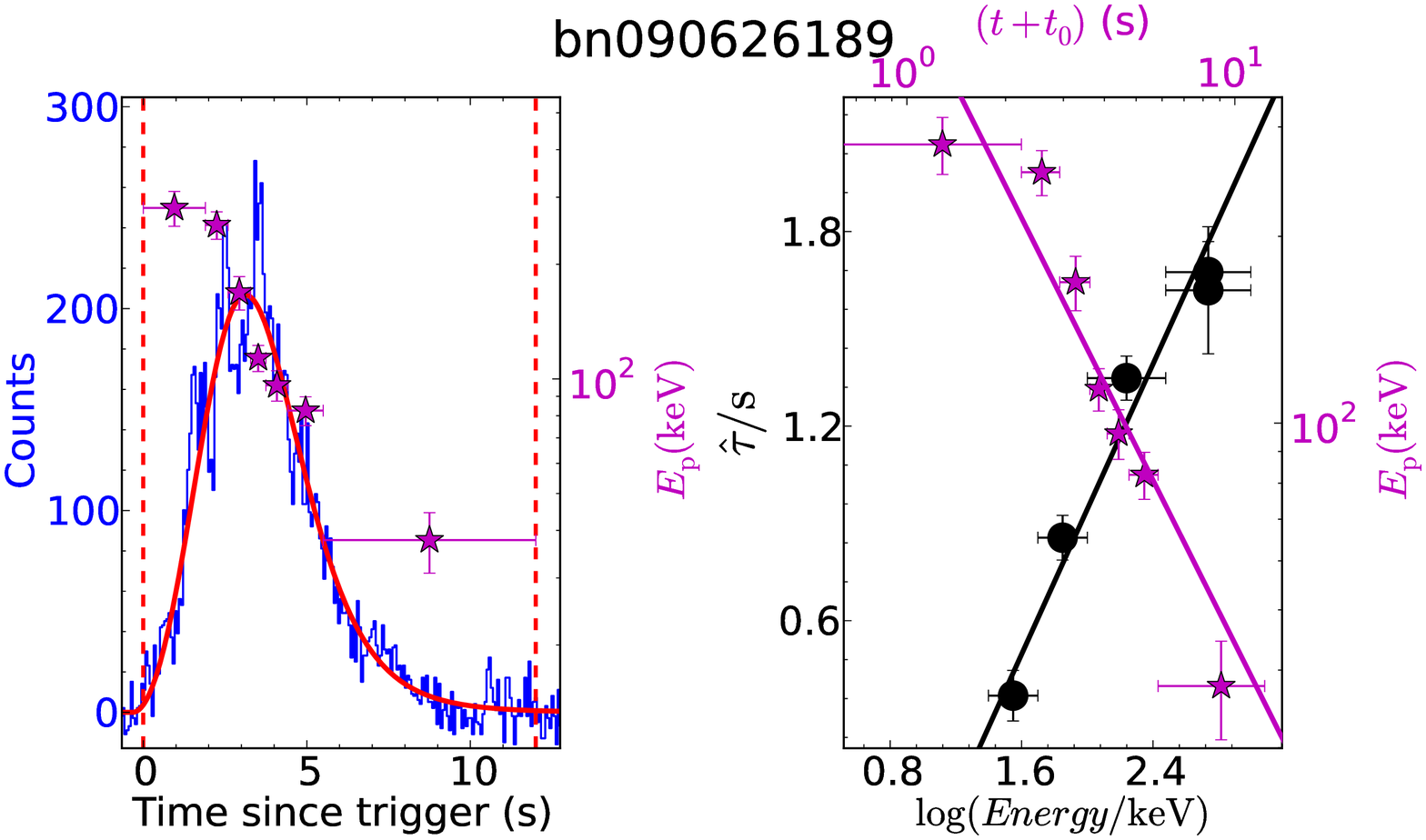} \\
\includegraphics[origin=c,angle=0,scale=1,width=0.500\textwidth,height=0.220\textheight]{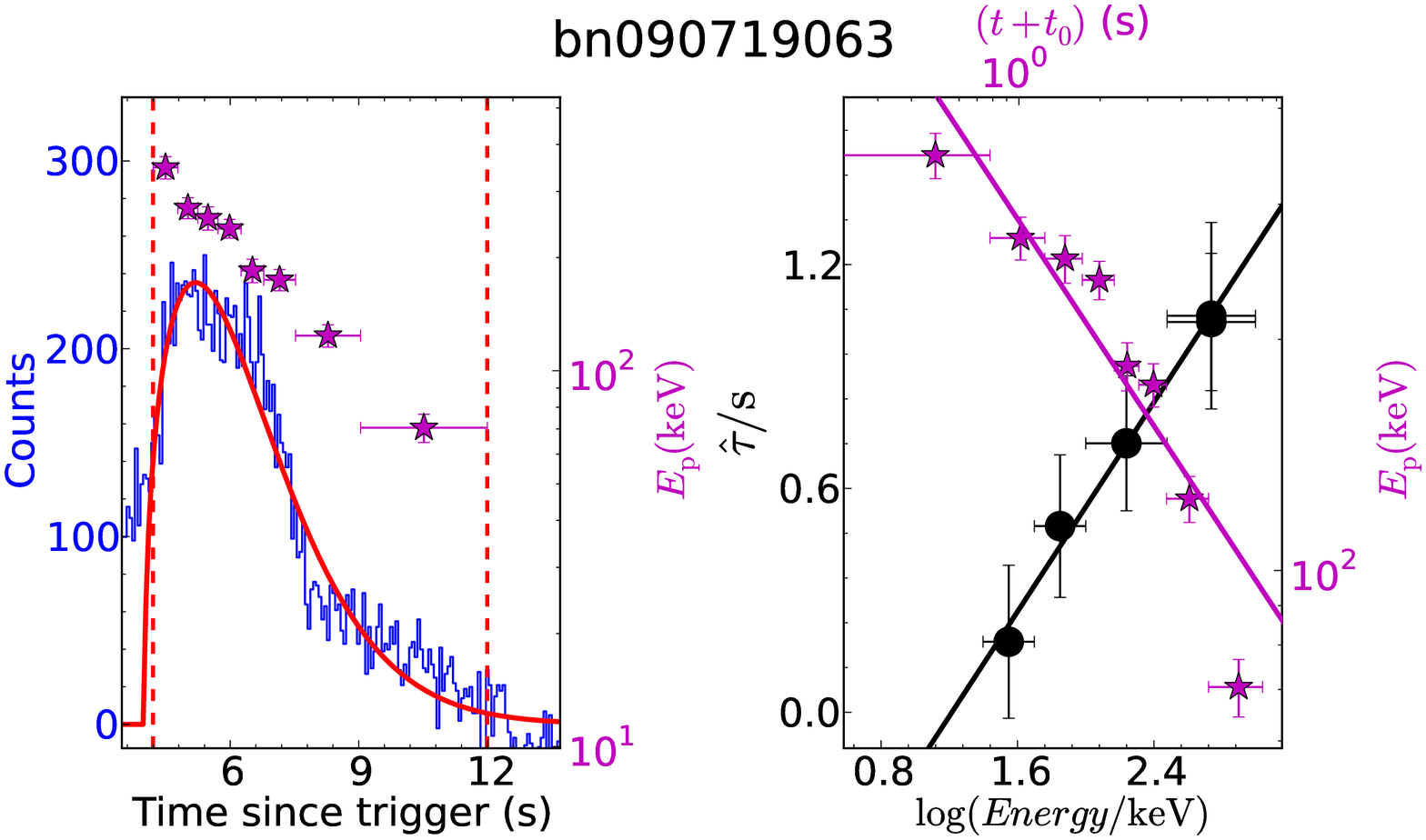} &
\includegraphics[origin=c,angle=0,scale=1,width=0.500\textwidth,height=0.22\textheight]{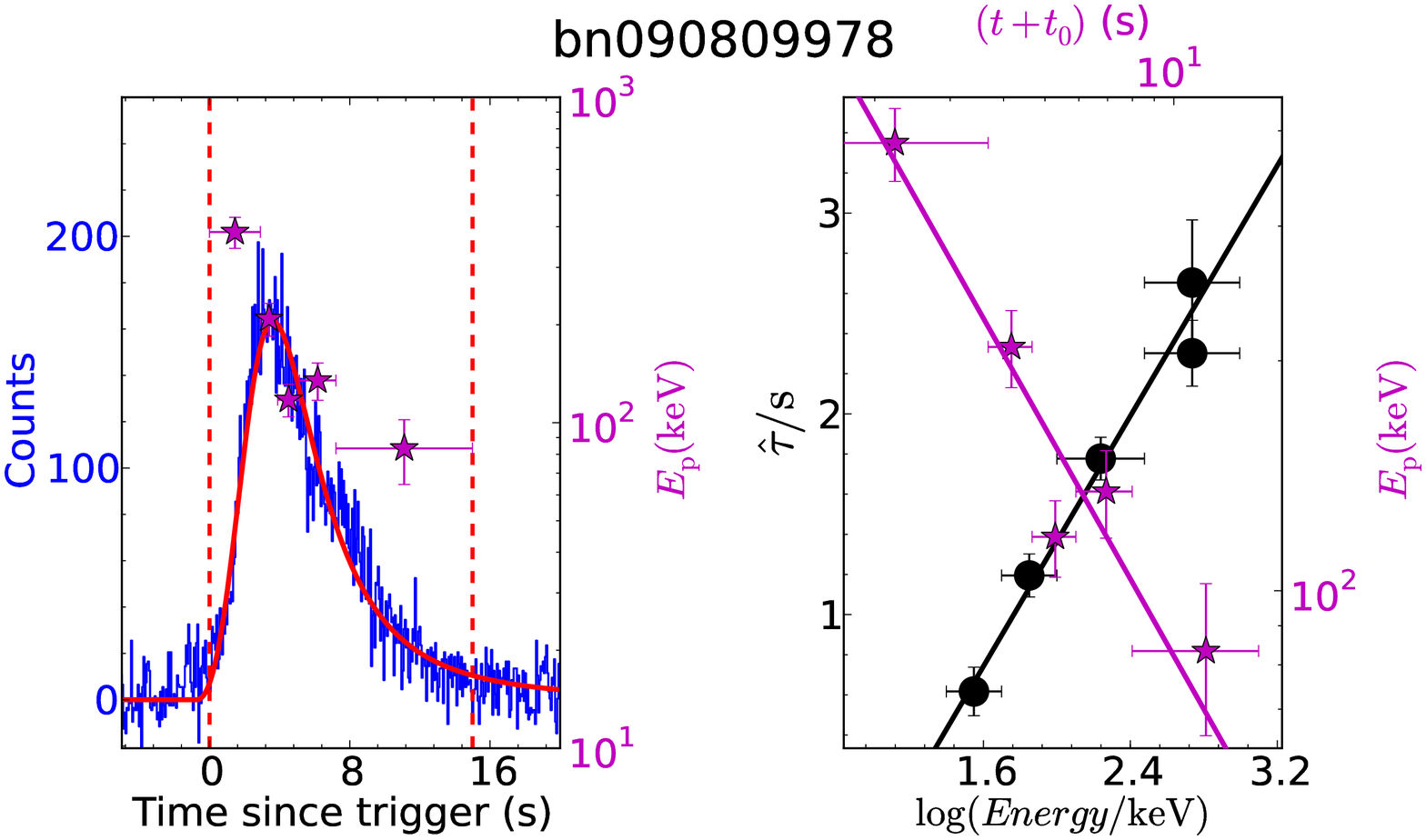} \\
\includegraphics[origin=c,angle=0,scale=1,width=0.500\textwidth,height=0.22\textheight]{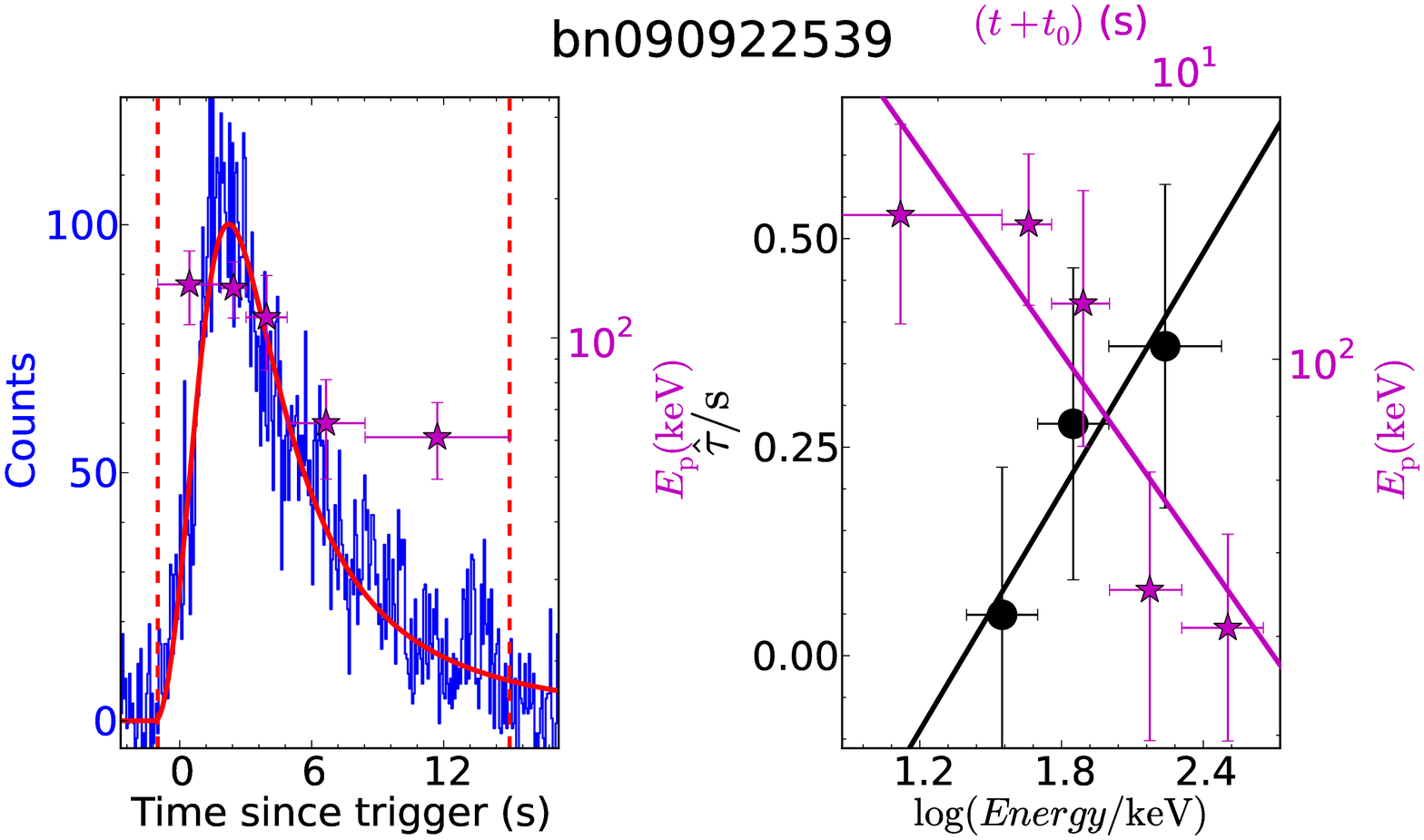} &
\includegraphics[origin=c,angle=0,scale=1,width=0.500\textwidth,height=0.22\textheight]{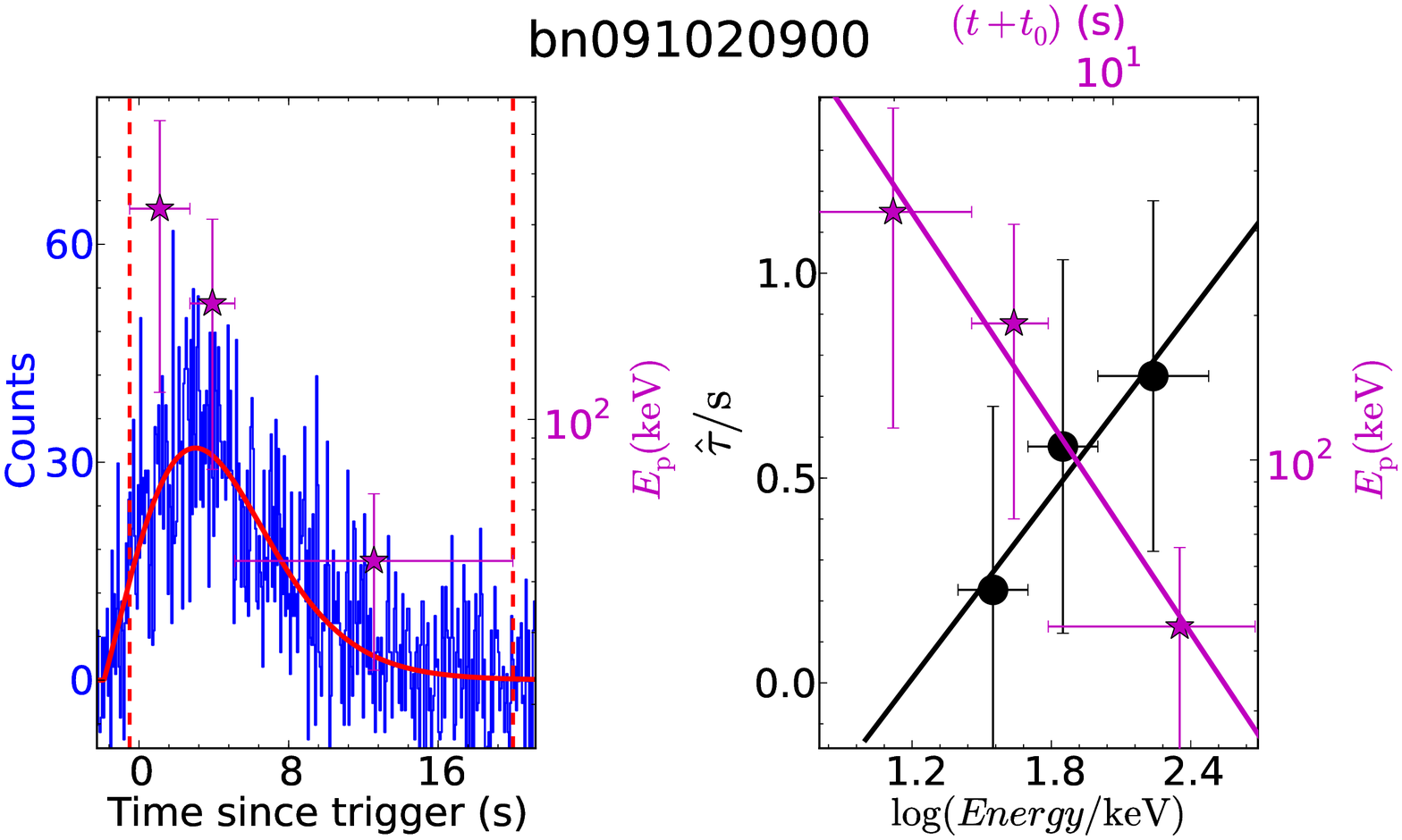} \\
\end{tabular}
\hfill
\vspace{4em}
\center{Fig.~\ref{myfigA}---Continued.}
\end{figure*}

\clearpage %p3
\begin{figure*}
\vspace{-4em}\hspace{-4em}
\centering
\begin{tabular}{cc}
\includegraphics[origin=c,angle=0,scale=1,width=0.500\textwidth,height=0.220\textheight]{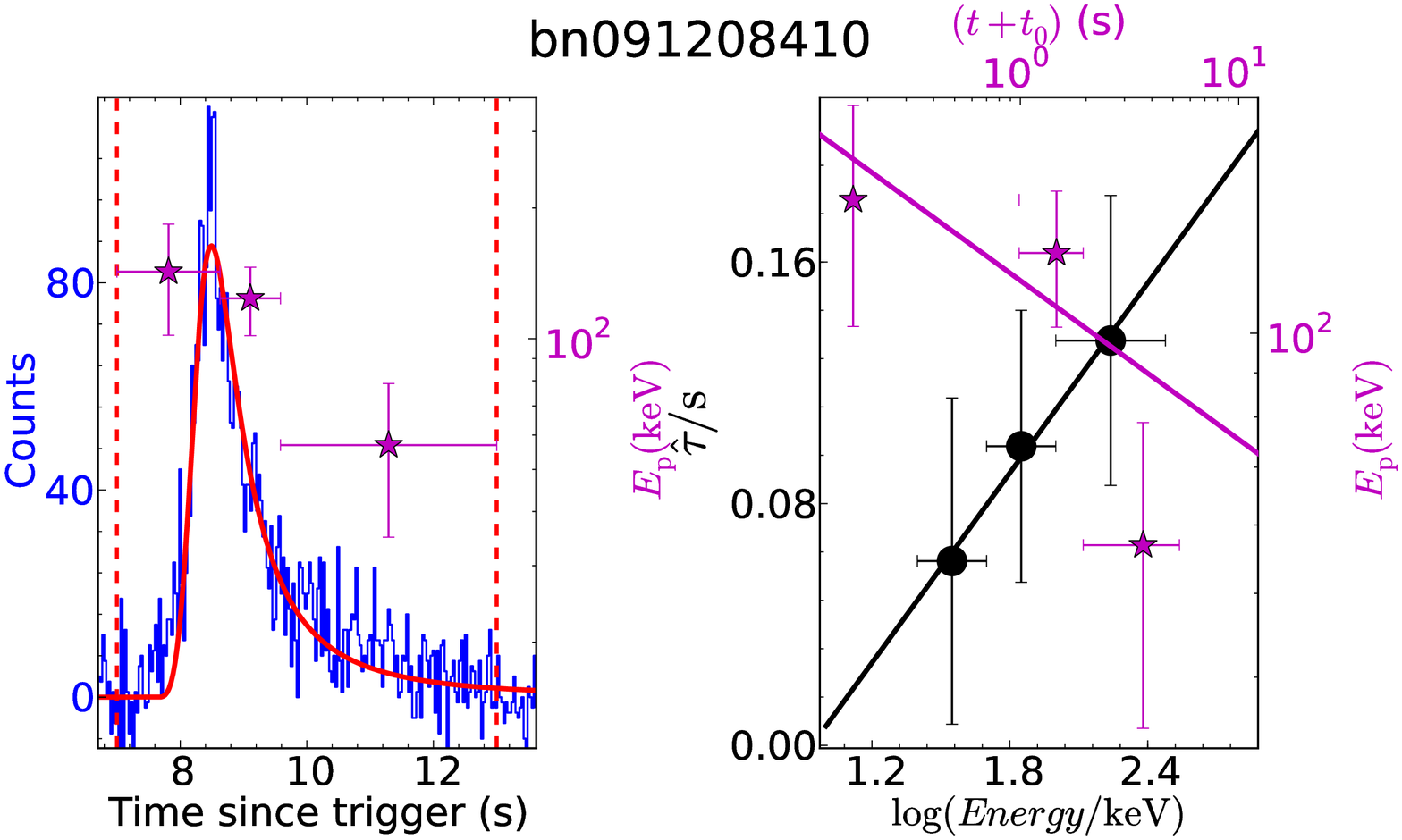} &
\includegraphics[origin=c,angle=0,scale=1,width=0.500\textwidth,height=0.23\textheight]{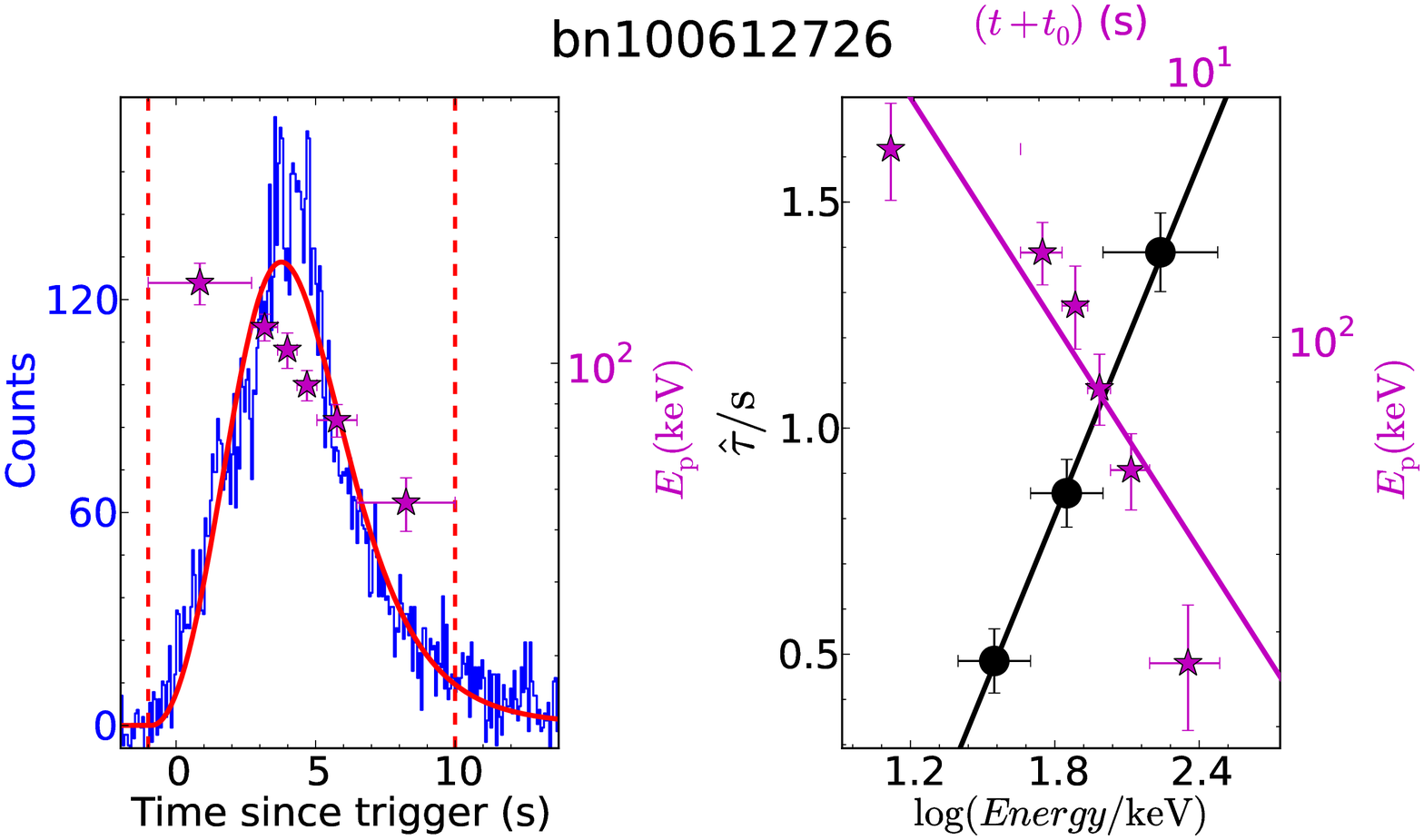} \\
\includegraphics[origin=c,angle=0,scale=1,width=0.500\textwidth,height=0.230\textheight]{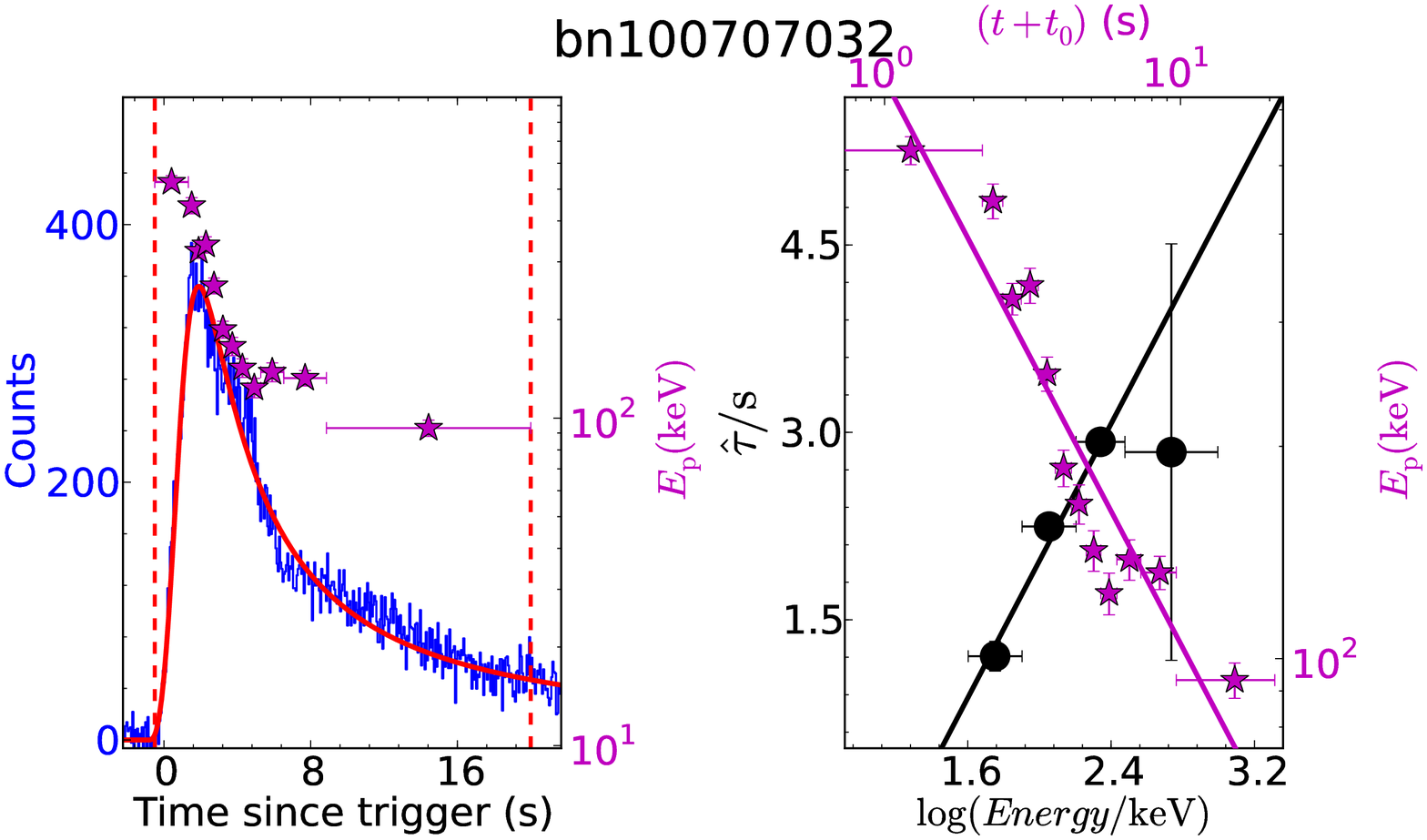} &
\includegraphics[origin=c,angle=0,scale=1,width=0.500\textwidth,height=0.23\textheight]{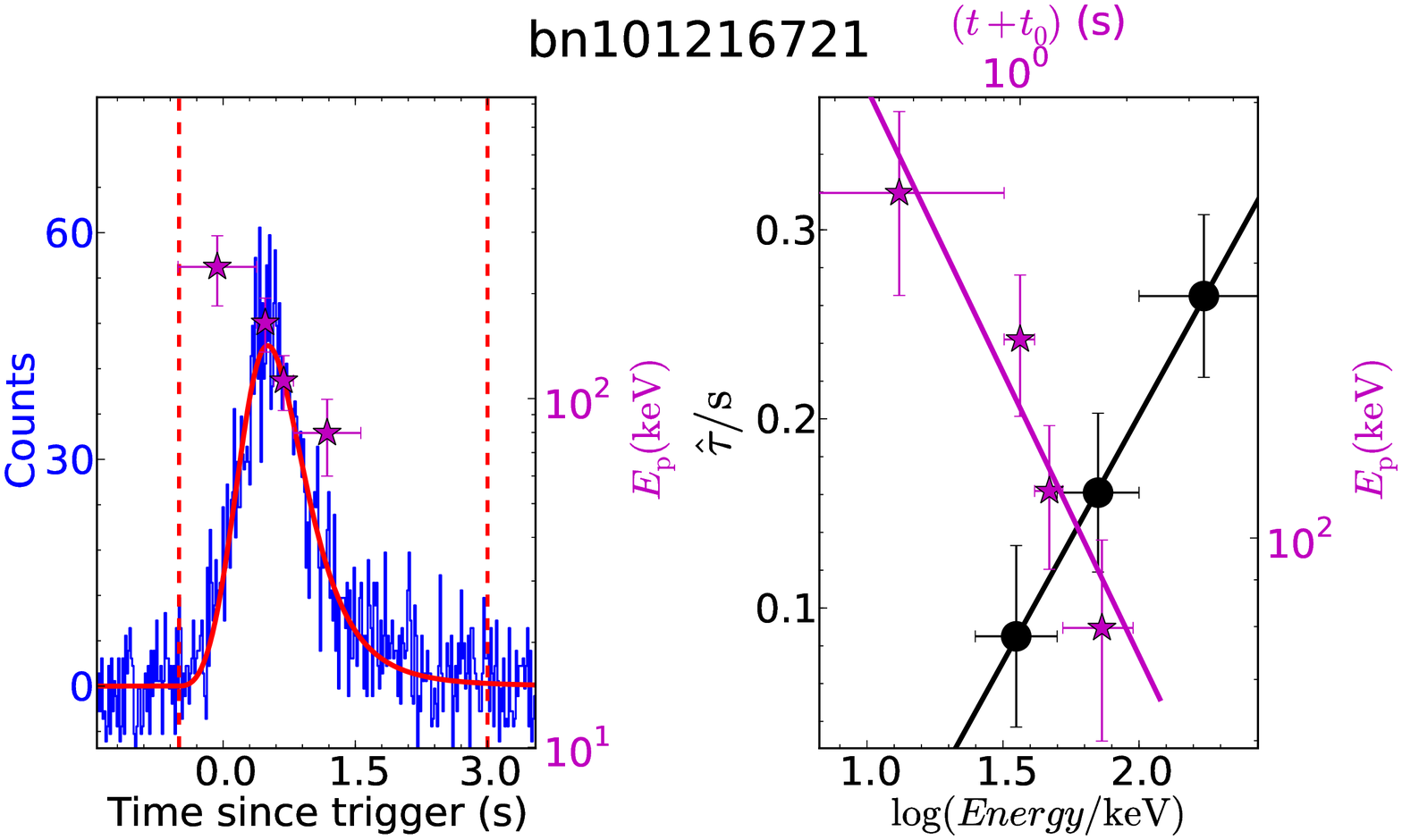} \\
\includegraphics[origin=c,angle=0,scale=1,width=0.500\textwidth,height=0.230\textheight]{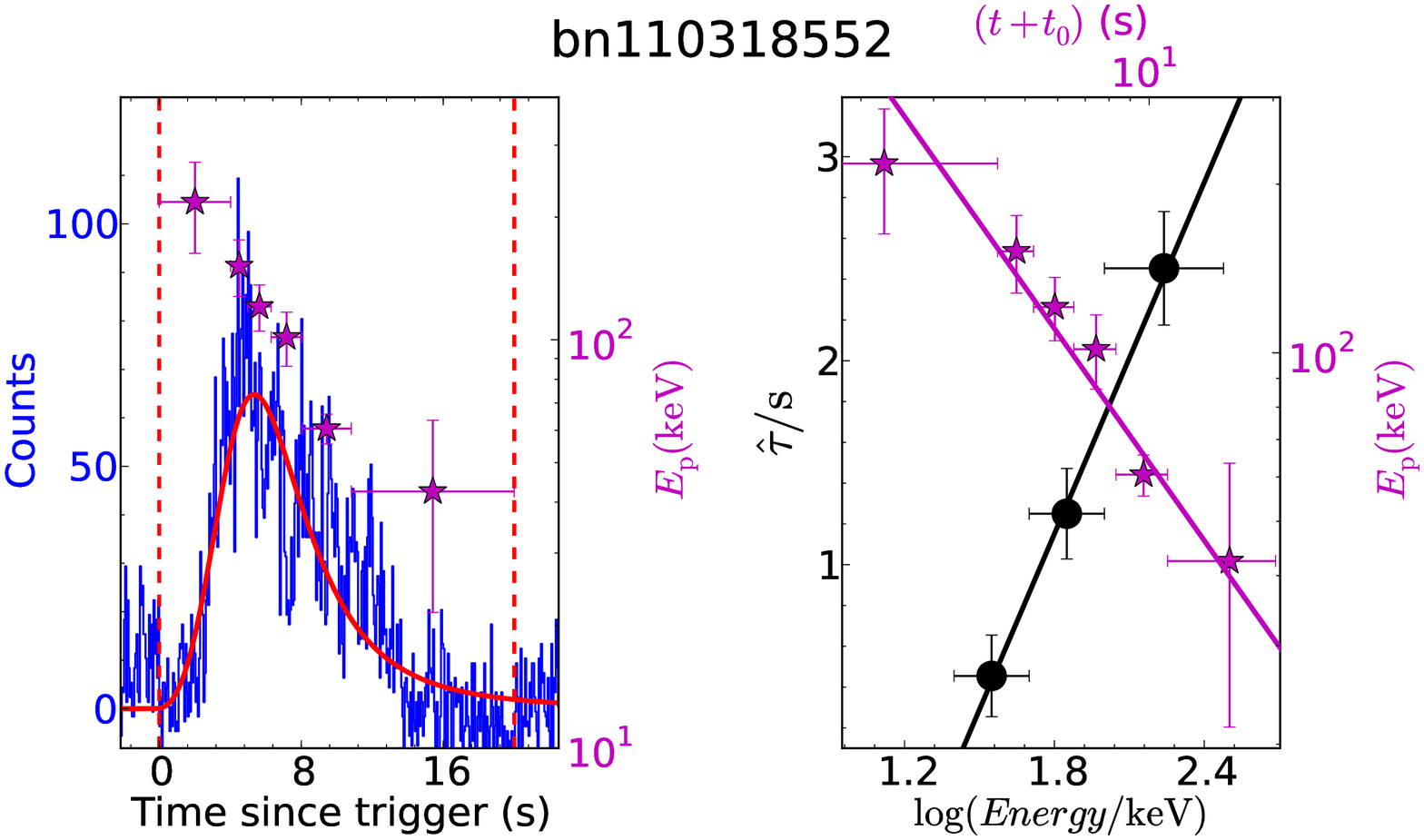} &
\includegraphics[origin=c,angle=0,scale=1,width=0.500\textwidth,height=0.22\textheight]{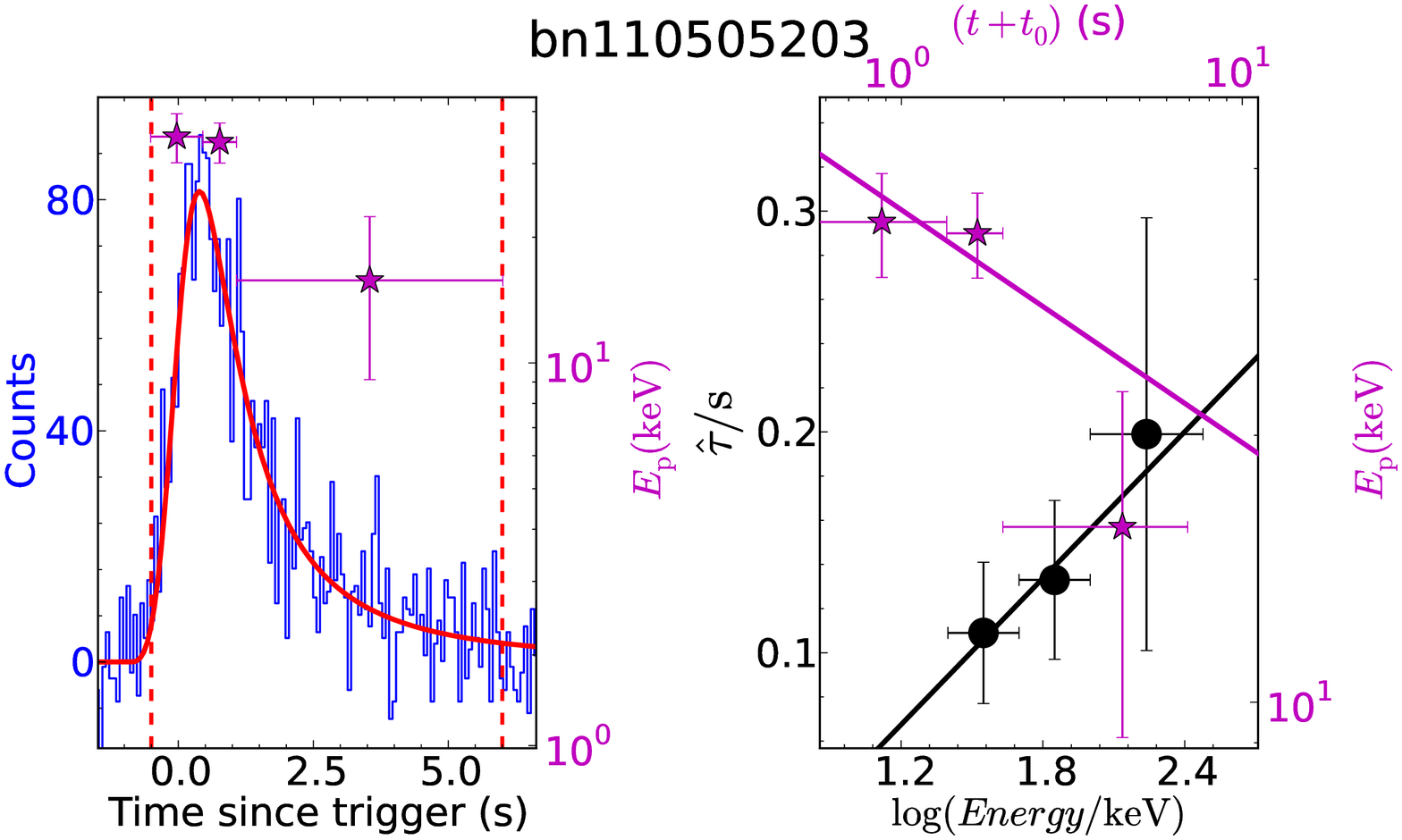} \\
\includegraphics[origin=c,angle=0,scale=1,width=0.500\textwidth,height=0.230\textheight]{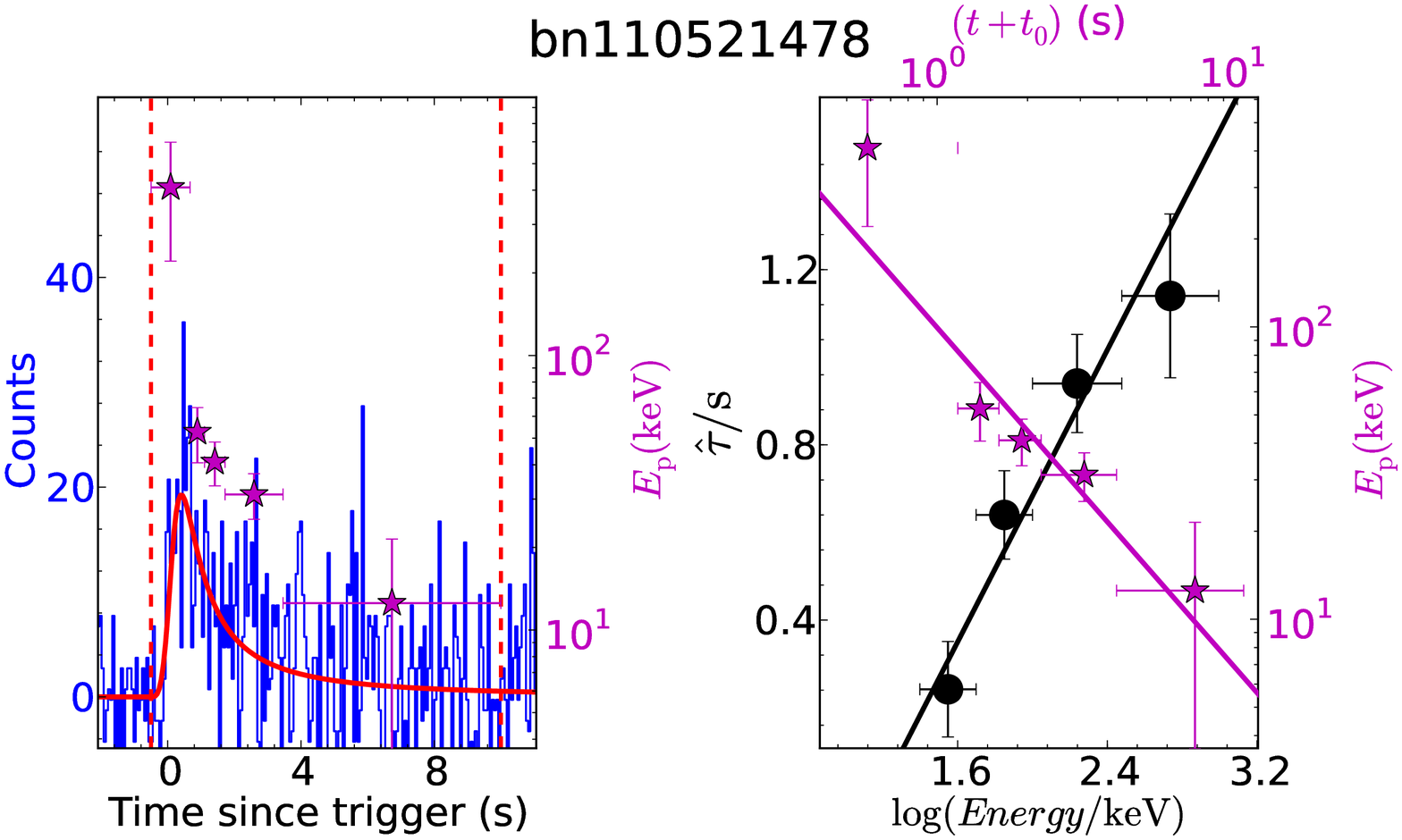} &
\includegraphics[origin=c,angle=0,scale=1,width=0.500\textwidth,height=0.23\textheight]{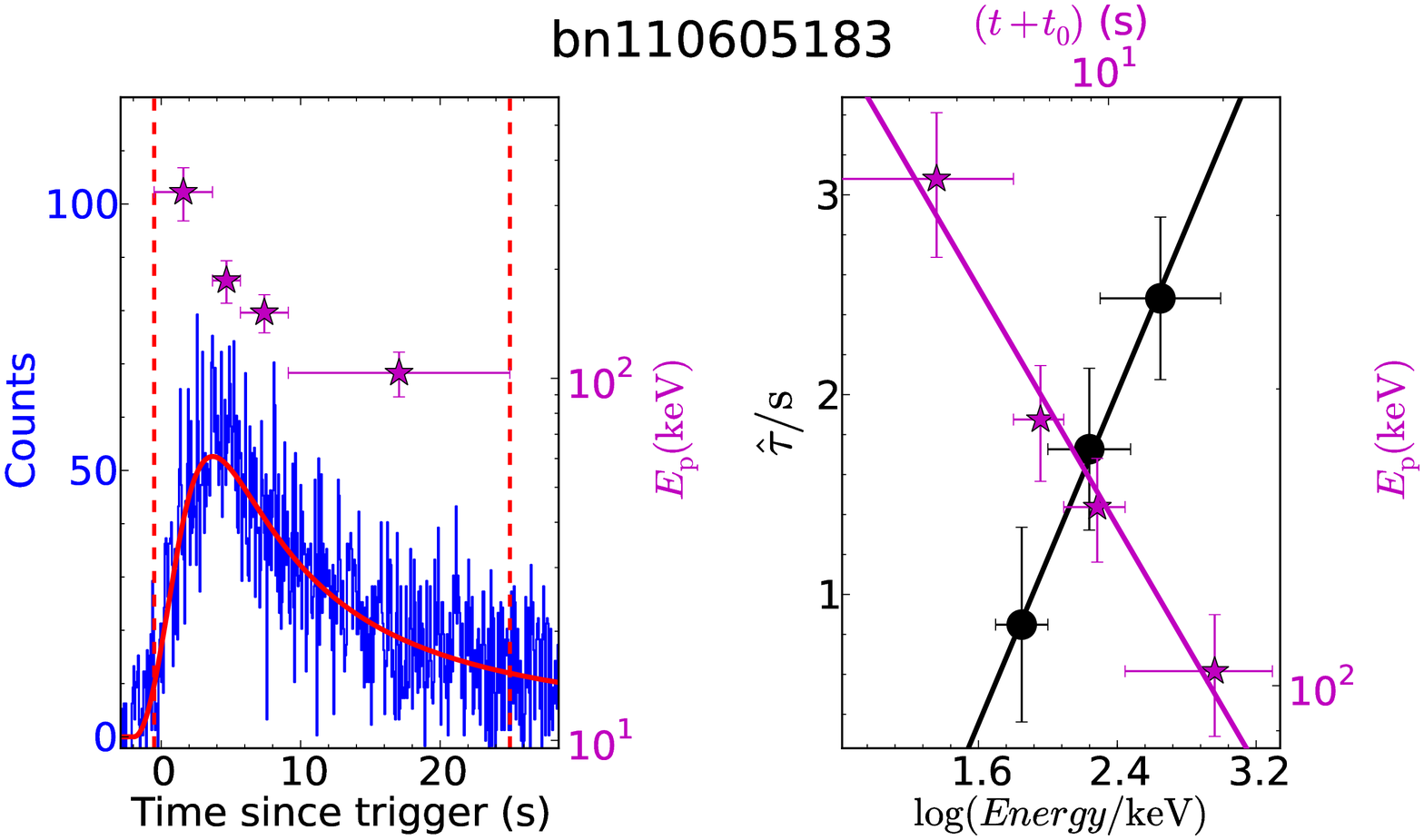} \\
\end{tabular}
\hfill
\vspace{4em}
\center{Fig.~\ref{myfigA}---Continued.}
\end{figure*}

\clearpage %p4
\begin{figure*}
\vspace{-4em}\hspace{-4em}
\centering
\begin{tabular}{cc}
\includegraphics[origin=c,angle=0,scale=1,width=0.500\textwidth,height=0.230\textheight]{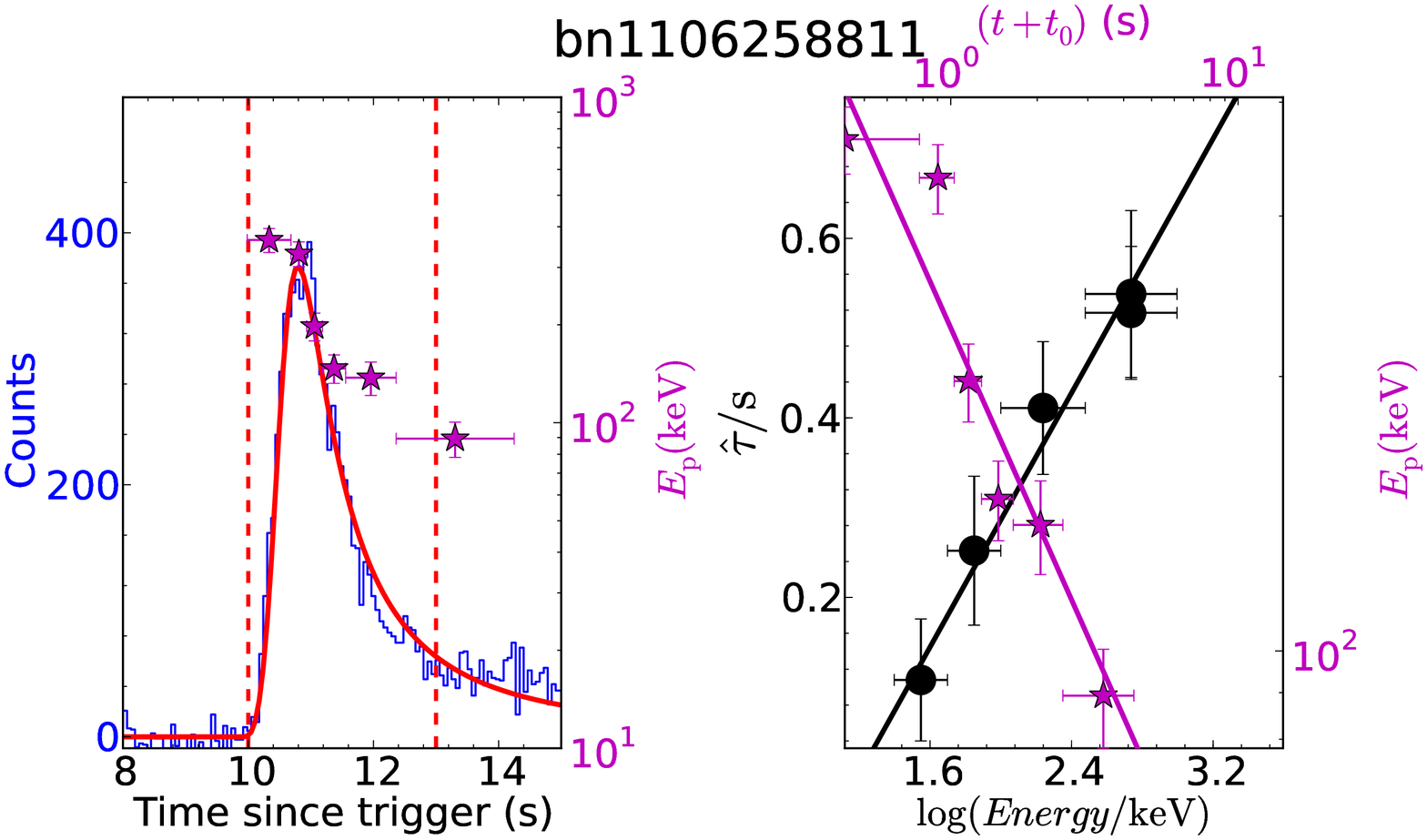} &
\includegraphics[origin=c,angle=0,scale=1,width=0.500\textwidth,height=0.230\textheight]{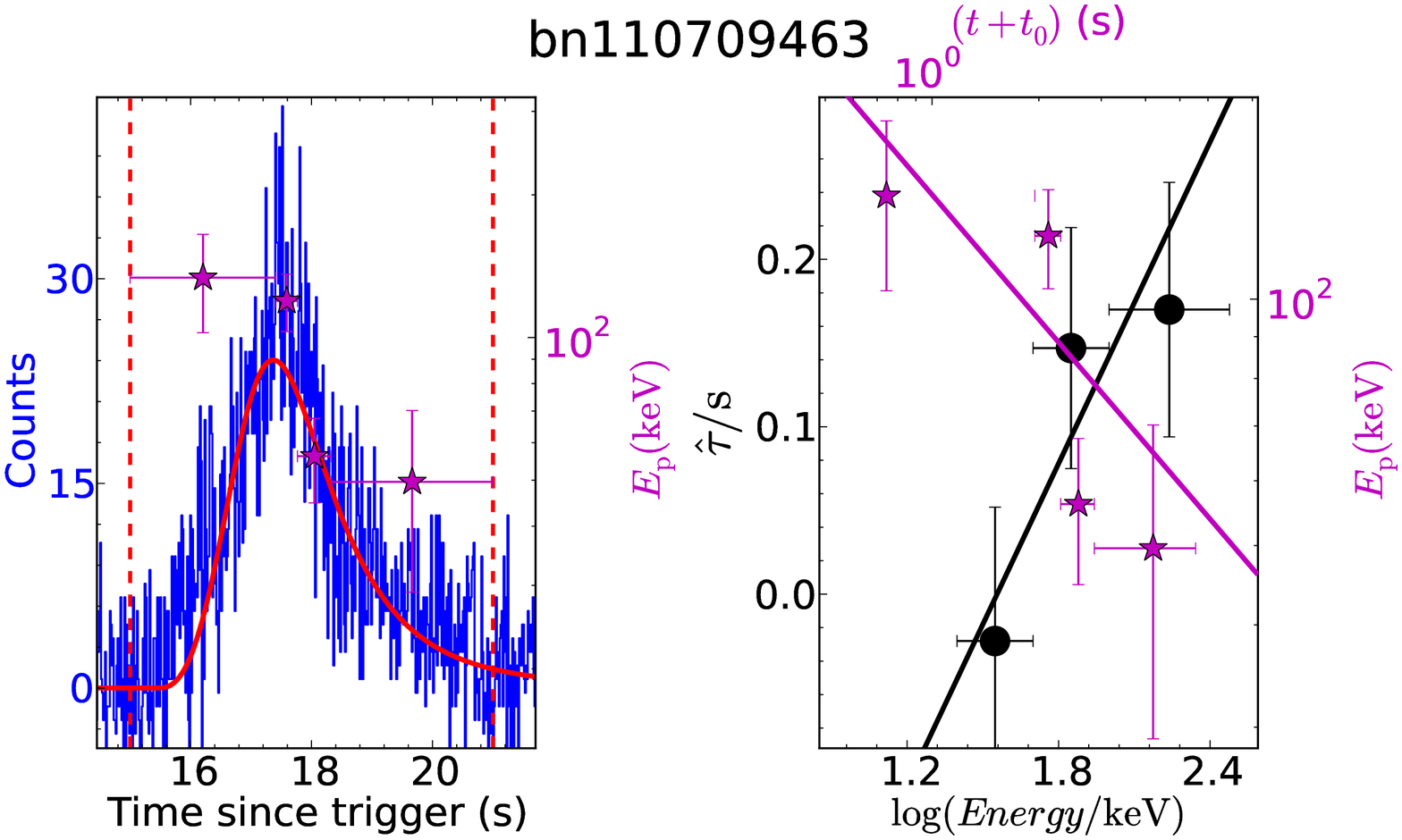} \\
\includegraphics[origin=c,angle=0,scale=1,width=0.500\textwidth,height=0.23\textheight]{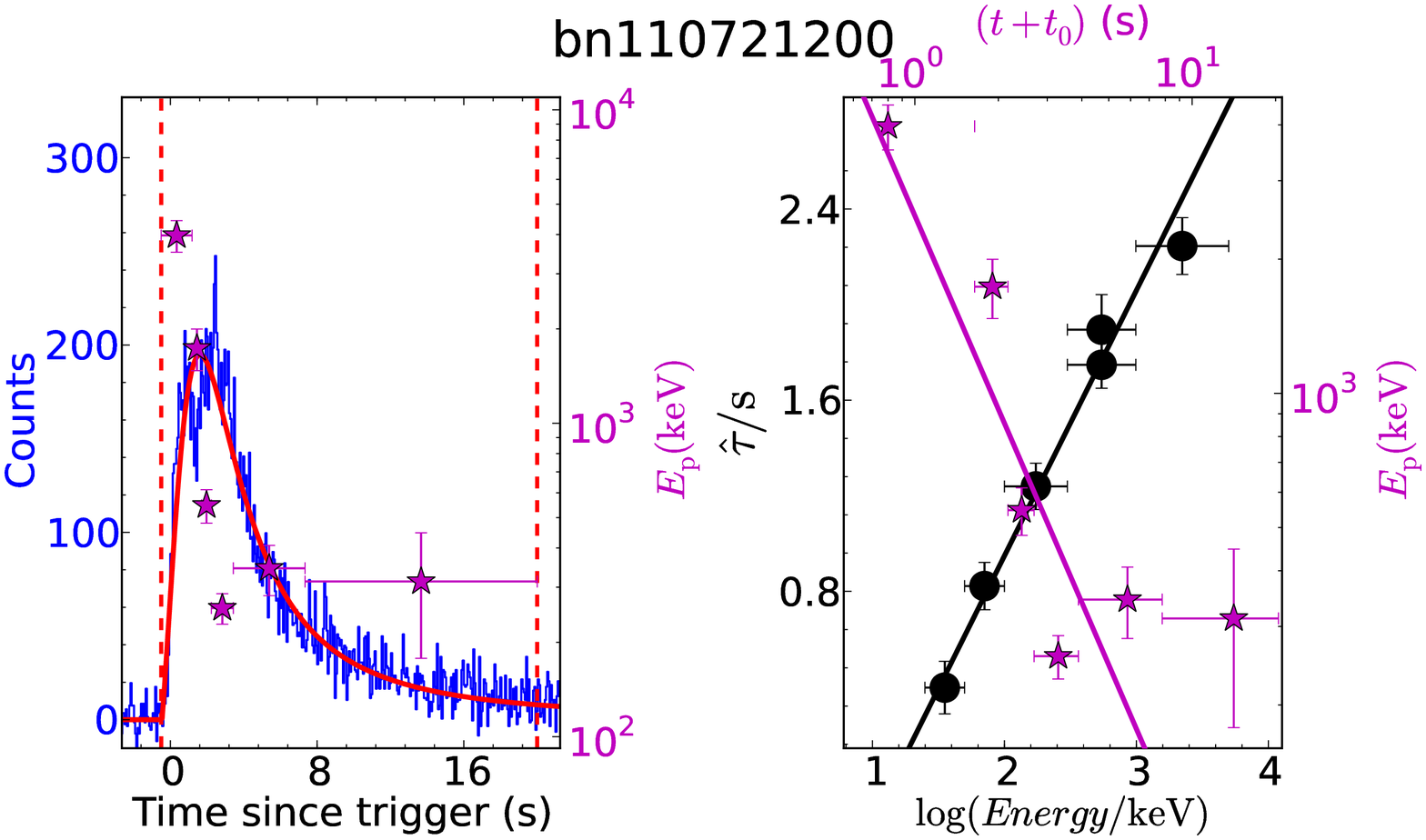} &
\includegraphics[origin=c,angle=0,scale=1,width=0.500\textwidth,height=0.23\textheight]{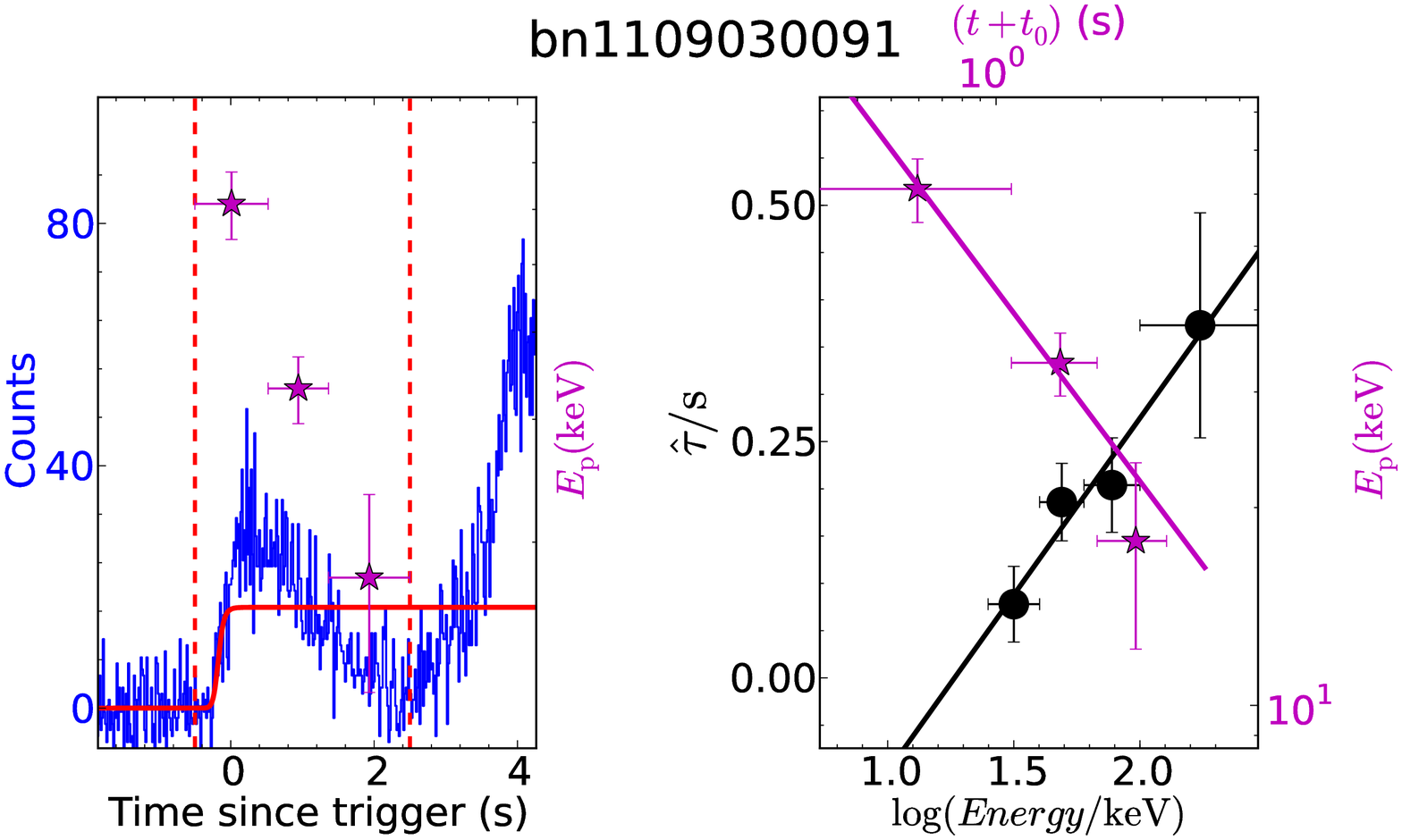} \\
\includegraphics[origin=c,angle=0,scale=1,width=0.500\textwidth,height=0.23\textheight]{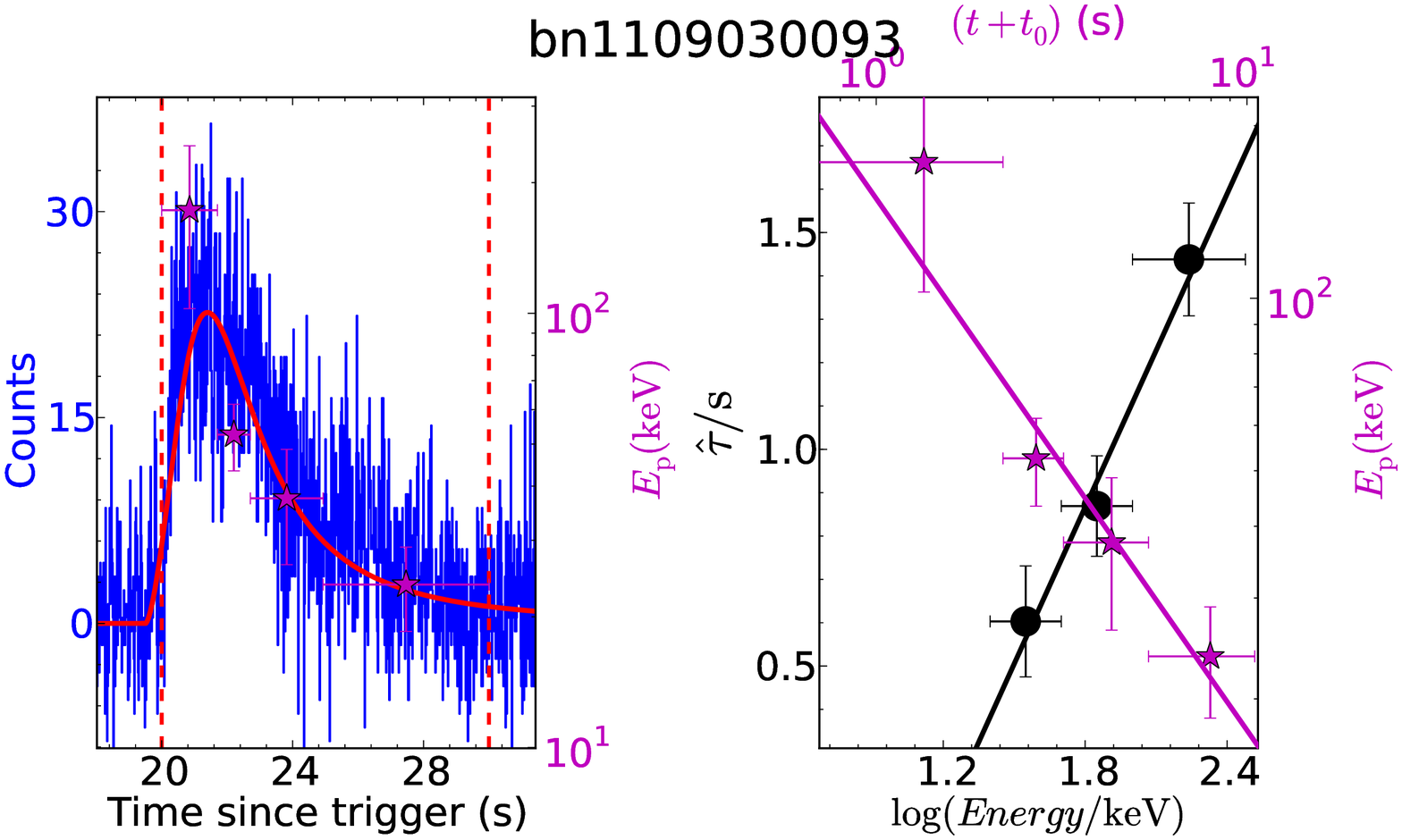} &
\includegraphics[origin=c,angle=0,scale=1,width=0.500\textwidth,height=0.23\textheight]{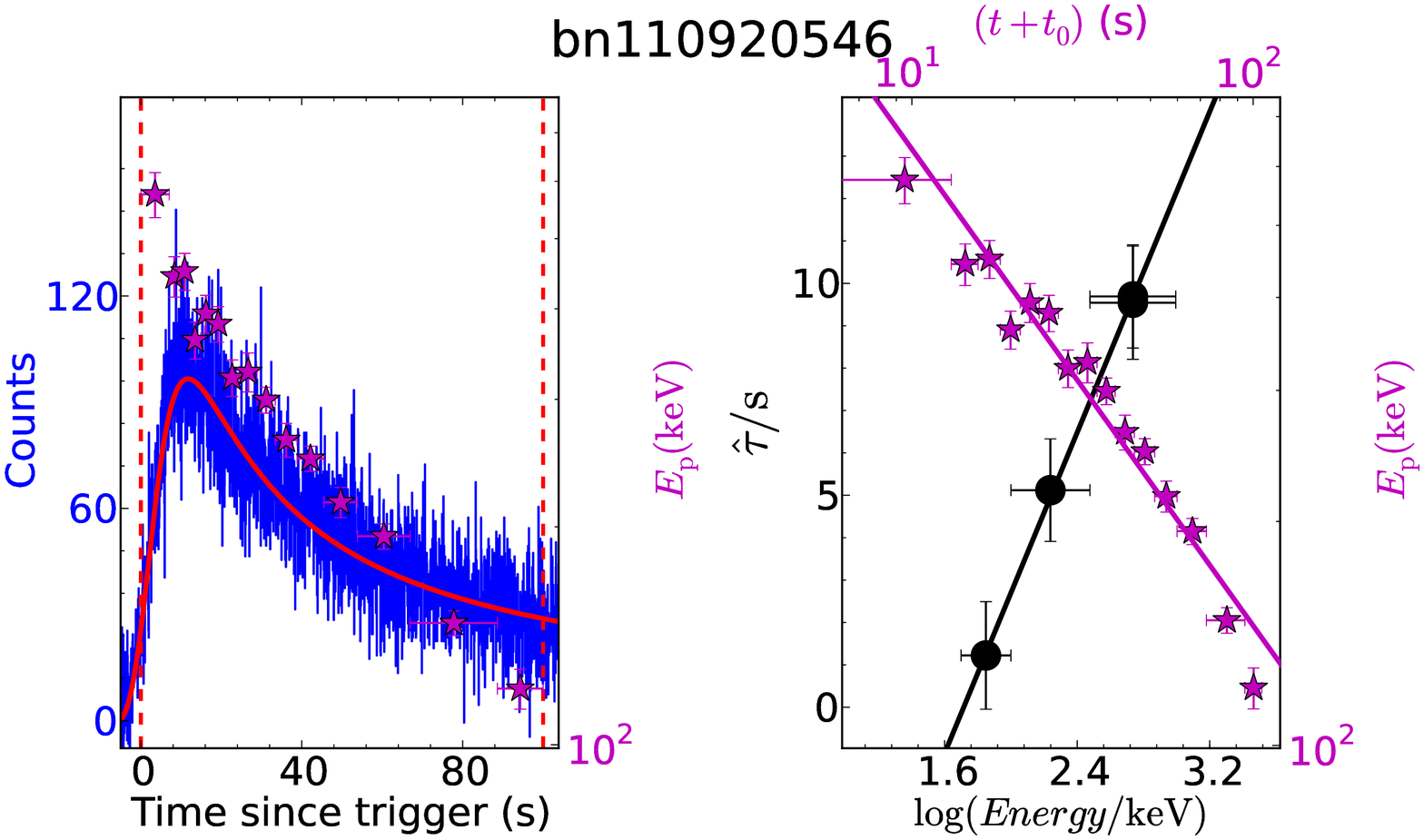} \\
\includegraphics[origin=c,angle=0,scale=1,width=0.500\textwidth,height=0.230\textheight]{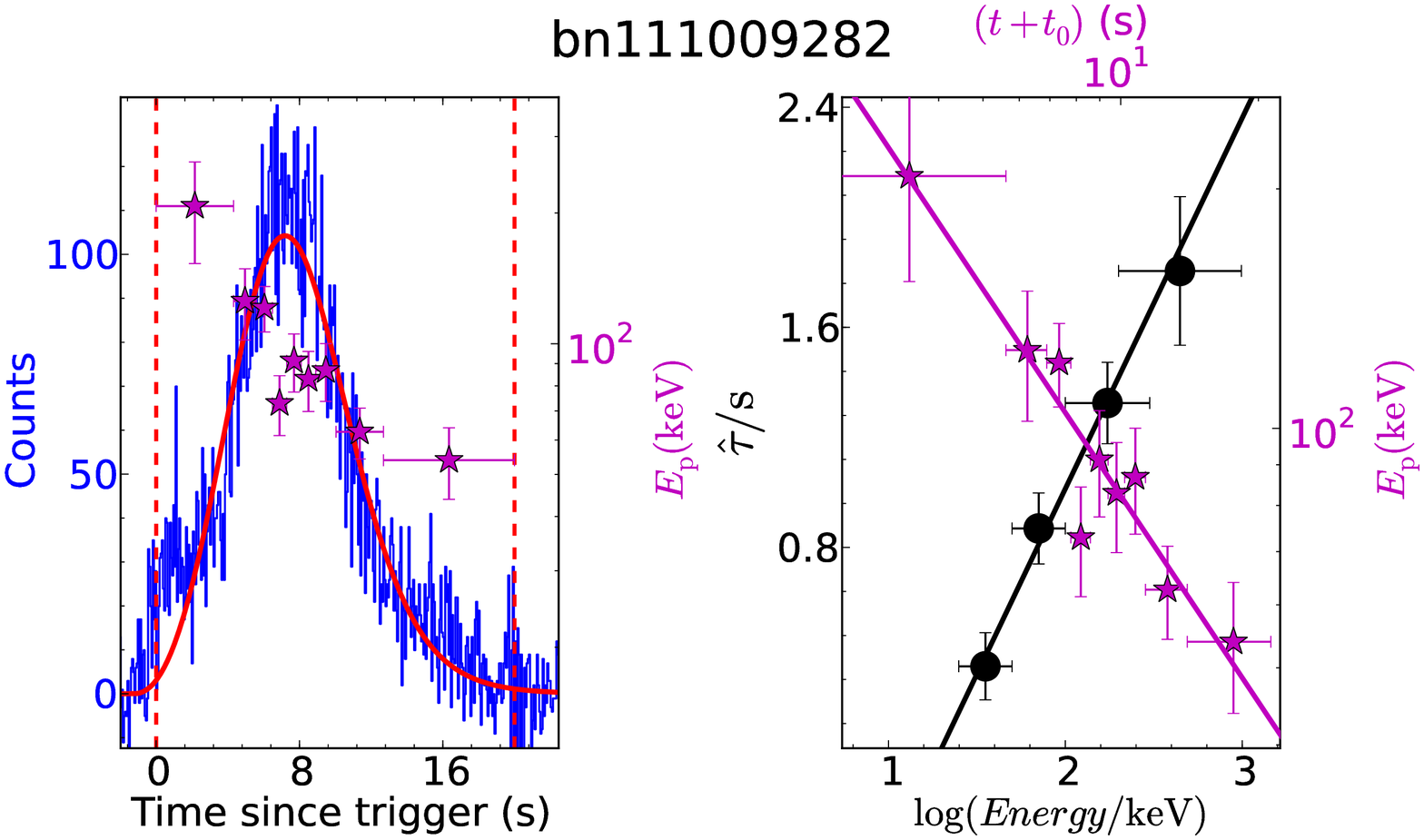} &
\includegraphics[origin=c,angle=0,scale=1,width=0.500\textwidth,height=0.230\textheight]{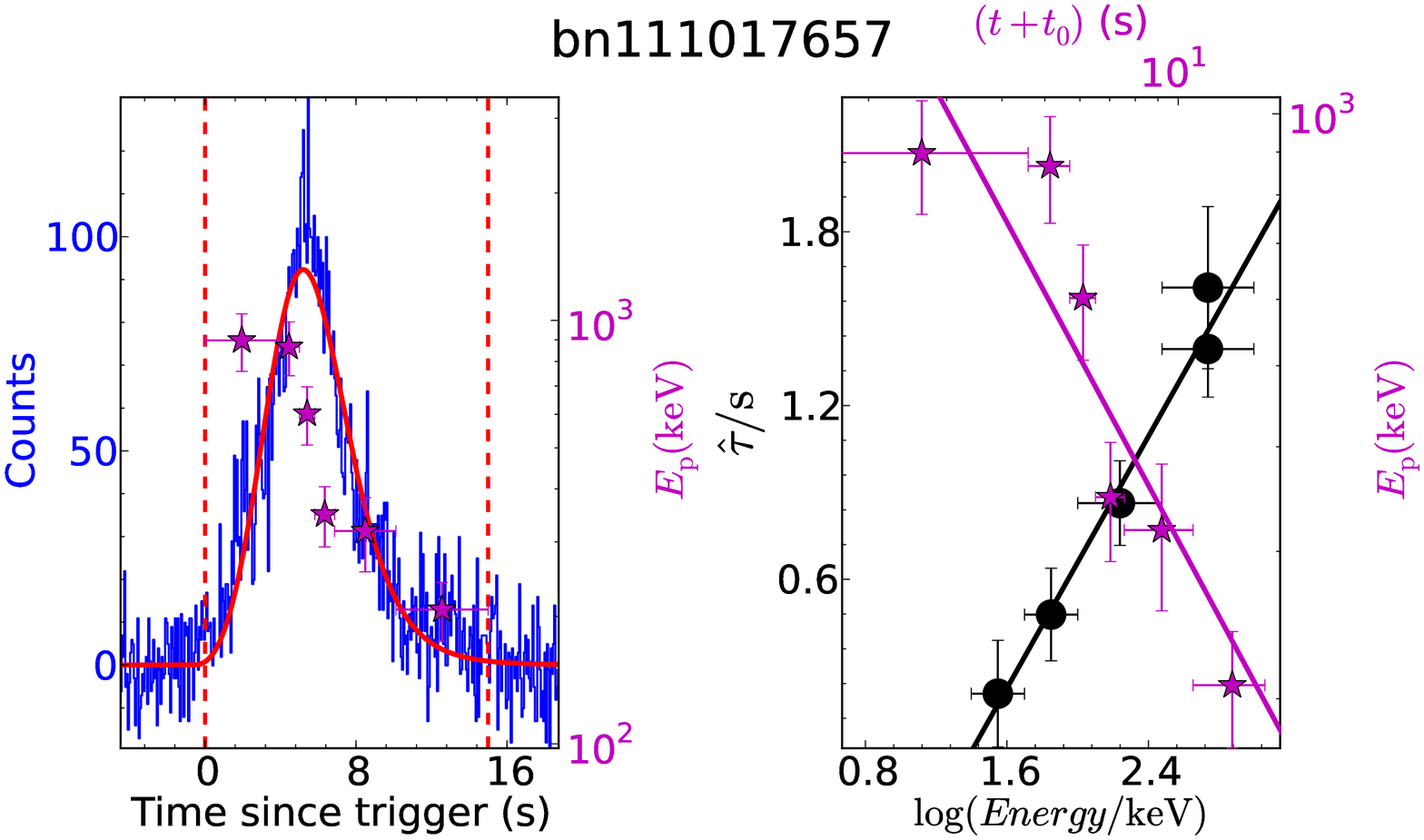} \\
\end{tabular}
\hfill
\vspace{4em}
\center{Fig.~\ref{myfigA}---Continued.}
\end{figure*}

\clearpage %p5
\begin{figure*}
\vspace{-4em}\hspace{-4em}
\centering
\begin{tabular}{cc}
\includegraphics[origin=c,angle=0,scale=1,width=0.500\textwidth,height=0.23\textheight]{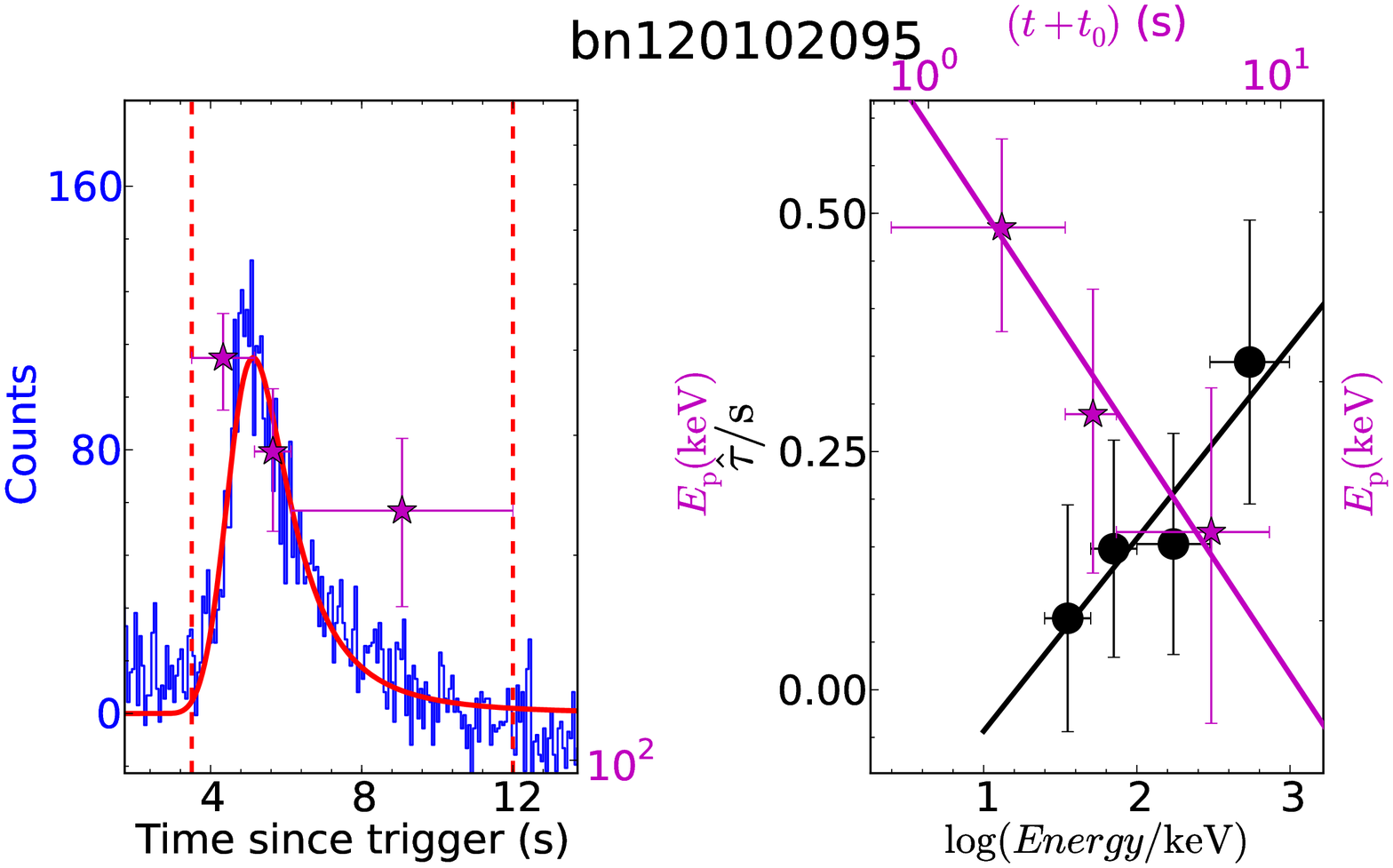} &
\includegraphics[origin=c,angle=0,scale=1,width=0.500\textwidth,height=0.23\textheight]{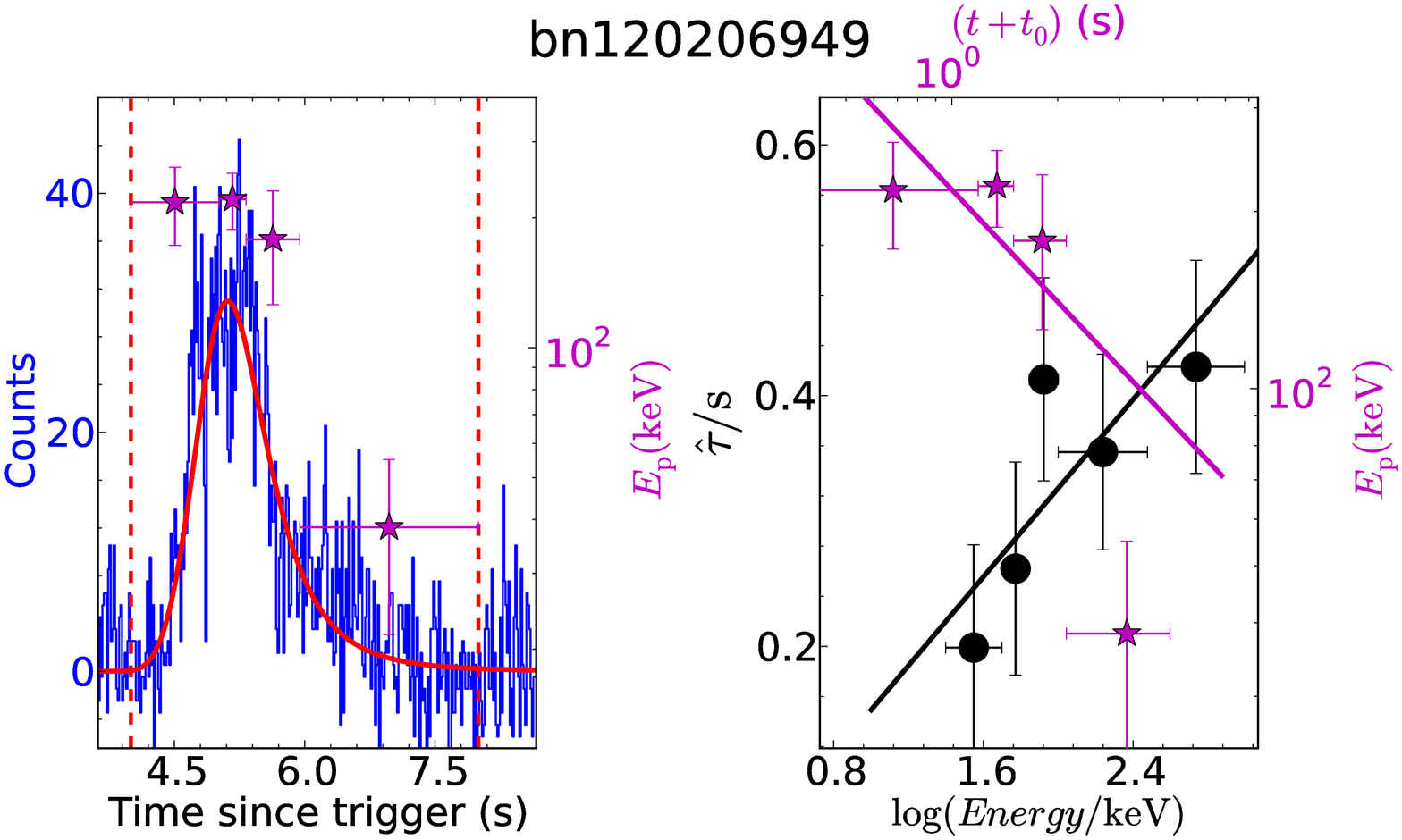} \\
\includegraphics[origin=c,angle=0,scale=1,width=0.500\textwidth,height=0.230\textheight]{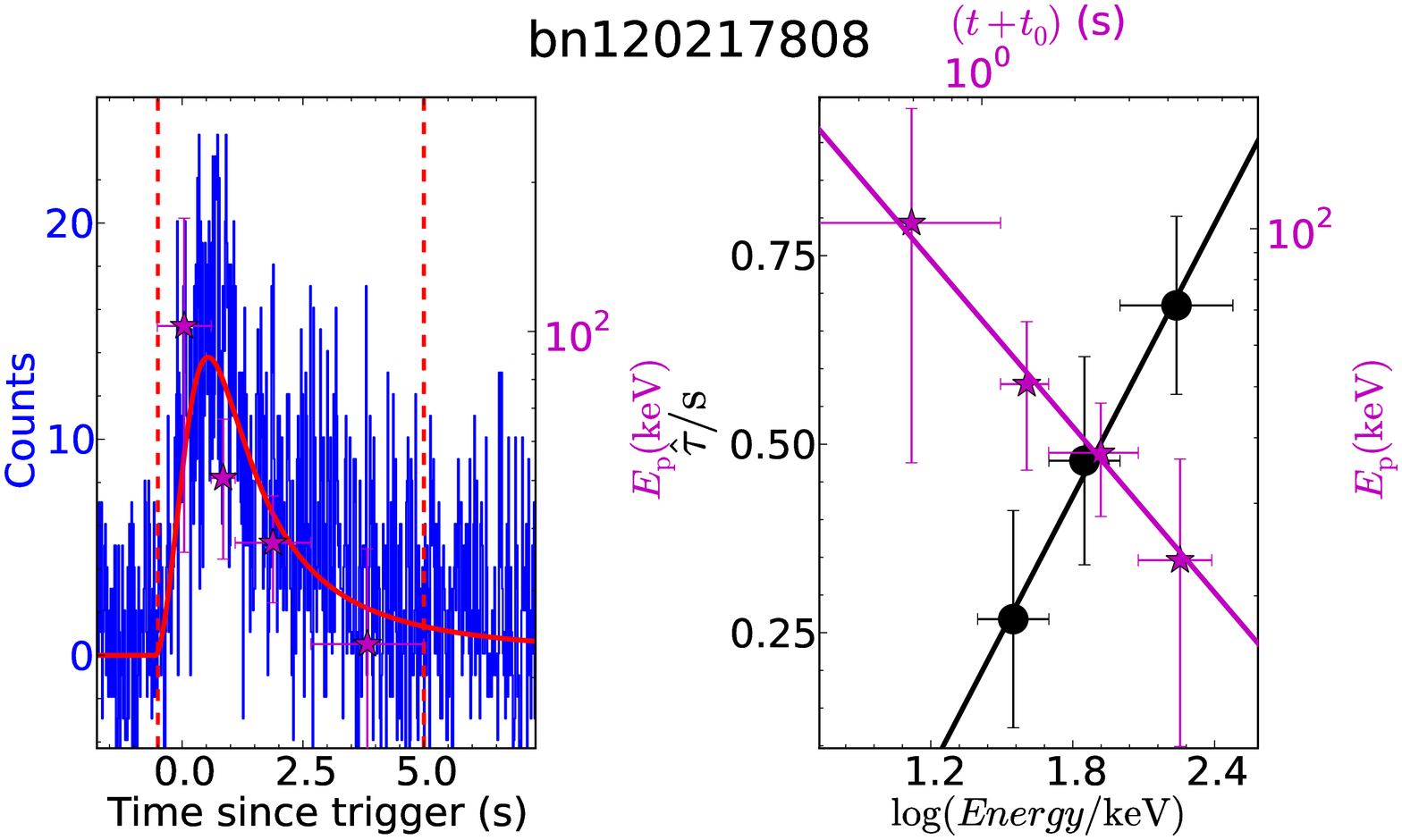} &
\includegraphics[origin=c,angle=0,scale=1,width=0.500\textwidth,height=0.23\textheight]{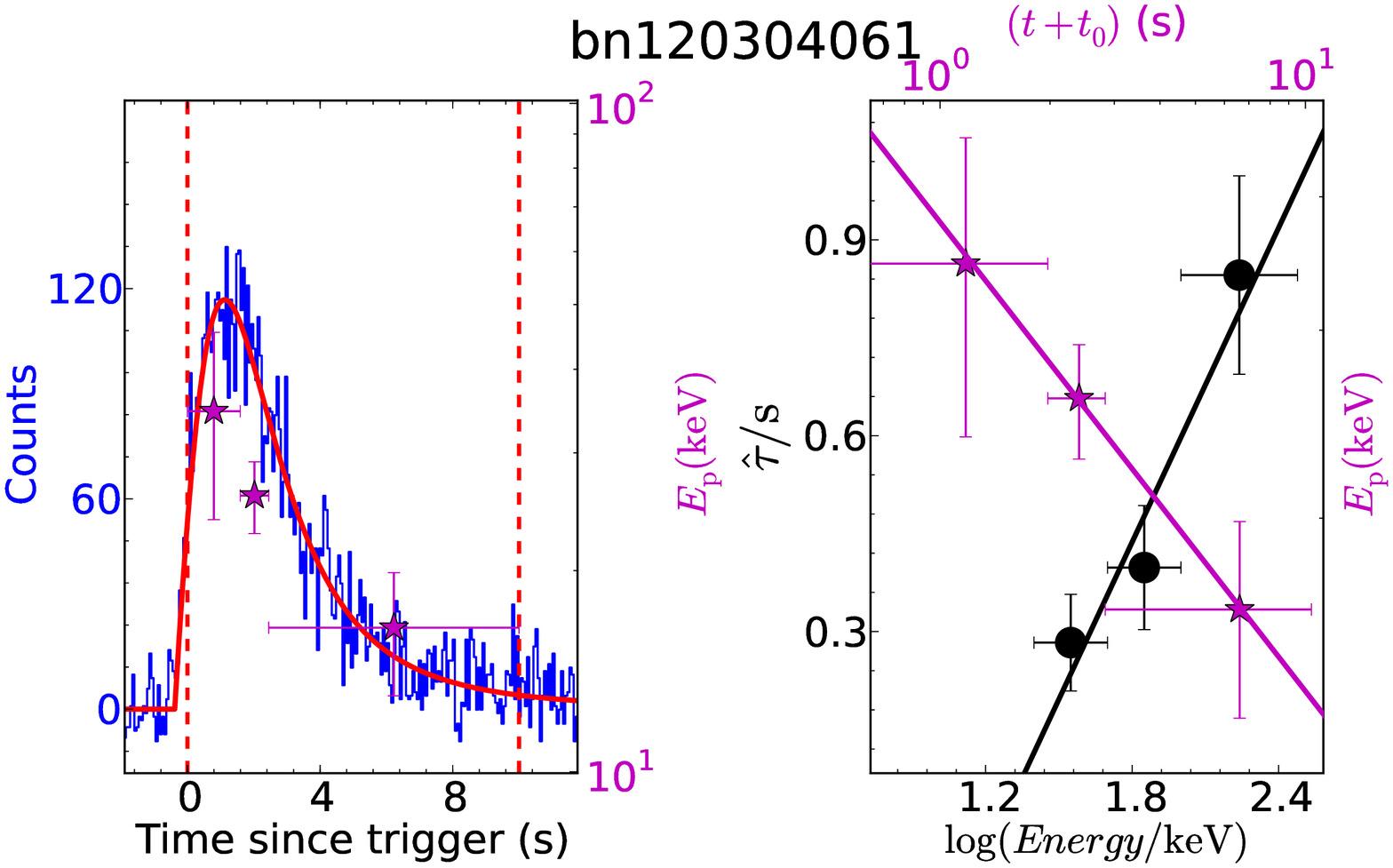} \\
\includegraphics[origin=c,angle=0,scale=1,width=0.500\textwidth,height=0.23\textheight]{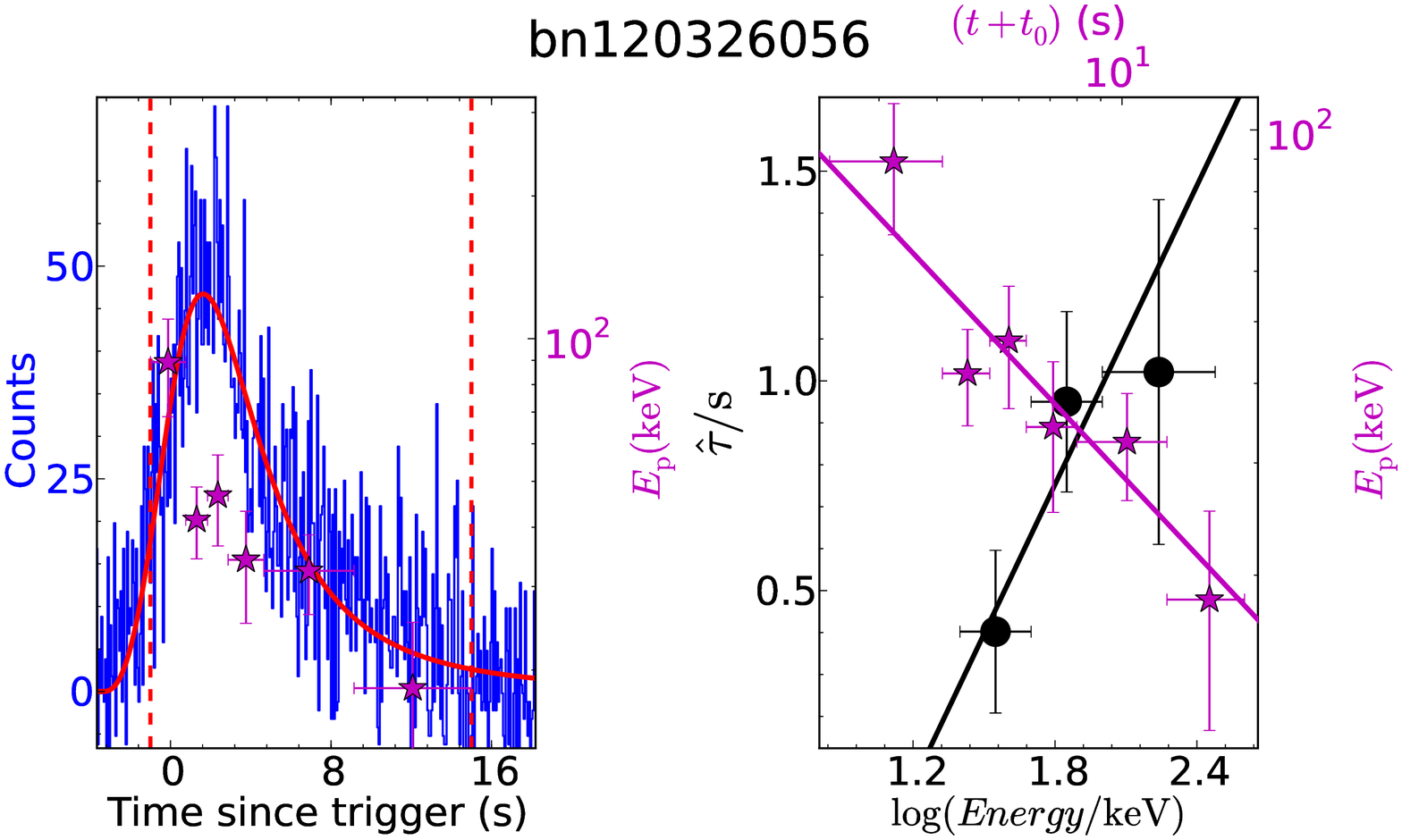} &
\includegraphics[origin=c,angle=0,scale=1,width=0.500\textwidth,height=0.22\textheight]{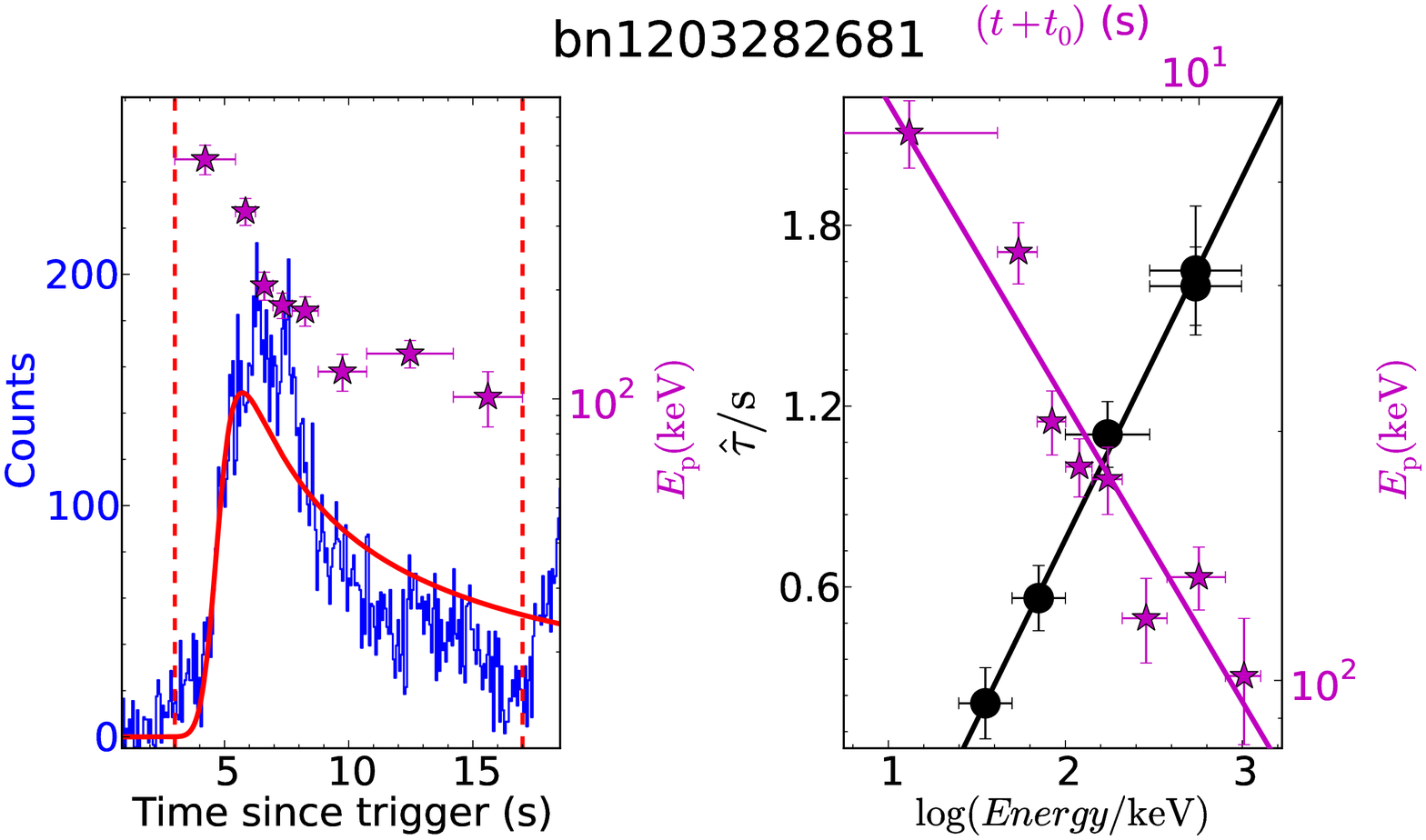} \\
\includegraphics[origin=c,angle=0,scale=1,width=0.500\textwidth,height=0.230\textheight]{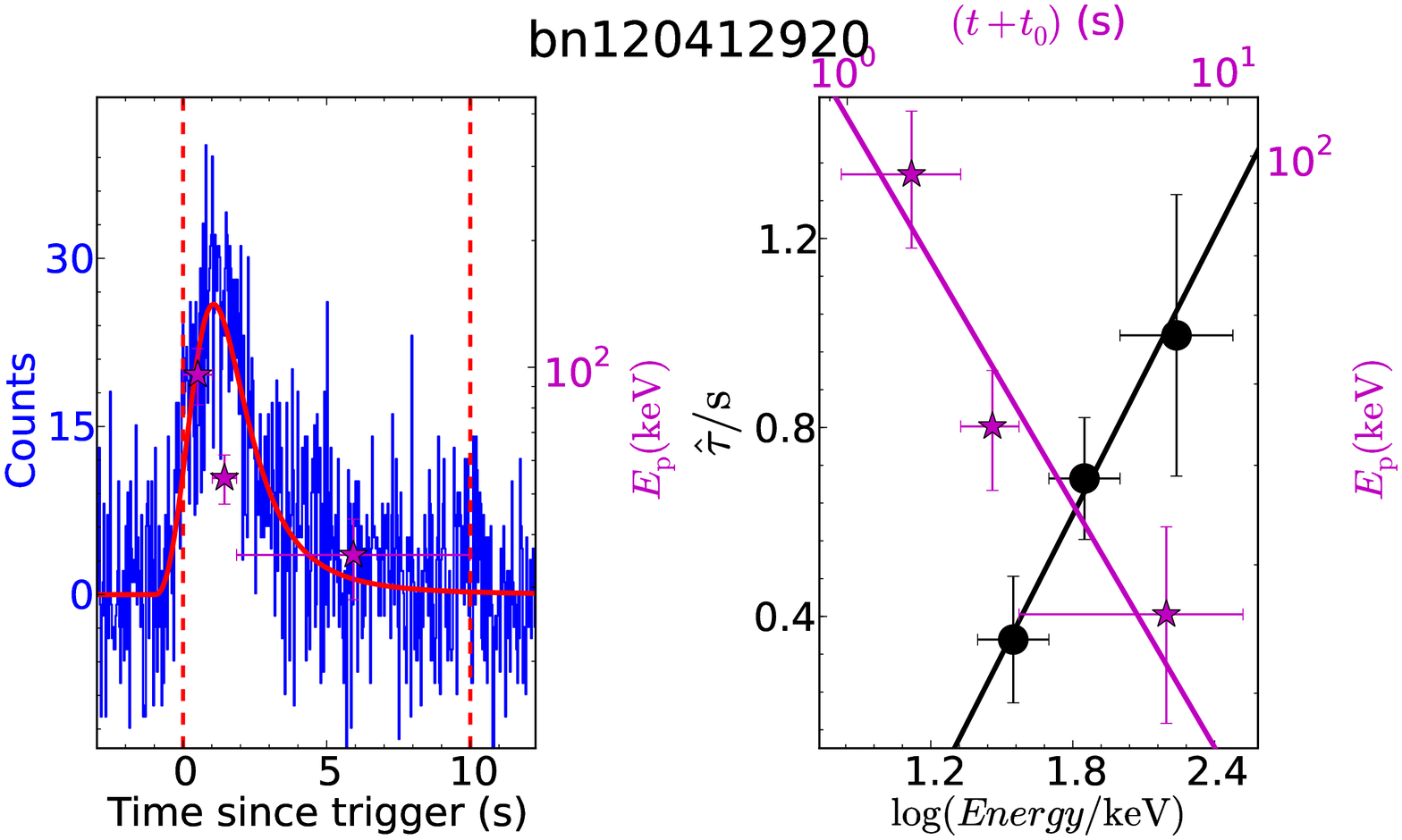} &
\includegraphics[origin=c,angle=0,scale=1,width=0.500\textwidth,height=0.230\textheight]{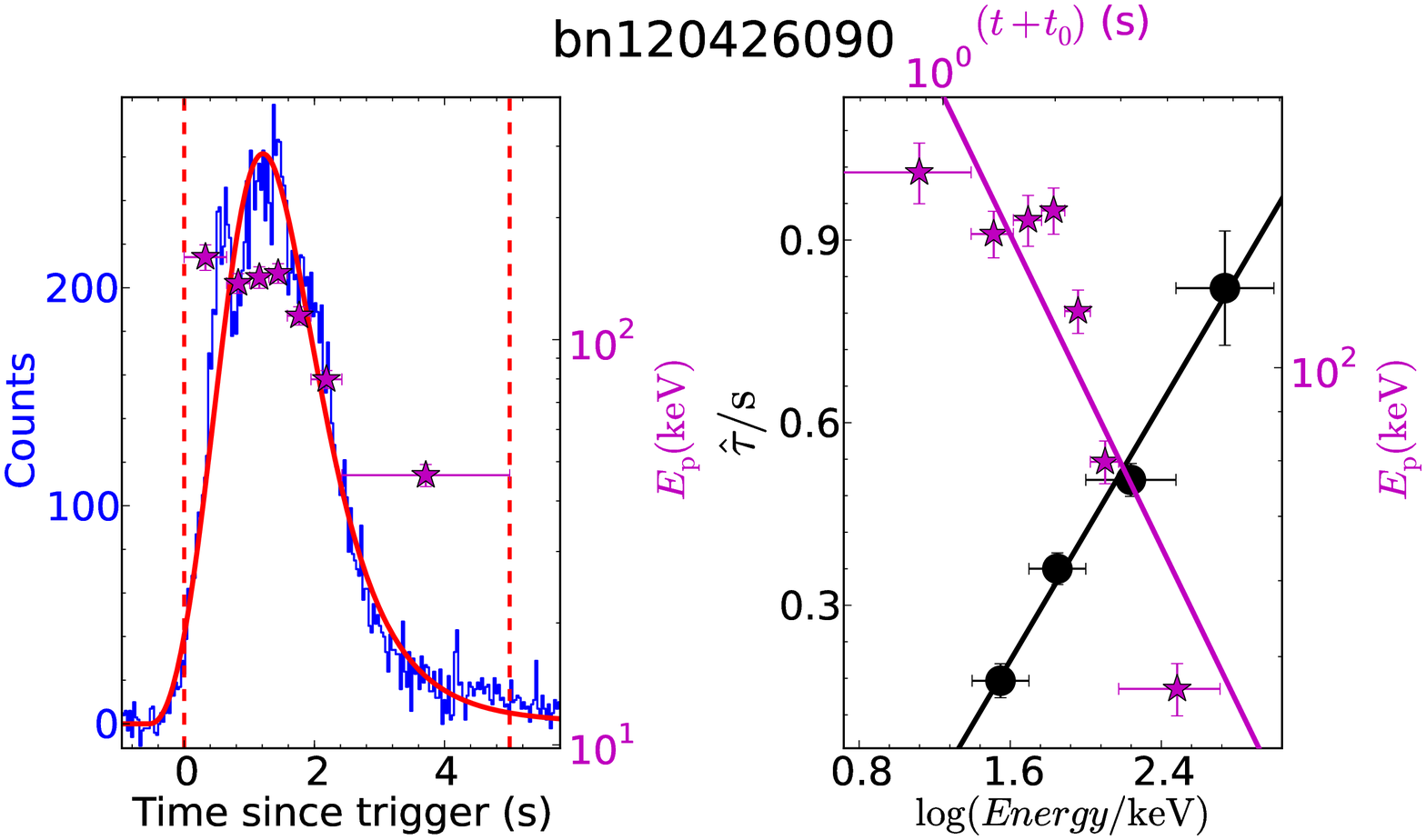} \\
\end{tabular}
\hfill
\vspace{4em}
\center{Fig.~\ref{myfigA}---Continued.}
\end{figure*}

\clearpage %p6
\begin{figure*}
\vspace{-4em}\hspace{-4em}
\centering
\begin{tabular}{cc}
\includegraphics[origin=c,angle=0,scale=1,width=0.500\textwidth,height=0.23\textheight]{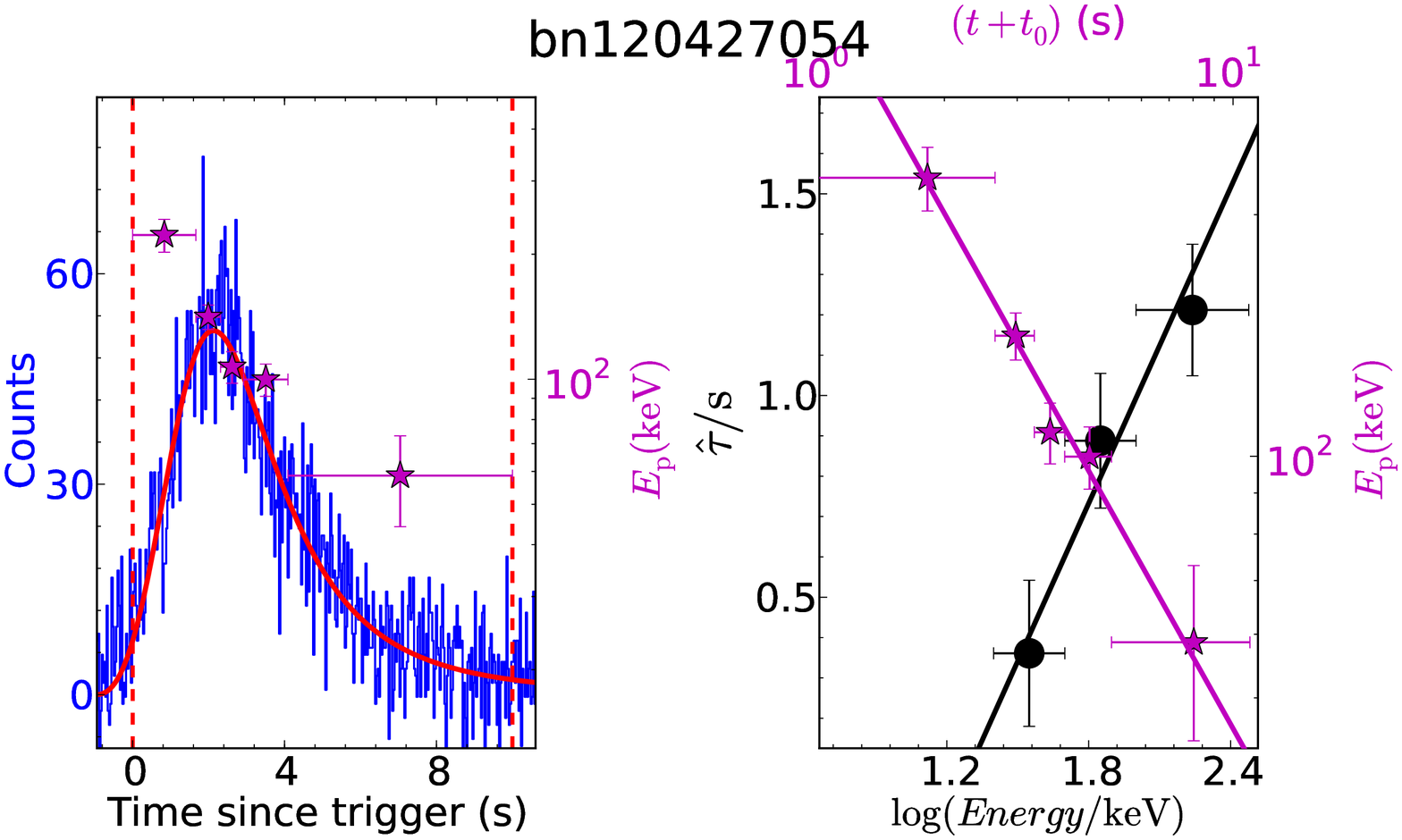} &
\includegraphics[origin=c,angle=0,scale=1,width=0.500\textwidth,height=0.23\textheight]{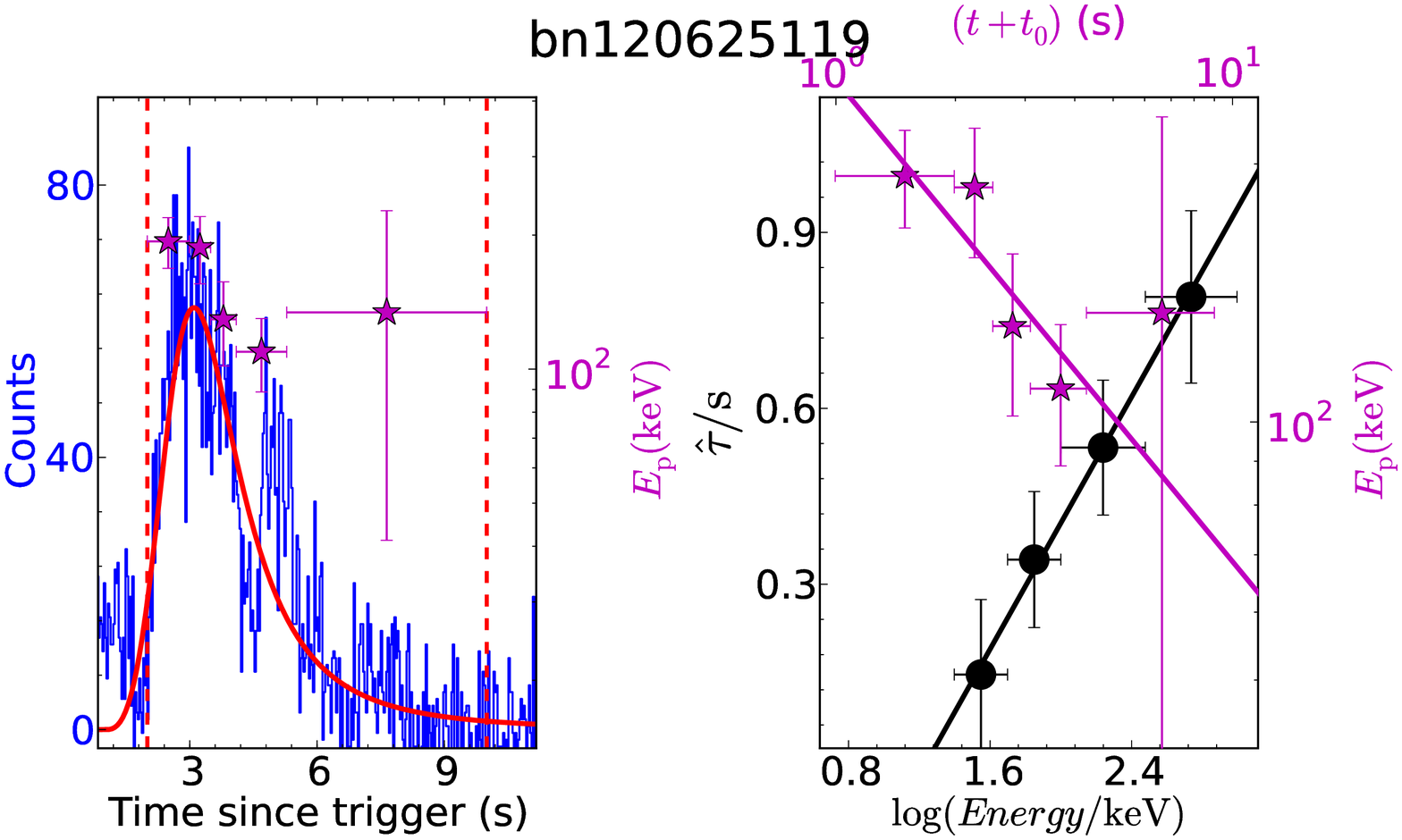} \\
\includegraphics[origin=c,angle=0,scale=1,width=0.500\textwidth,height=0.230\textheight]{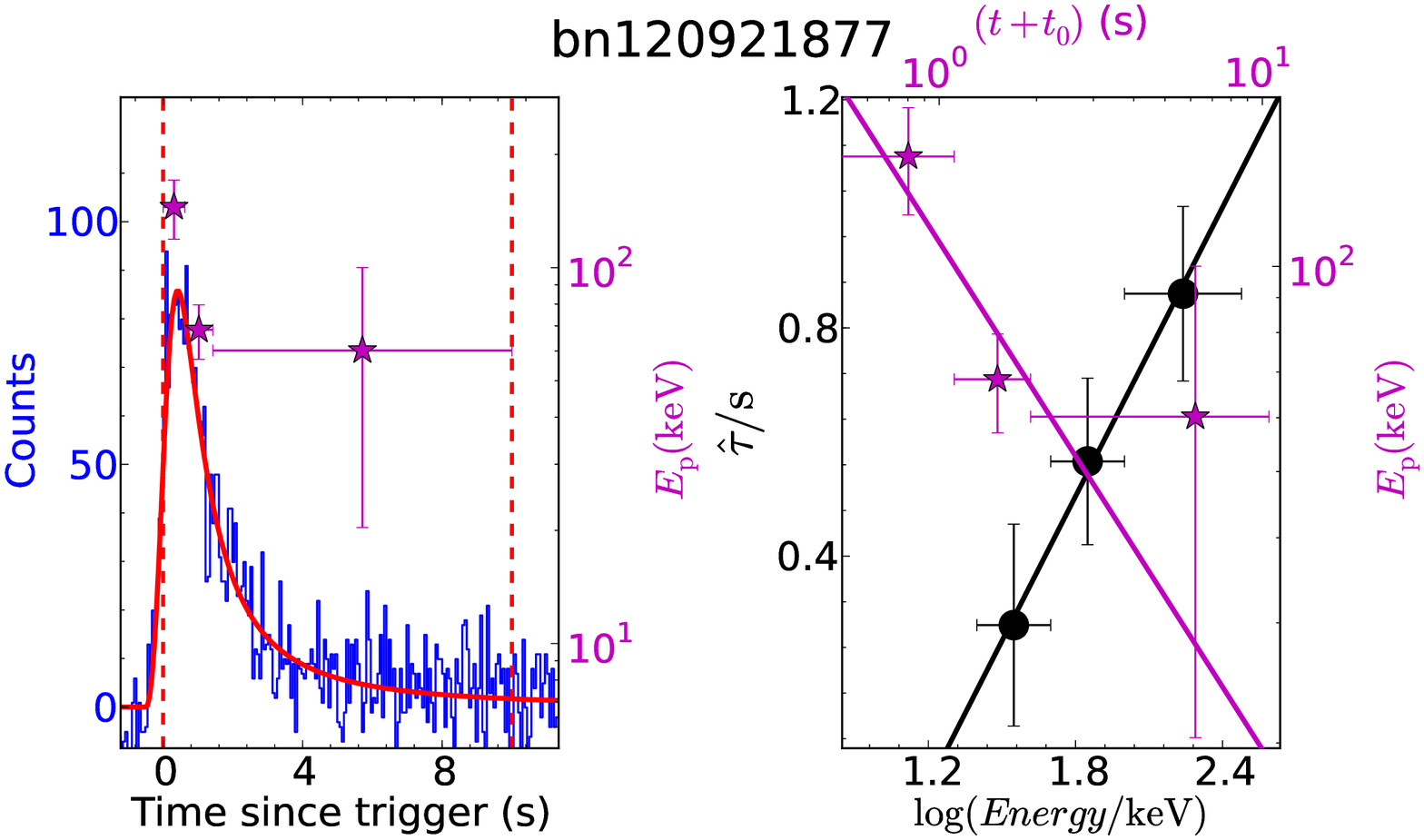} &
\includegraphics[origin=c,angle=0,scale=1,width=0.500\textwidth,height=0.23\textheight]{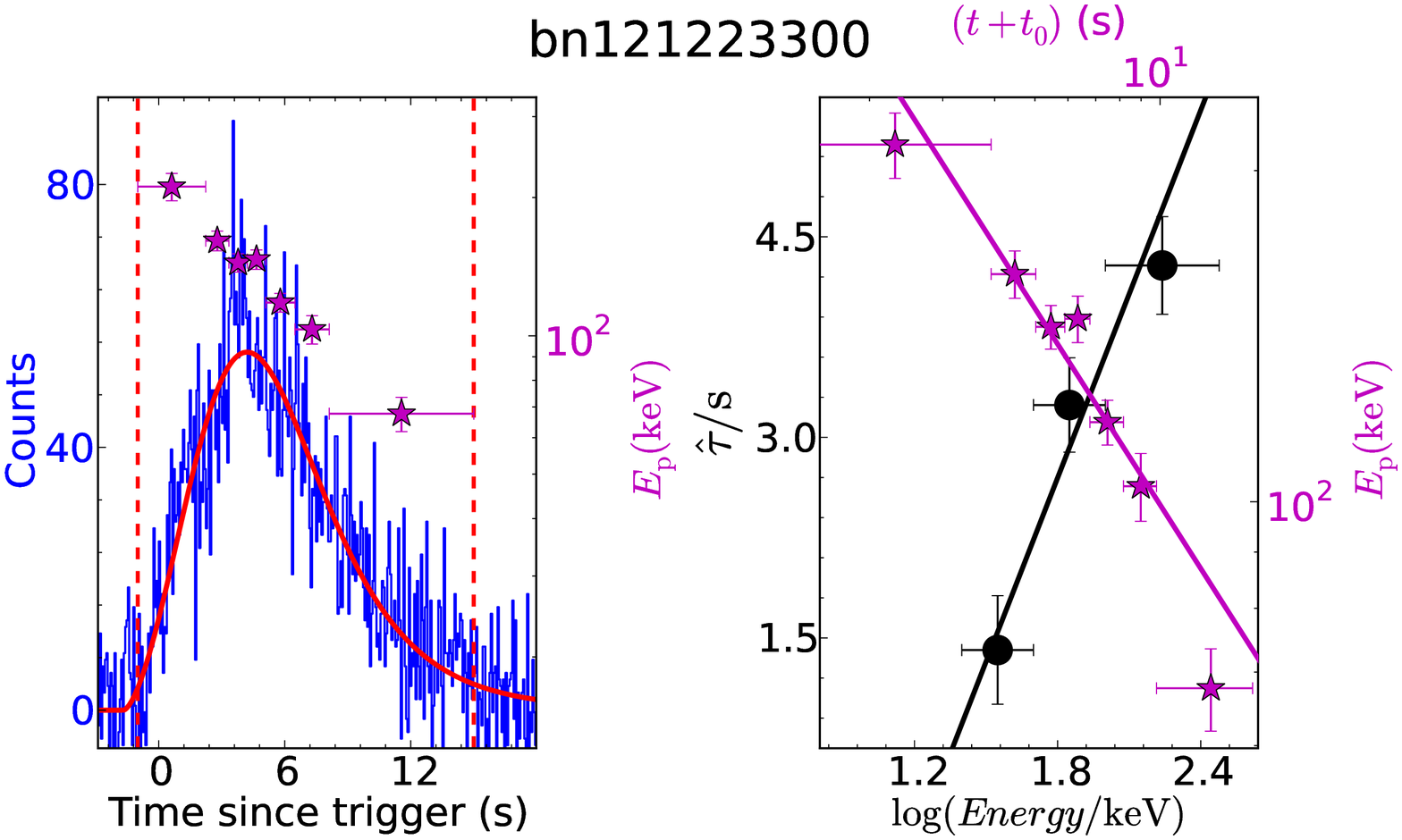} \\
\includegraphics[origin=c,angle=0,scale=1,width=0.500\textwidth,height=0.23\textheight]{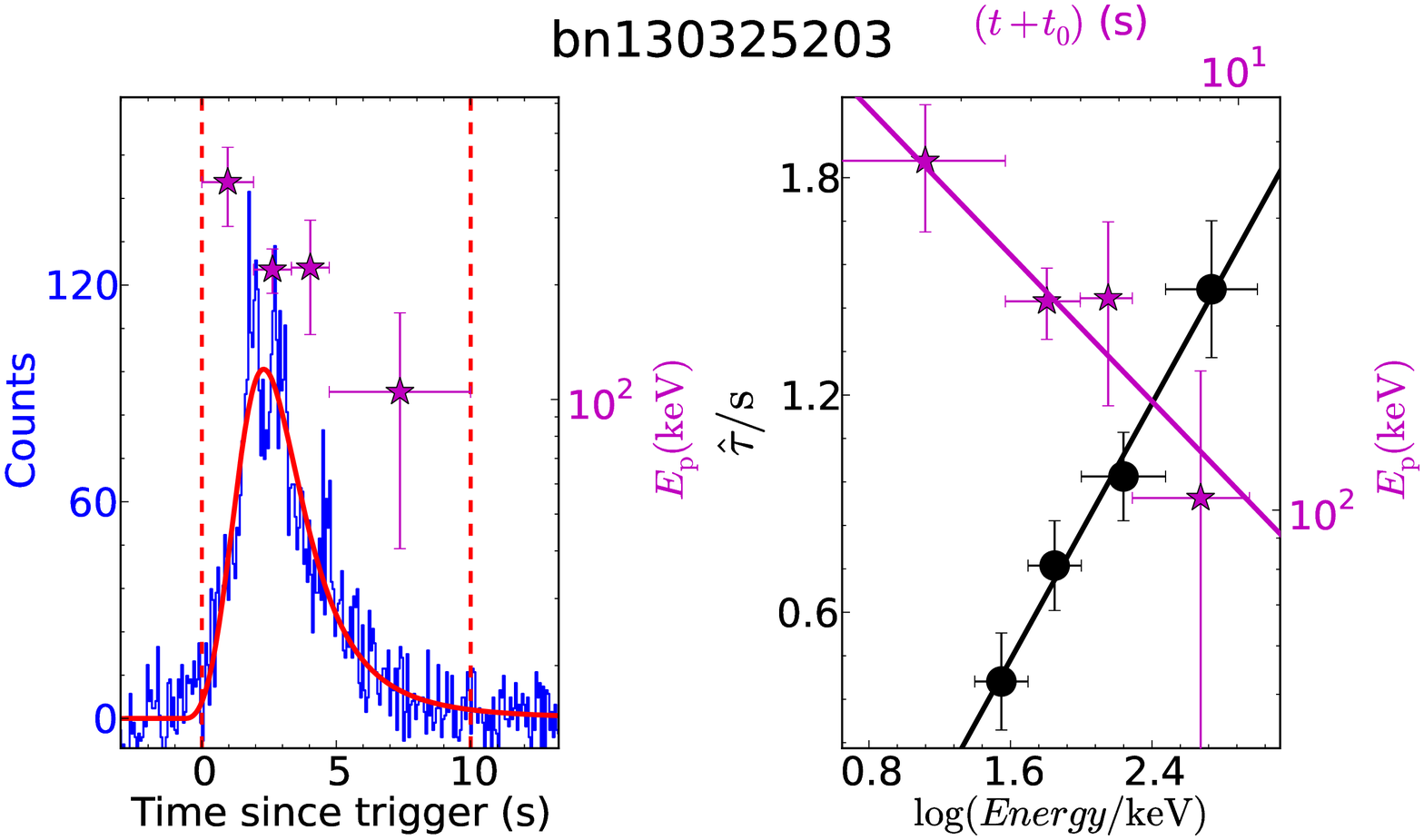} &
\includegraphics[origin=c,angle=0,scale=1,width=0.500\textwidth,height=0.230\textheight]{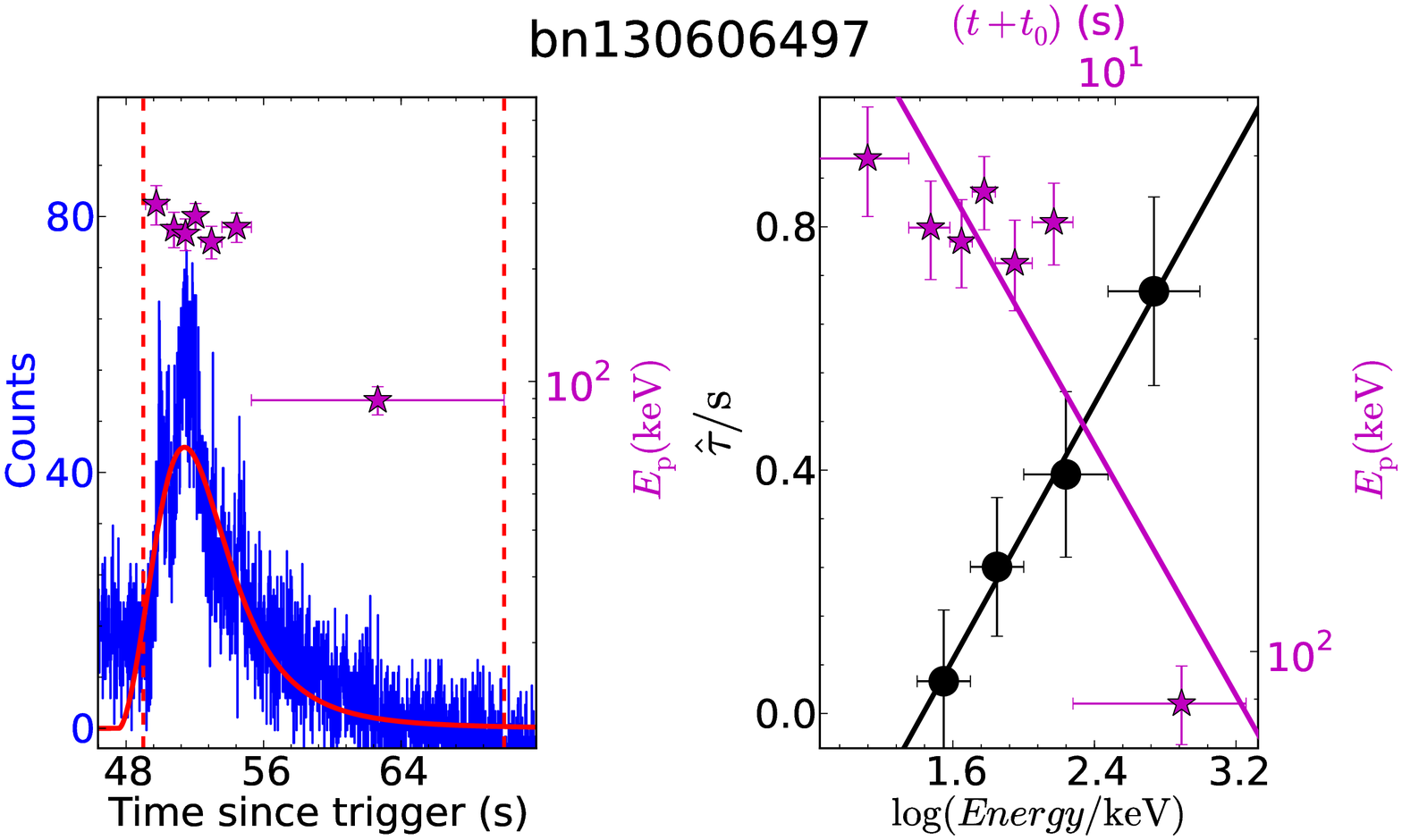} \\
\includegraphics[origin=c,angle=0,scale=1,width=0.500\textwidth,height=0.230\textheight]{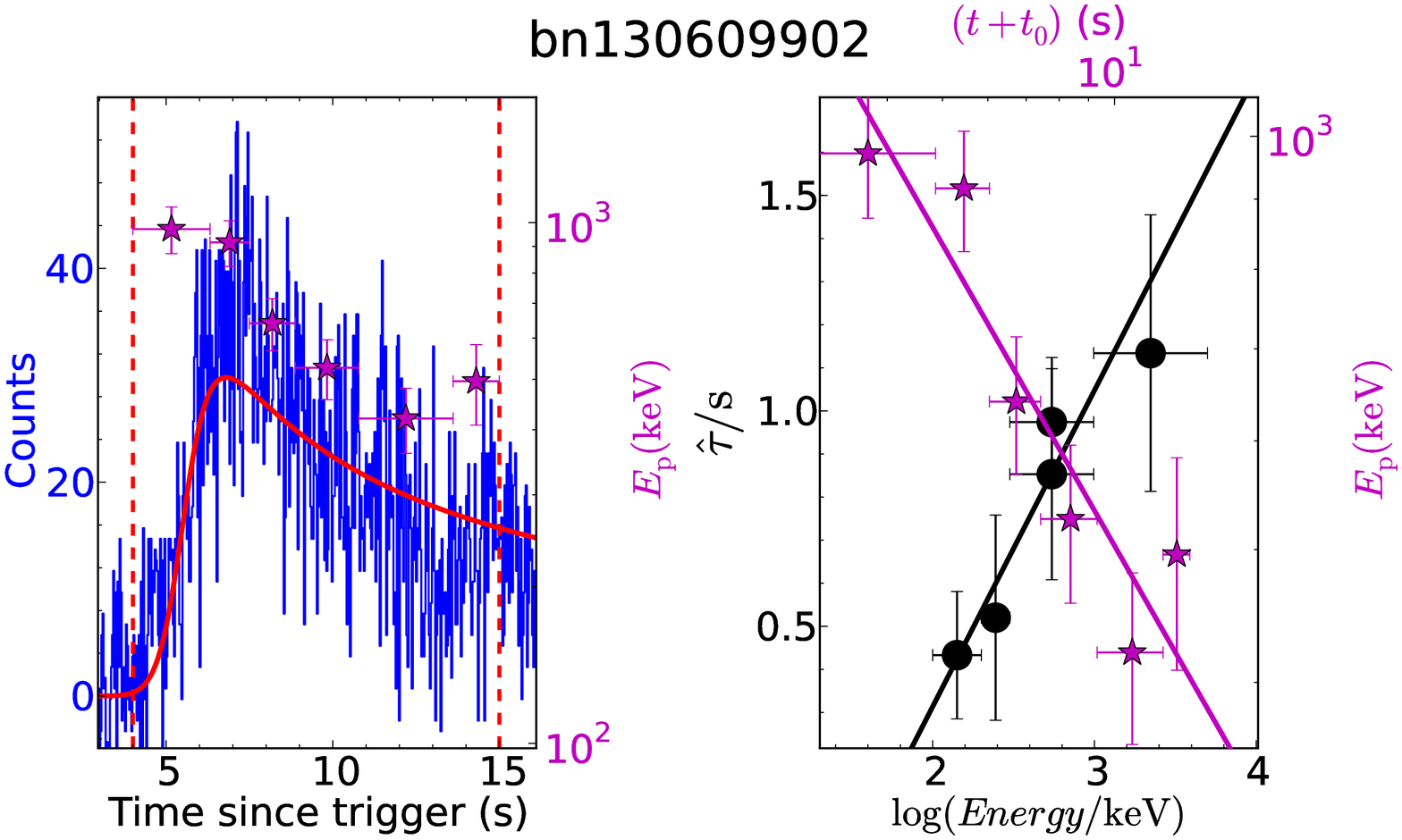} &
\includegraphics[origin=c,angle=0,scale=1,width=0.500\textwidth,height=0.230\textheight]{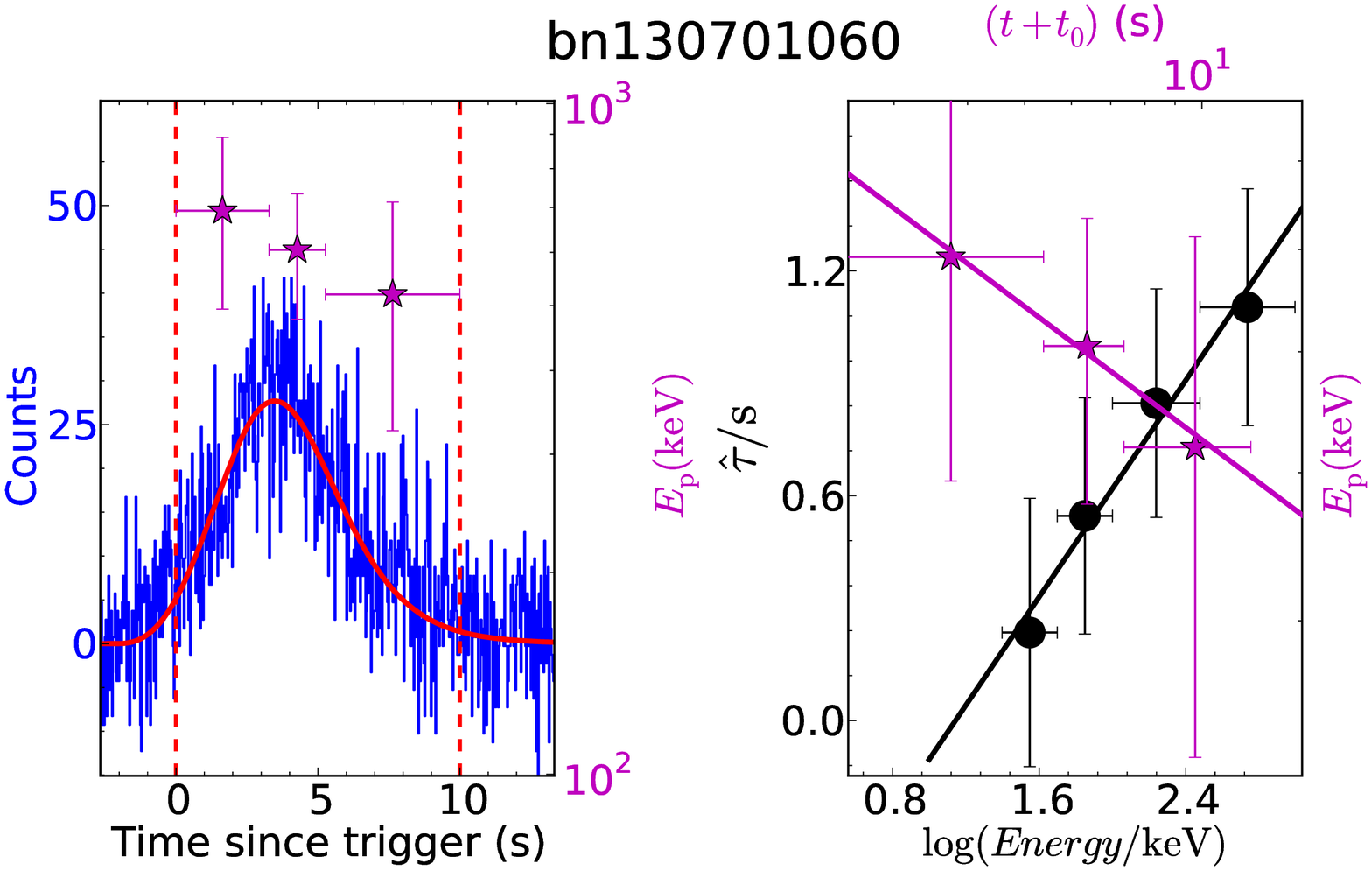} \\
\end{tabular}
\hfill
\vspace{4em}
\center{Fig.~\ref{myfigA}---Continued.}
\end{figure*}

\clearpage %7

\begin{figure*}
\vspace{-4em}\hspace{-4em}
\centering
\begin{tabular}{cc}
\includegraphics[origin=c,angle=0,scale=1,width=0.500\textwidth,height=0.230\textheight]{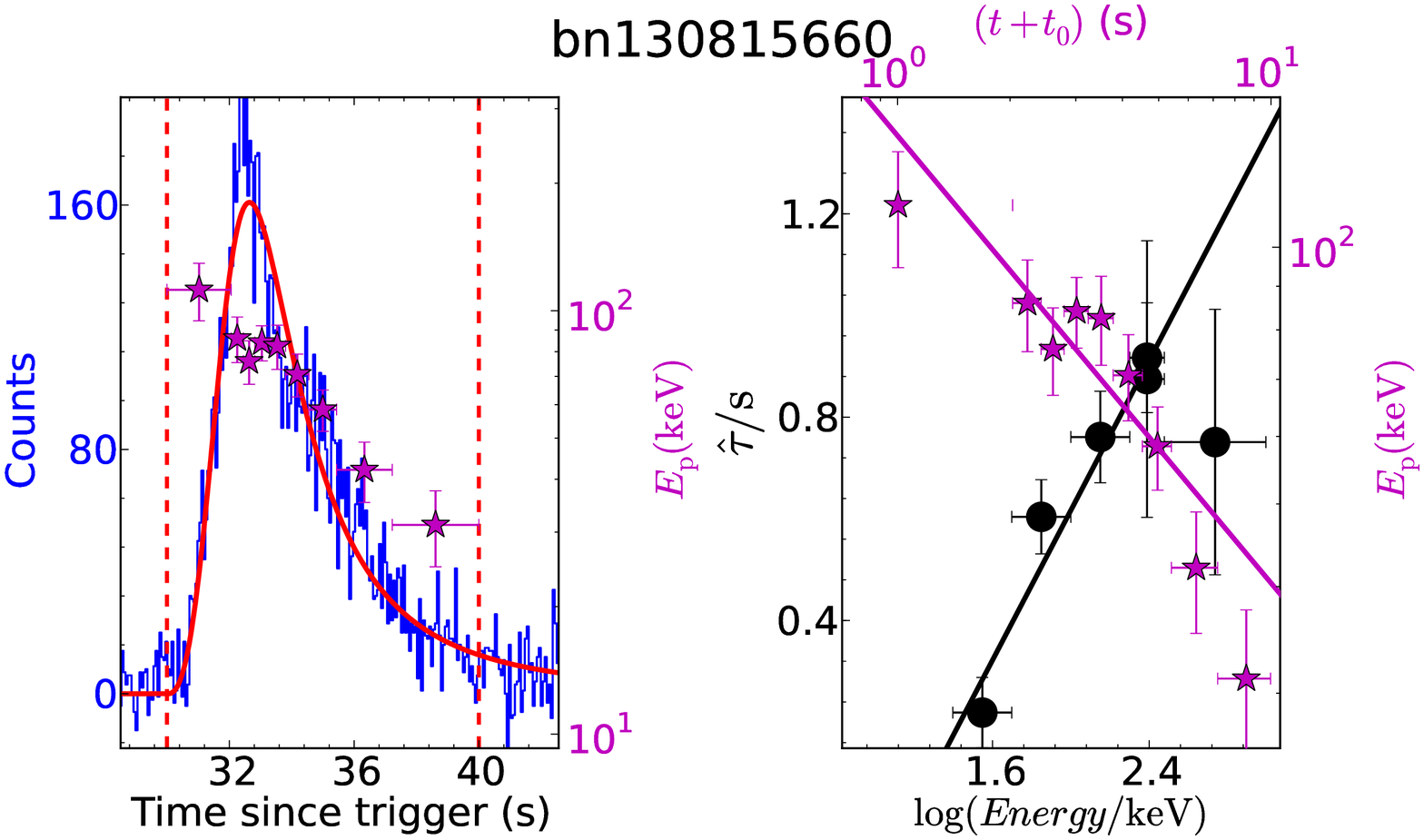} &
\includegraphics[origin=c,angle=0,scale=1,width=0.500\textwidth,height=0.23\textheight]{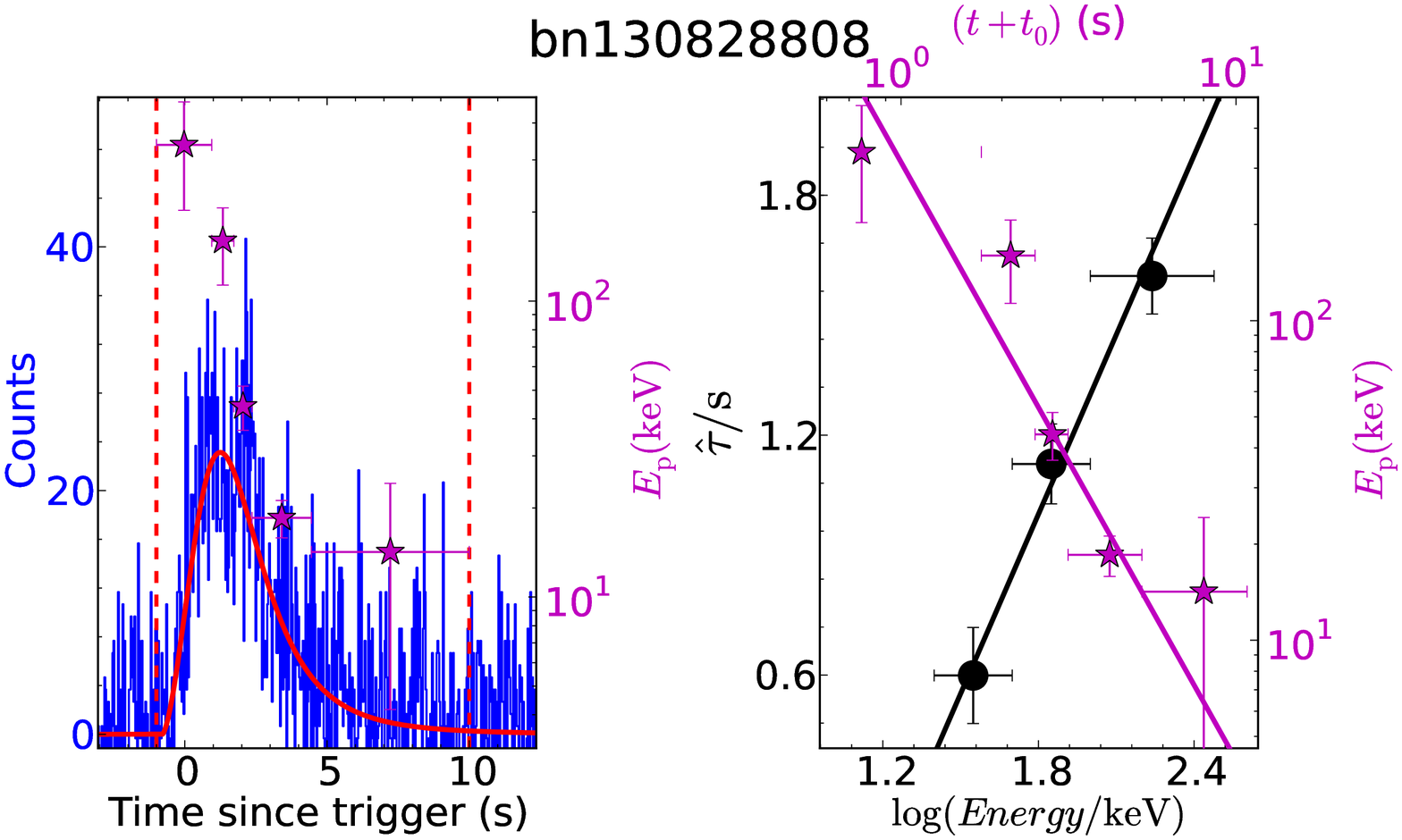} \\
\includegraphics[origin=c,angle=0,scale=1,width=0.500\textwidth,height=0.23\textheight]{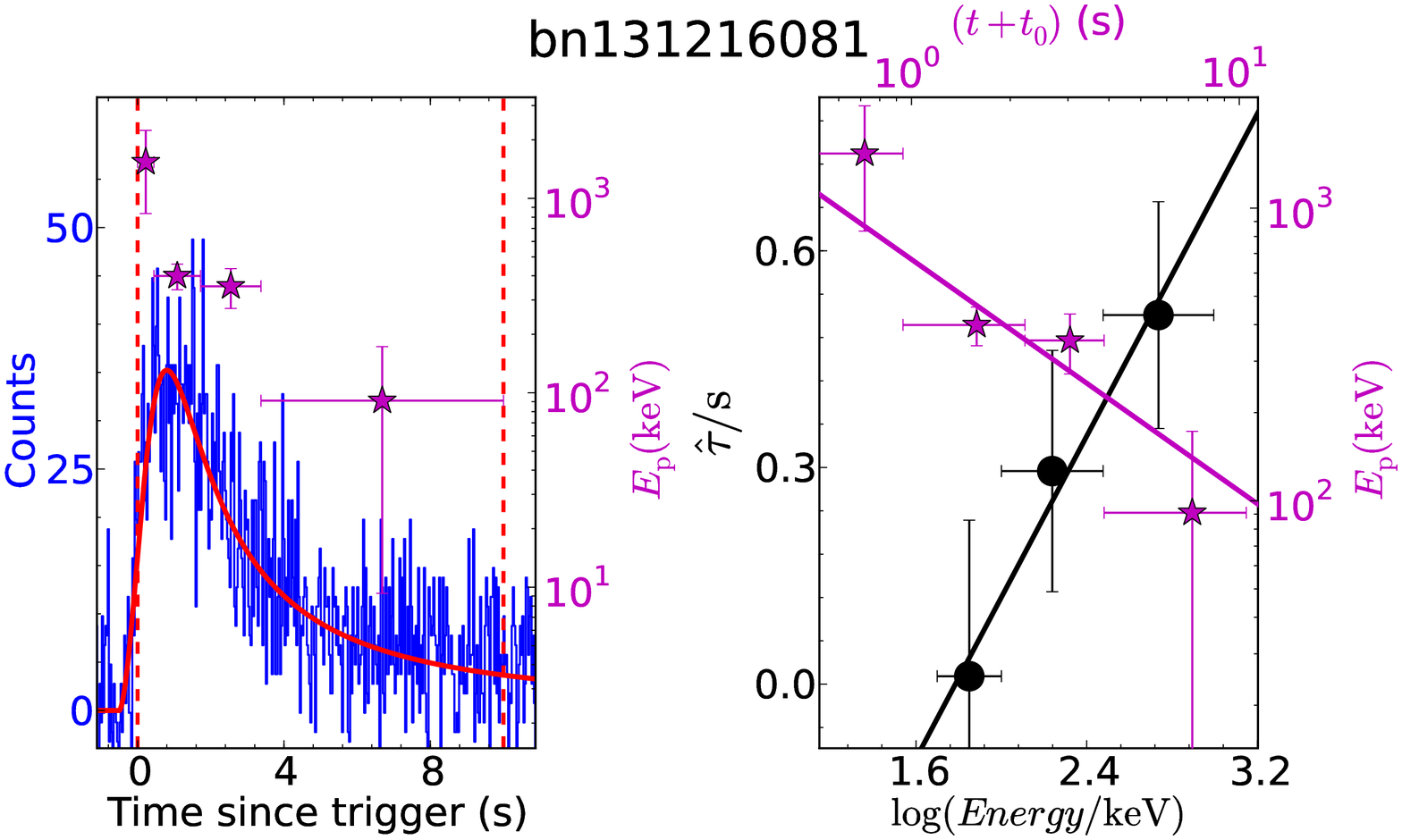} &
\includegraphics[origin=c,angle=0,scale=1,width=0.500\textwidth,height=0.23\textheight]{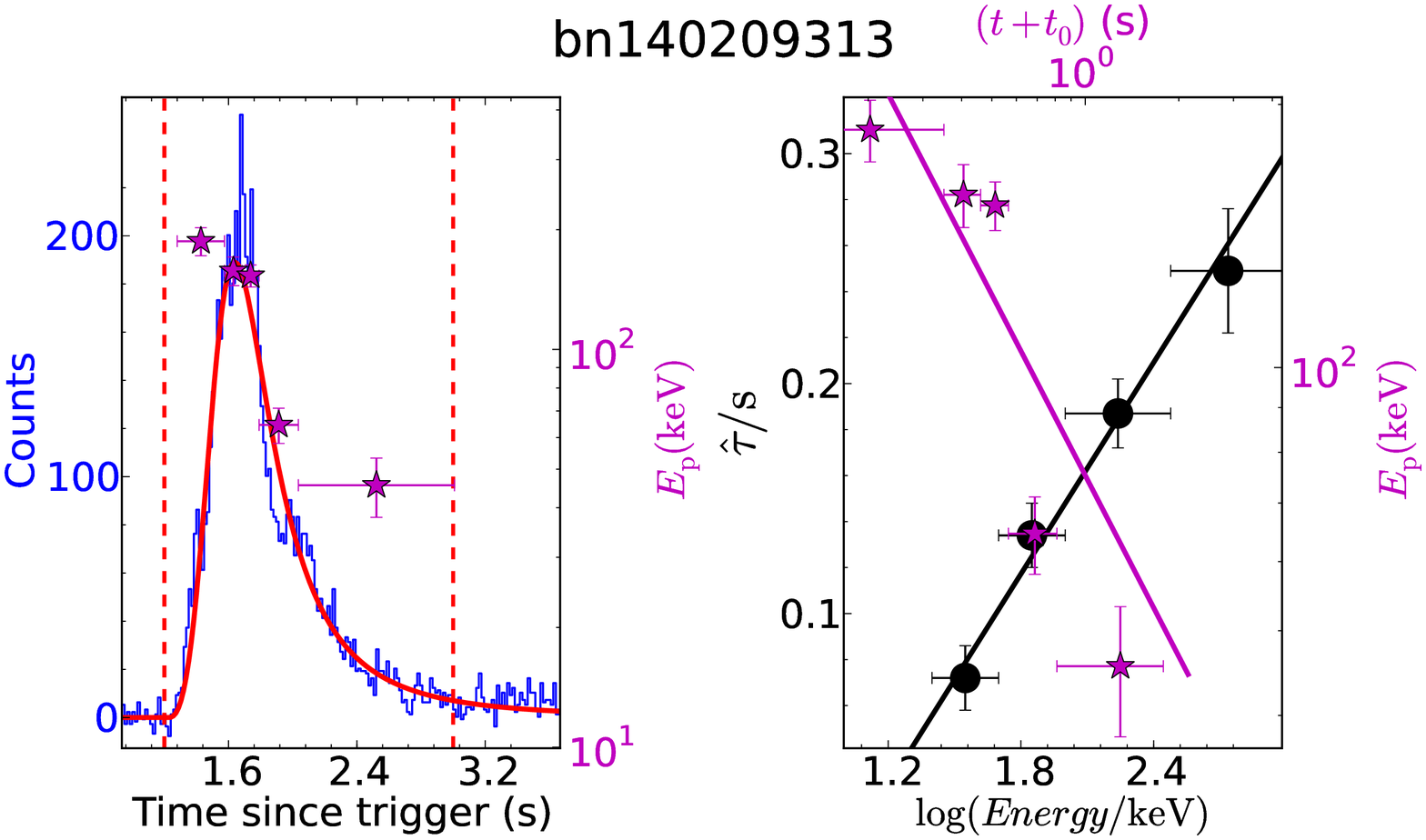} \\
\includegraphics[origin=c,angle=0,scale=1,width=0.500\textwidth,height=0.230\textheight]{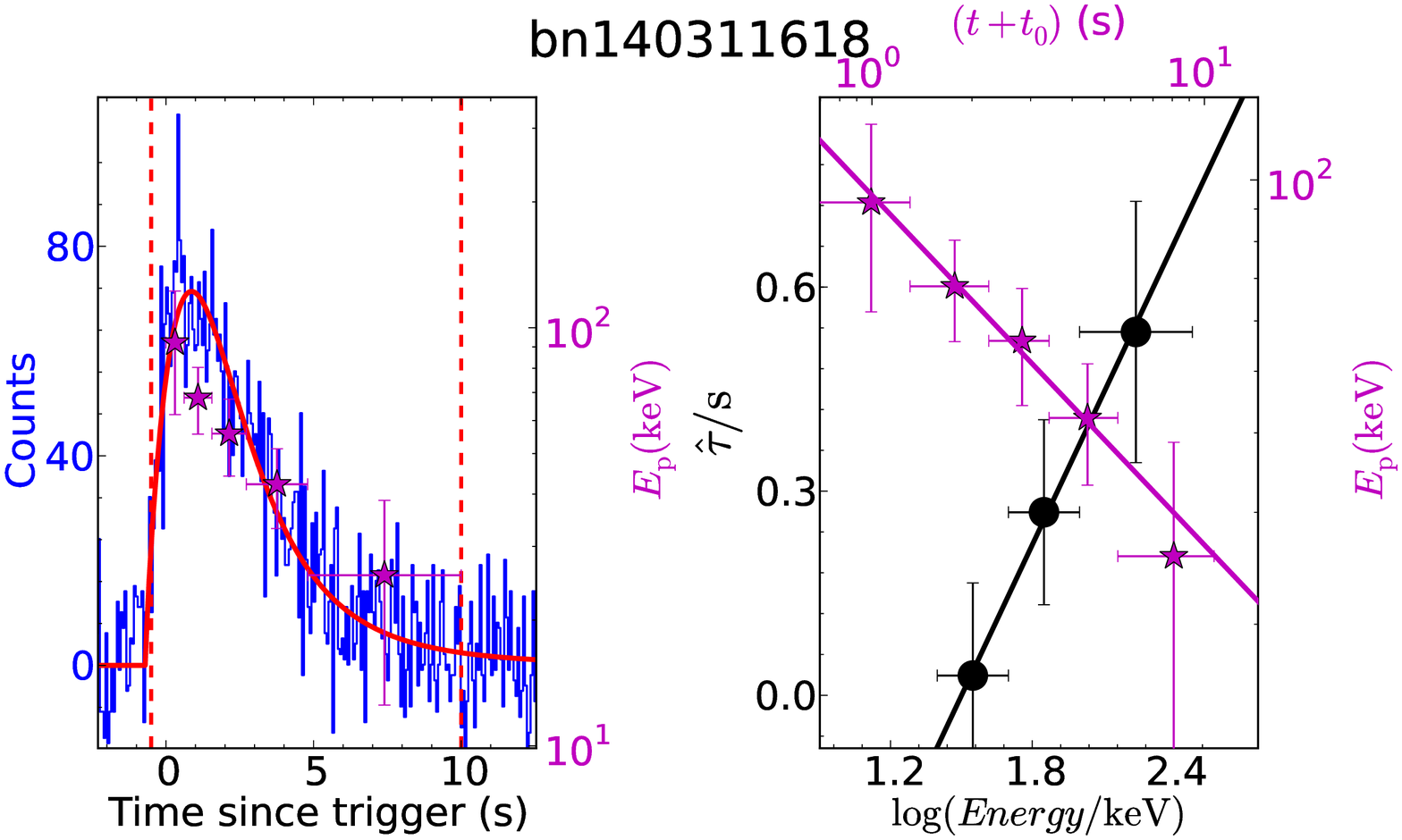} &
\includegraphics[origin=c,angle=0,scale=1,width=0.500\textwidth,height=0.19\textheight]{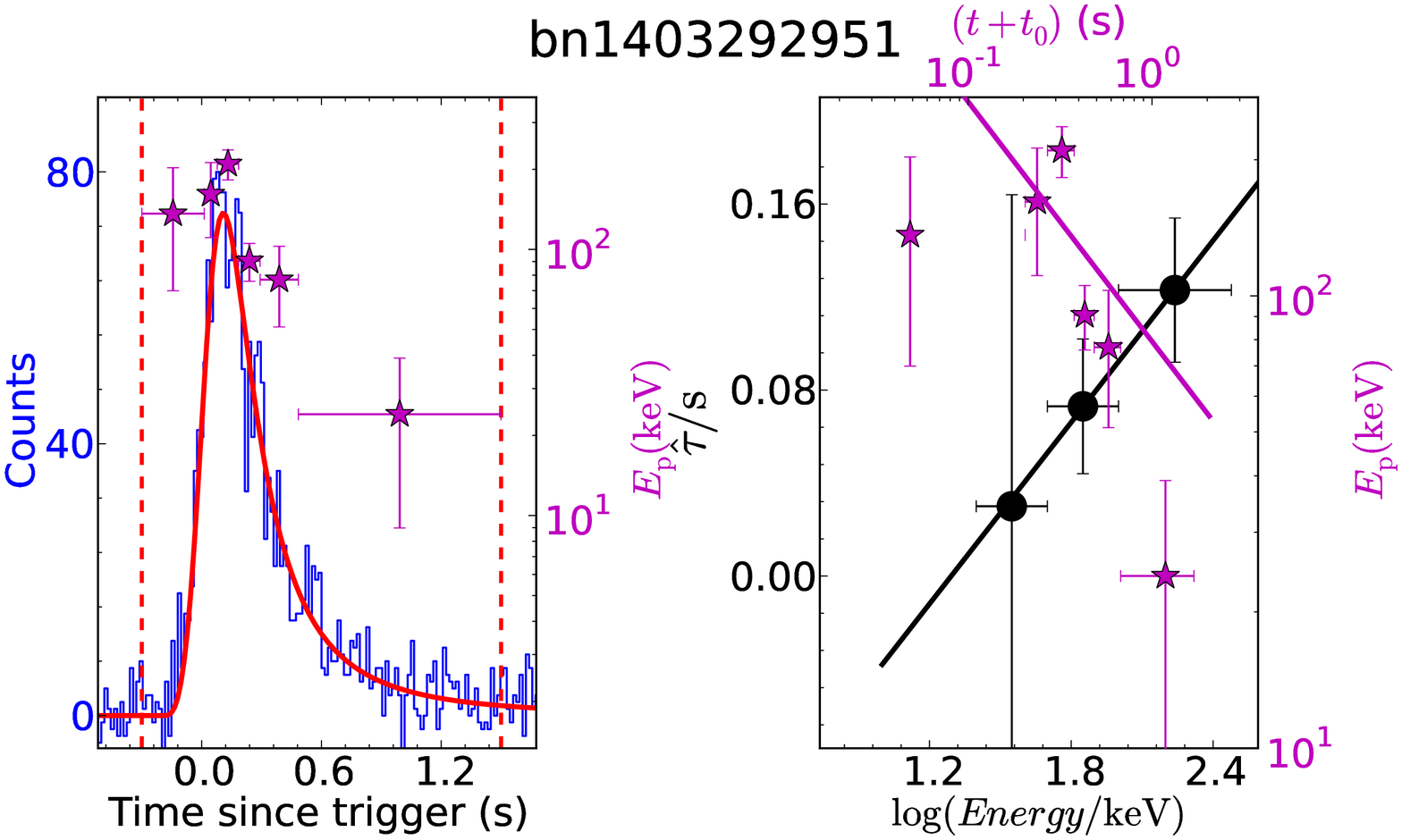} \\
\includegraphics[origin=c,angle=0,scale=1,width=0.500\textwidth,height=0.230\textheight]{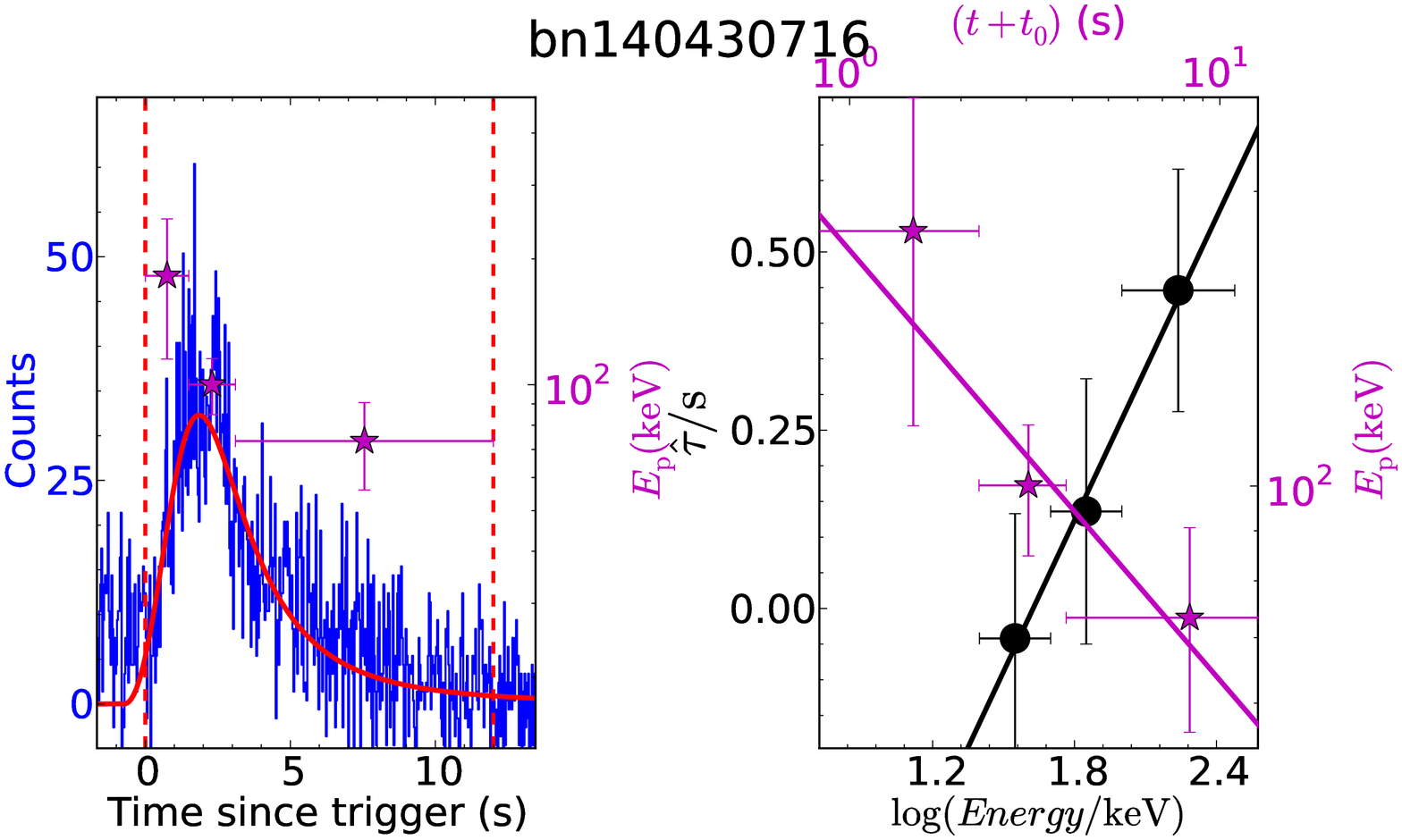} &
\includegraphics[origin=c,angle=0,scale=1,width=0.500\textwidth,height=0.23\textheight]{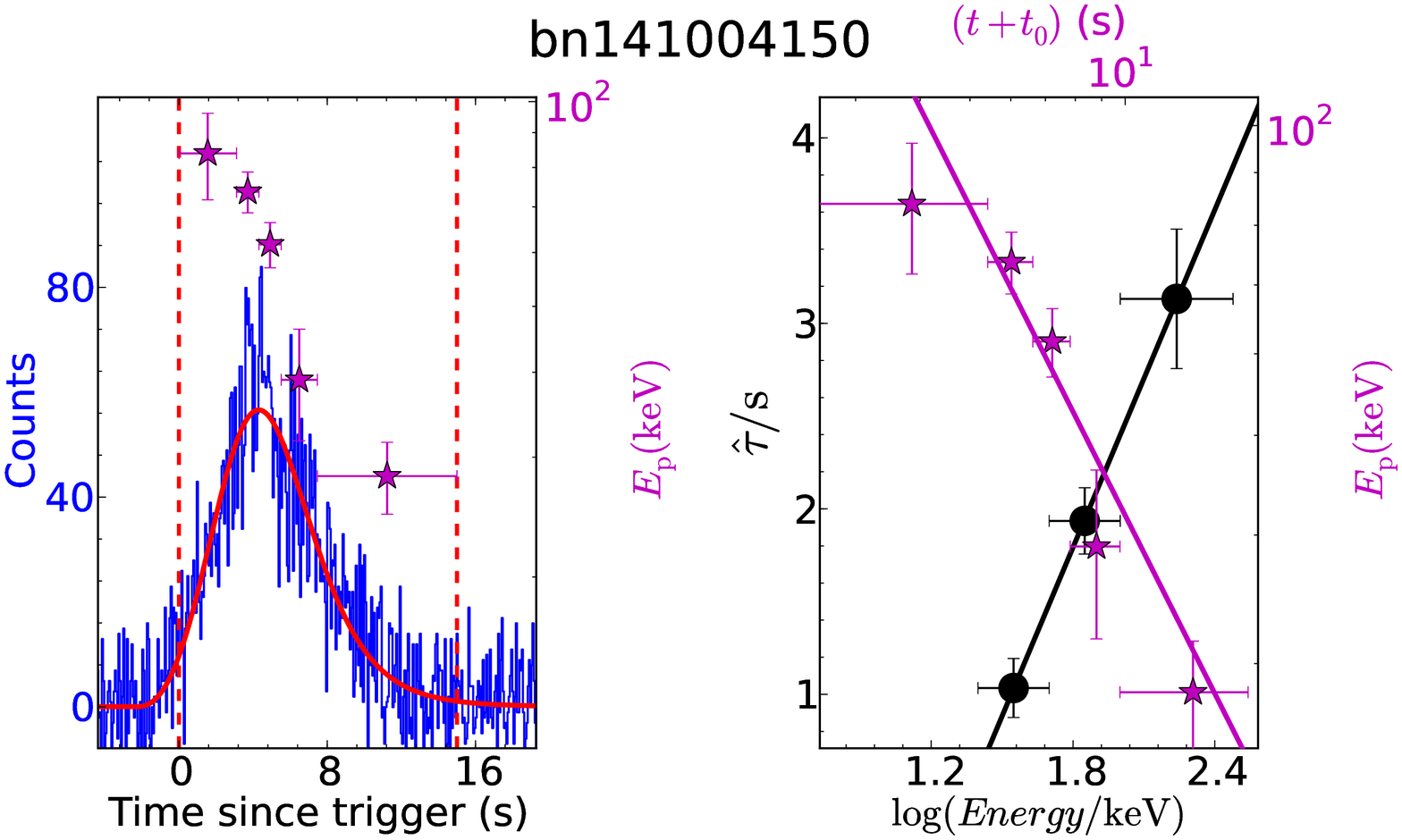} \\
\end{tabular}
\hfill
\vspace{4em}
\center{Fig. \ref{myfigA}---Continued.}
\end{figure*}

\clearpage %8

\begin{figure*}
\vspace{-4em}\hspace{-4em}
\centering
\begin{tabular}{cc}
\includegraphics[origin=c,angle=0,scale=1,width=0.500\textwidth,height=0.230\textheight]{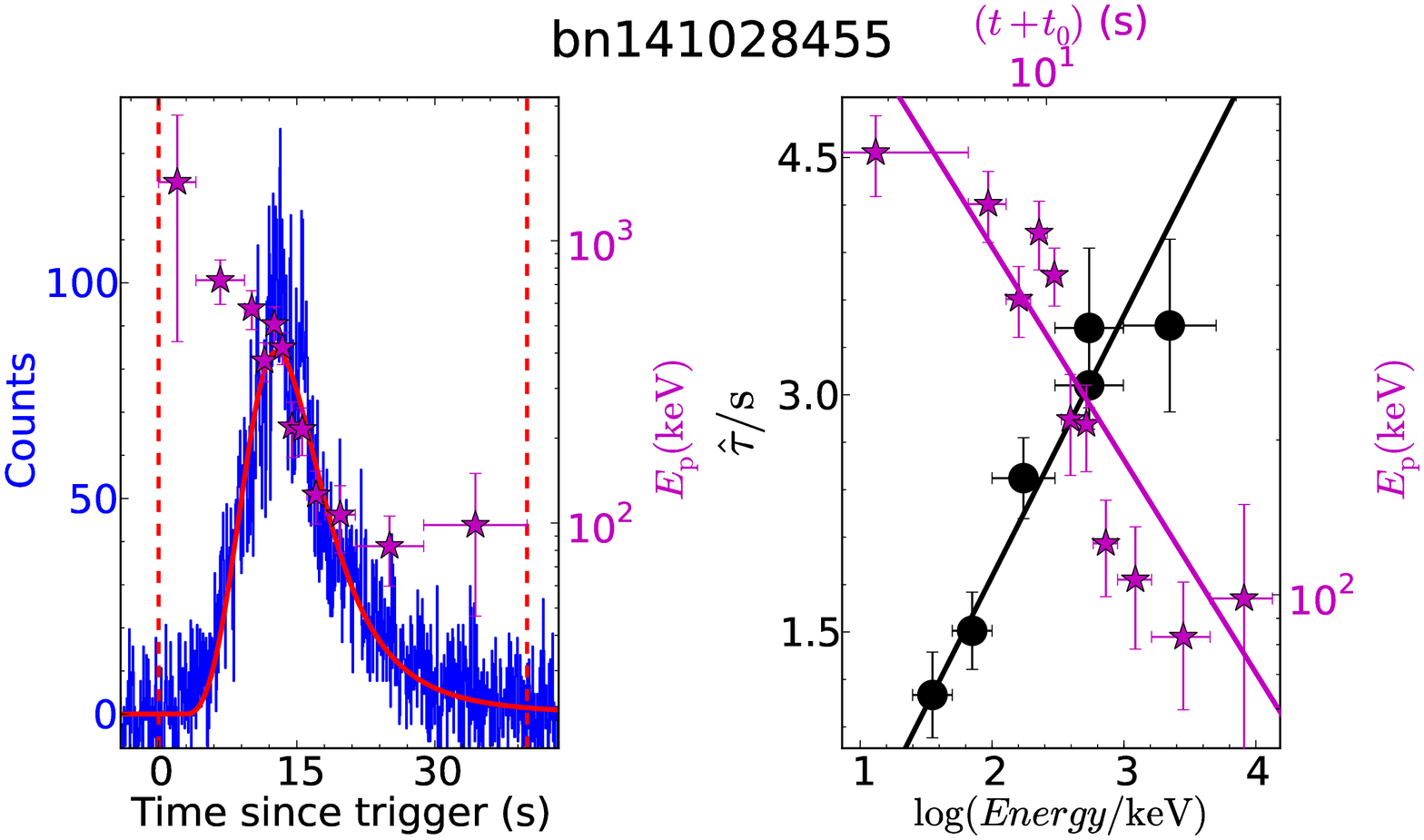} &
\includegraphics[origin=c,angle=0,scale=1,width=0.500\textwidth,height=0.23\textheight]{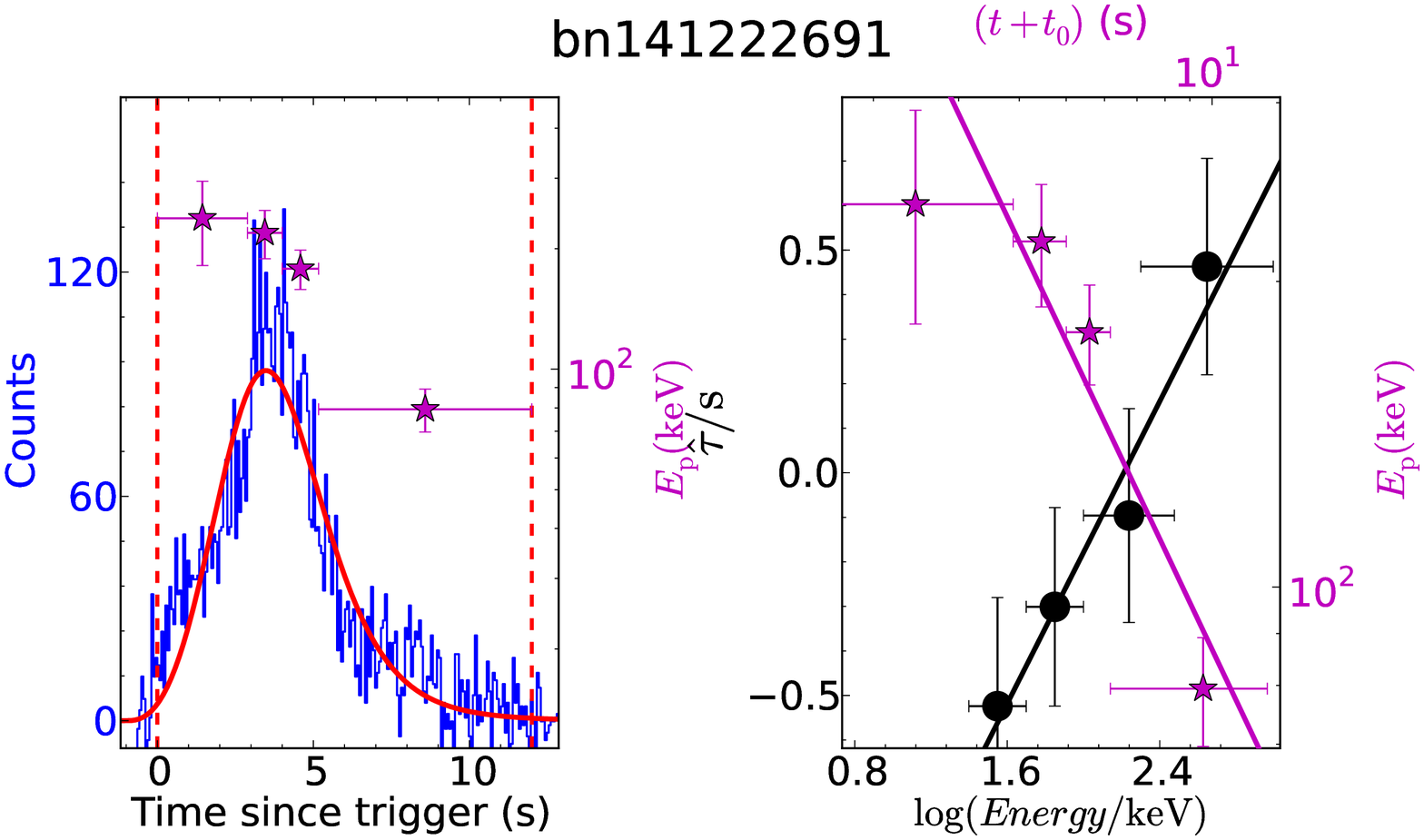} \\
\includegraphics[origin=c,angle=0,scale=1,width=0.500\textwidth,height=0.23\textheight]{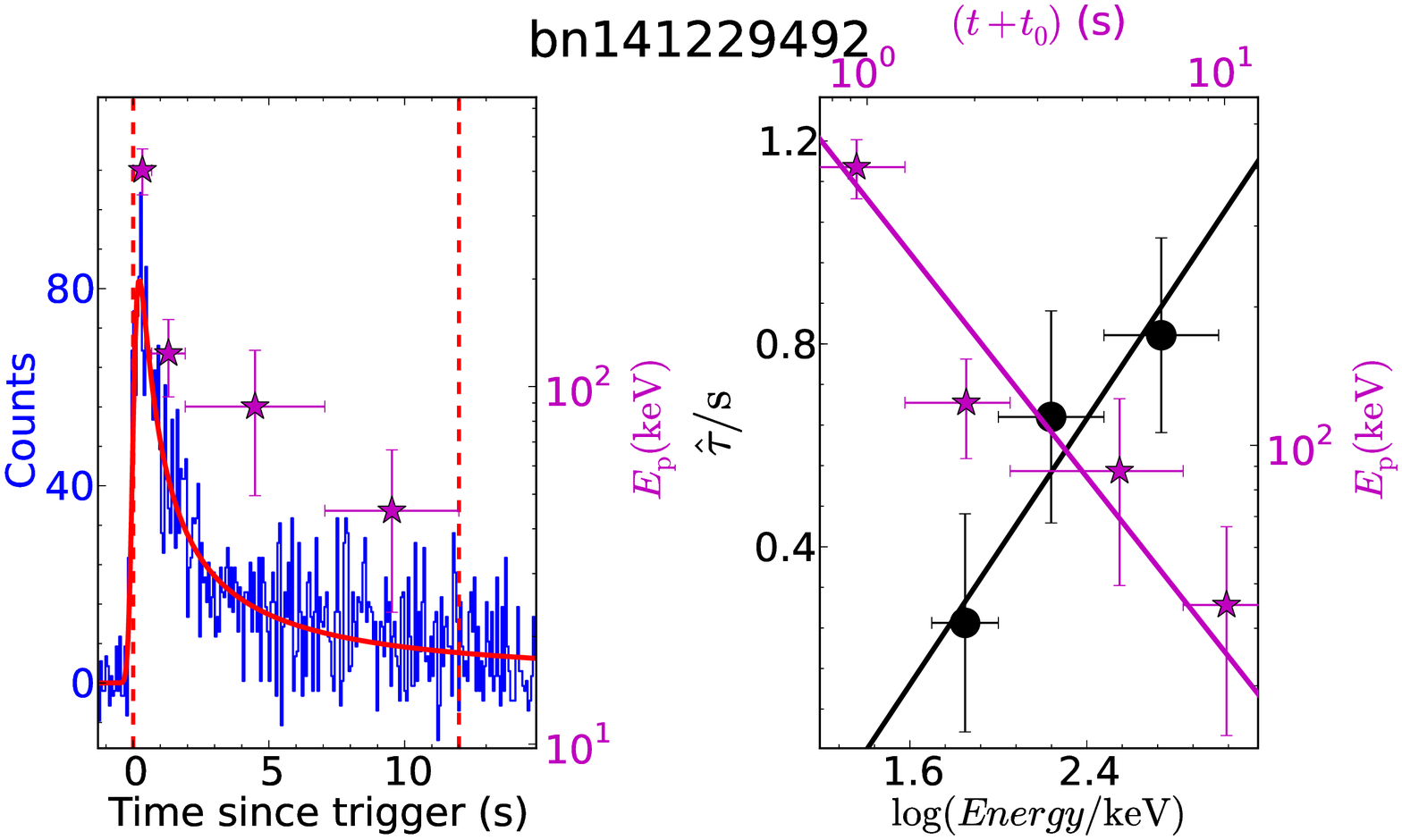} &
\includegraphics[origin=c,angle=0,scale=1,width=0.500\textwidth,height=0.23\textheight]{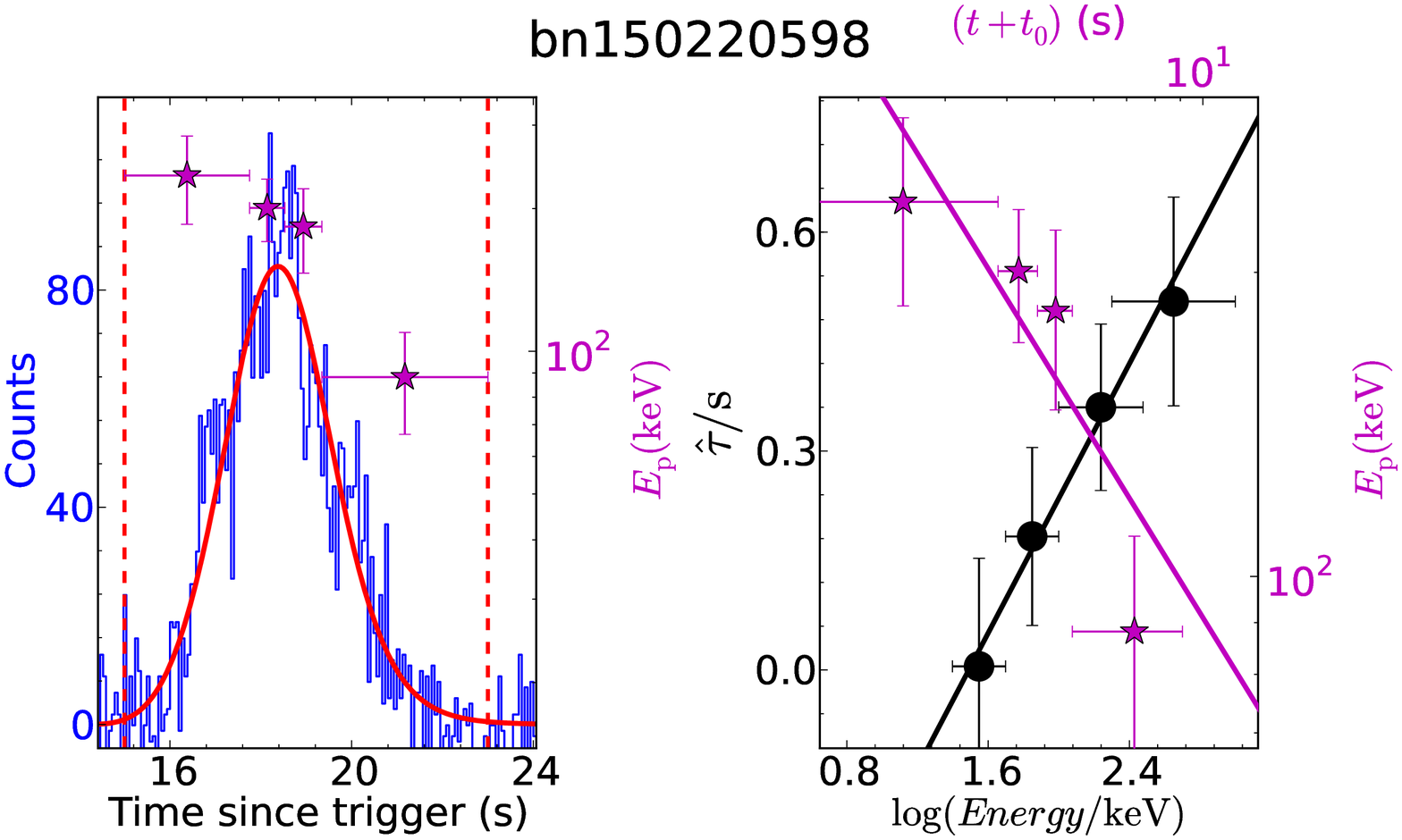} \\
\includegraphics[origin=c,angle=0,scale=1,width=0.500\textwidth,height=0.230\textheight]{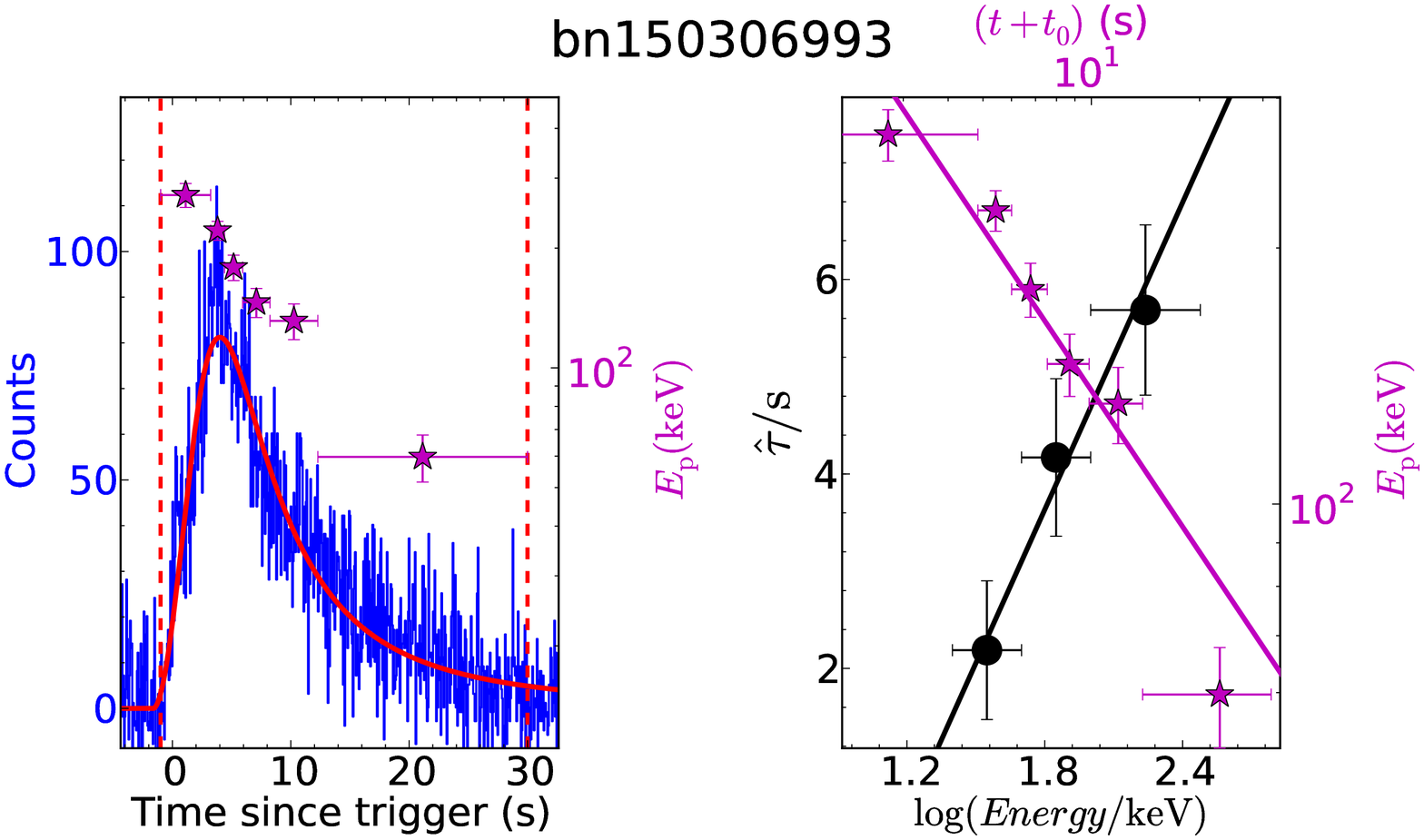} &
\includegraphics[origin=c,angle=0,scale=1,width=0.500\textwidth,height=0.22\textheight]{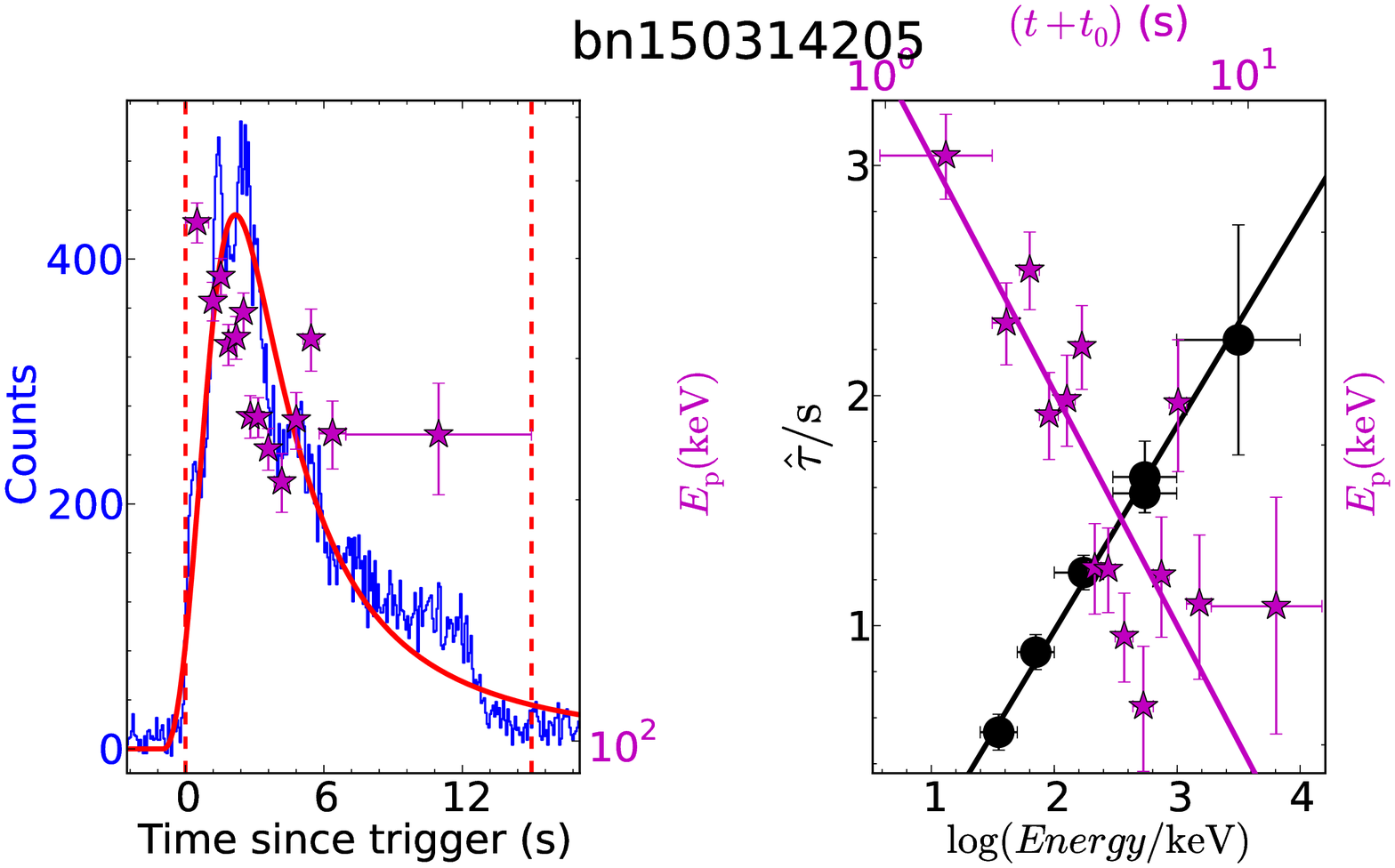} \\
\includegraphics[origin=c,angle=0,scale=1,width=0.500\textwidth,height=0.23\textheight]{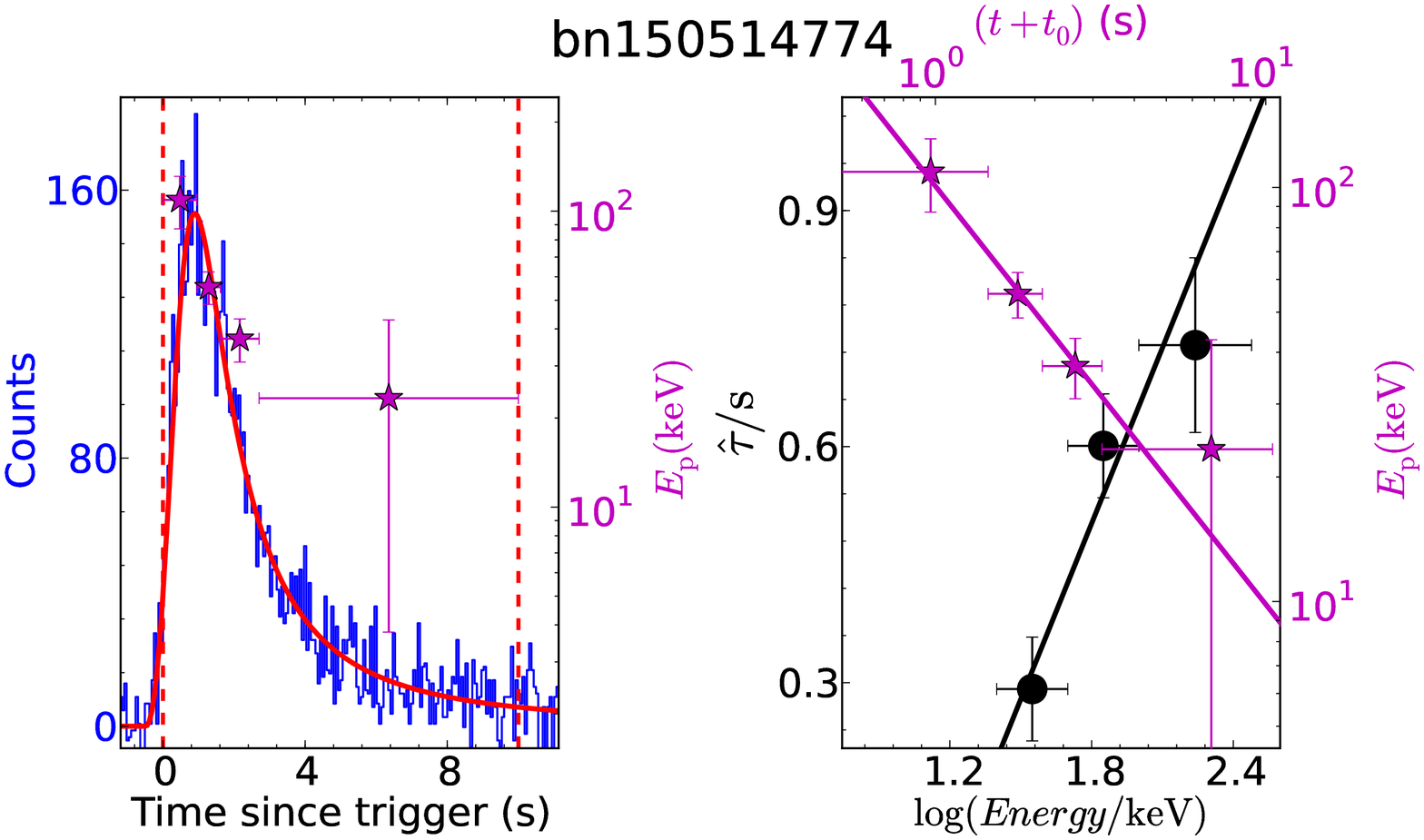} &
\includegraphics[origin=c,angle=0,scale=1,width=0.500\textwidth,height=0.23\textheight]{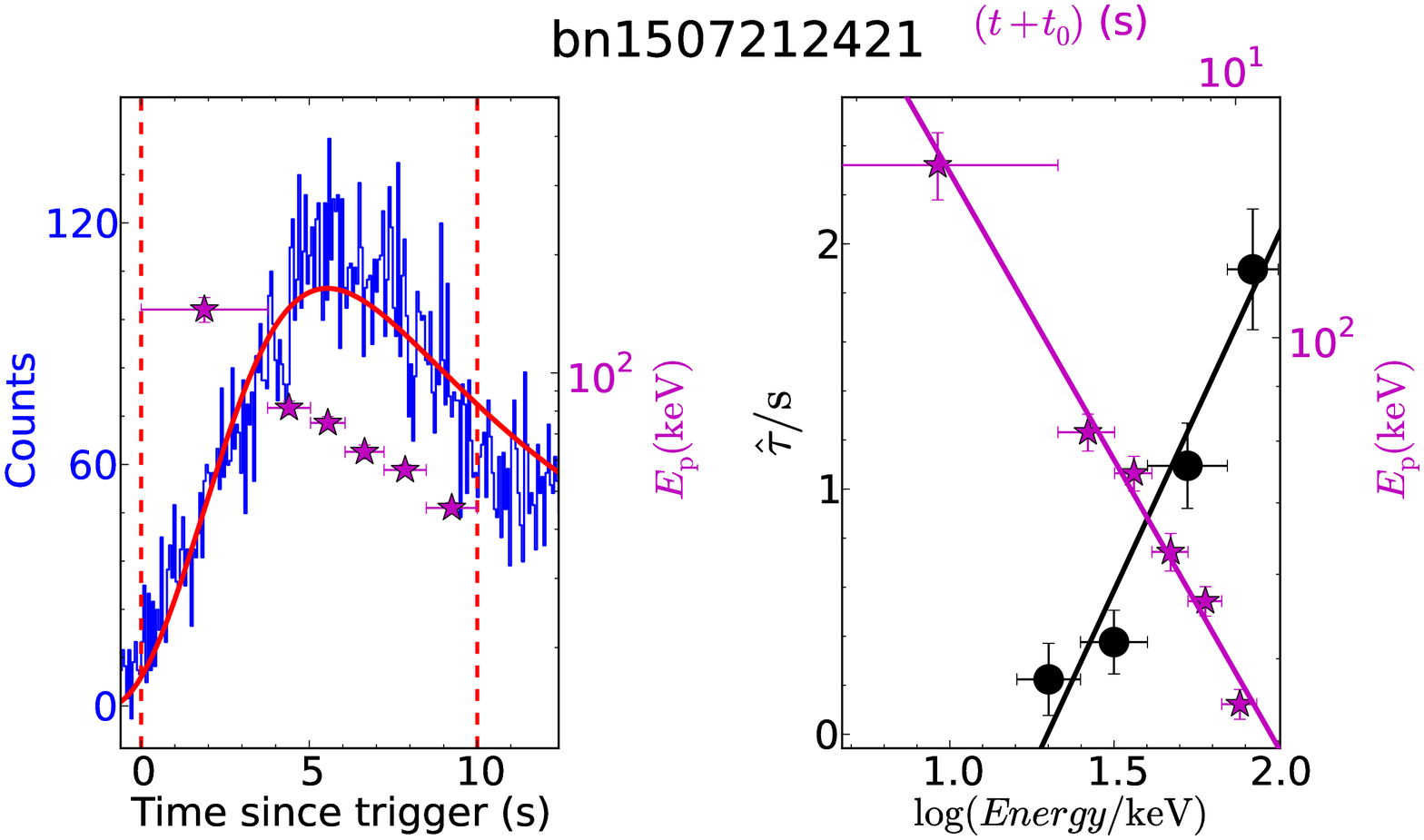} \\
\end{tabular}
\hfill
\vspace{4em}
\center{Fig.~\ref{myfigA}---Continued.}
\end{figure*}

%%%%%%%%%%%%%%%%%%%%%%%%%%%%%%%%%%%%%%%%%%%%%%%%%%%%%%%%%%%%%%%%%%%%%%%%%%%%%%%%%%%%%%%%%%%%%%%%%%%%%%%%%%%%%%%%%%%%
%%%%%%%%%%%%%%%%%%%%%%%0%%%%%%%%%%%%%%%%%%%%%%%%%%%%%%%%%%%%%%%%%%%%%%%%%%%%%%%%%%%%%%%%%%%%%%%%%%%%%%%%%%%%%%%%%%%%%
%%%%% Figure 2
\clearpage
\begin{figure*}
\vspace{-4em}\hspace{-4em}
\centering
\begin{tabular}{cc}
\includegraphics[origin=c,angle=0,scale=1,width=0.500\textwidth,height=0.22\textheight]{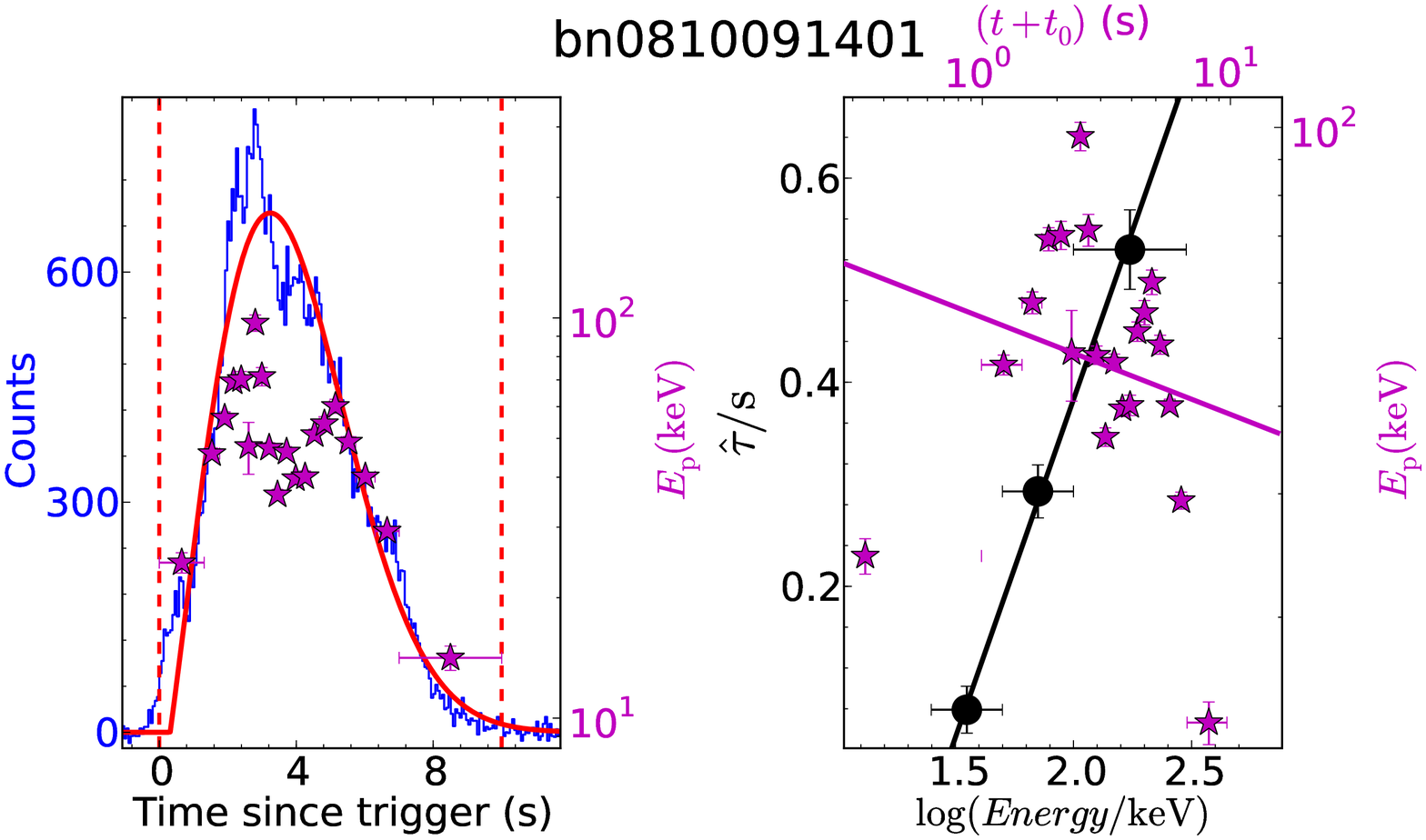} &
\includegraphics[origin=c,angle=0,scale=1,width=0.500\textwidth,height=0.220\textheight]{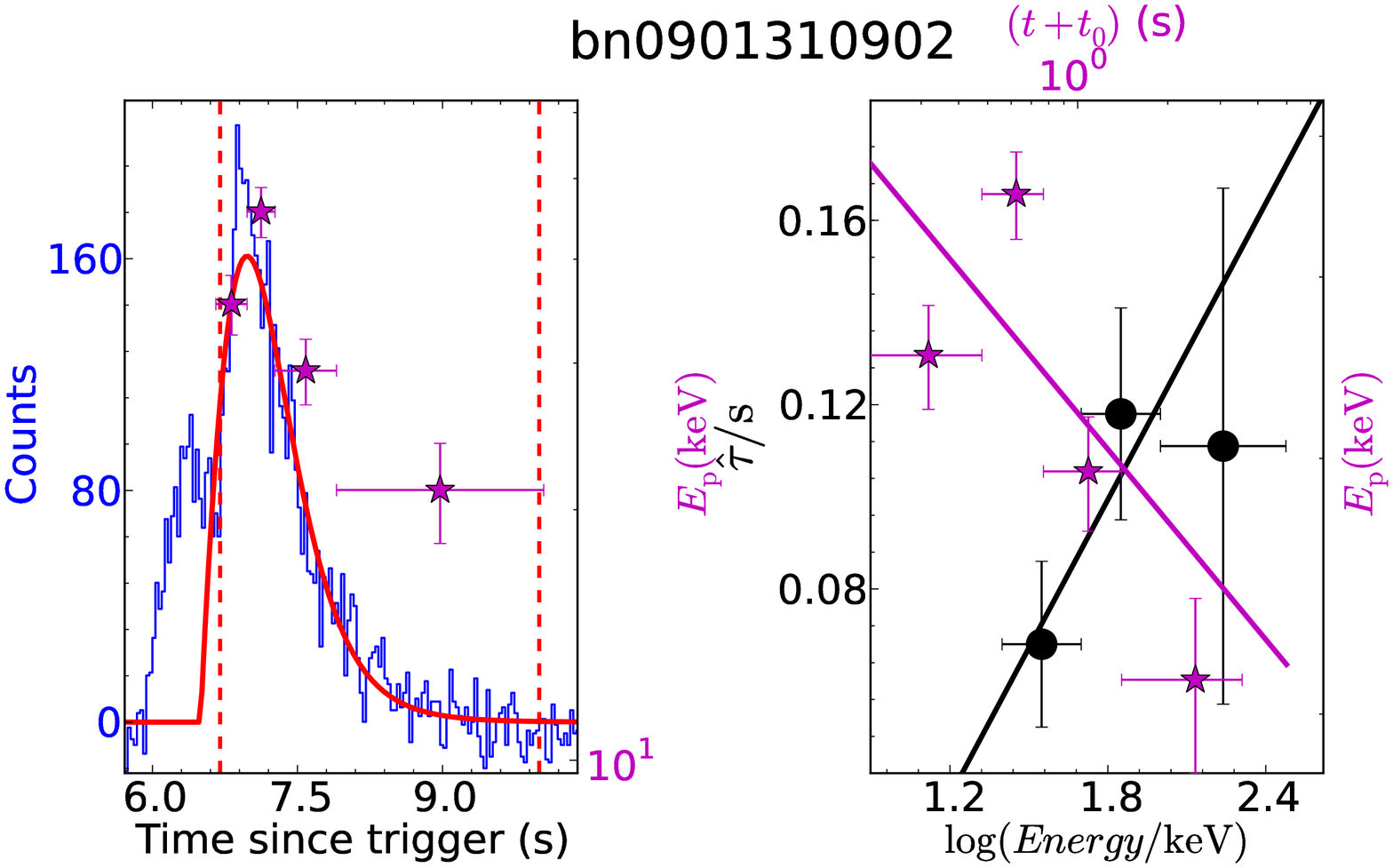} \\
\includegraphics[origin=c,angle=0,scale=1,width=0.500\textwidth,height=0.22\textheight]{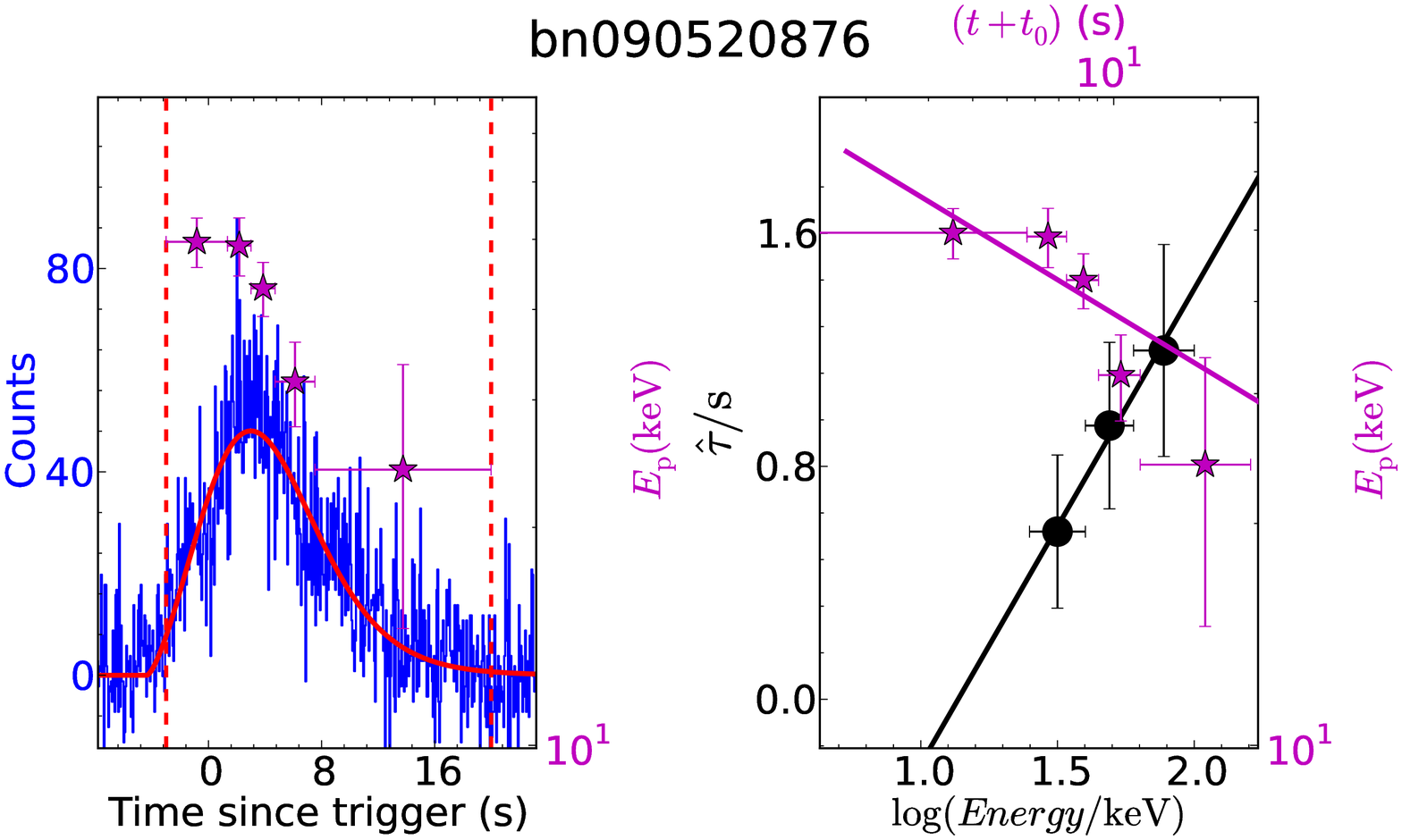} &
\includegraphics[origin=c,angle=0,scale=1,width=0.500\textwidth,height=0.22\textheight]{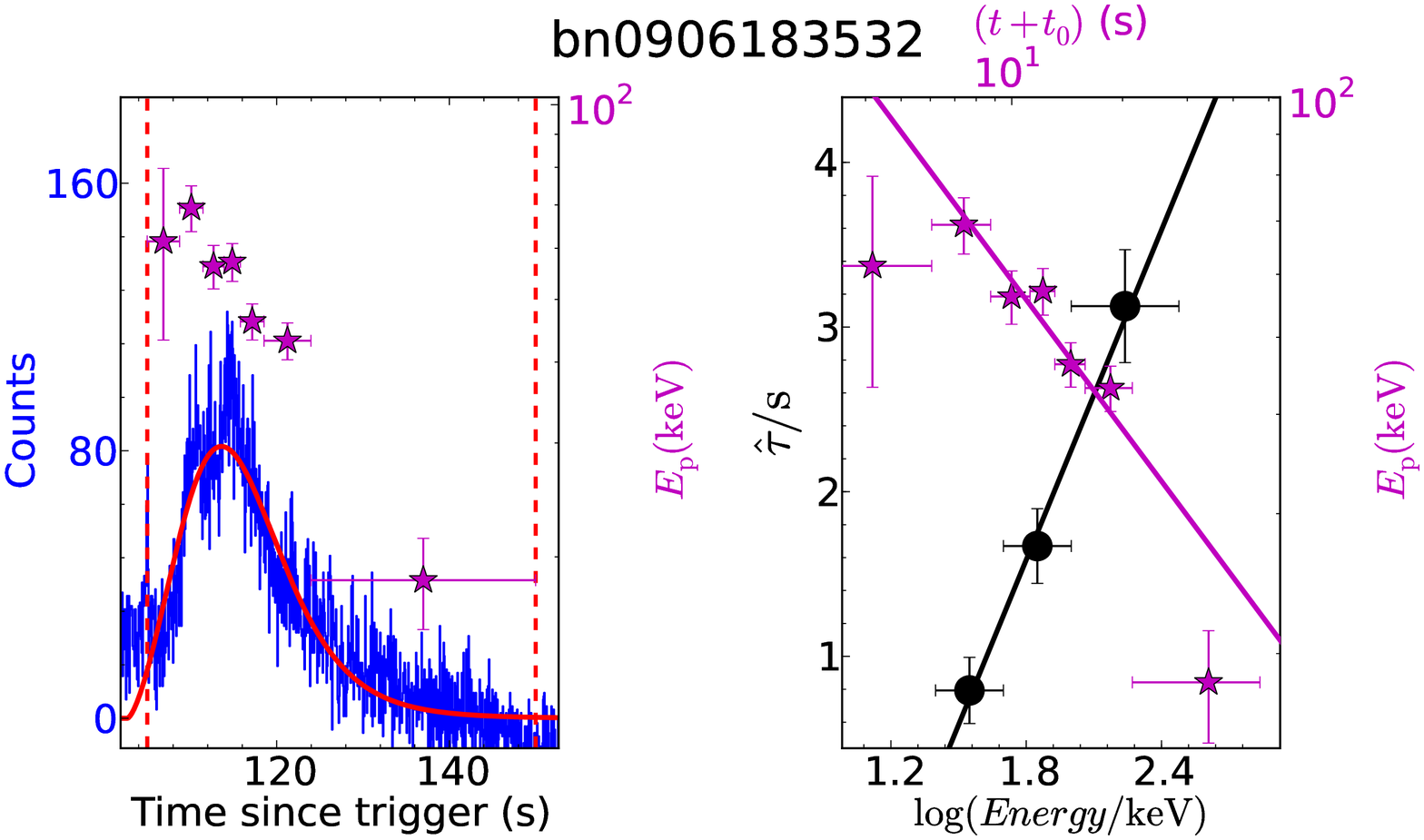} \\
\includegraphics[origin=c,angle=0,scale=1,width=0.500\textwidth,height=0.220\textheight]{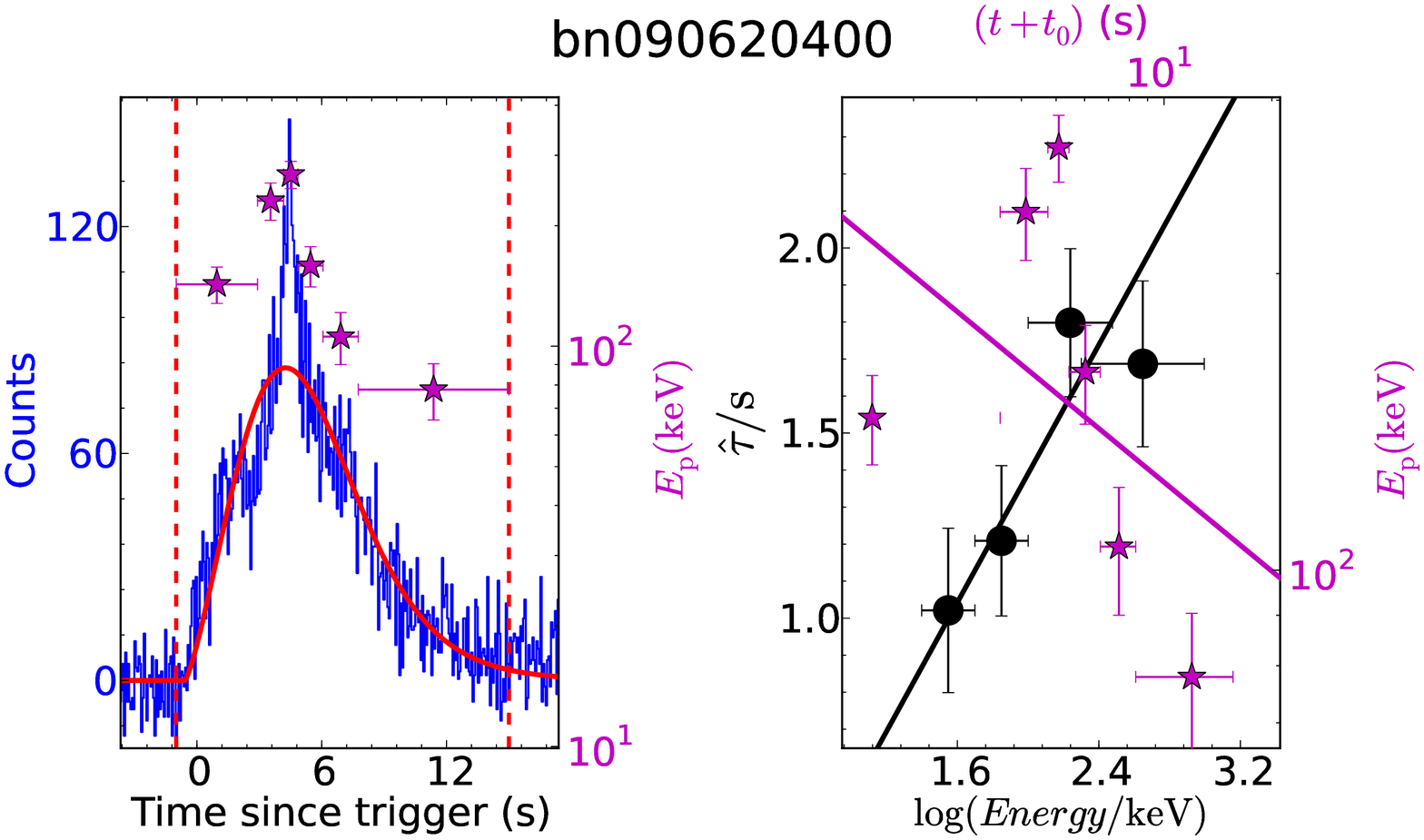} &
\includegraphics[origin=c,angle=0,scale=1,width=0.500\textwidth,height=0.22\textheight]{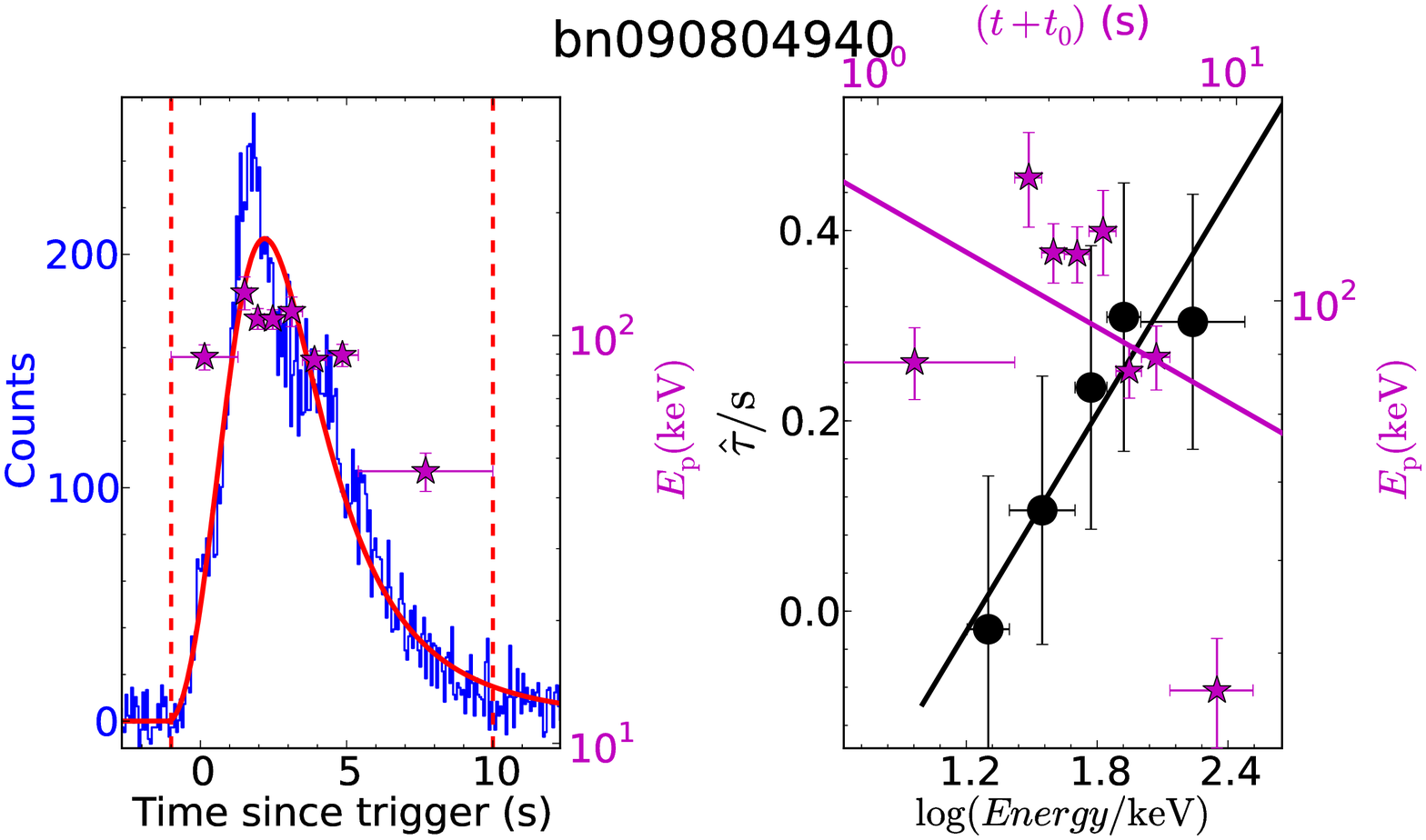} \\
\includegraphics[origin=c,angle=0,scale=1,width=0.500\textwidth,height=0.22\textheight]{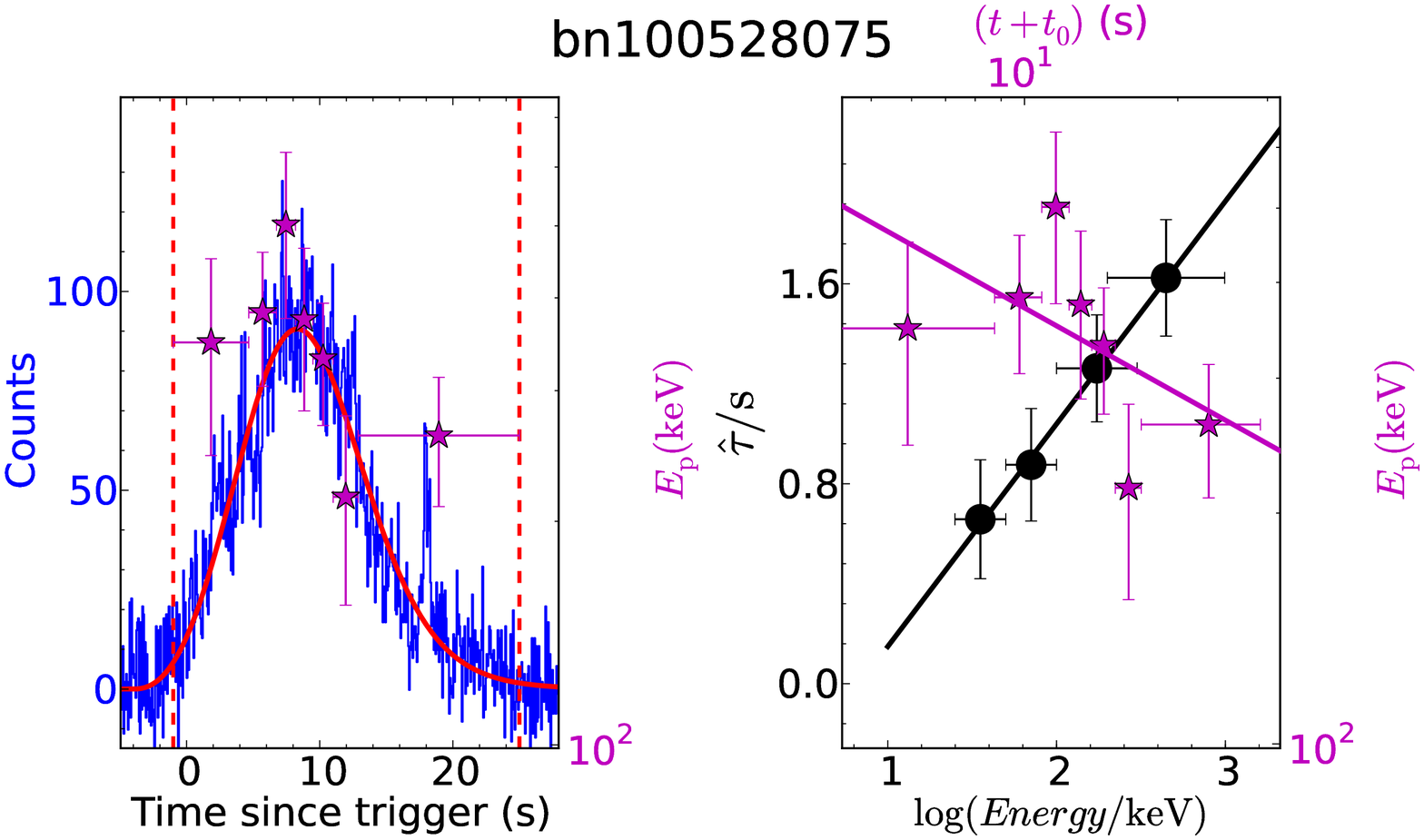} &
\includegraphics[origin=c,angle=0,scale=1,width=0.500\textwidth,height=0.22\textheight]{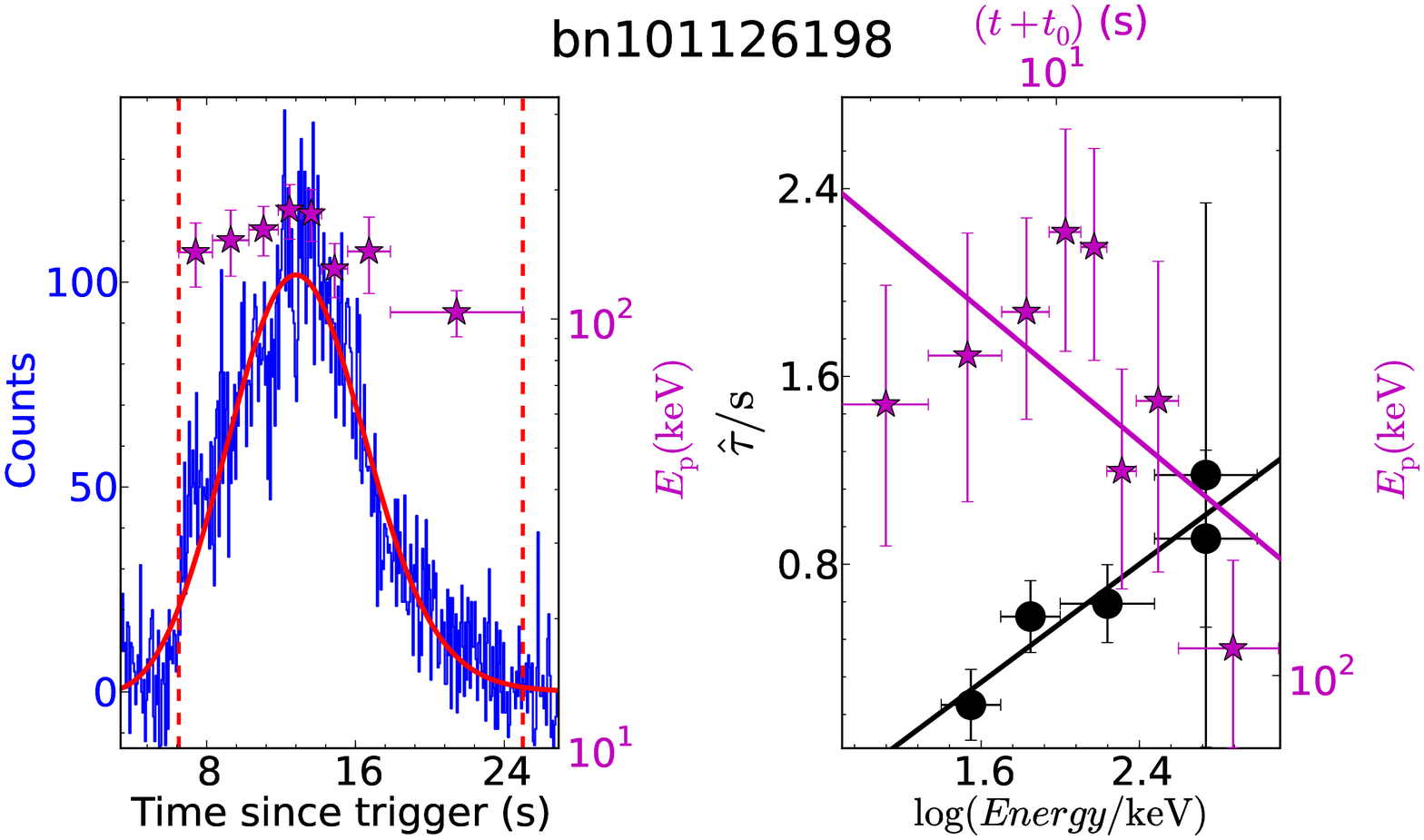} \\
\end{tabular}
\hfill
\vspace{4em}
\caption{Our selected pulses with tracking spectral evolution.
The meanings of the symbols and lines are the same as those in Figure~\ref{myfigA}.}\label{myfigB}
\end{figure*}

\clearpage %p2

\begin{figure*}
\vspace{-4em}\hspace{-4em}
\centering
\begin{tabular}{cc}
\includegraphics[origin=c,angle=0,scale=1,width=0.500\textwidth,height=0.22\textheight]{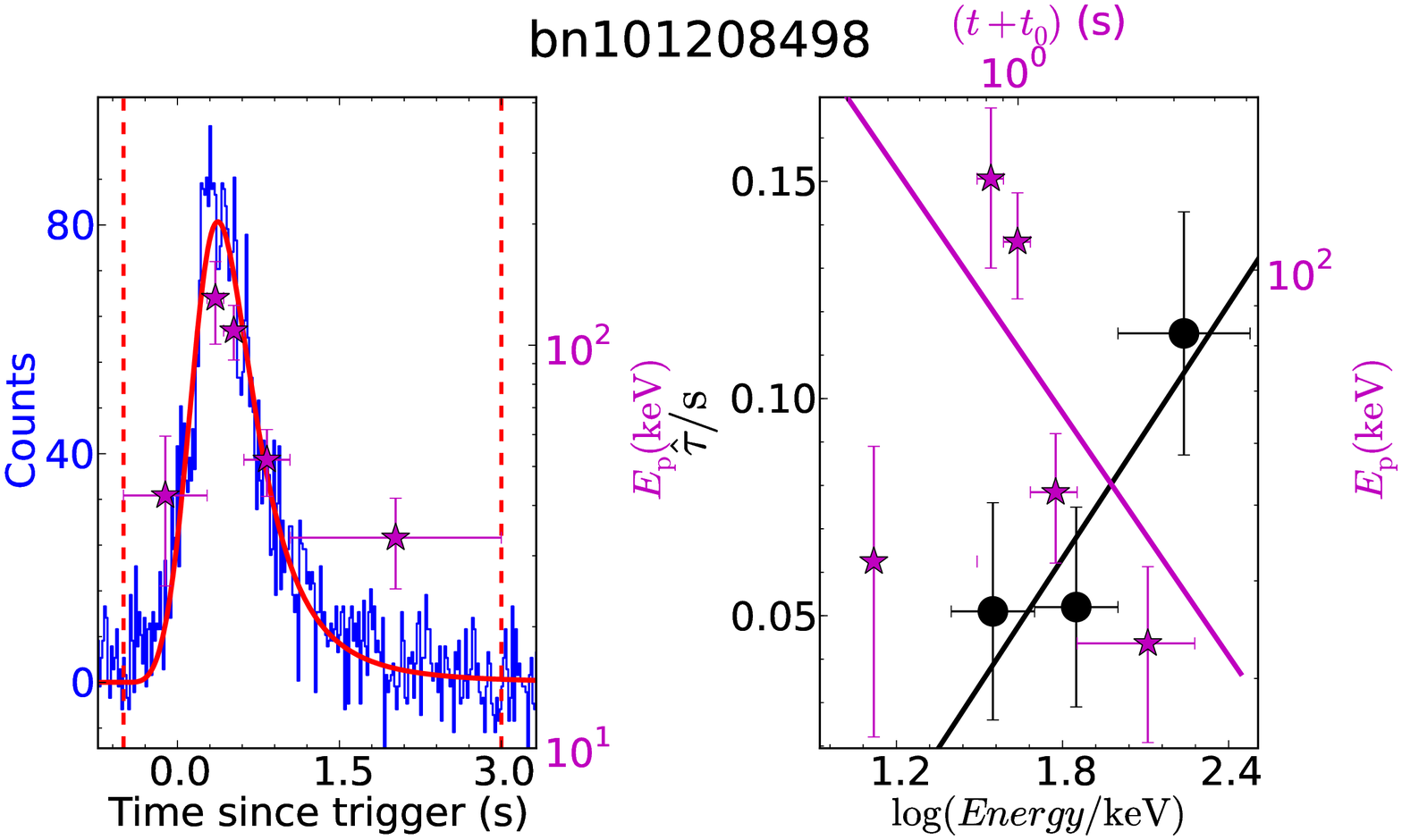} &
\includegraphics[origin=c,angle=0,scale=1,width=0.500\textwidth,height=0.22\textheight]{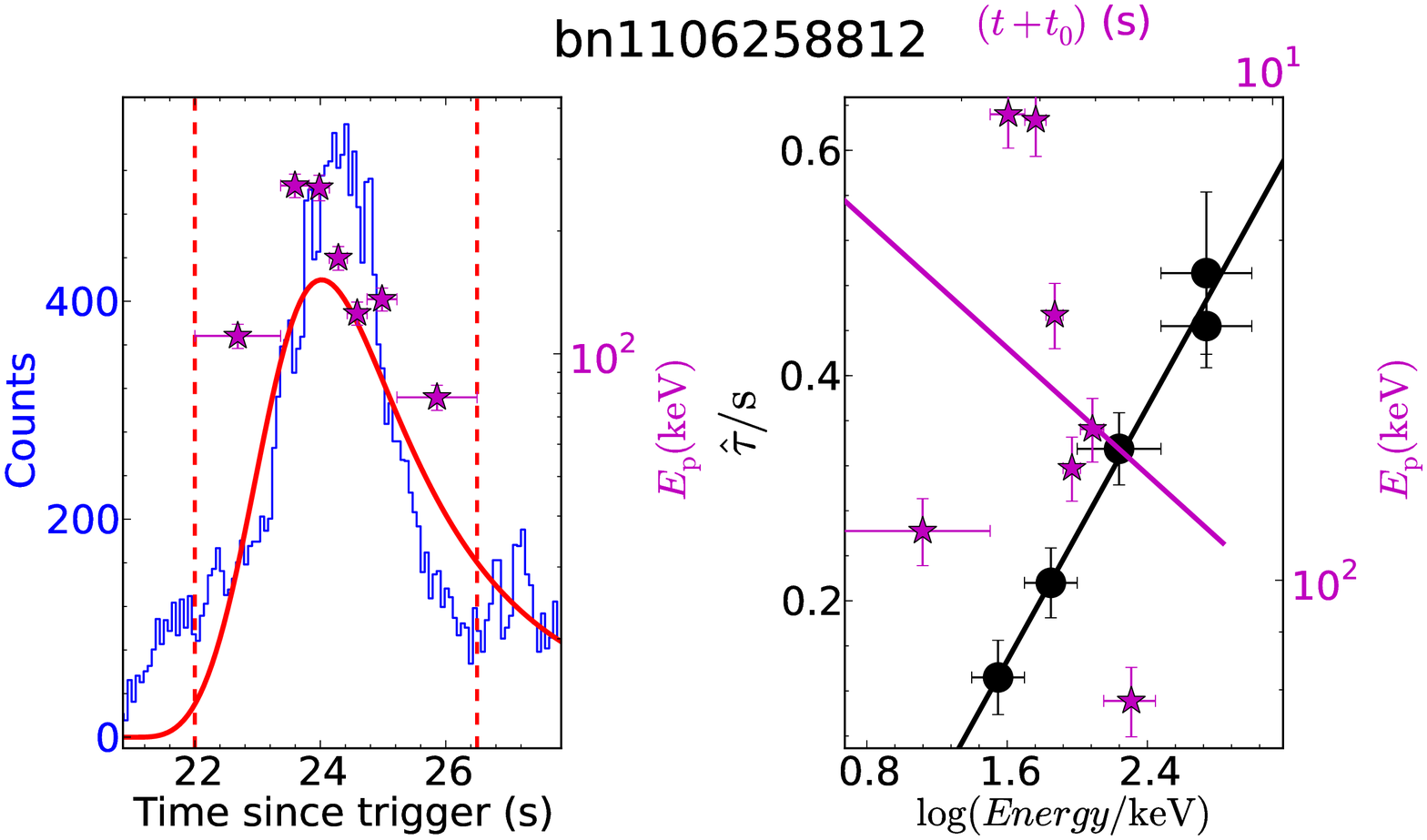} \\
\includegraphics[origin=c,angle=0,scale=1,width=0.500\textwidth,height=0.22\textheight]{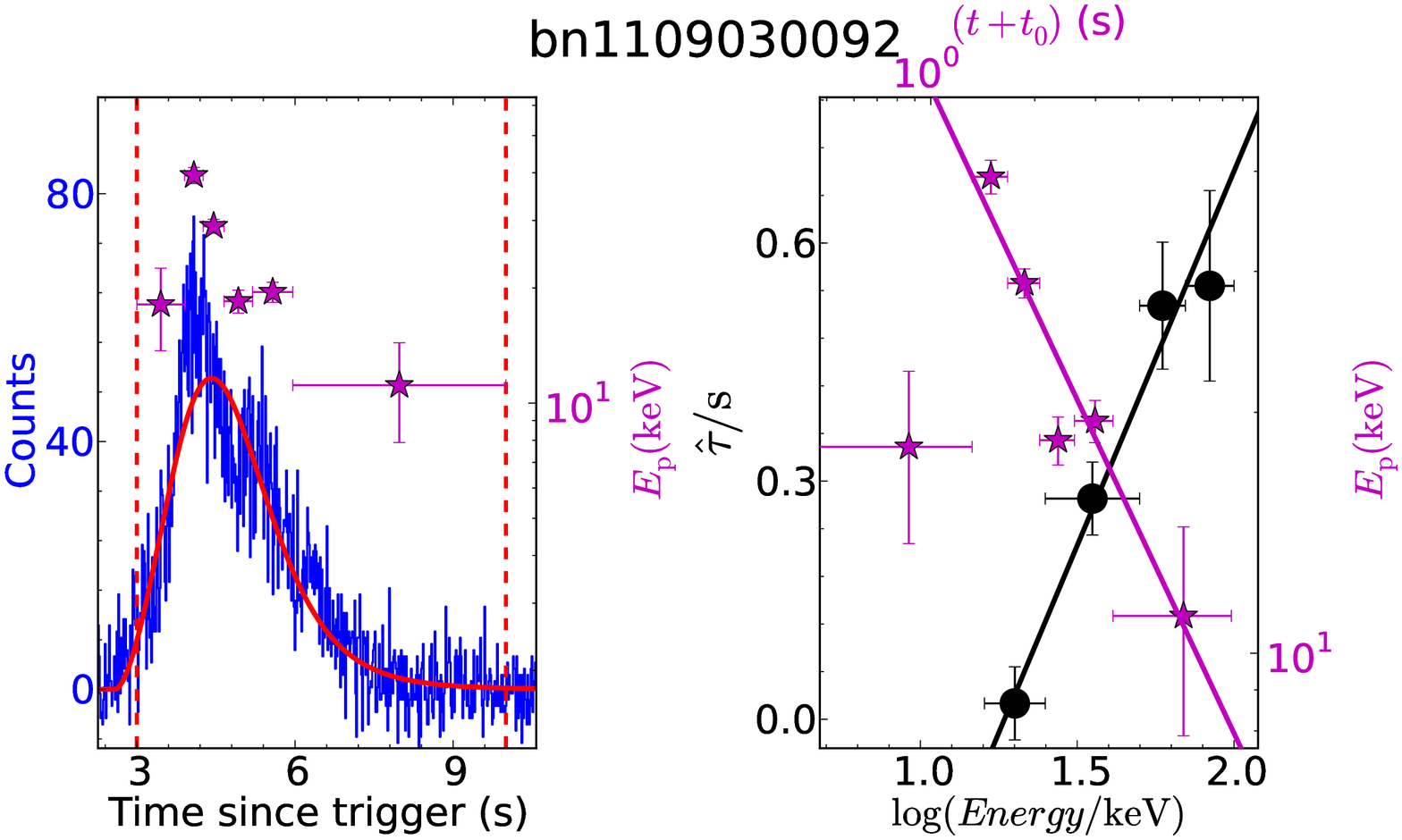} &
\includegraphics[origin=c,angle=0,scale=1,width=0.500\textwidth,height=0.22\textheight]{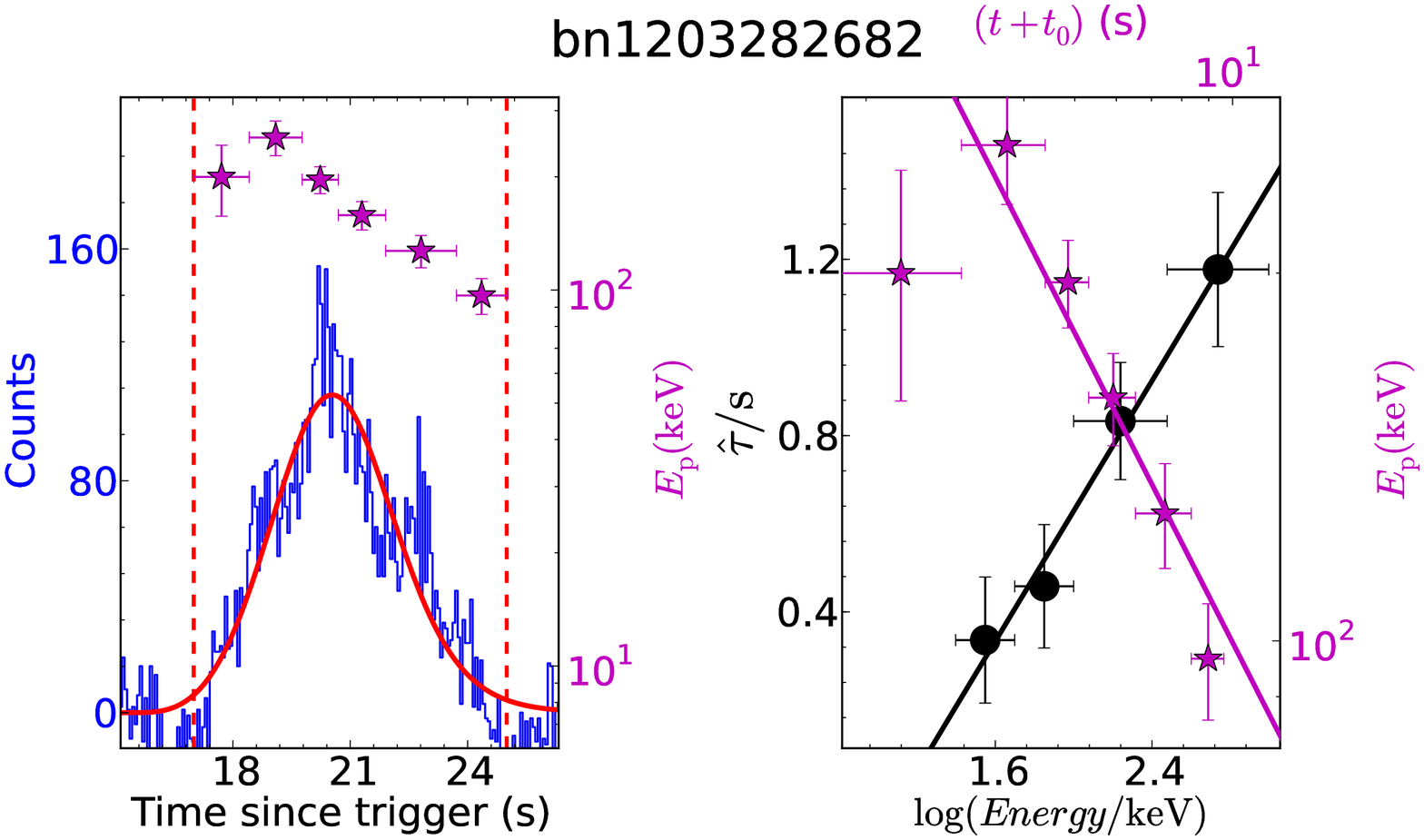} \\
\includegraphics[origin=c,angle=0,scale=1,width=0.500\textwidth,height=0.22\textheight]{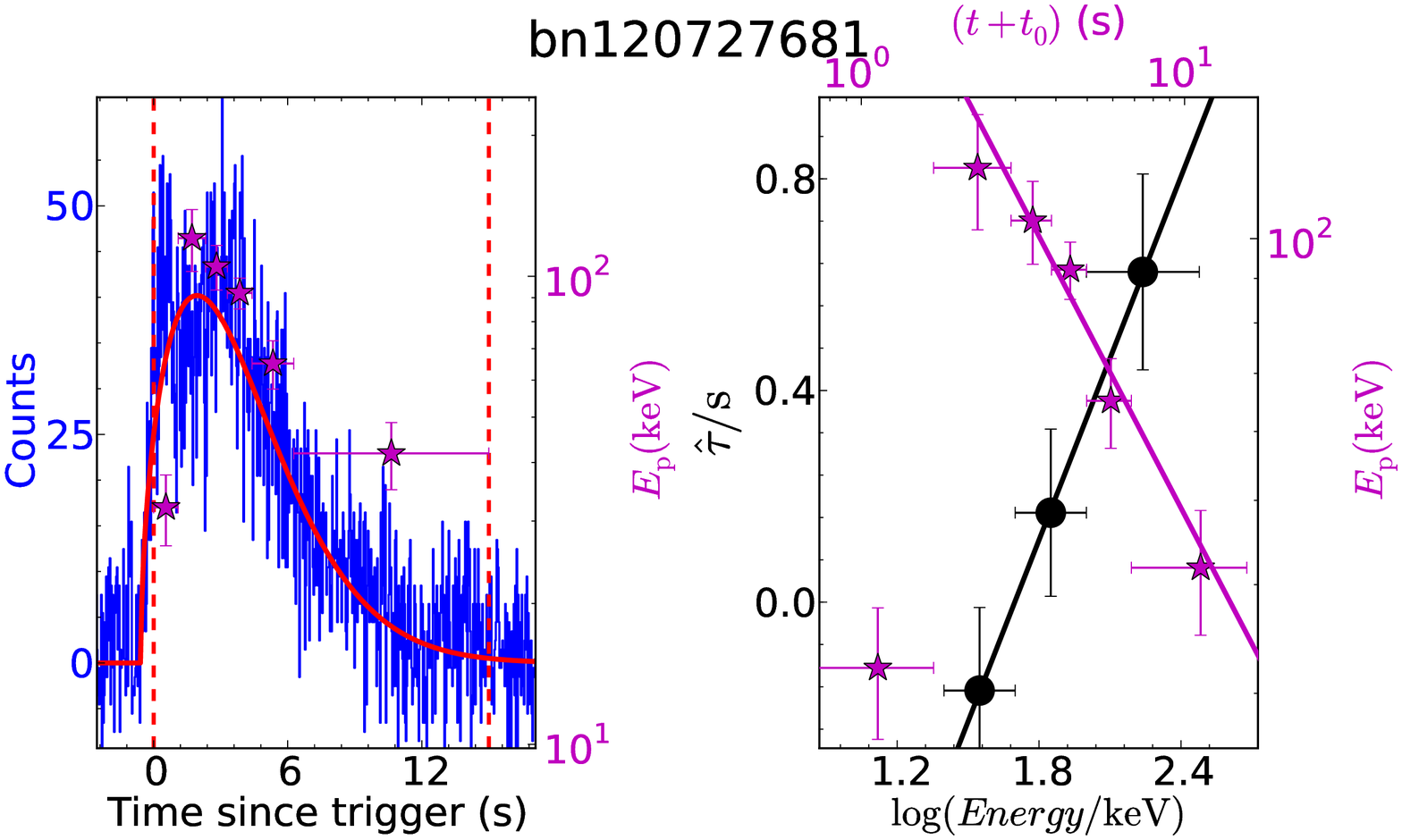} &
\includegraphics[origin=c,angle=0,scale=1,width=0.500\textwidth,height=0.22\textheight]{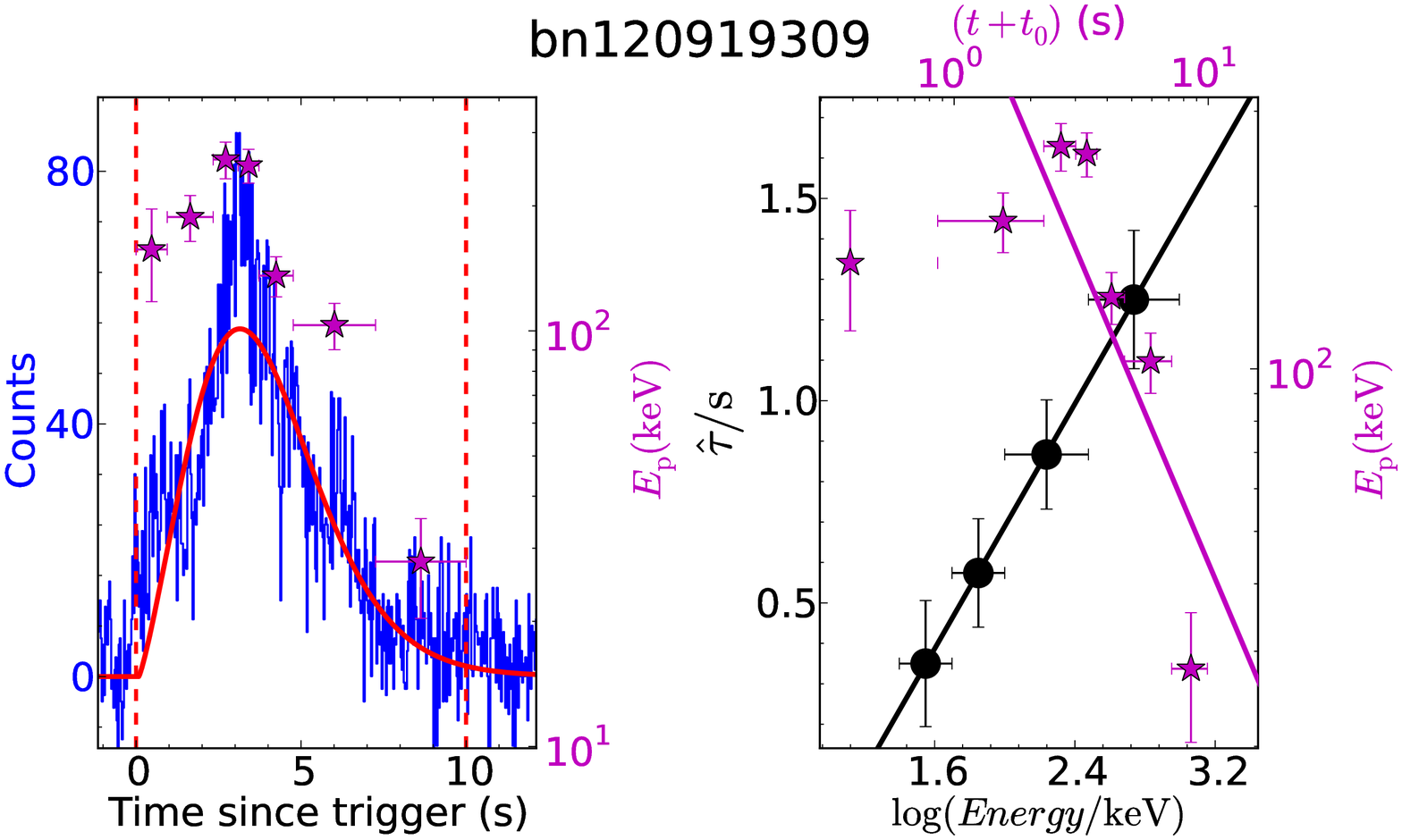} \\
\includegraphics[origin=c,angle=0,scale=1,width=0.500\textwidth,height=0.23\textheight]{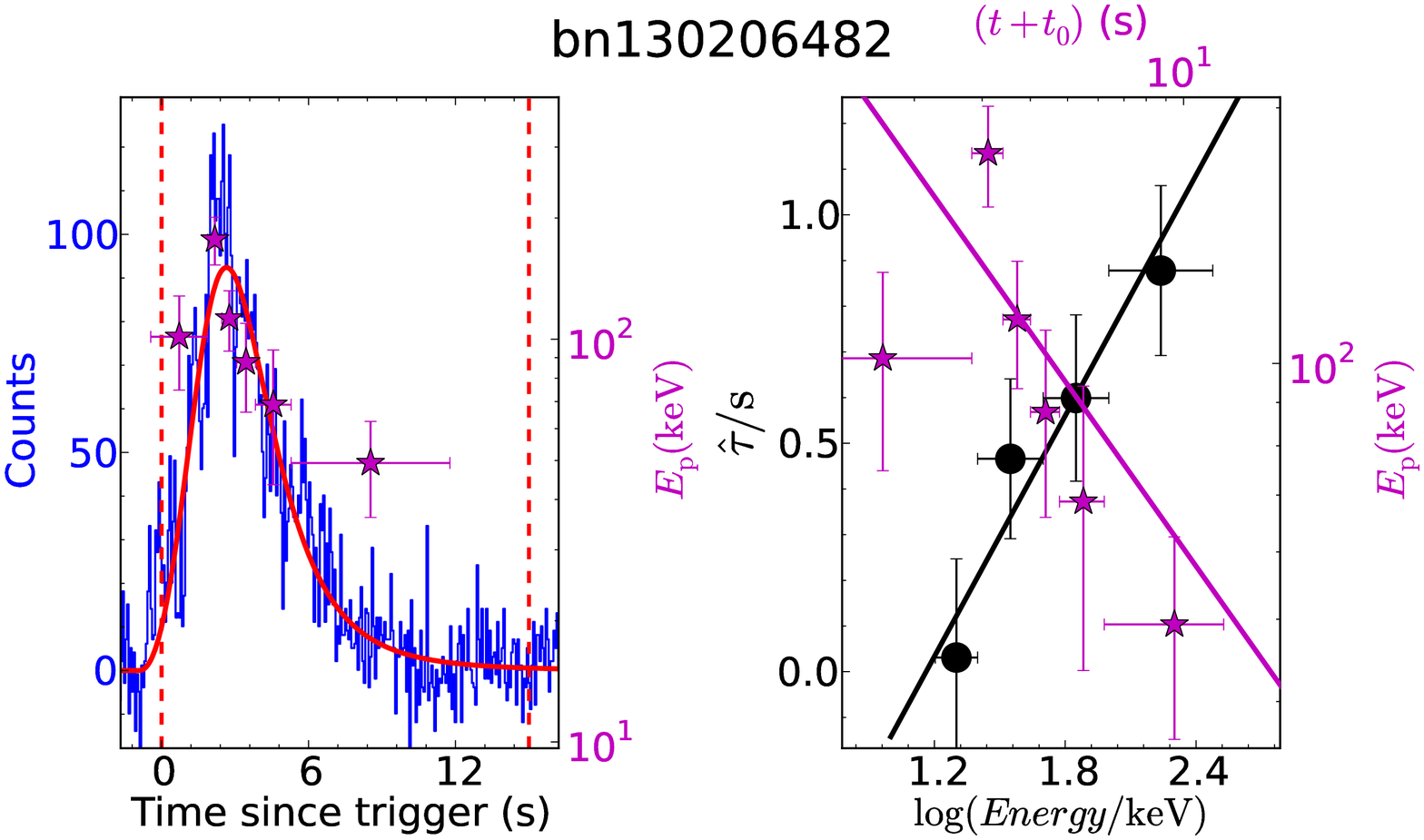} &
\includegraphics[origin=c,angle=0,scale=1,width=0.500\textwidth,height=0.22\textheight]{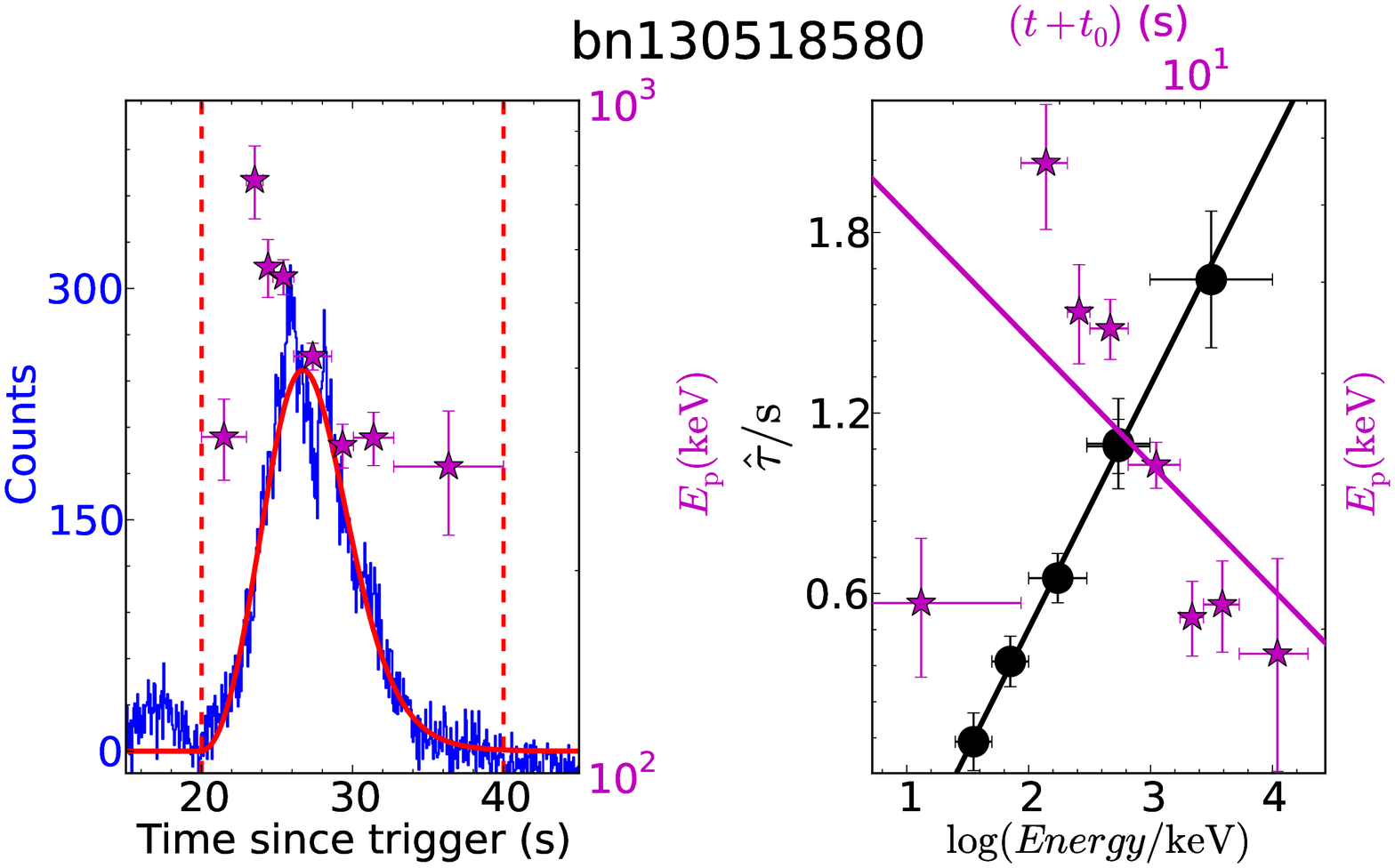} \\
\end{tabular}
\hfill
\vspace{4em}
\center{Fig.~\ref{myfigB}---Continued.}
\end{figure*}

\clearpage %p3

\begin{figure*}
\vspace{-4em}\hspace{-4em}
\centering
\begin{tabular}{cc}
\includegraphics[origin=c,angle=0,scale=1,width=0.500\textwidth,height=0.22\textheight]{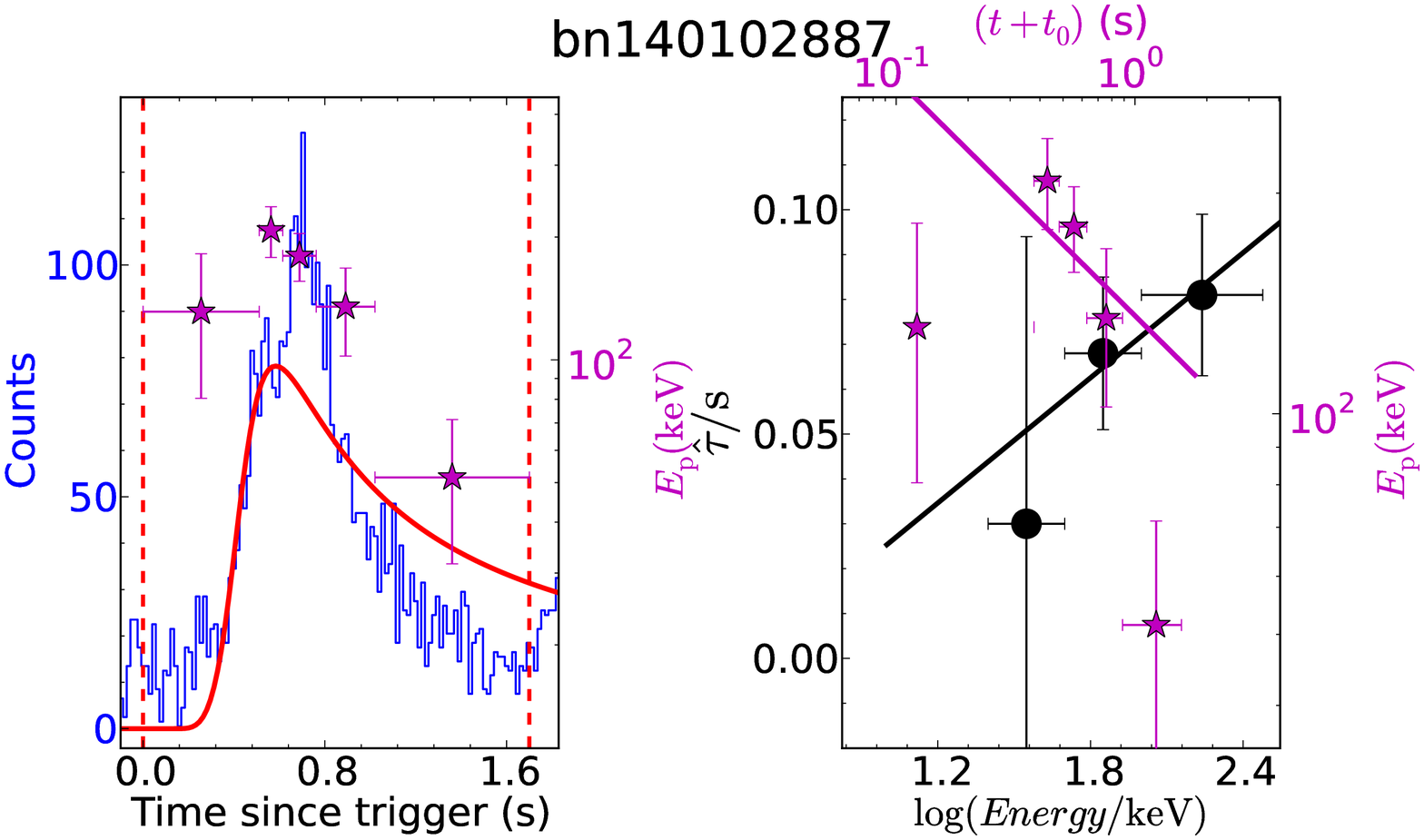} &
\includegraphics[origin=c,angle=0,scale=1,width=0.500\textwidth,height=0.19\textheight]{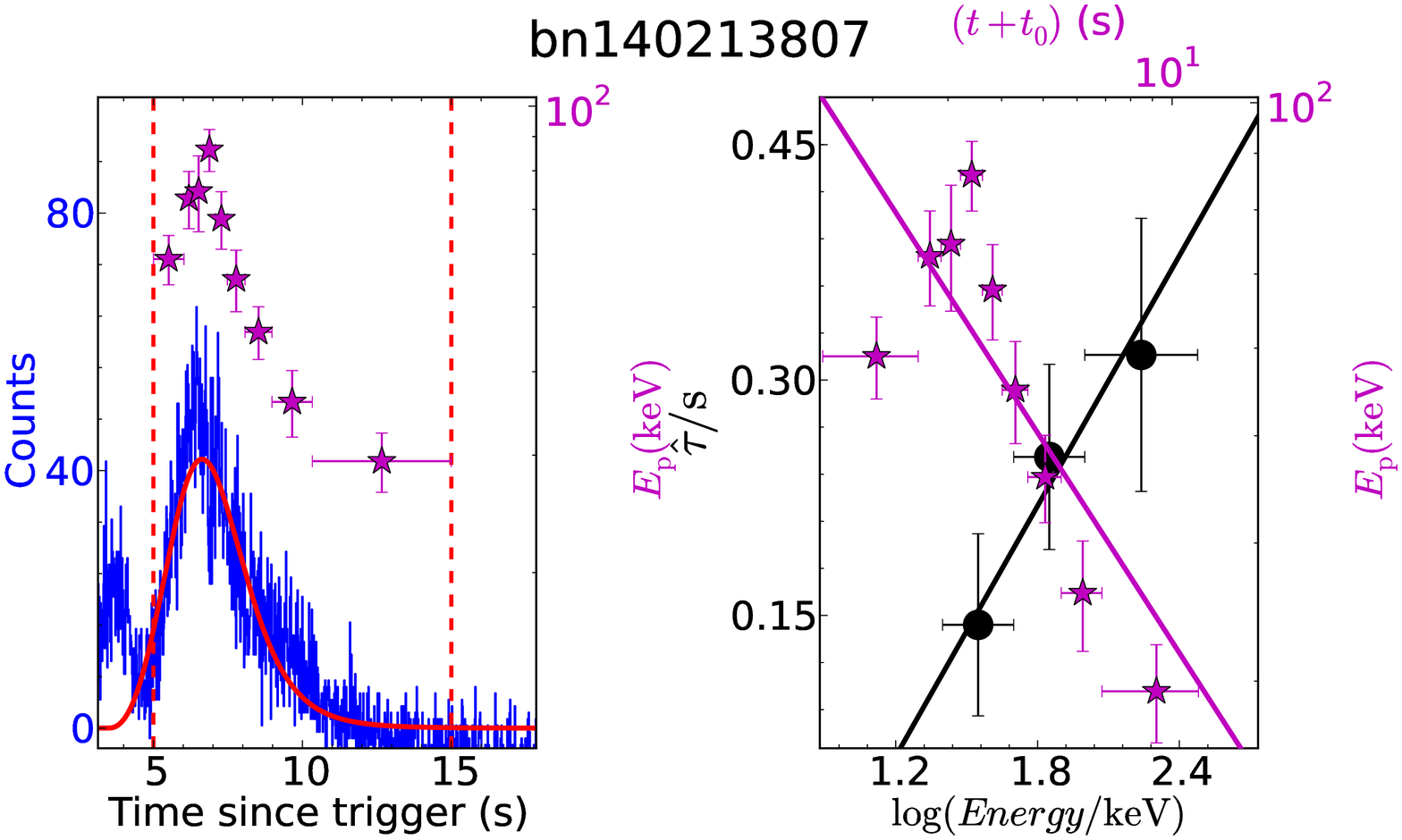} \\
\includegraphics[origin=c,angle=0,scale=1,width=0.500\textwidth,height=0.19\textheight]{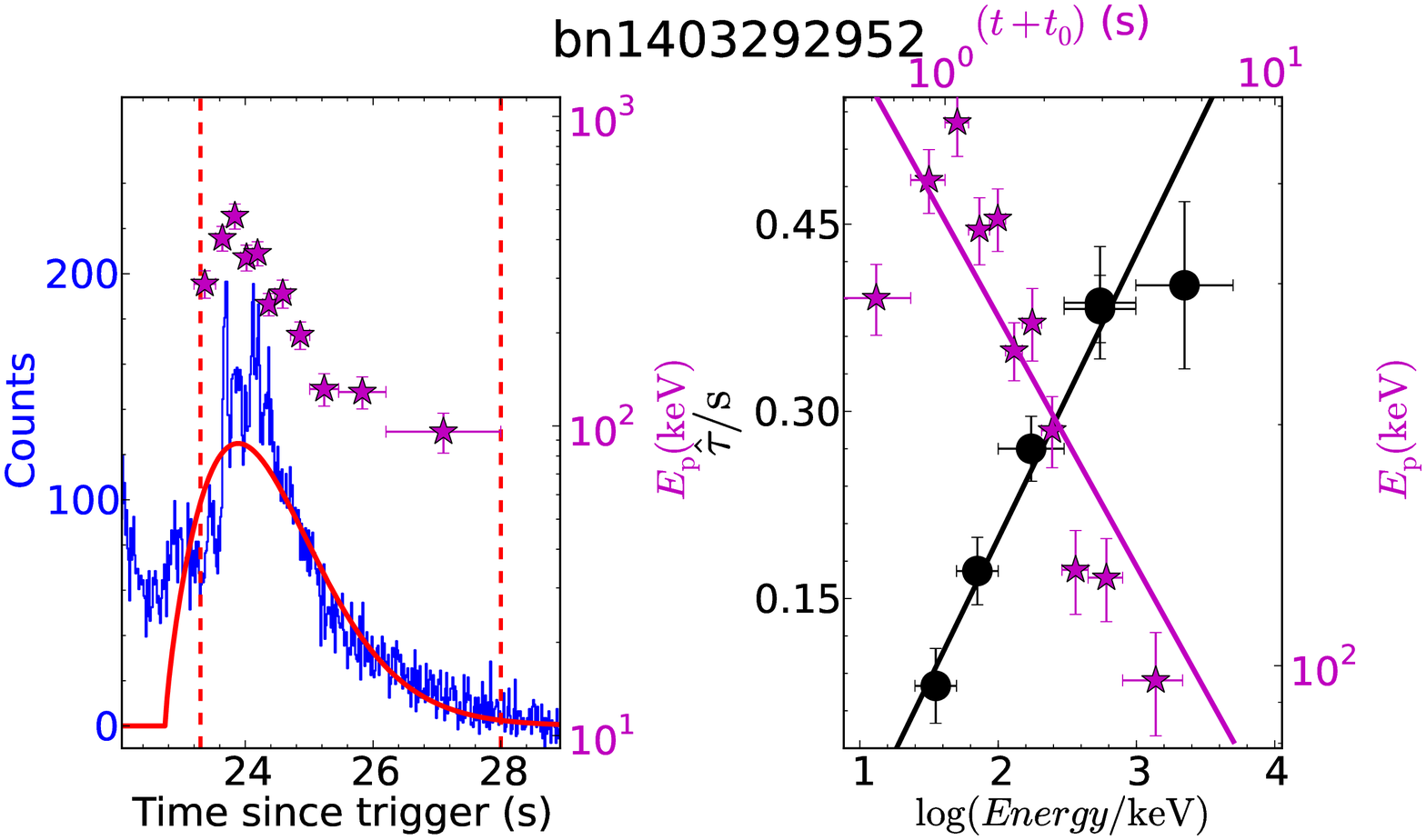} &
\includegraphics[origin=c,angle=0,scale=1,width=0.500\textwidth,height=0.19\textheight]{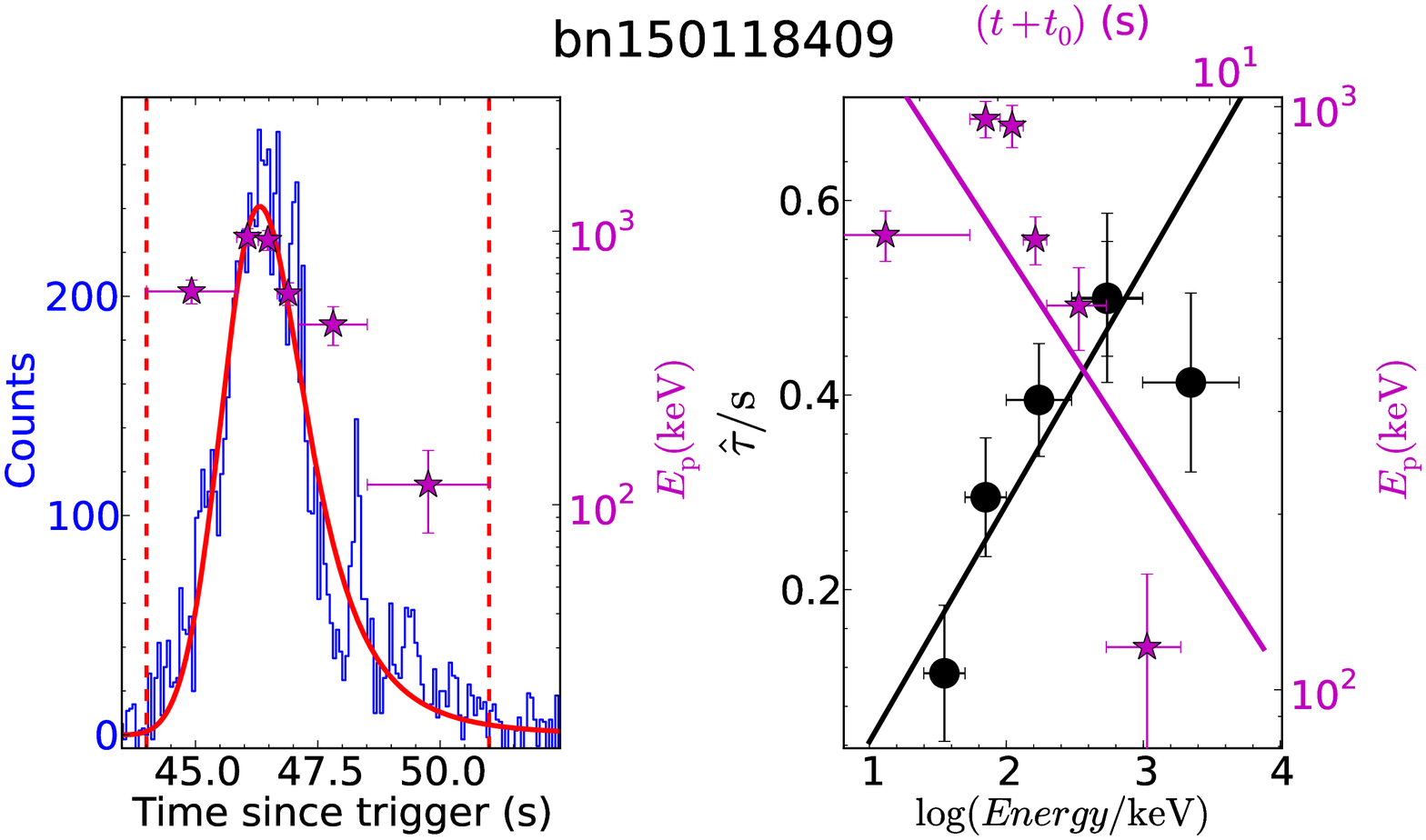} \\
\includegraphics[origin=c,angle=0,scale=1,width=0.500\textwidth,height=0.19\textheight]{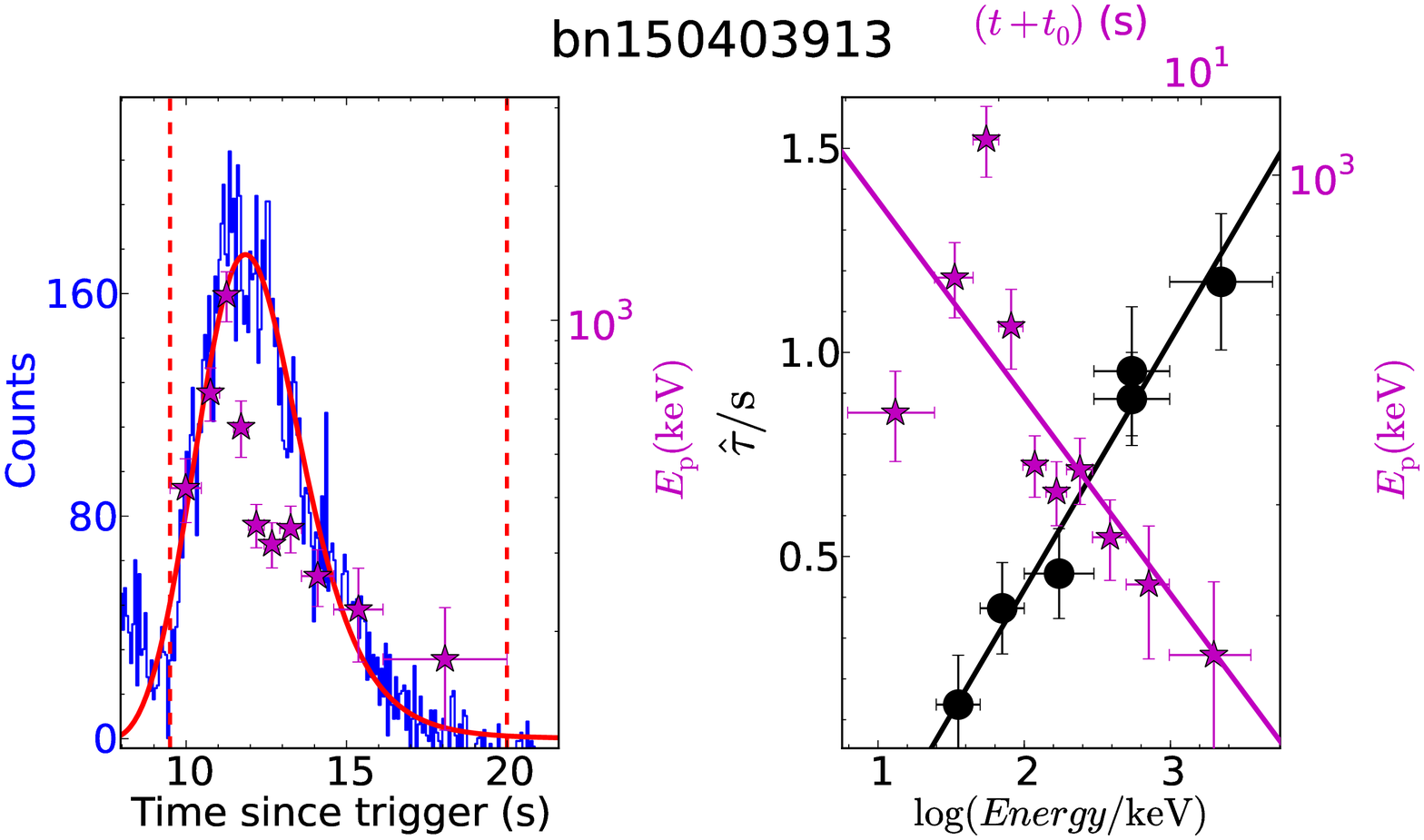} &
\includegraphics[origin=c,angle=0,scale=1,width=0.500\textwidth,height=0.19\textheight]{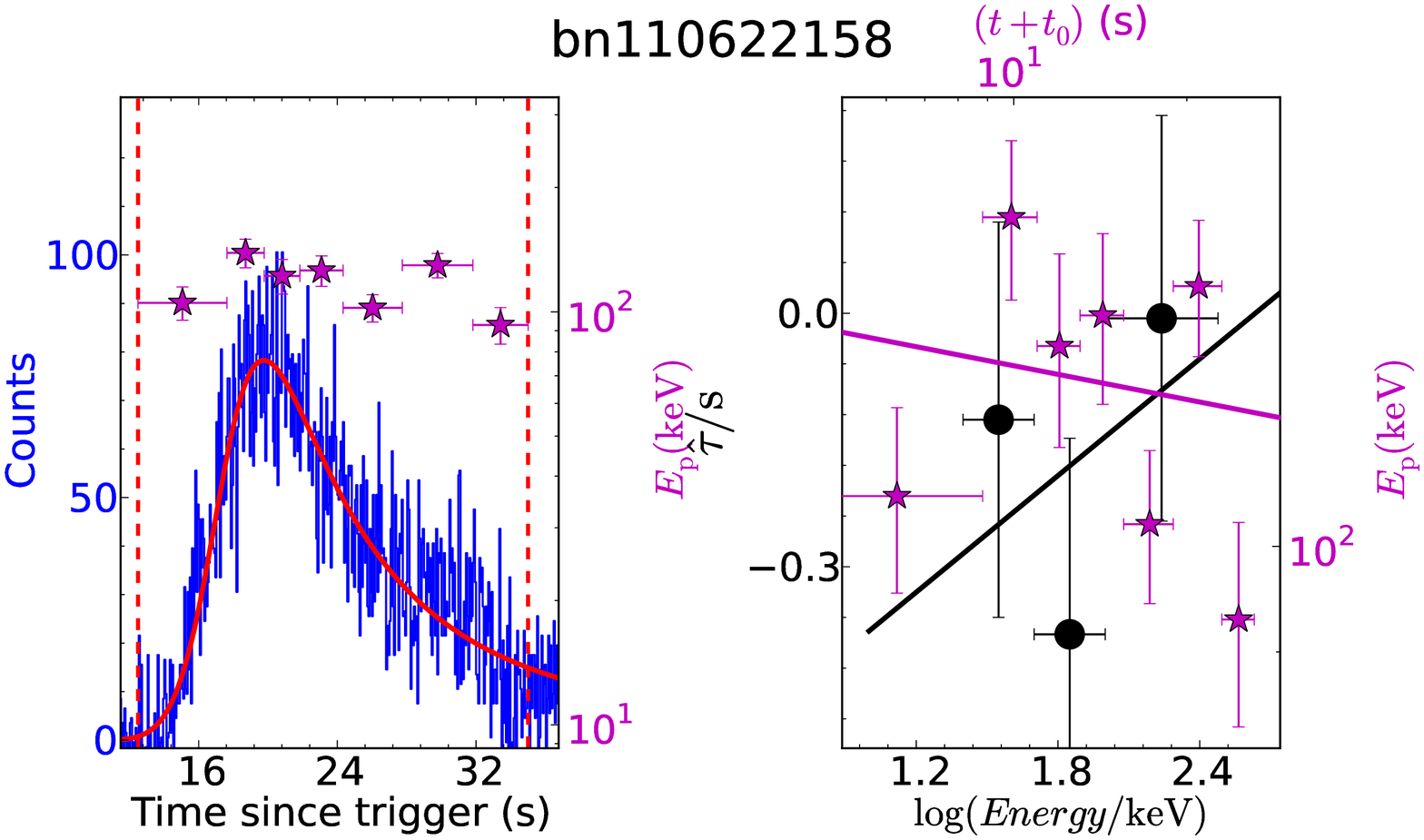} \\
\includegraphics[origin=c,angle=0,scale=1,width=0.500\textwidth,height=0.19\textheight]{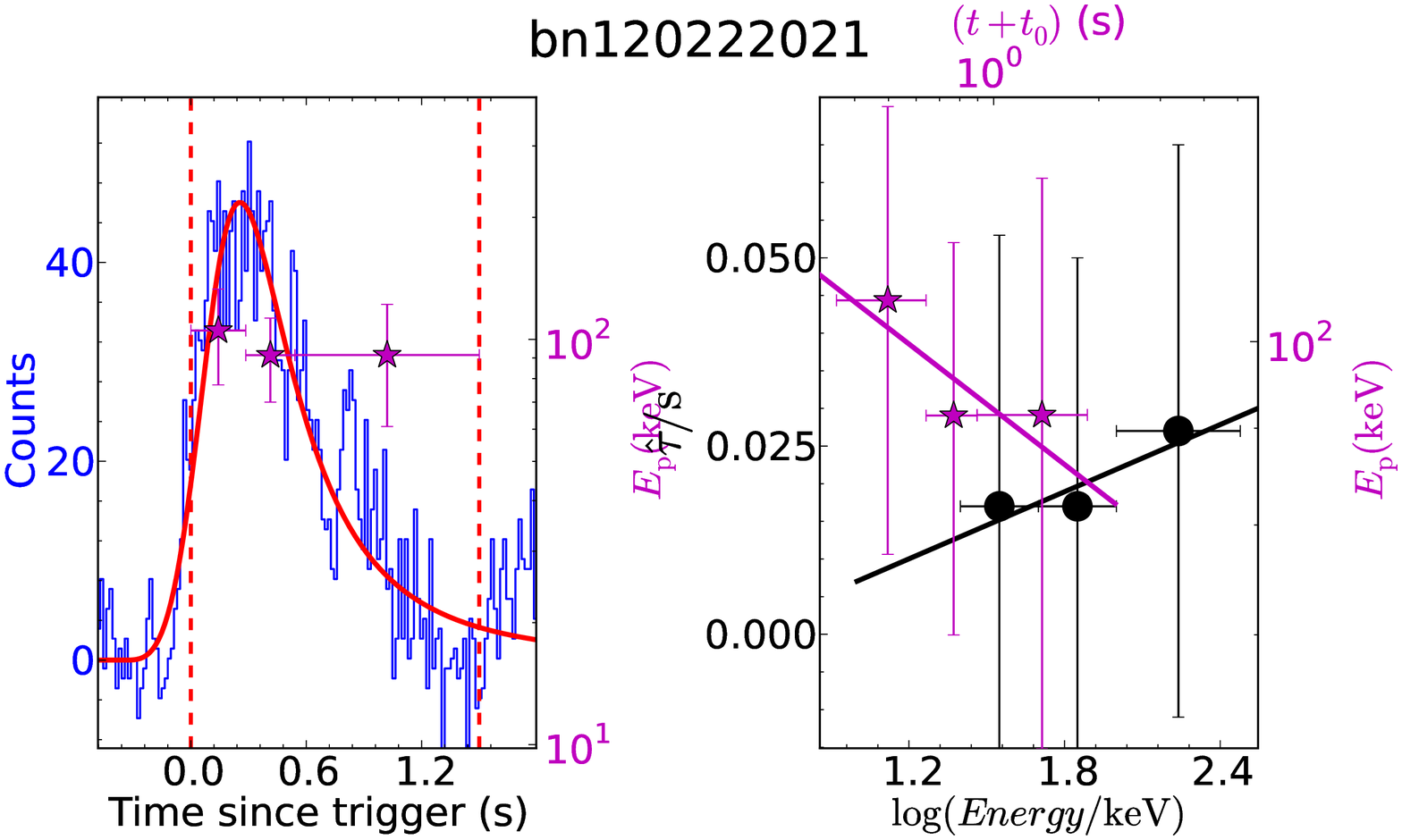} &
\includegraphics[origin=c,angle=0,scale=1,width=0.500\textwidth,height=0.19\textheight]{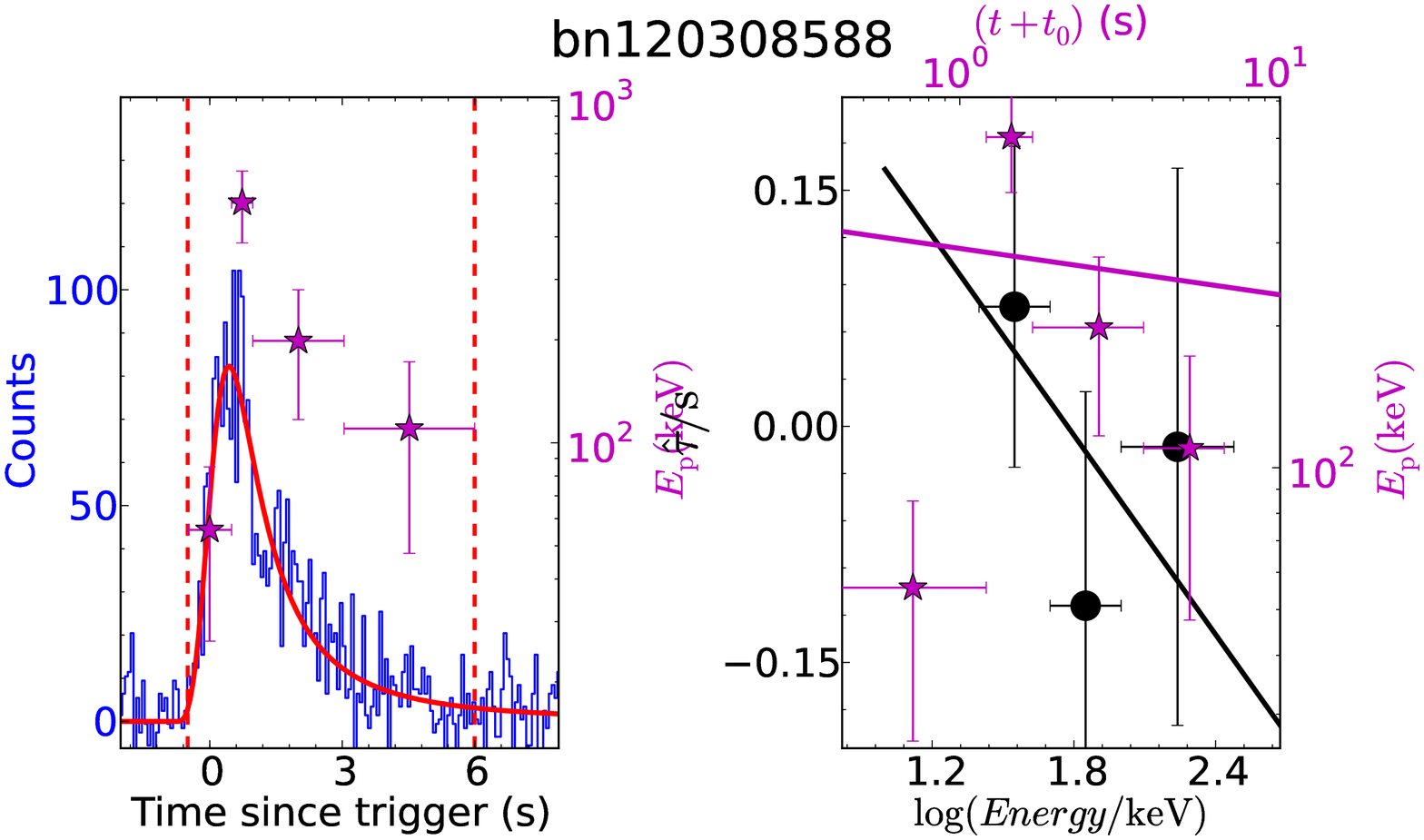} \\
\end{tabular}
\hfill
\vspace{4em}
\center{Fig.~\ref{myfigB}---Continued.}
\end{figure*}

\clearpage %p4

\begin{figure*}
\vspace{-4em}\hspace{-4em}
\centering
\begin{tabular}{cc}
\includegraphics[origin=c,angle=0,scale=1,width=0.500\textwidth,height=0.19\textheight]{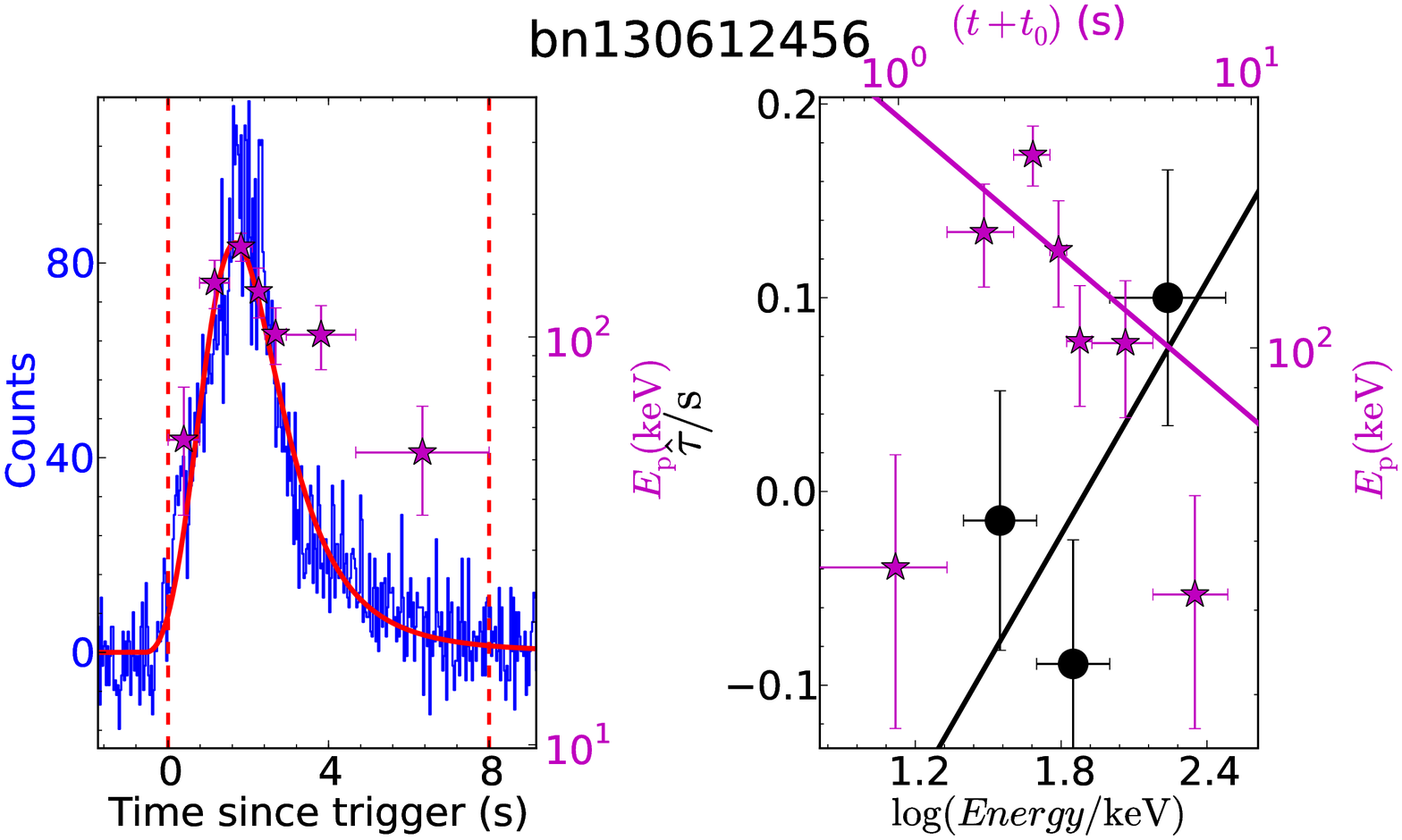} &
\includegraphics[origin=c,angle=0,scale=1,width=0.500\textwidth,height=0.19\textheight]{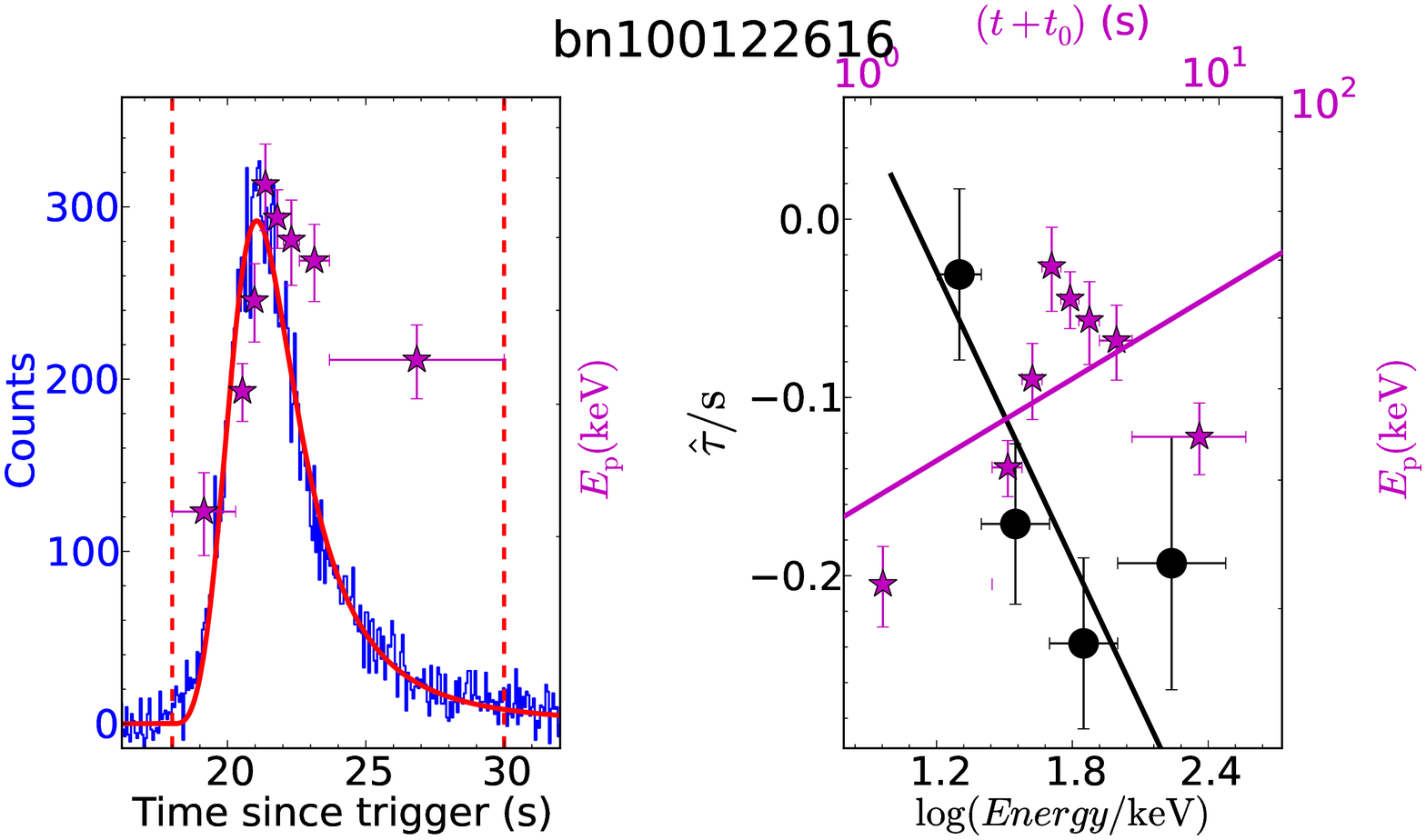} \\
\includegraphics[origin=c,angle=0,scale=1,width=0.500\textwidth,height=0.19\textheight]{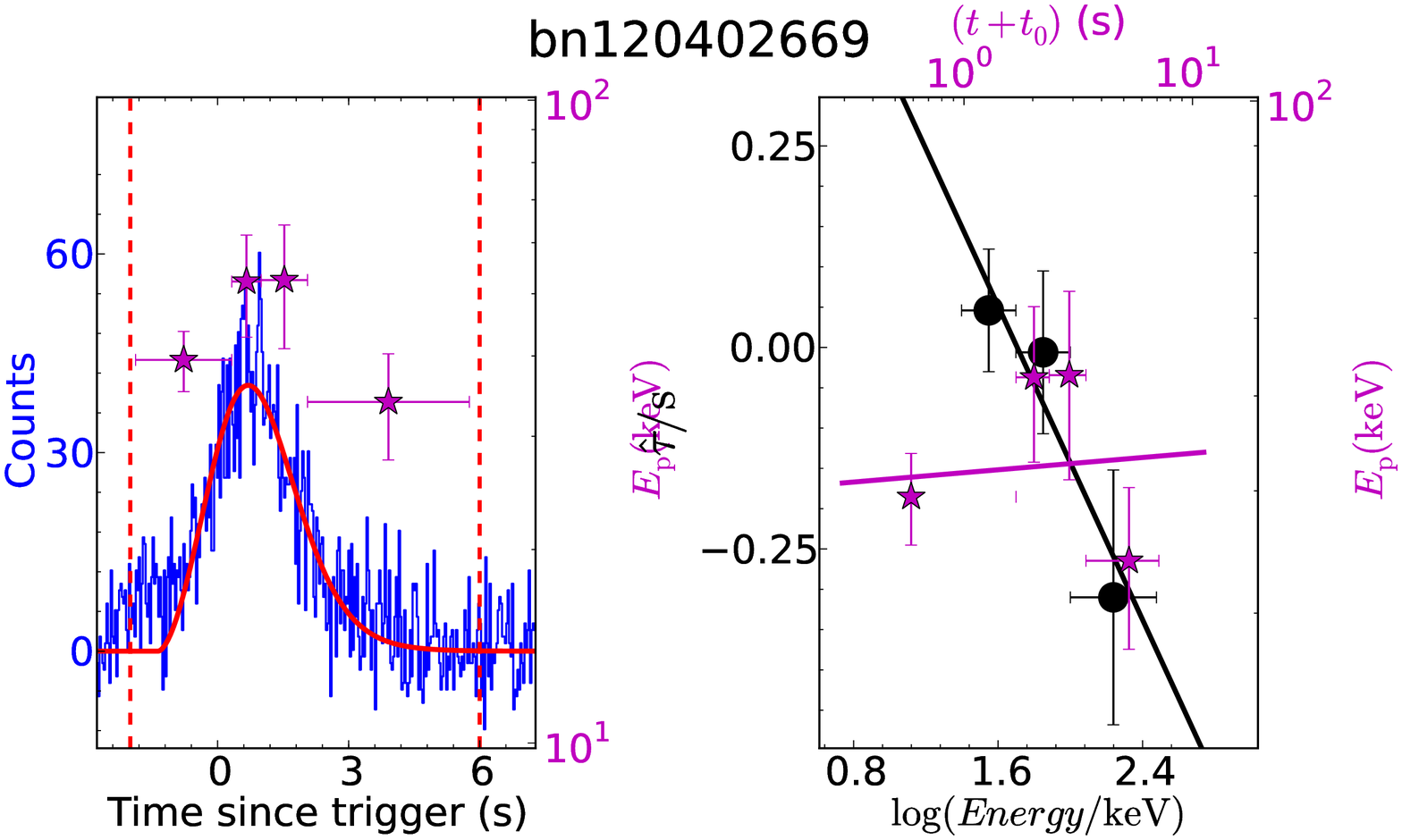} &
\includegraphics[origin=c,angle=0,scale=1,width=0.500\textwidth,height=0.19\textheight]{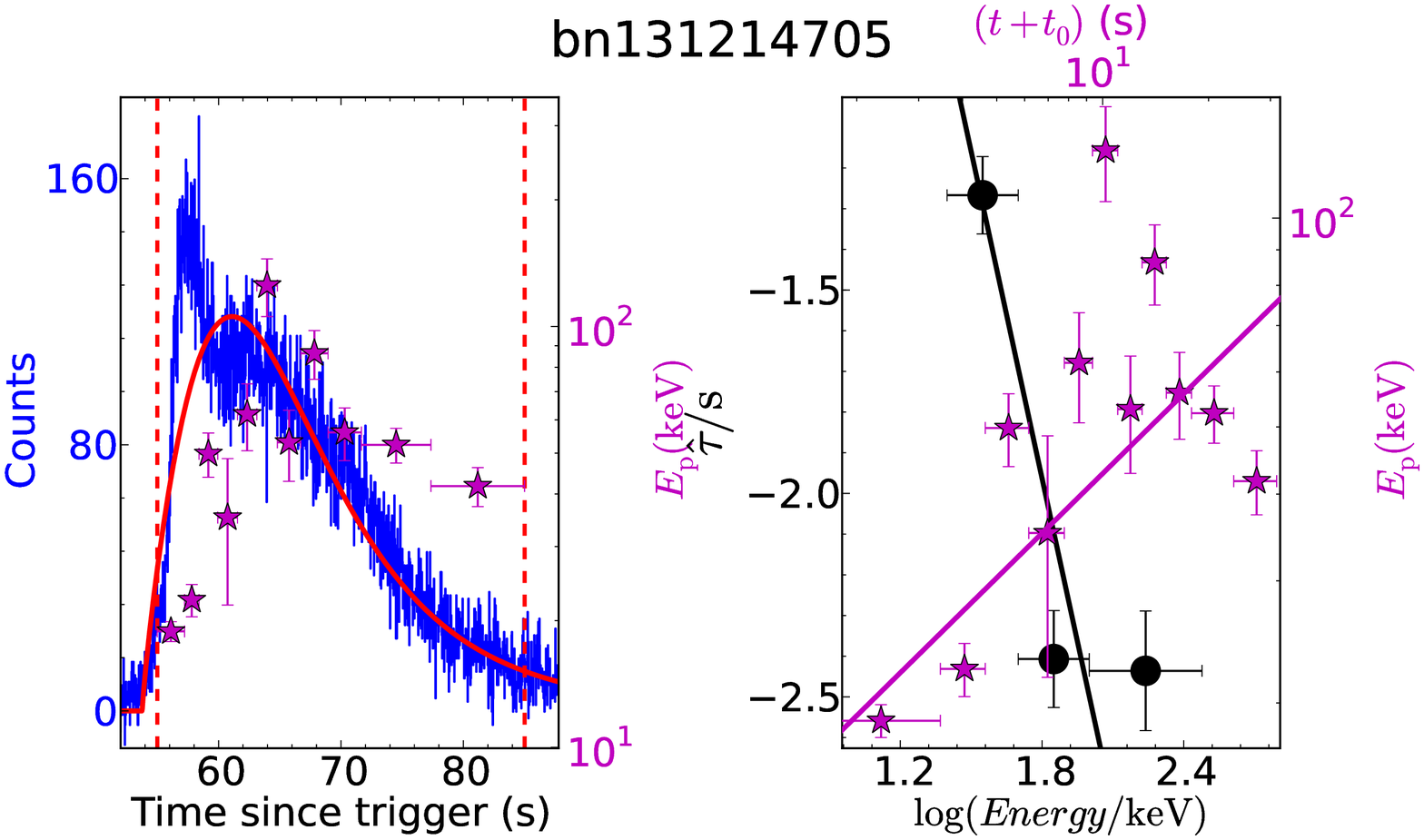} \\
\end{tabular}
\hfill
\vspace{4em}
\center{Fig.~\ref{myfigB}---Continued.}
\end{figure*}

%%%%%%%%%%%%%%%%%%%%%%%%%%%%%%%%%%%%%%%%%%%%%%%%%%%%%%%%%%%%%%%%%%%%%%%%%%%%%%
%%%%%%%%%%%%%%%%%%%%%%%%%%%%%%%%%%%%%%%%%%%%%%%%%%%%%%%%%%%%%%%%%%%%%%%%%%%%%%
%%%%%  Figure 3
\clearpage
\begin{figure*}\label{Distribution_kEktau}
\vspace{-2em}\hspace{-2em}
\centering
\begin{tabular}{cc}
\includegraphics[origin=c,angle=0,scale=1,width=0.48\textwidth,height=0.30\textheight]{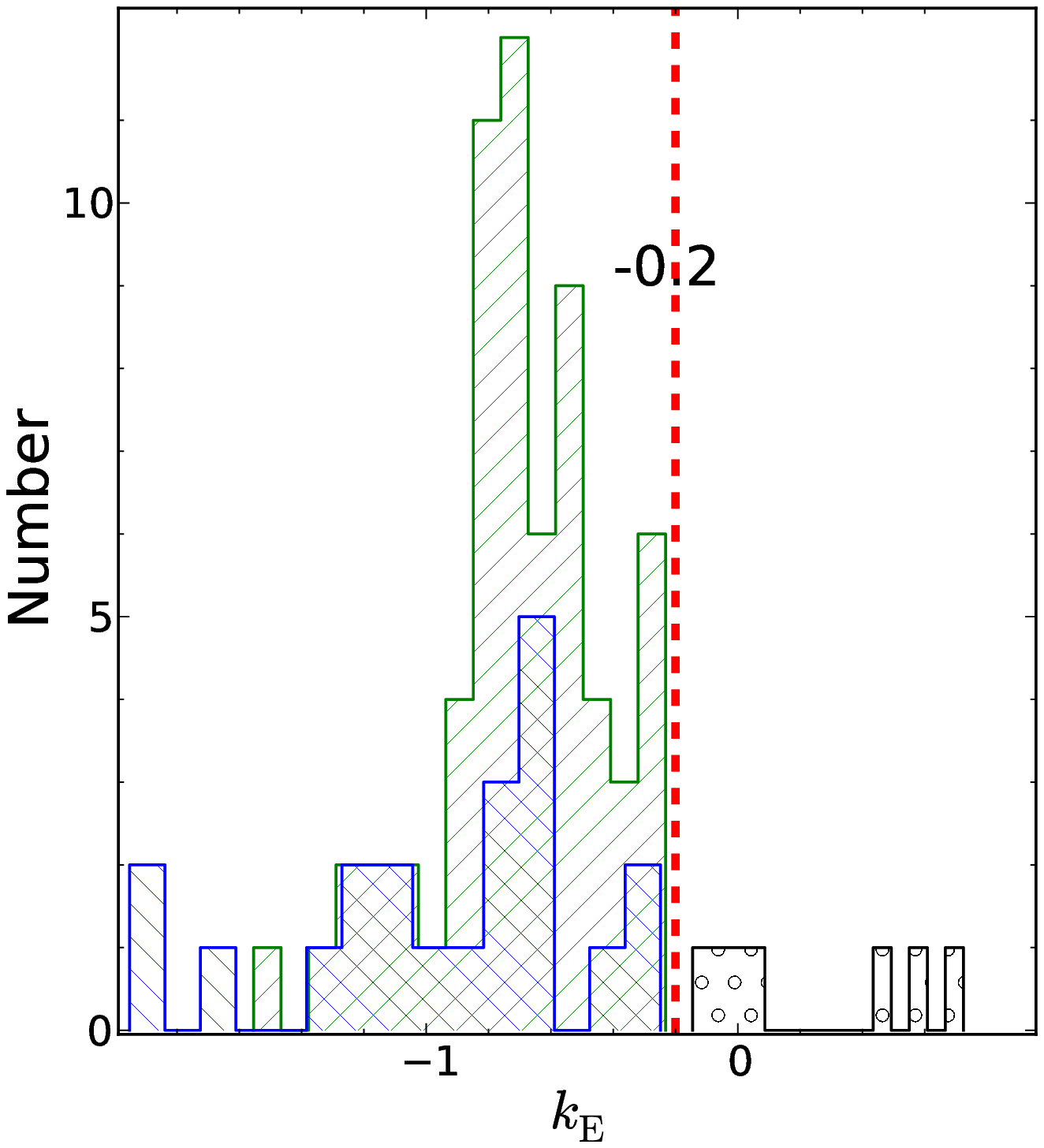} &
\includegraphics[origin=c,angle=0,scale=1,width=0.48\textwidth,height=0.30\textheight]{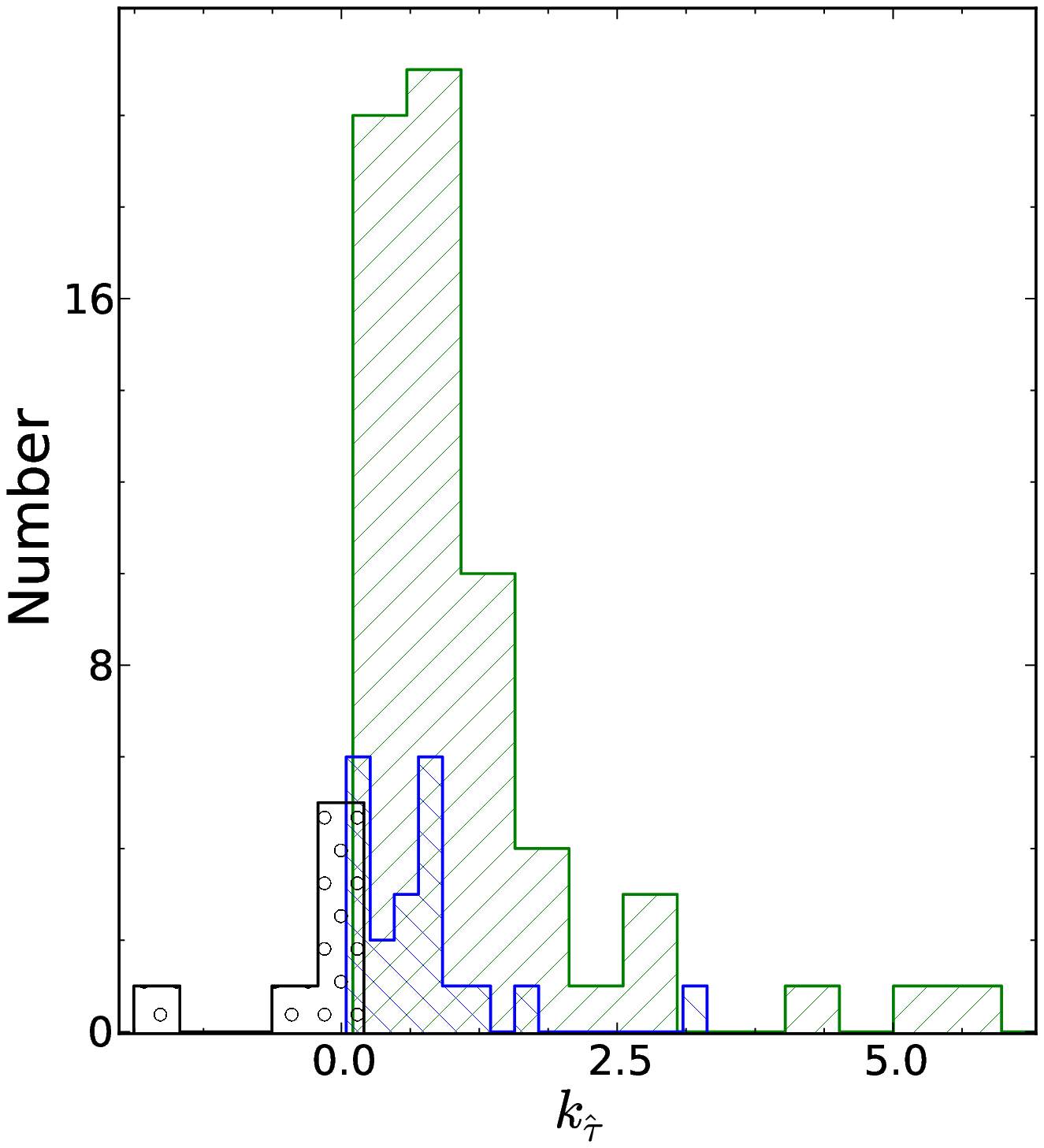} \\
\end{tabular}
\hfill
\vspace{4em}
\caption{Left panel: $k_{\rm E}$ distribution,
where the red vertical dashed line marks $k_{\rm E}=-0.2$.
Right panel: $k_{\hat{\tau}}$ distribution, where the green ``$/$'', blue ``$\setminus$'', and black ``$o$''
hatches represent the data from the H2S pulses, H2S-dominated-tracking pulses,
and tracking spectral evolution pulses. }\label{myfigC}
\end{figure*}

%%%%%%%%%%%%%%%%%%%%%%%%%%%%%%%%%%%%%%%%%%%%%%%%%%%%%%%%%%%%%%%%%%%%%%%%%%%%%%
%%%%%%%%%%%%%%%%%%%%%%%%%%%%%%%%%%%%%%%%%%%%%%%%%%%%%%%%%%%%%%%%%%%%%%%%%%%%%%
%%  Figure 11

\clearpage
\begin{figure*}
\vspace{-4em}\hspace{-4em}
\centering
\includegraphics[origin=c,angle=0,scale=1,width=0.60\textwidth,height=0.50\textheight]{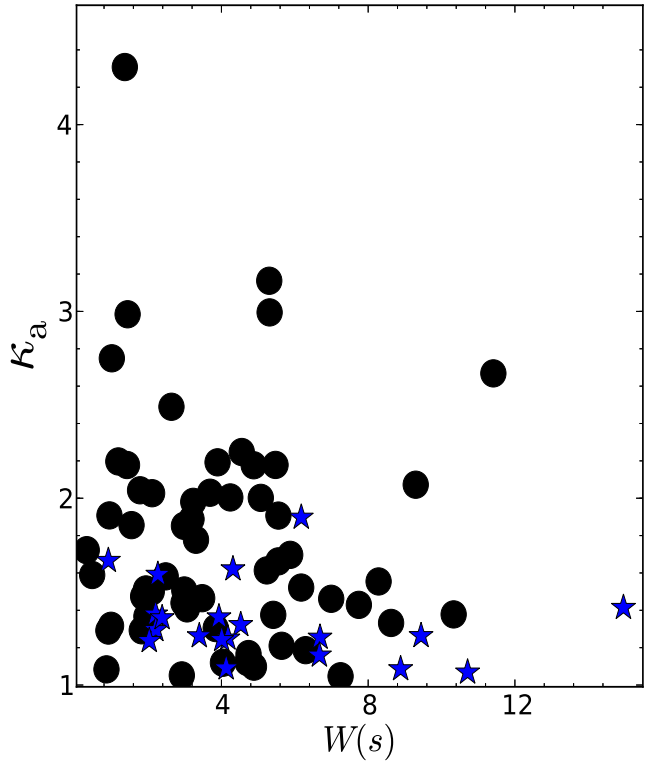}
\hfill
\vspace{4em}
\caption{\bf Relationship between the pulse asymmetry $\kappa_a$ and its width $W$ for the H2S pulses (black solid circles) and the tracking pulses (blue stars) in our sample.}\label{myfigK}
\end{figure*}

%%%%%%%%%%%%%%%%%%%%%%%%%%%%%%%%%%%%%0%%%%%%%%%%%%%%%%%%%%%%%%%%%%%%%%%%%%%%%%
%%%  Figure 4
\clearpage
\begin{figure*}\label{3pairs-relations}
\vspace{-4em}\hspace{-4em}
\centering
\begin{tabular}{ccc}
\includegraphics[origin=c,angle=0,scale=1,width=0.30\textwidth,height=0.30\textheight]{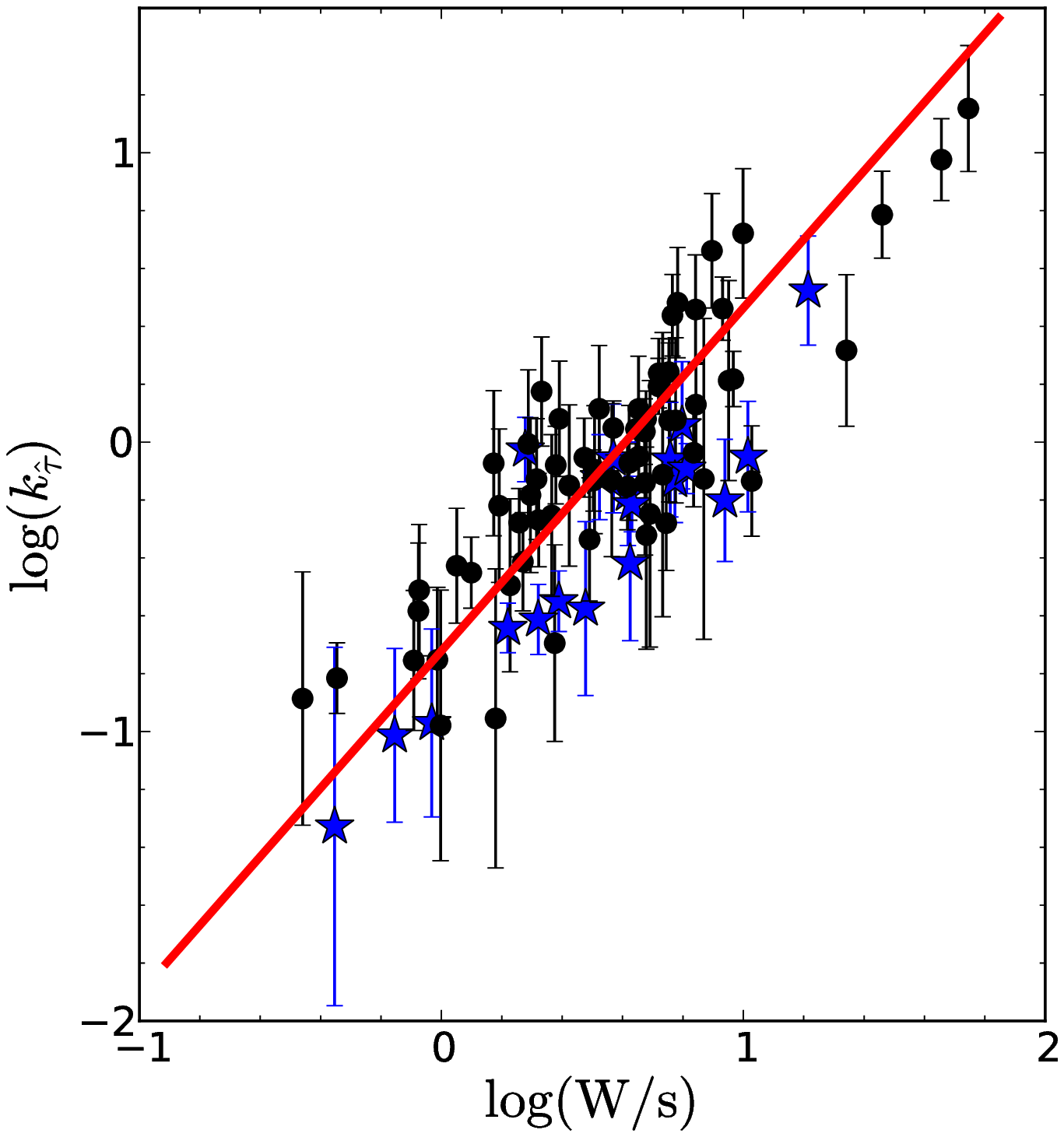} &
\includegraphics[origin=c,angle=0,scale=1,width=0.30\textwidth,height=0.30\textheight]{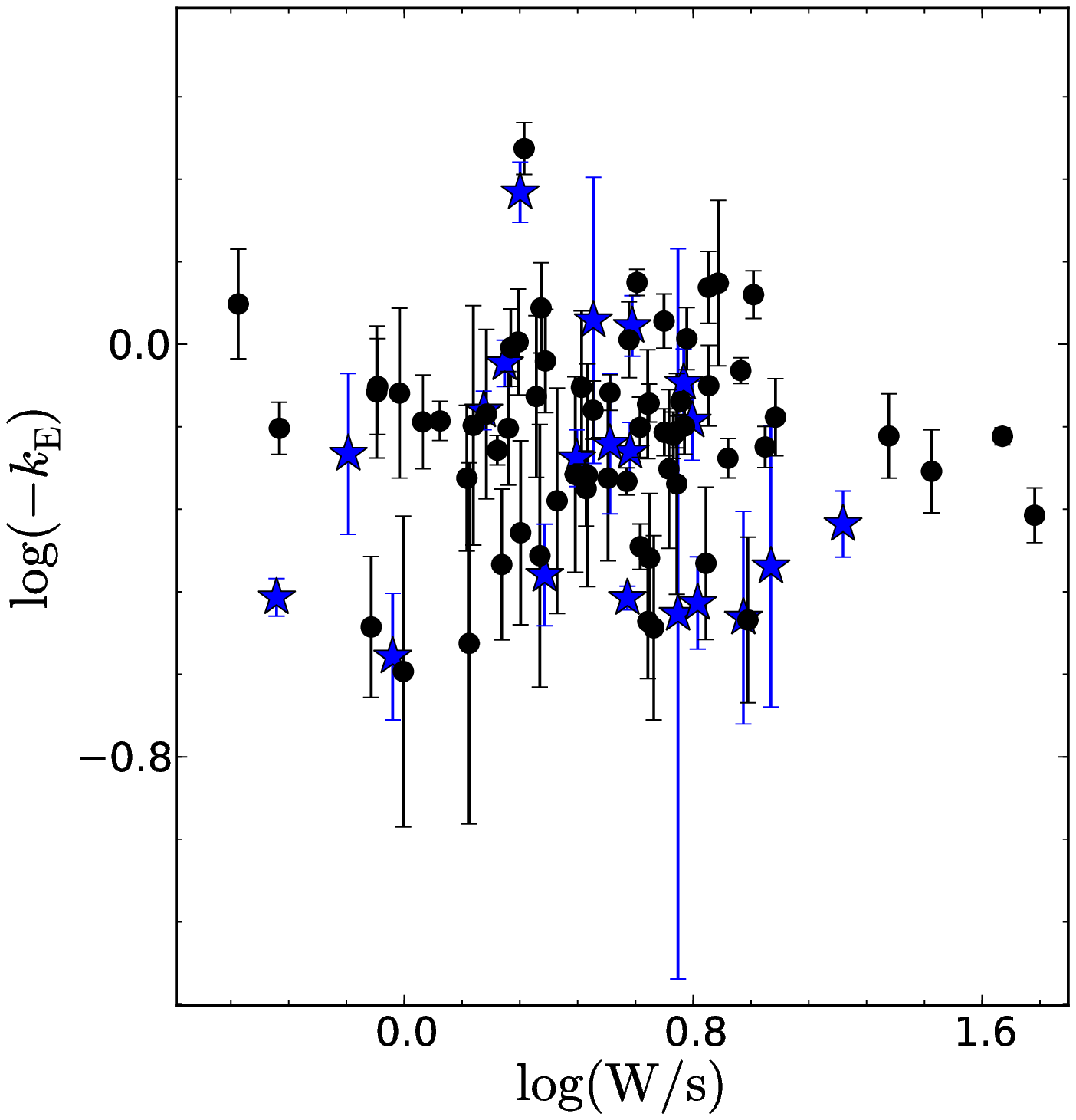} &
\includegraphics[origin=c,angle=0,scale=1,width=0.30\textwidth,height=0.30\textheight]{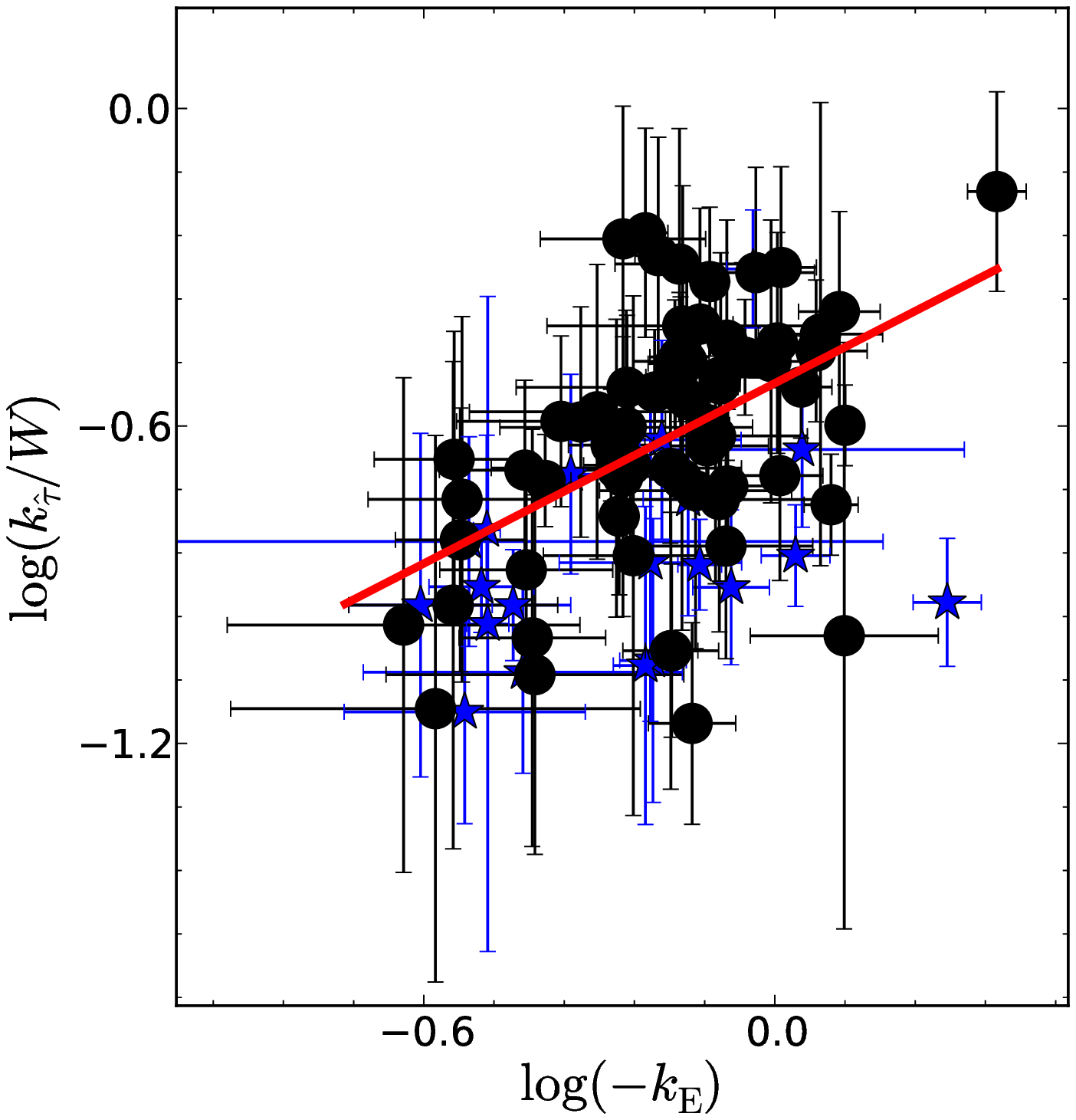} \\
\end{tabular}
\hfill
\vspace{4em}
\caption{Relations of $k_{\hat{\tau}}$-$W$ (left panel), $k_{E}$-$W$ (middle panel),
and $k_{\hat{\tau}}/W$-$k_{E}$ (right panel) for H2S pulses (black ``$\bullet$'') and H2S-dominated-tracking pulses (blue ``$\star$'').
The red solid lines in the left and right panels represent the best fitting for the data of the H2S pulses and H2S-dominated-tracking pulses.}\label{myfigD}
\end{figure*}

%%%%%%%%%%%%%%%%%%%%%%%%%%%%%%%%%%%%%%0%%%%%%%%%%%%%%%%%%%%%%%%%%%%%%%%%%%%%%%%
%%%  Figure 5
\clearpage
\begin{figure*}\label{lag31-w-tauE}
\vspace{-4em}\hspace{-4em}
\centering
\begin{tabular}{cc}
\includegraphics[origin=c,angle=0,scale=1,width=0.48\textwidth,height=0.30\textheight]{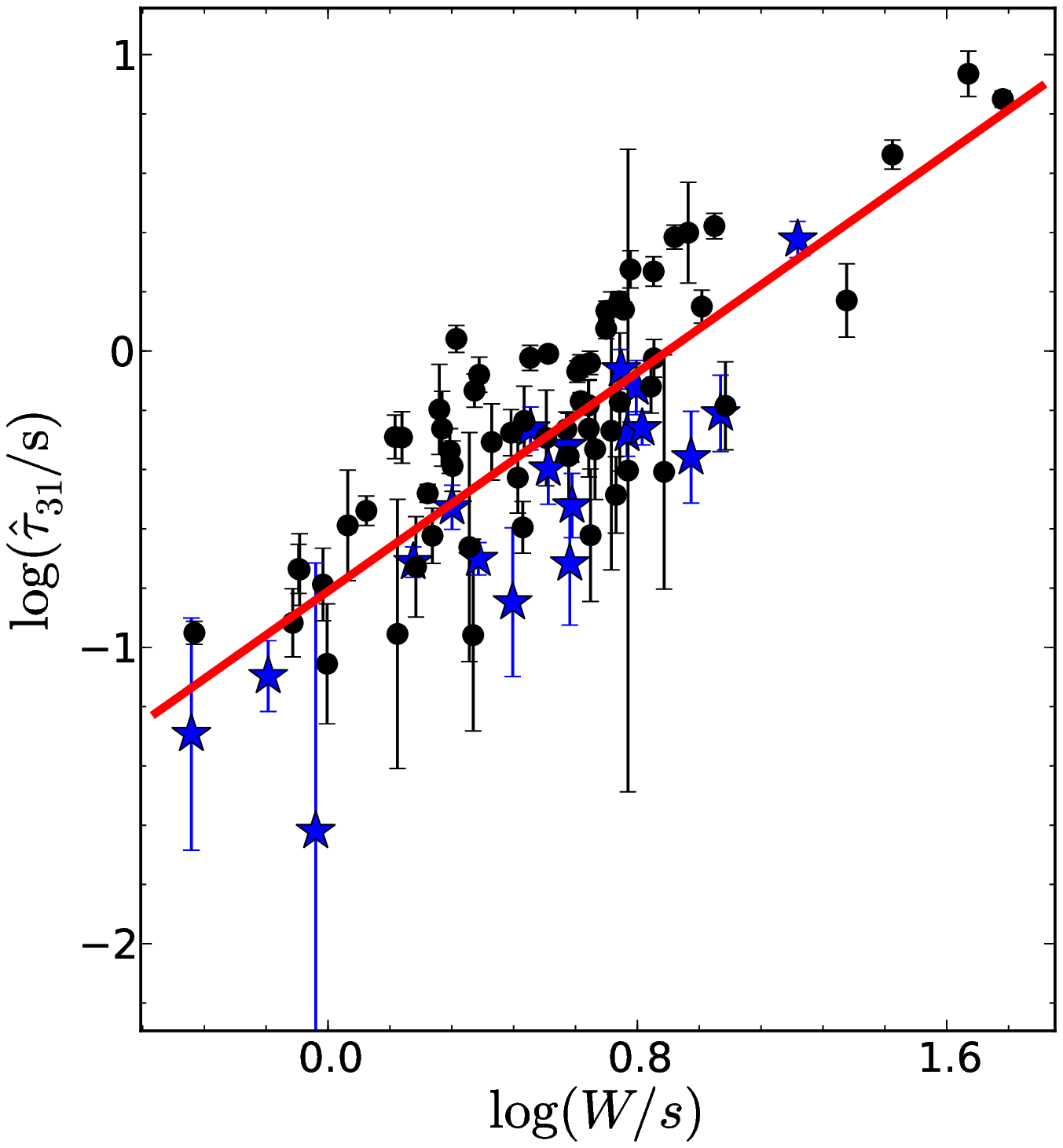} &
\includegraphics[origin=c,angle=0,scale=1,width=0.48\textwidth,height=0.30\textheight]{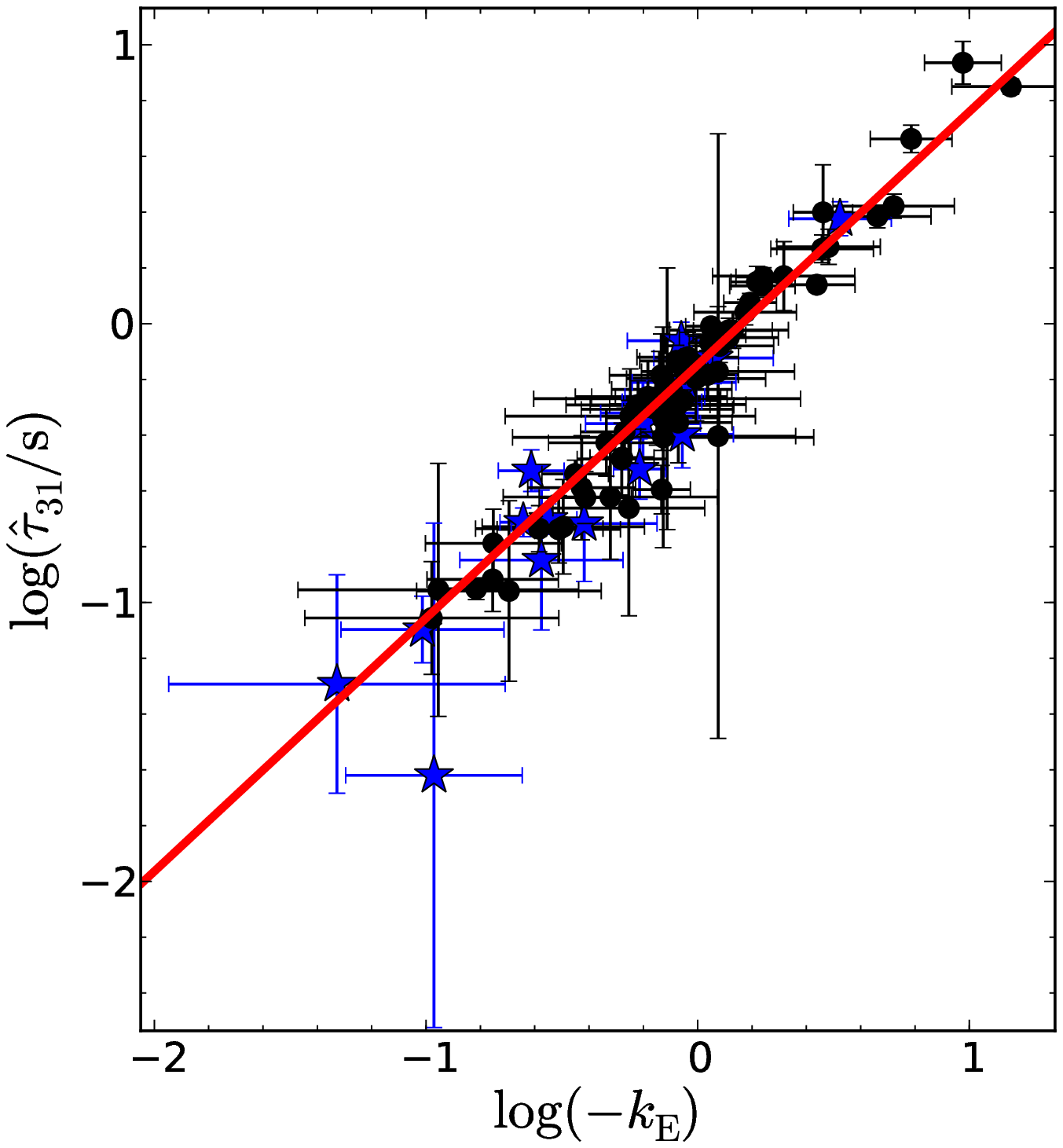} \\
\end{tabular}
\hfill
\vspace{4em}
\caption{Relations of ${\hat{\tau}}_{31}$-$W$ and ${\hat{\tau}}_{31}$-$k_{E}$ for the H2S pulses (black ``$\bullet$'')
and the H2S-dominated-tracking pulses (blue ``$\star$'').
The red solid lines represent the best fitting to the H2S pulses and H2S-dominated-tracking pulses.}\label{myfigE}
\end{figure*}

%%%%%%%%%%%%%%%%%%%%%%%%%%%%%%%%%%%%%%%%%%%%%%%%%%%%%%%%%%%%%%%%%%%%%%%%%%%%%
%%%%%%%%%%%%%%%%%%%%%%%%%%%%%%%%%%%%%%%%%%%%%%%0%%%%%%%%%%%%%%%%%%%%%%%%%%%%%%
%%%%% Figure 6
\clearpage
\begin{figure*}
\vspace{-4em}\hspace{-4em}
\centering
\begin{tabular}{cc}
\includegraphics[origin=c,angle=0,scale=1,width=0.50\textwidth,height=0.20\textheight]{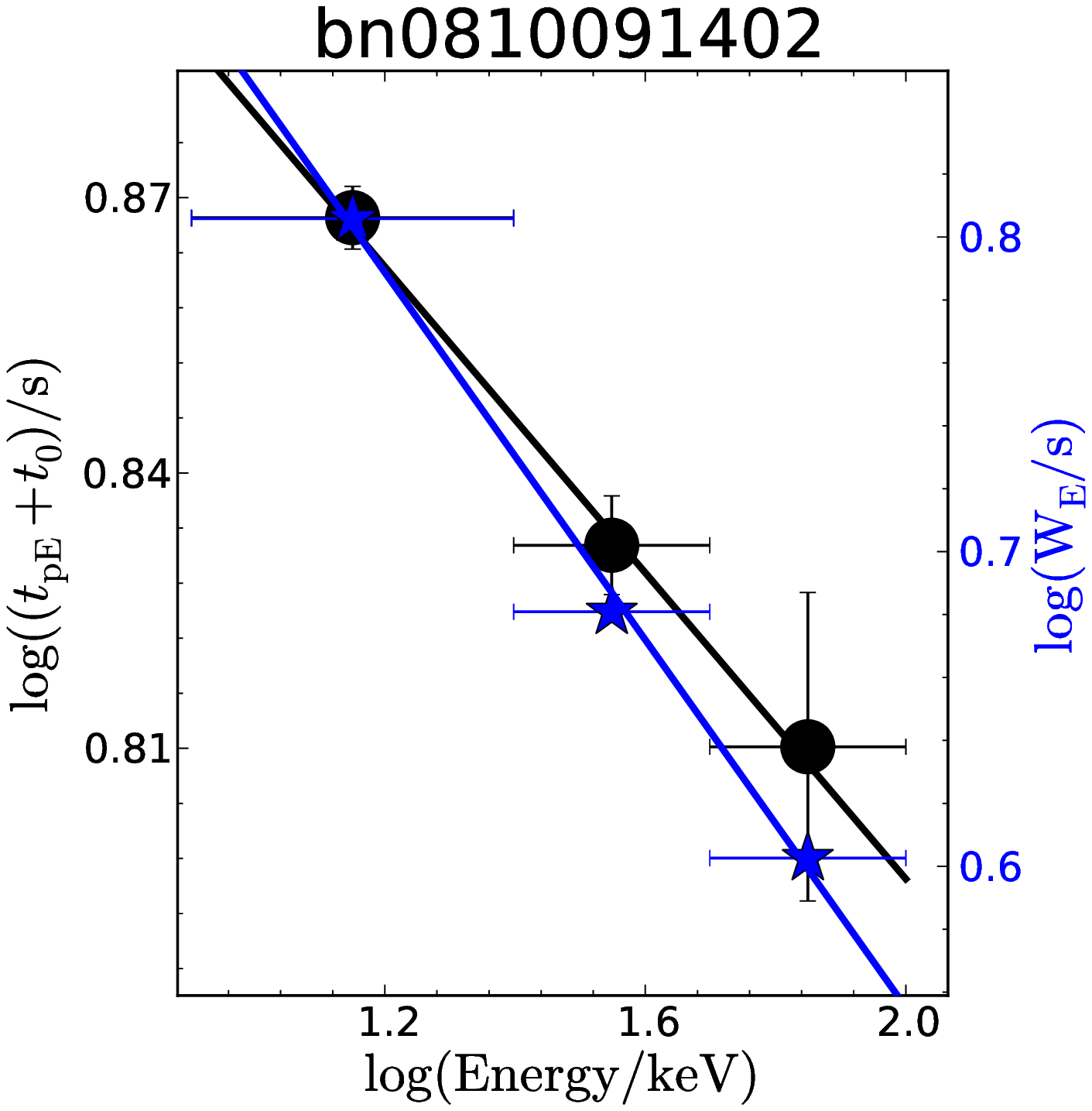} &
\includegraphics[origin=c,angle=0,scale=1,width=0.50\textwidth,height=0.20\textheight]{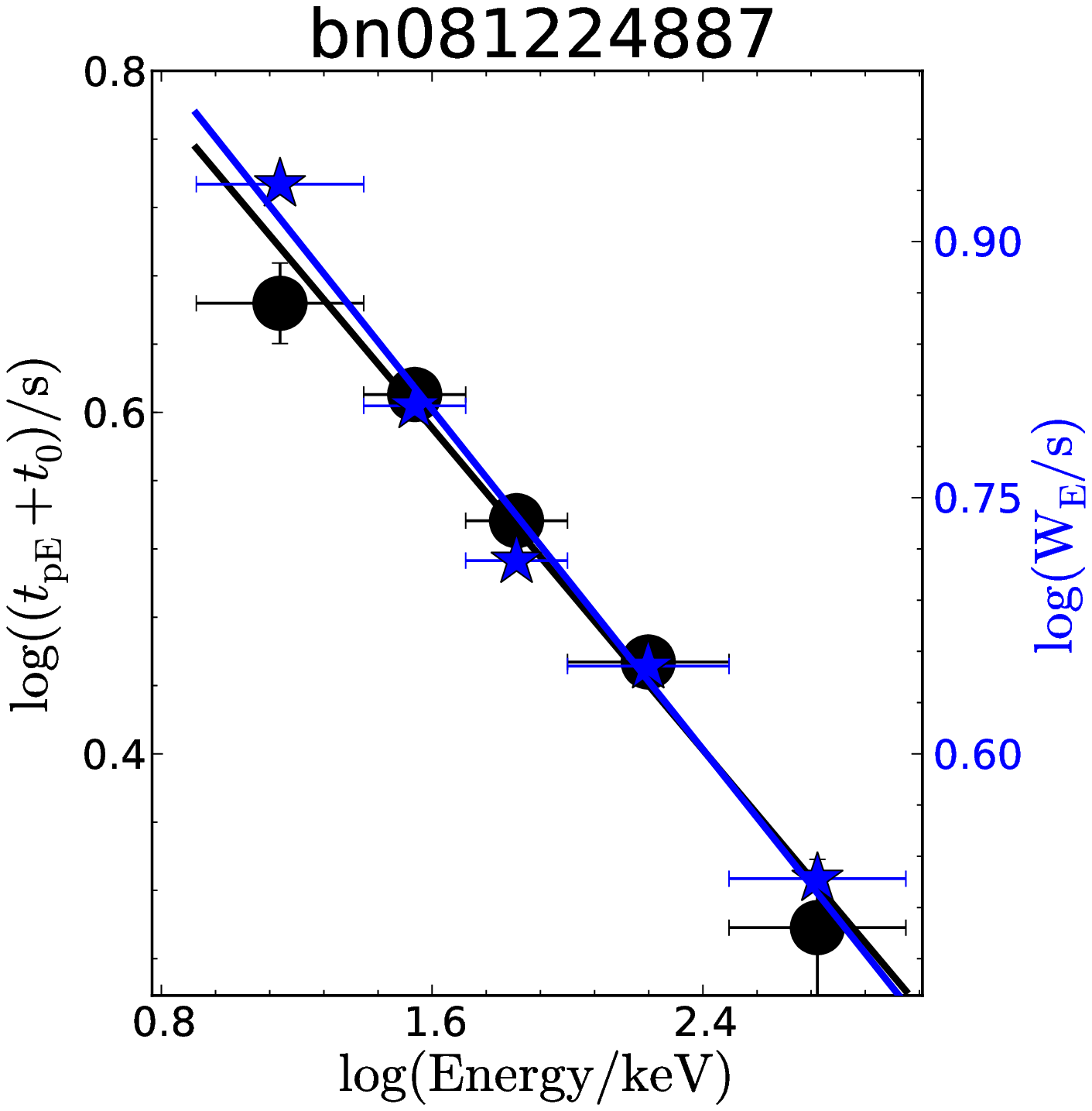} \\
\includegraphics[origin=c,angle=0,scale=1,width=0.50\textwidth,height=0.20\textheight]{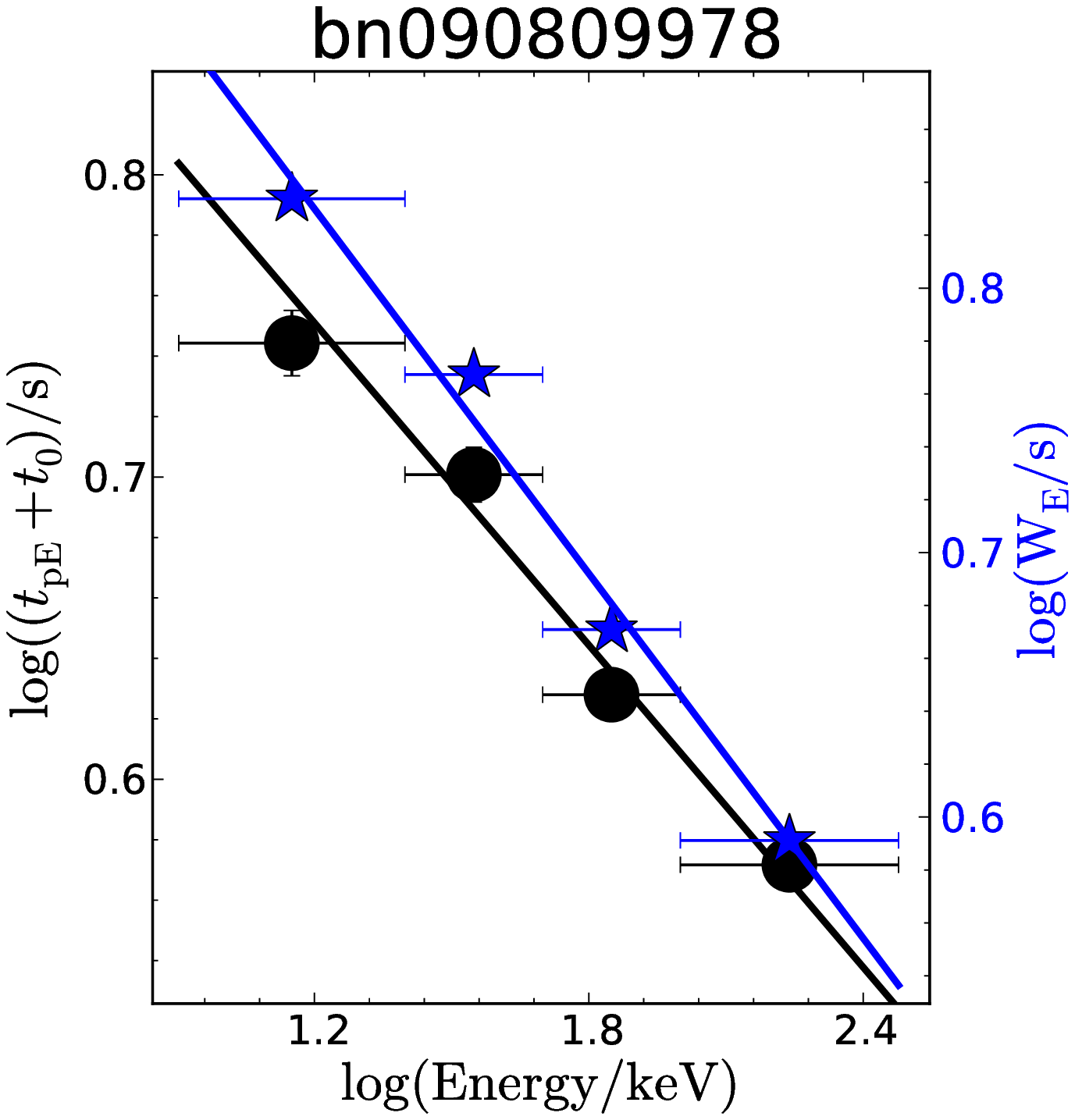} &
\includegraphics[origin=c,angle=0,scale=1,width=0.50\textwidth,height=0.20\textheight]{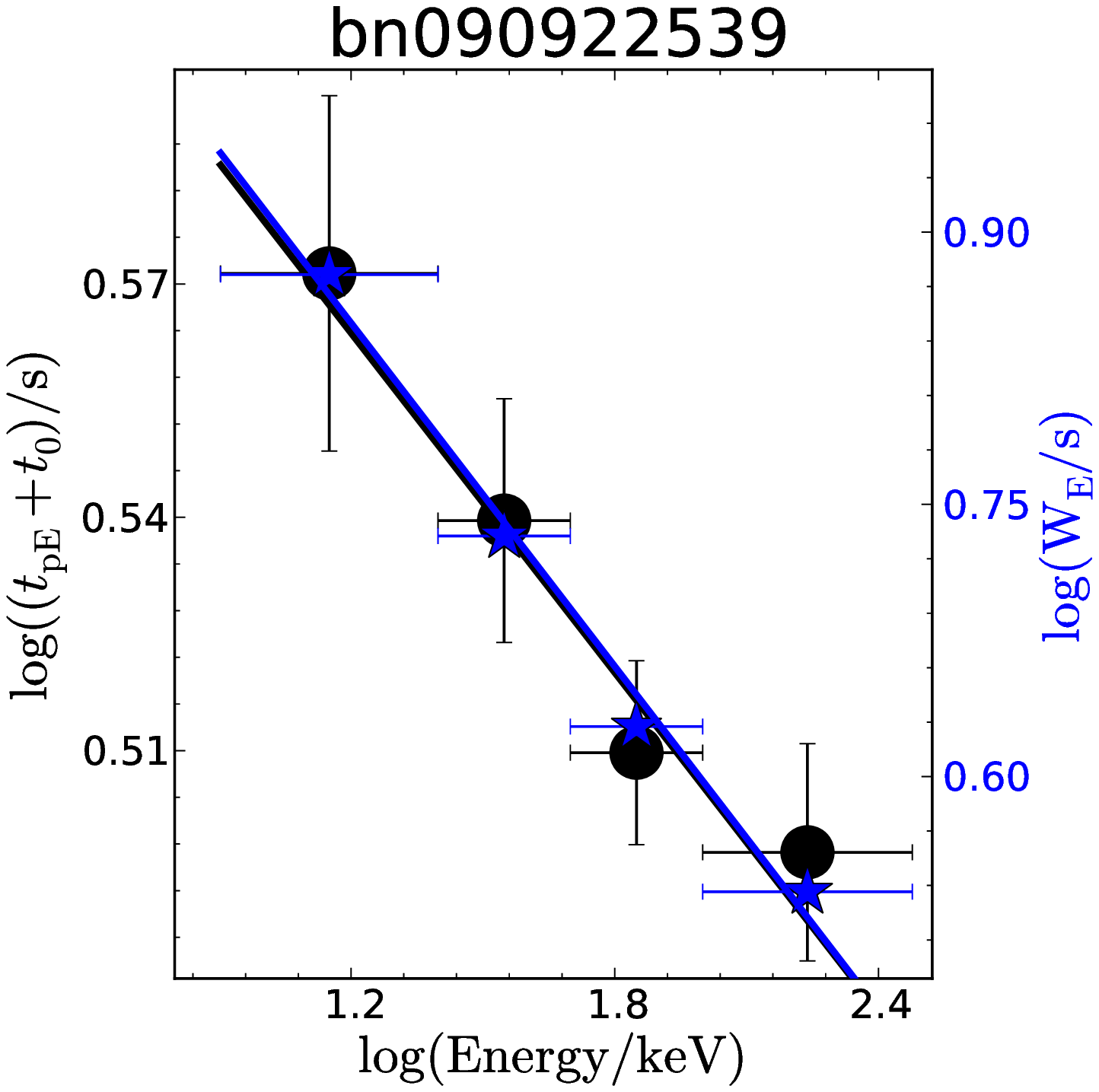} \\
\includegraphics[origin=c,angle=0,scale=1,width=0.50\textwidth,height=0.20\textheight]{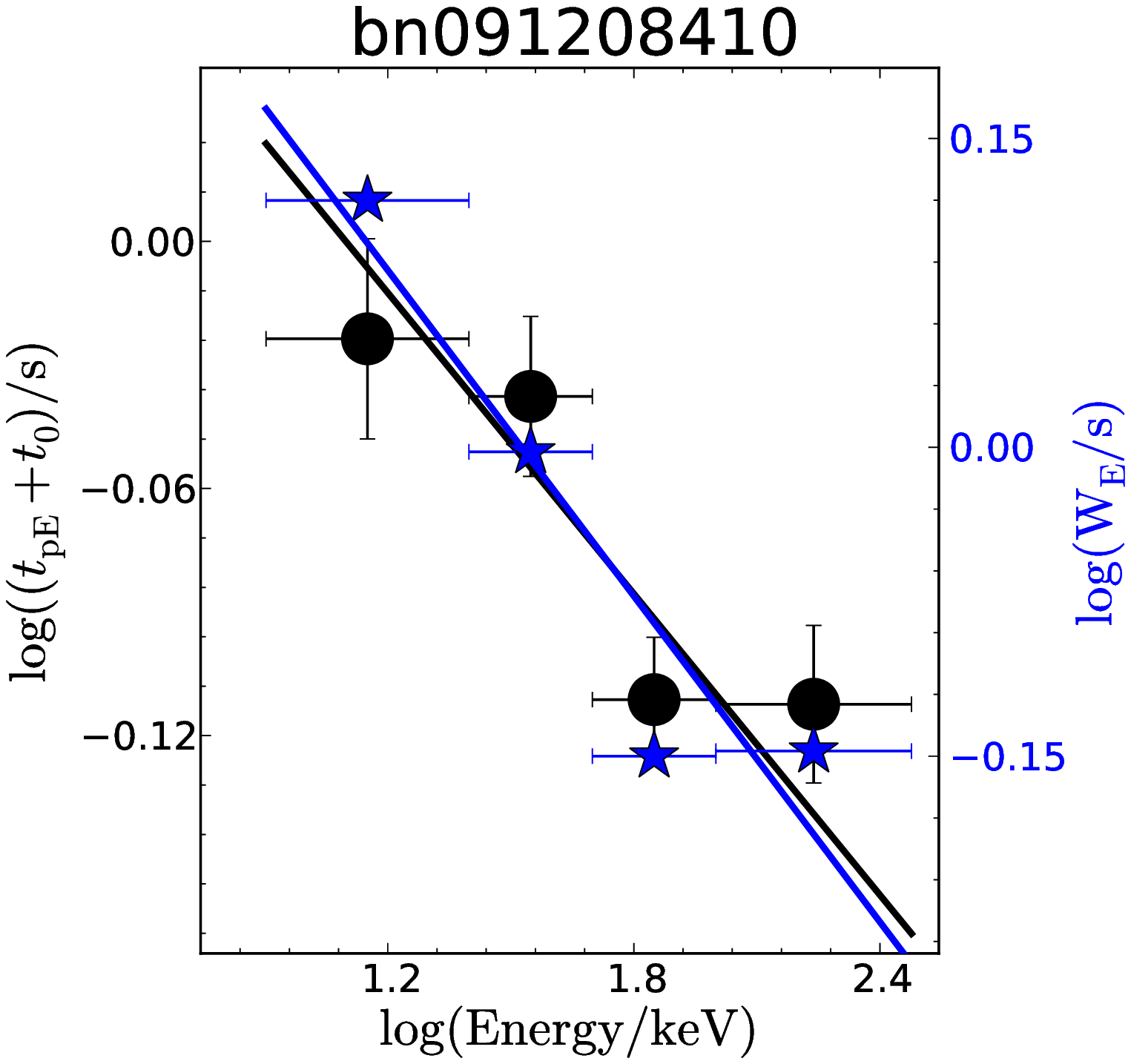} &
\includegraphics[origin=c,angle=0,scale=1,width=0.50\textwidth,height=0.20\textheight]{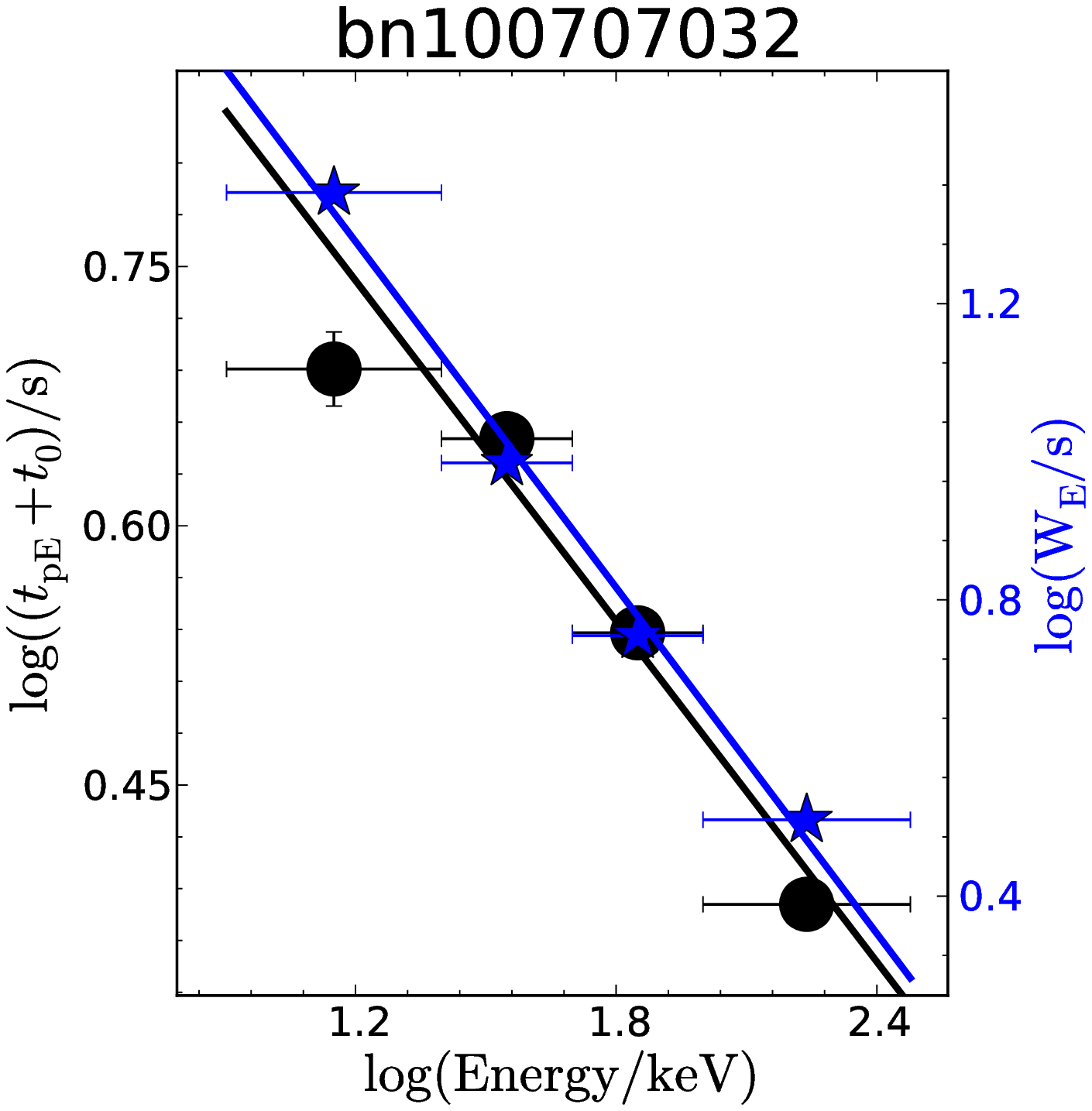} \\
\includegraphics[origin=c,angle=0,scale=1,width=0.50\textwidth,height=0.20\textheight]{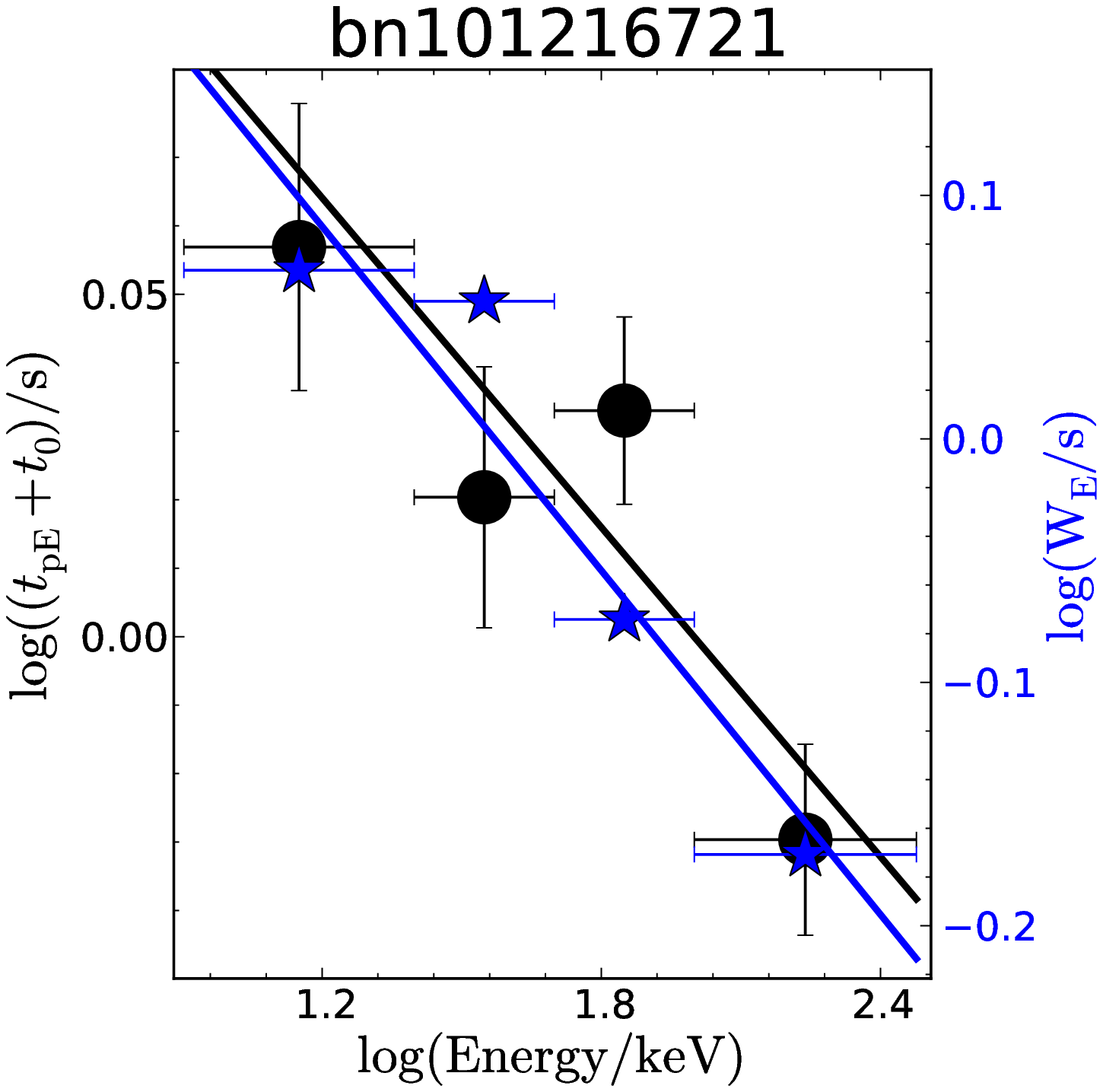} &
\includegraphics[origin=c,angle=0,scale=1,width=0.50\textwidth,height=0.20\textheight]{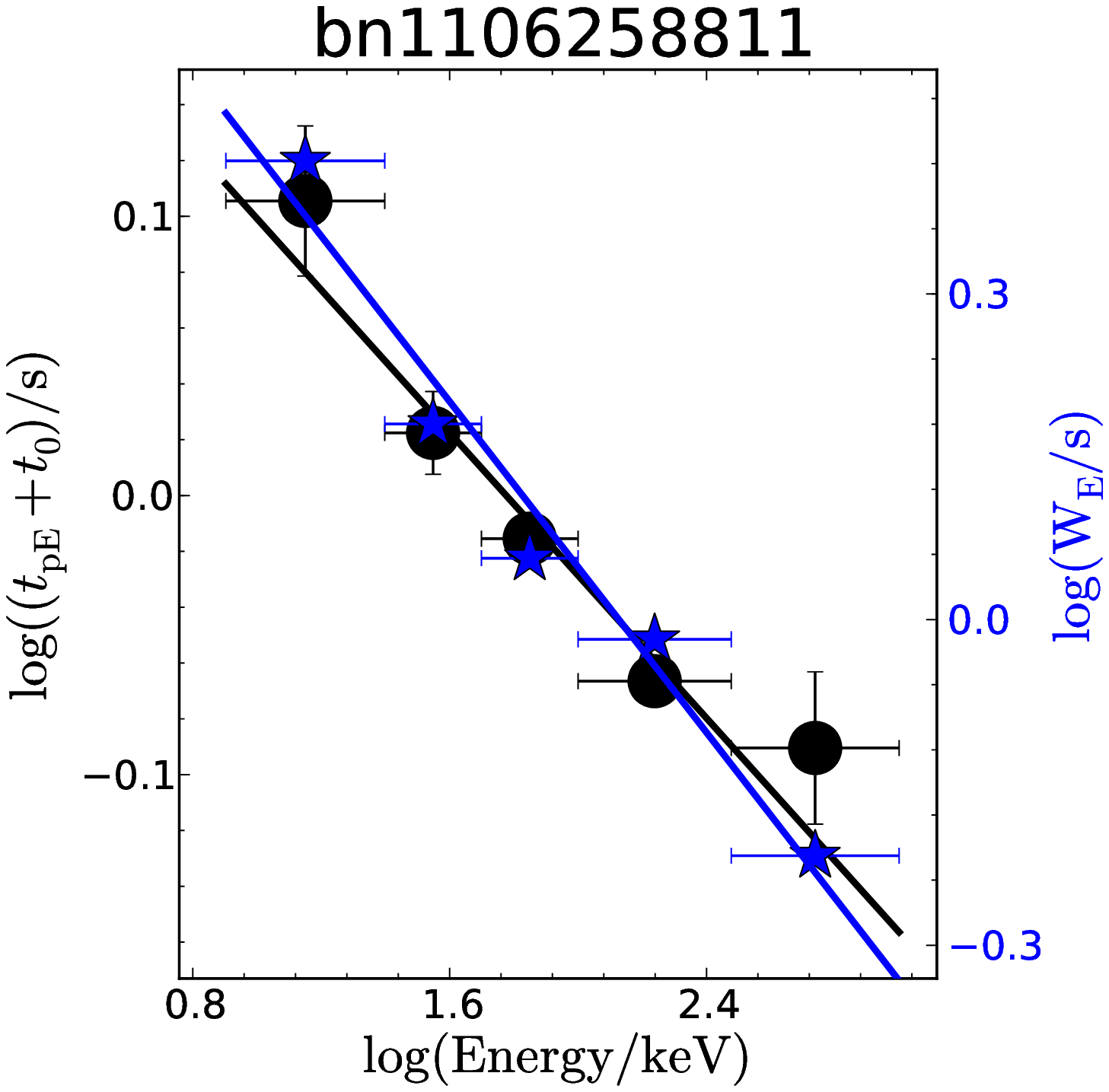} \\
\includegraphics[origin=c,angle=0,scale=1,width=0.50\textwidth,height=0.20\textheight]{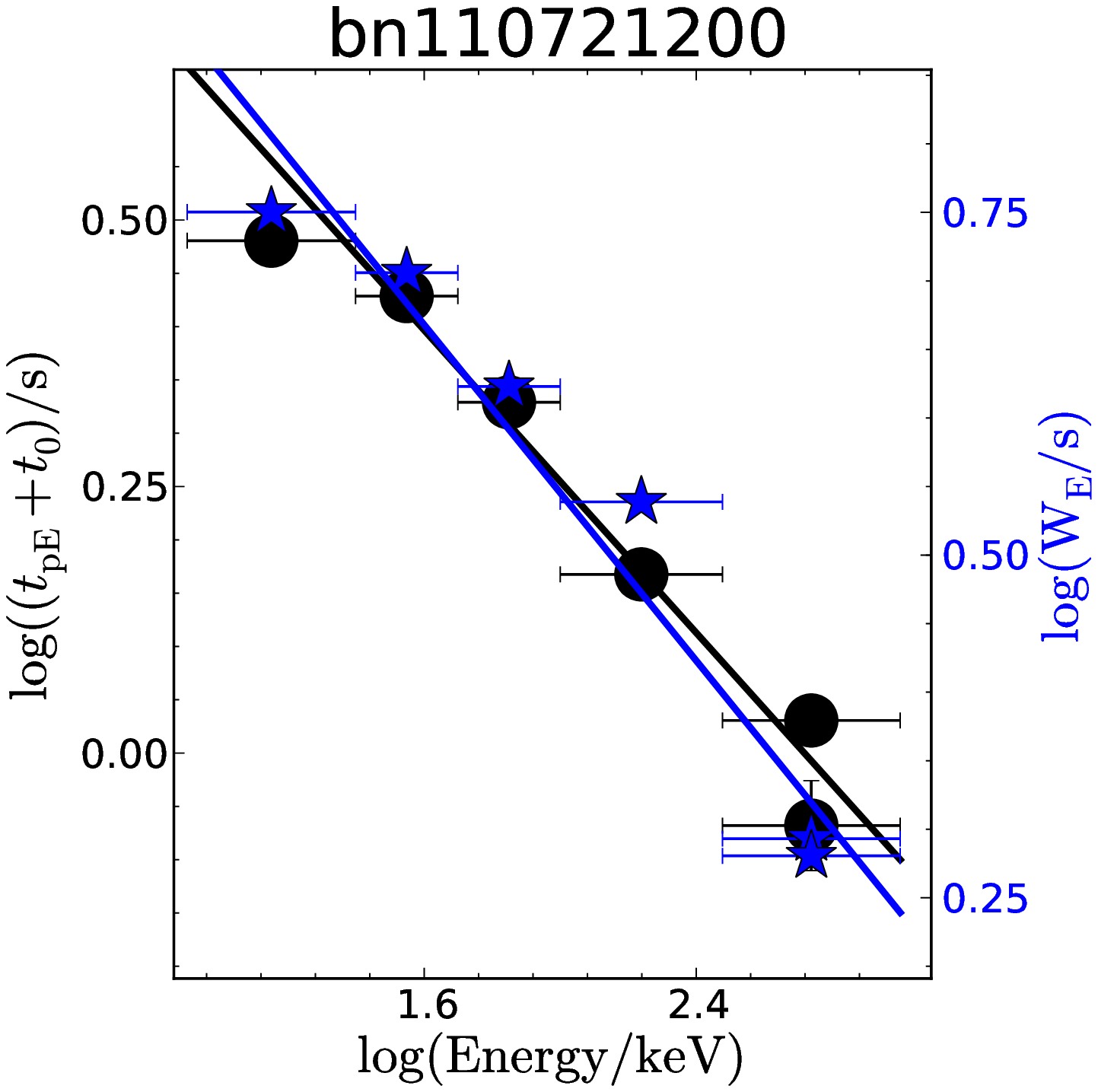} &
\includegraphics[origin=c,angle=0,scale=1,width=0.50\textwidth,height=0.20\textheight]{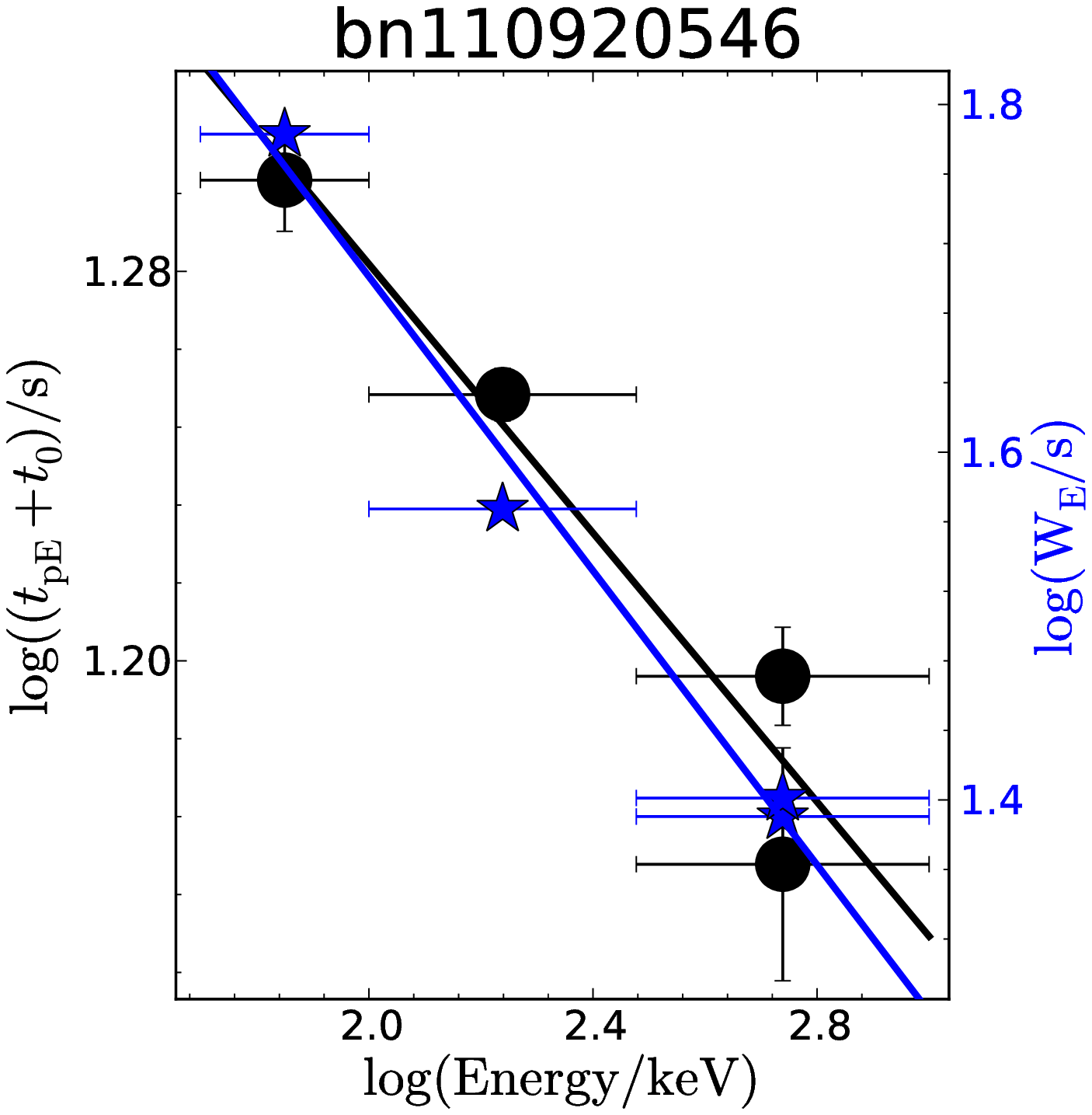} \\
\end{tabular}
\hfill
\vspace{4em}
\caption{Relations of $t_{\rm p_E}$-$E$ (black ``$\bullet$'', left $y$-axis)
and $W_{E}$-$E$ (blue ``$\star$'', right $y$-axis) for 30 pulses in our sample.
The two solid lines are the best linear fittings in logarithmic scale.}\label{myfigF}
\end{figure*}

\clearpage %p3

\begin{figure*}
\vspace{-4em}\hspace{-4em}
\centering
\begin{tabular}{cc}
\includegraphics[origin=c,angle=0,scale=1,width=0.50\textwidth,height=0.20\textheight]{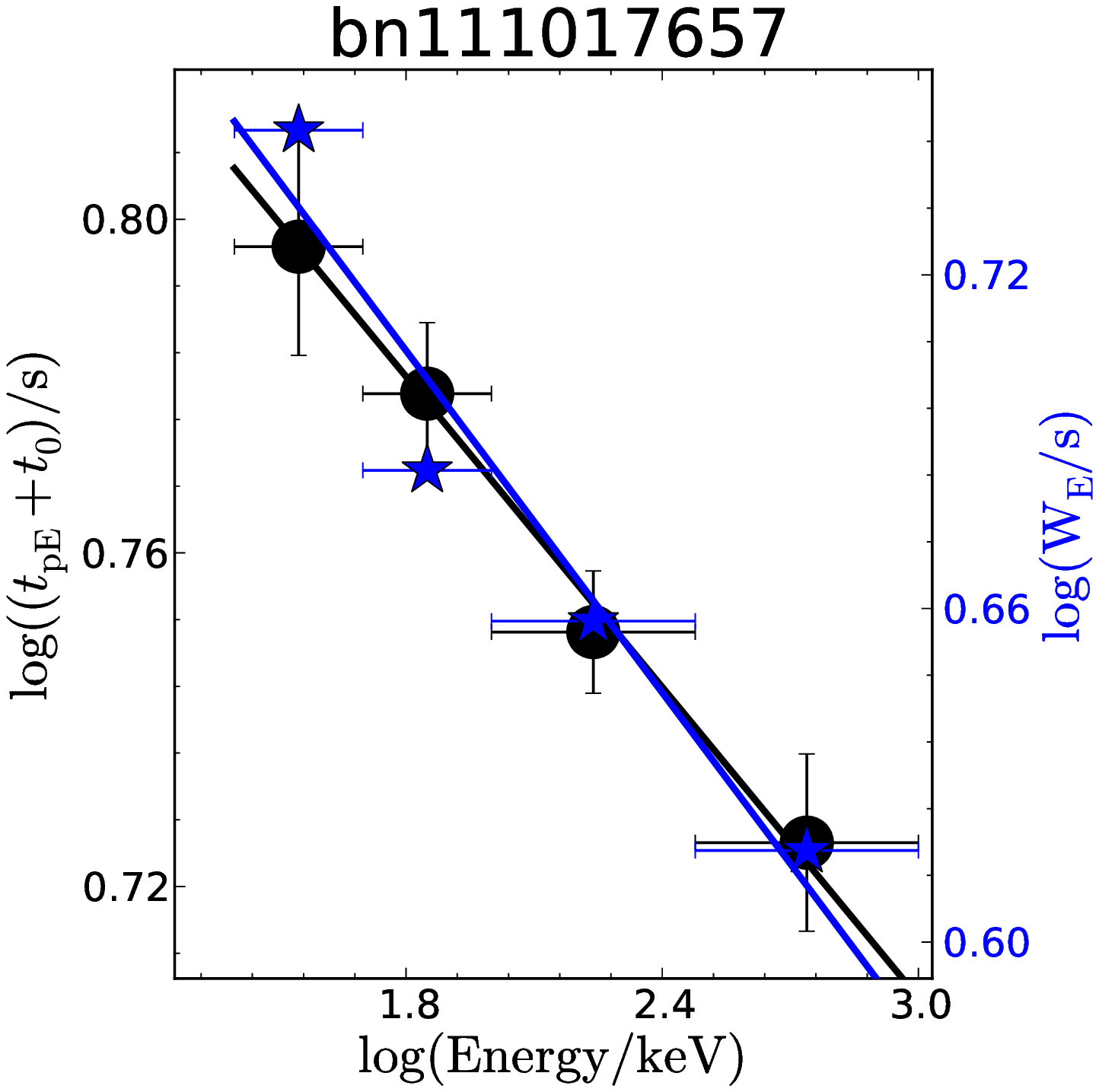} &
\includegraphics[origin=c,angle=0,scale=1,width=0.50\textwidth,height=0.20\textheight]{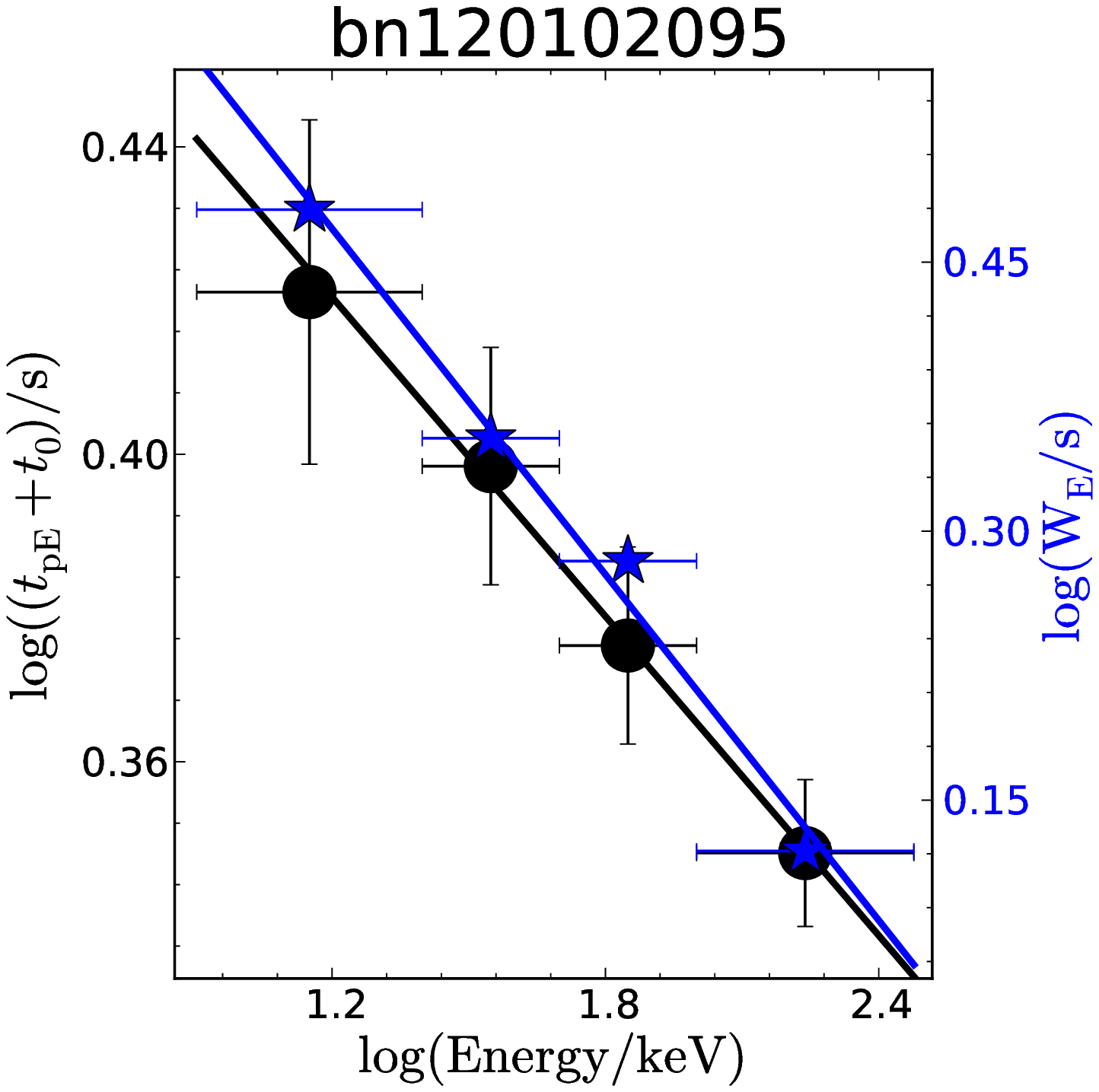} \\
\includegraphics[origin=c,angle=0,scale=1,width=0.50\textwidth,height=0.20\textheight]{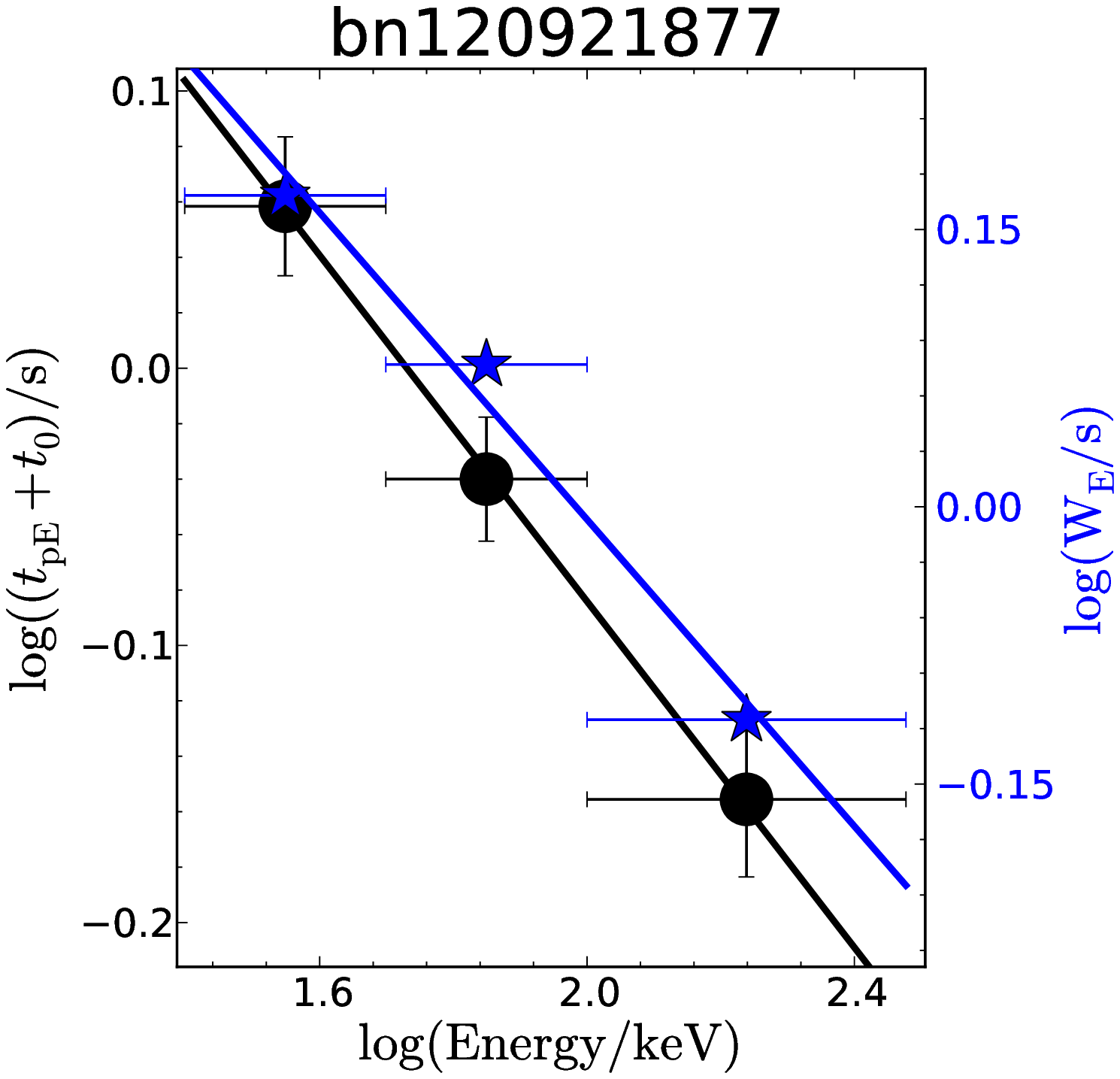} &
\includegraphics[origin=c,angle=0,scale=1,width=0.50\textwidth,height=0.20\textheight]{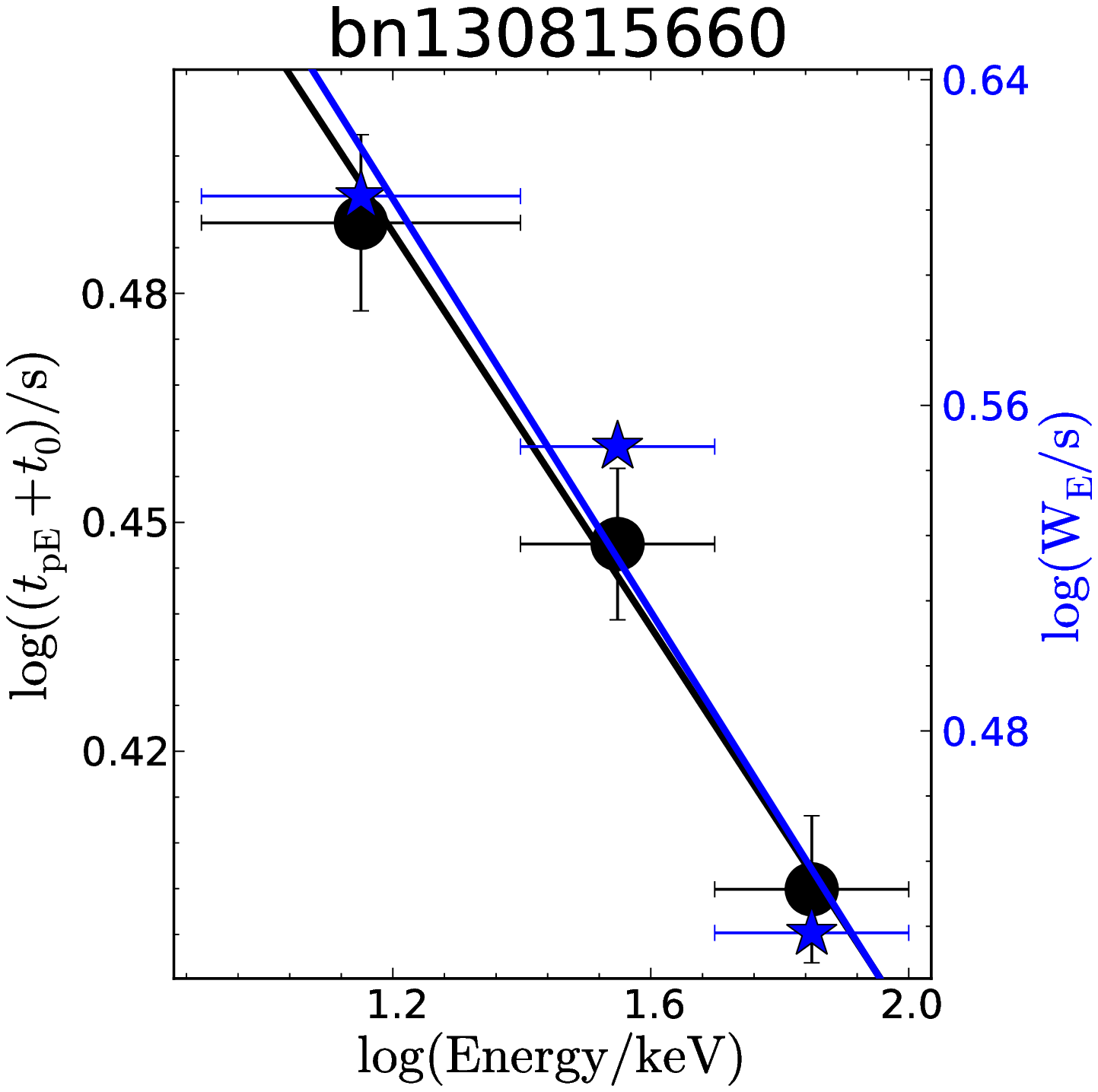} \\
\includegraphics[origin=c,angle=0,scale=1,width=0.50\textwidth,height=0.20\textheight]{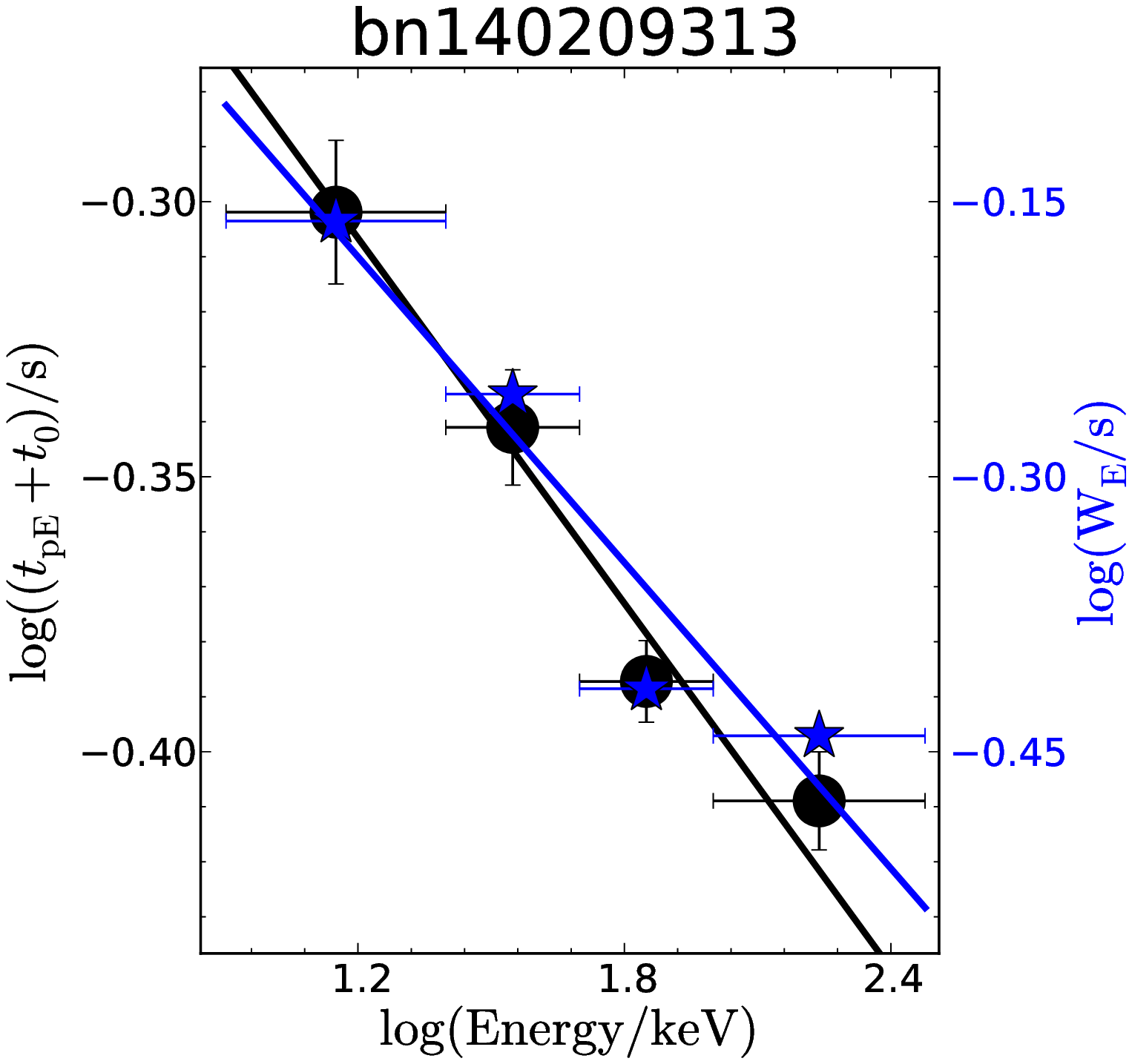} &
\includegraphics[origin=c,angle=0,scale=1,width=0.50\textwidth,height=0.20\textheight]{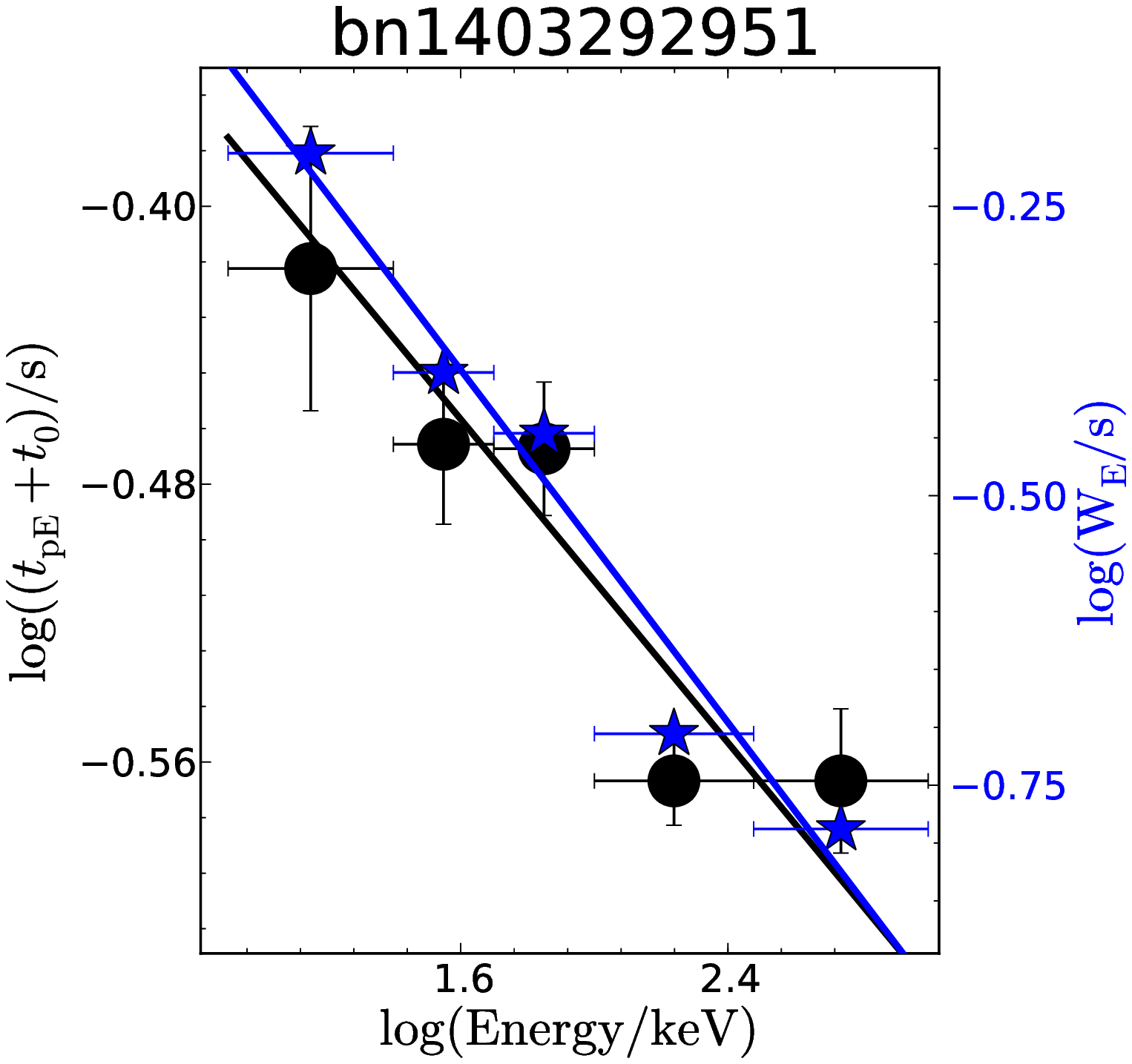} \\
\includegraphics[origin=c,angle=0,scale=1,width=0.50\textwidth,height=0.20\textheight]{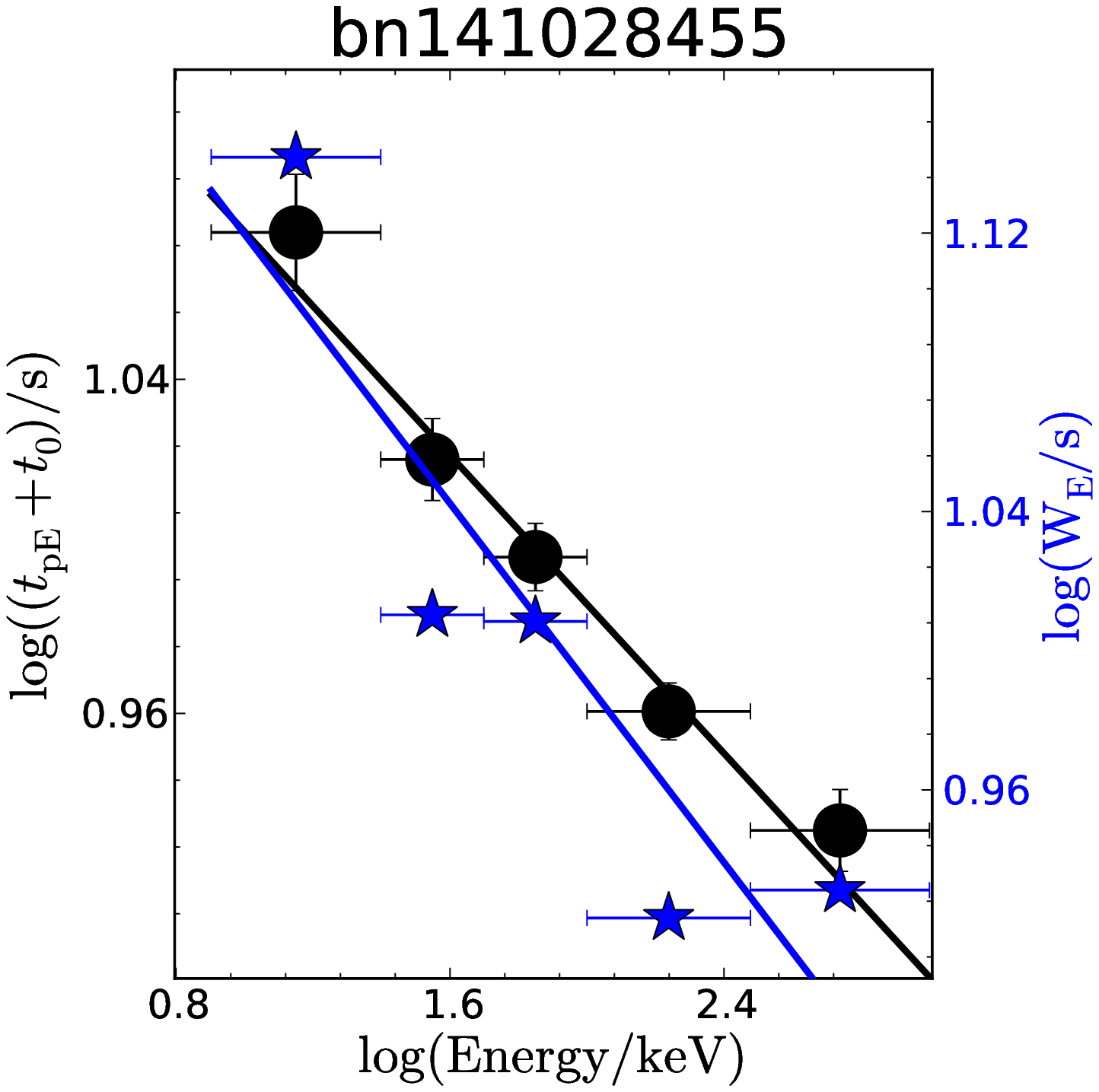} &
\includegraphics[origin=c,angle=0,scale=1,width=0.50\textwidth,height=0.20\textheight]{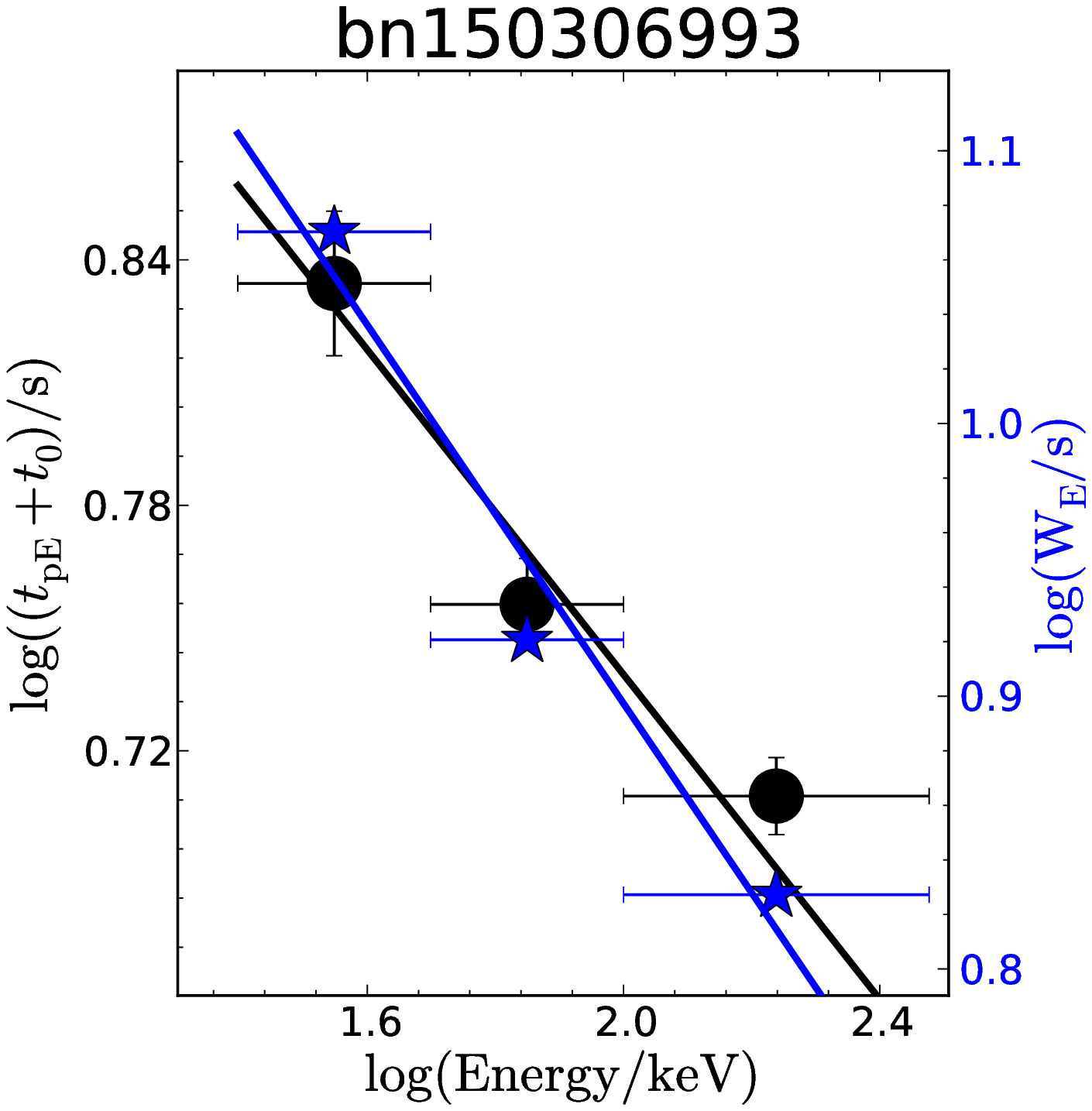} \\
\includegraphics[origin=c,angle=0,scale=1,width=0.50\textwidth,height=0.20\textheight]{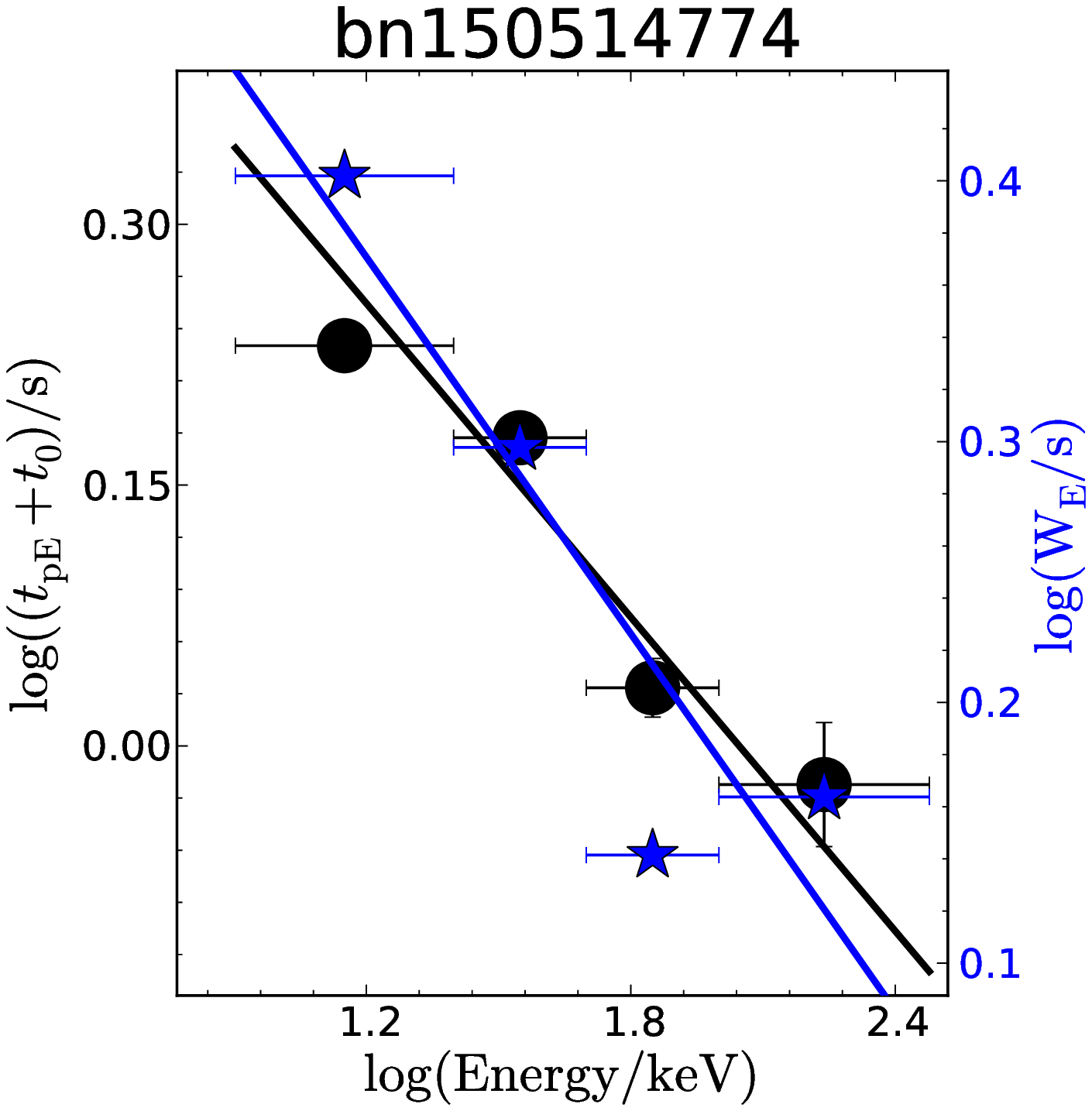} &
\includegraphics[origin=c,angle=0,scale=1,width=0.50\textwidth,height=0.20\textheight]{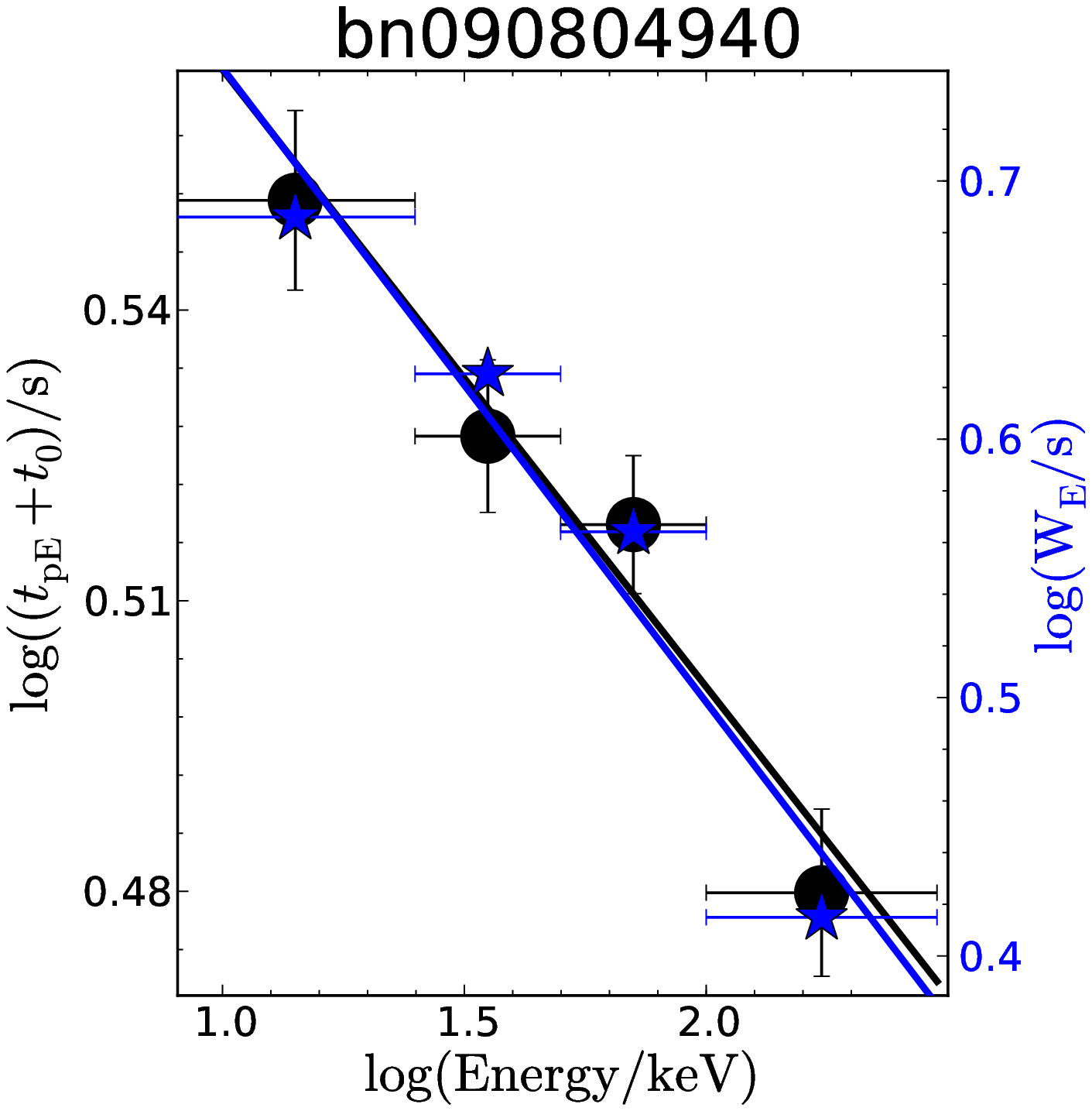} \\
\end{tabular}
\hfill
\vspace{4em}
\center{Fig. \ref{myfigF}---Continued.}
\end{figure*}

\clearpage %p5

\begin{figure*}
\vspace{-4em}\hspace{-4em}
\centering
\begin{tabular}{cc}
\includegraphics[origin=c,angle=0,scale=1,width=0.50\textwidth,height=0.20\textheight]{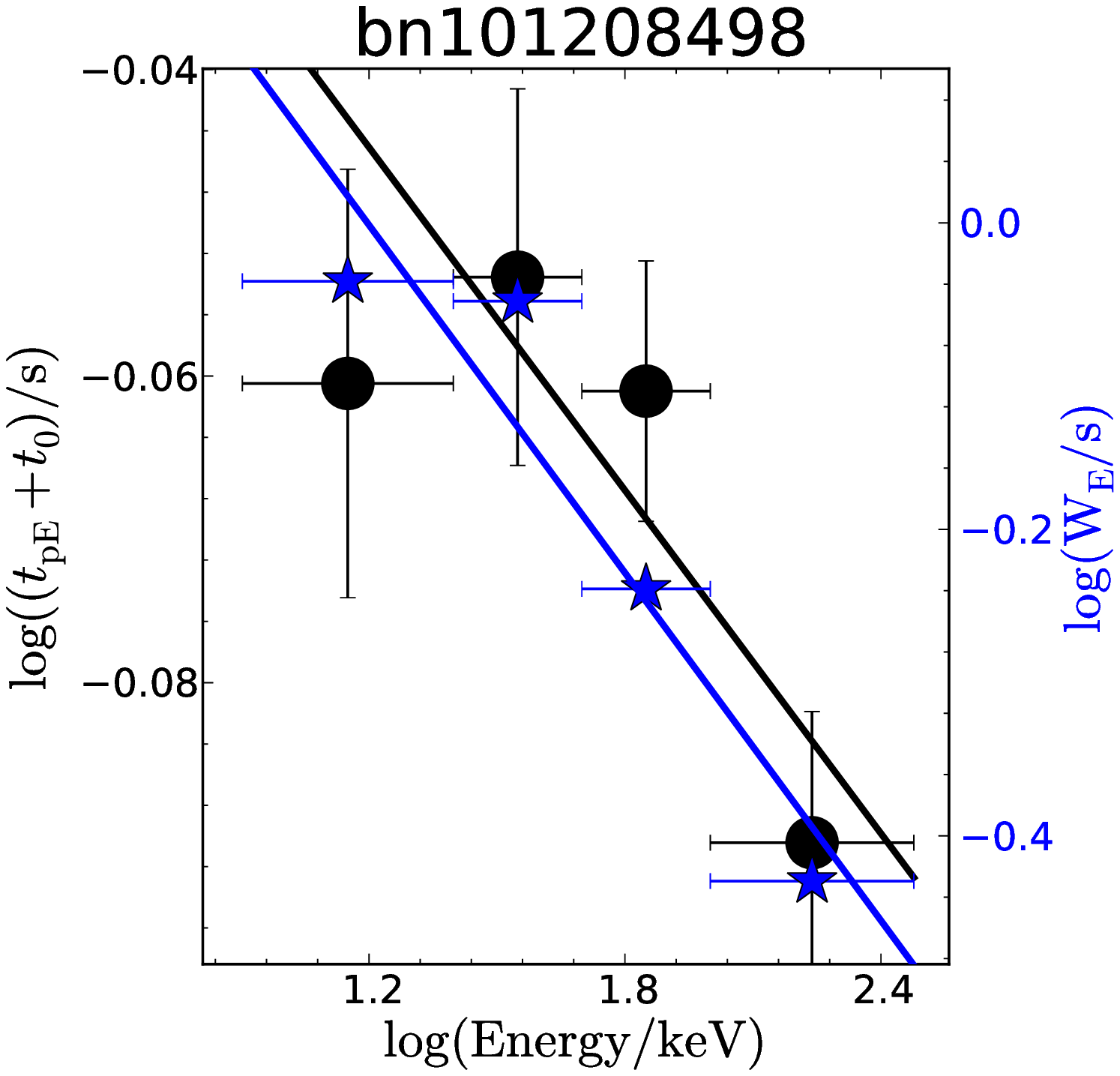} &
\includegraphics[origin=c,angle=0,scale=1,width=0.50\textwidth,height=0.20\textheight]{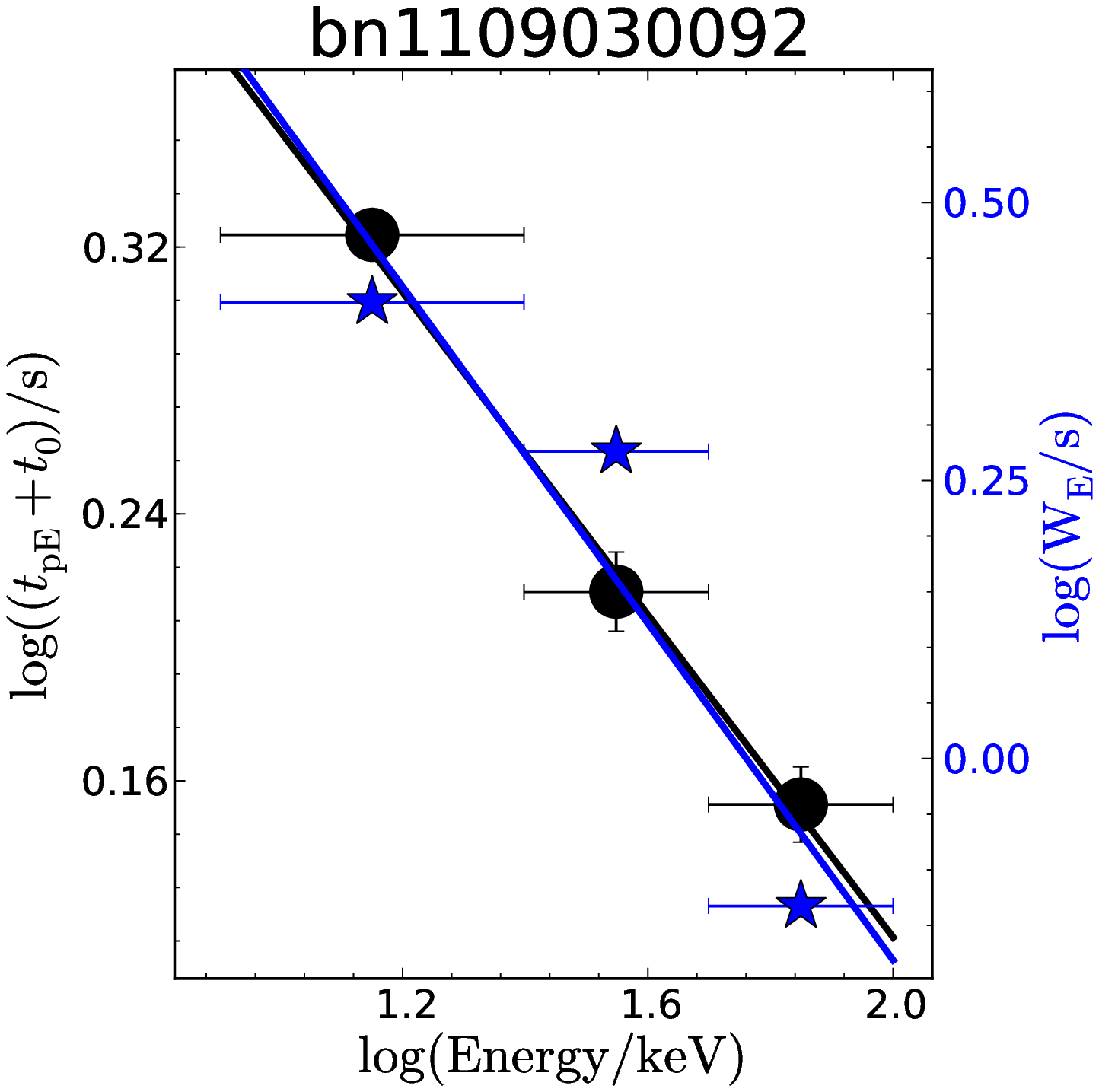} \\
\includegraphics[origin=c,angle=0,scale=1,width=0.50\textwidth,height=0.20\textheight]{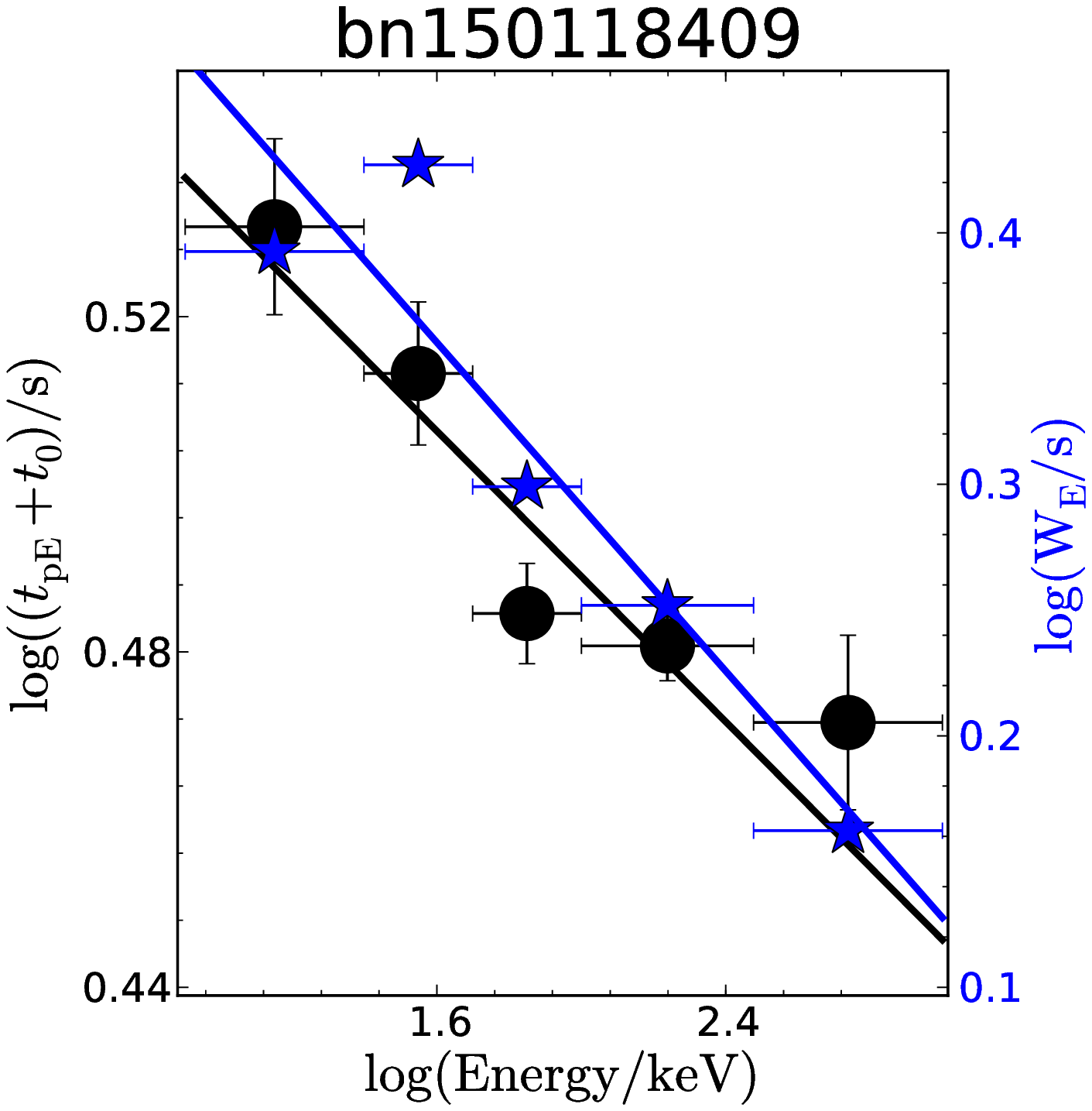} &
\includegraphics[origin=c,angle=0,scale=1,width=0.50\textwidth,height=0.20\textheight]{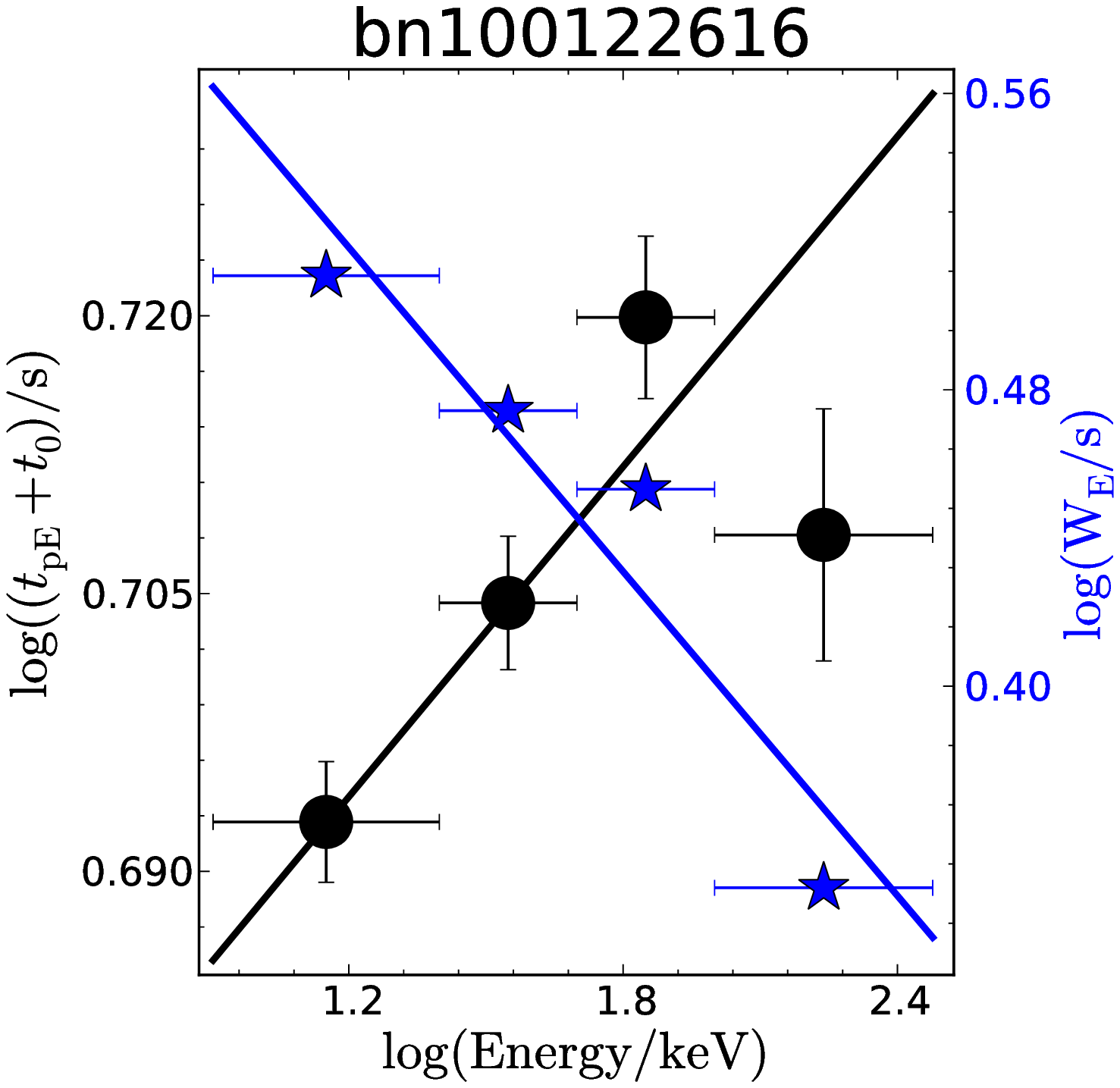} \\
\includegraphics[origin=c,angle=0,scale=1,width=0.50\textwidth,height=0.20\textheight]{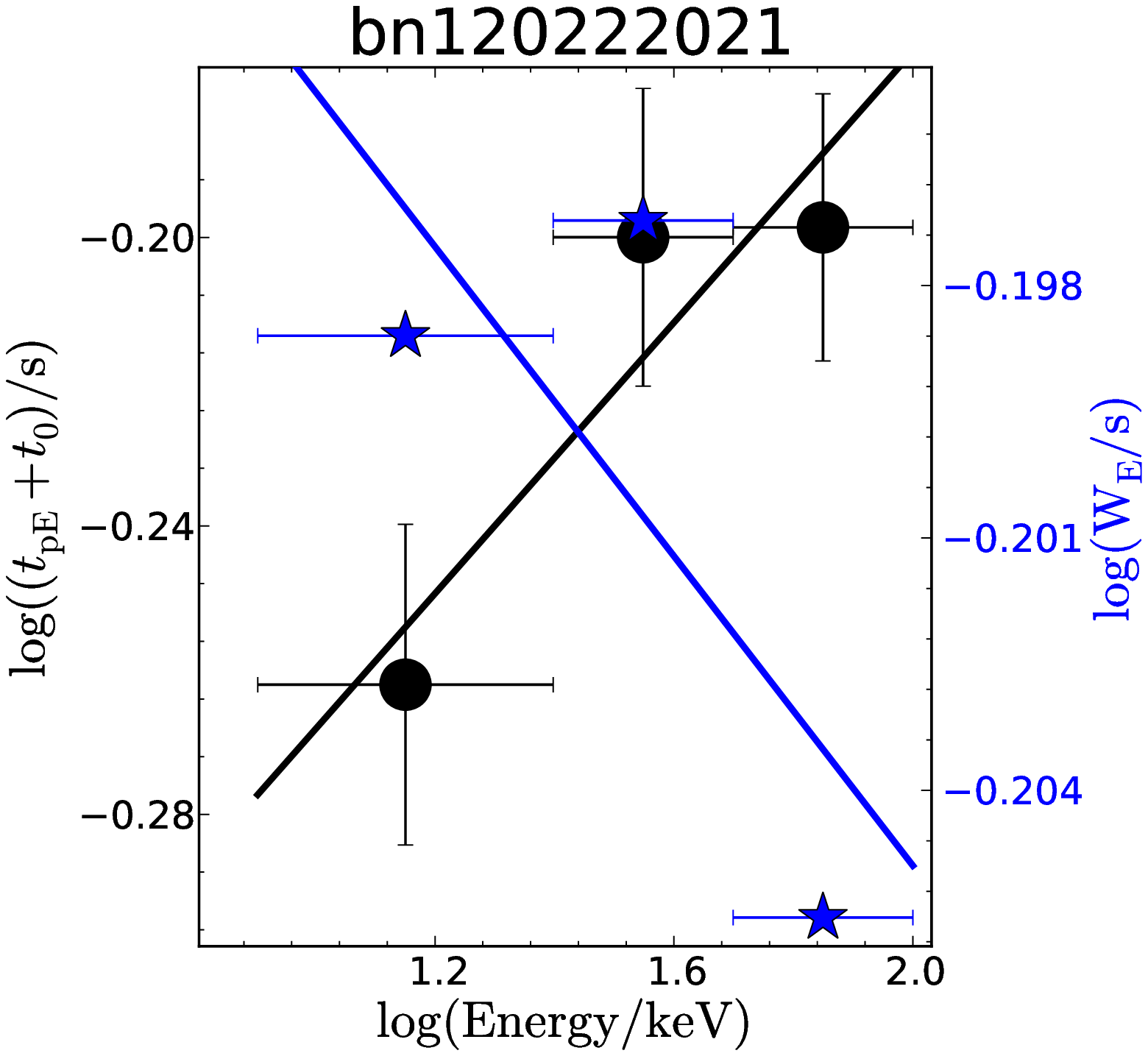} &
\includegraphics[origin=c,angle=0,scale=1,width=0.50\textwidth,height=0.20\textheight]{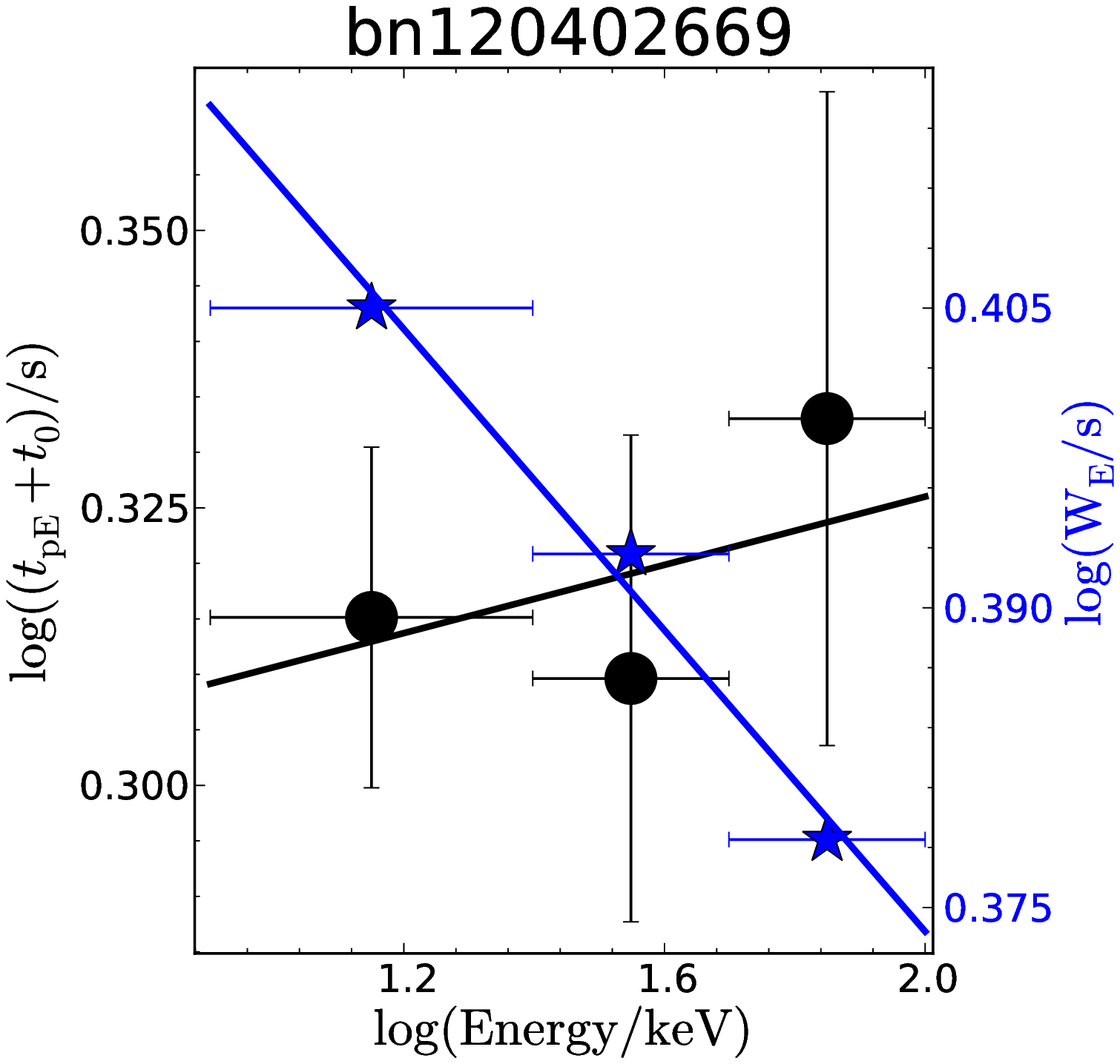} \\
\includegraphics[origin=c,angle=0,scale=1,width=0.50\textwidth,height=0.20\textheight]{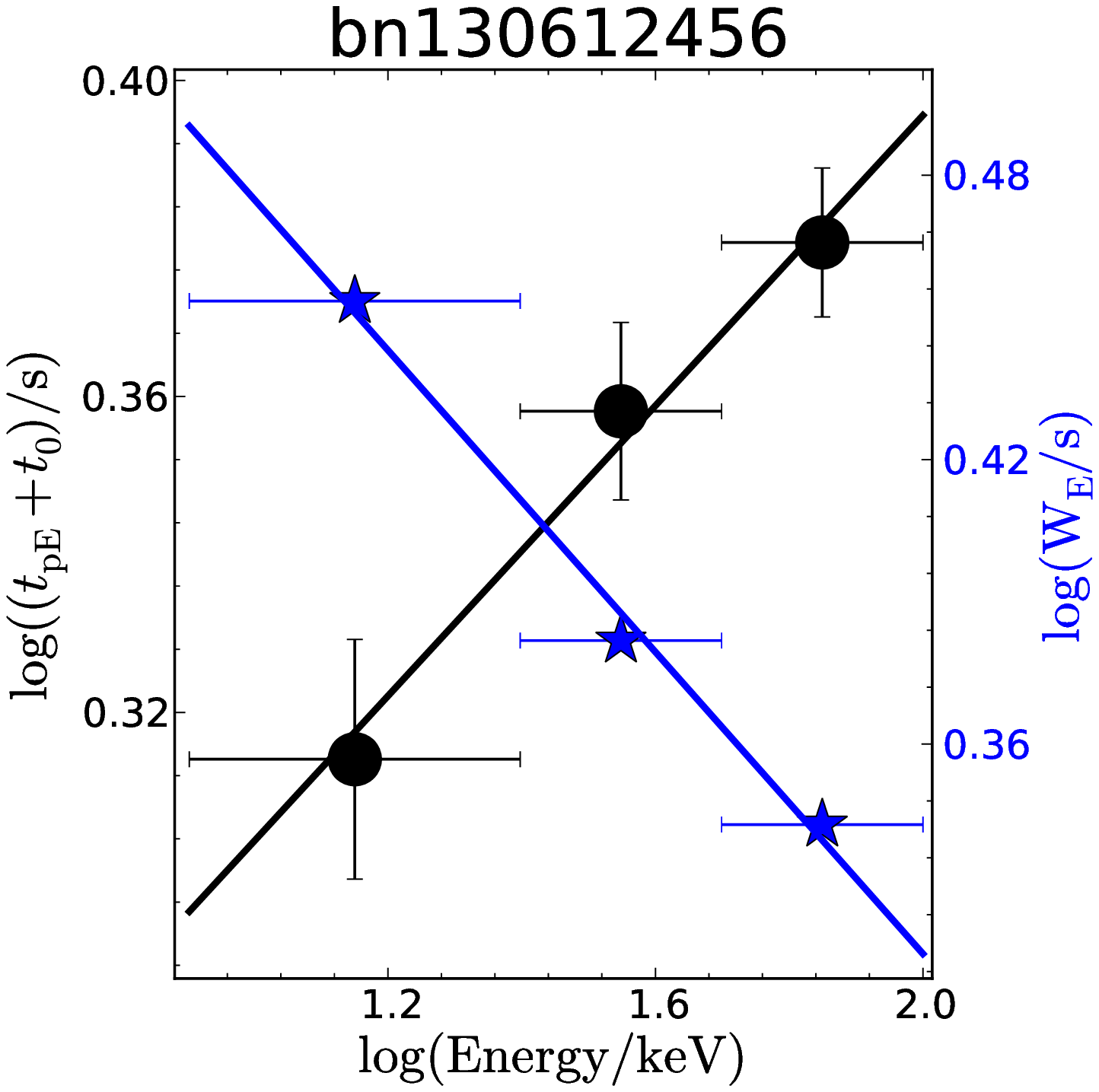} &
\includegraphics[origin=c,angle=0,scale=1,width=0.50\textwidth,height=0.20\textheight]{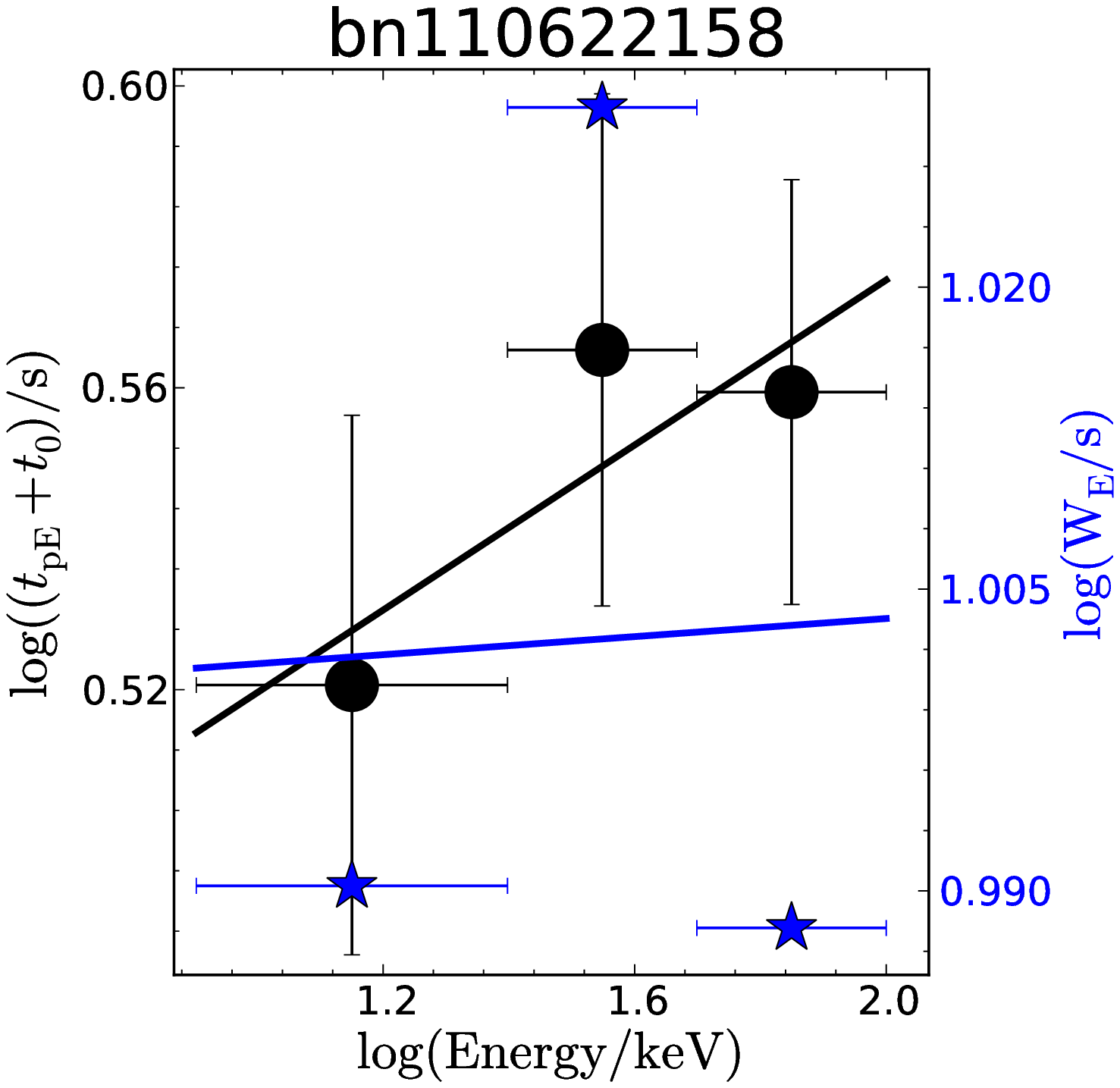} \\
\includegraphics[origin=c,angle=0,scale=1,width=0.50\textwidth,height=0.20\textheight]{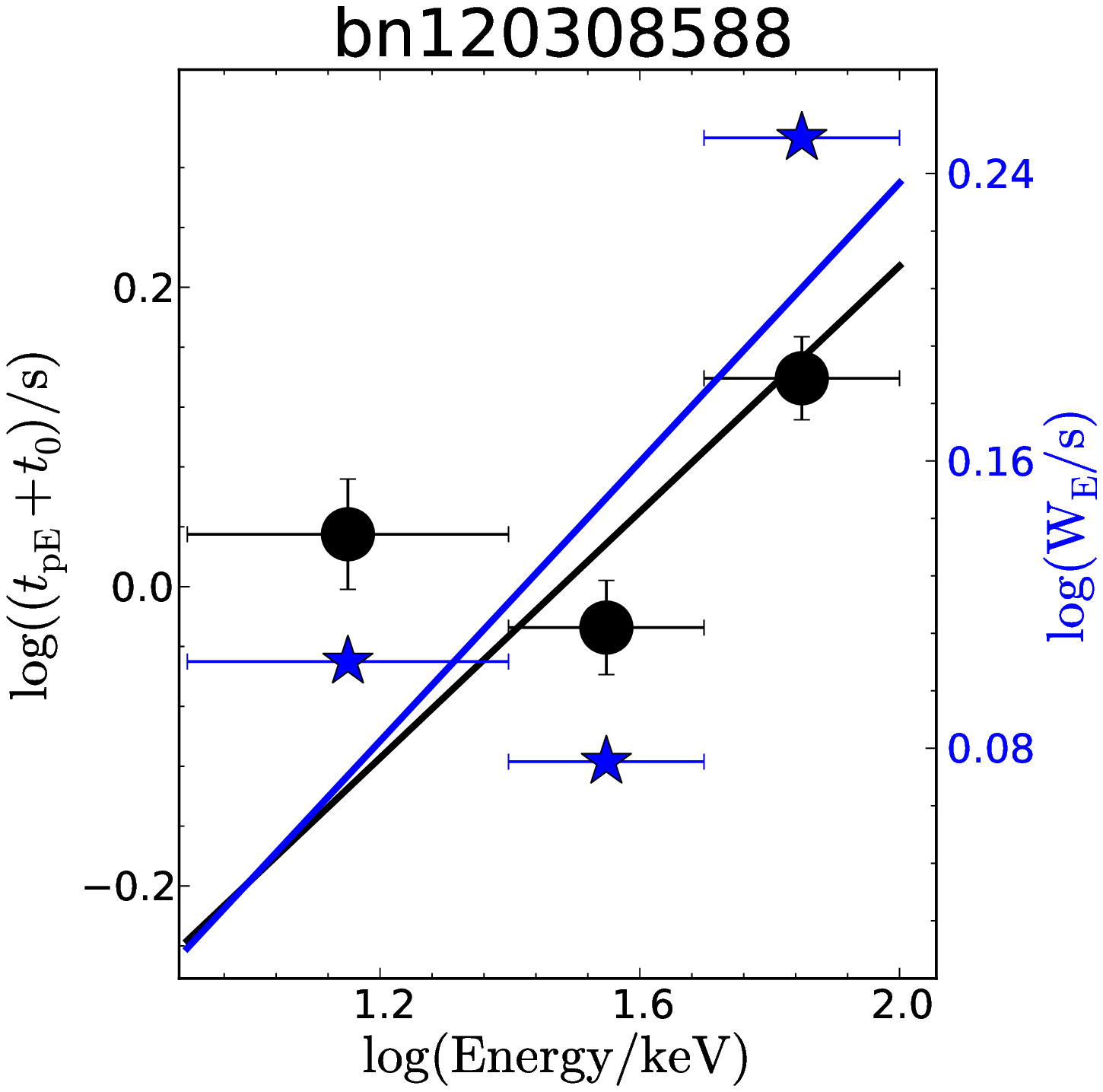} &
\includegraphics[origin=c,angle=0,scale=1,width=0.50\textwidth,height=0.20\textheight]{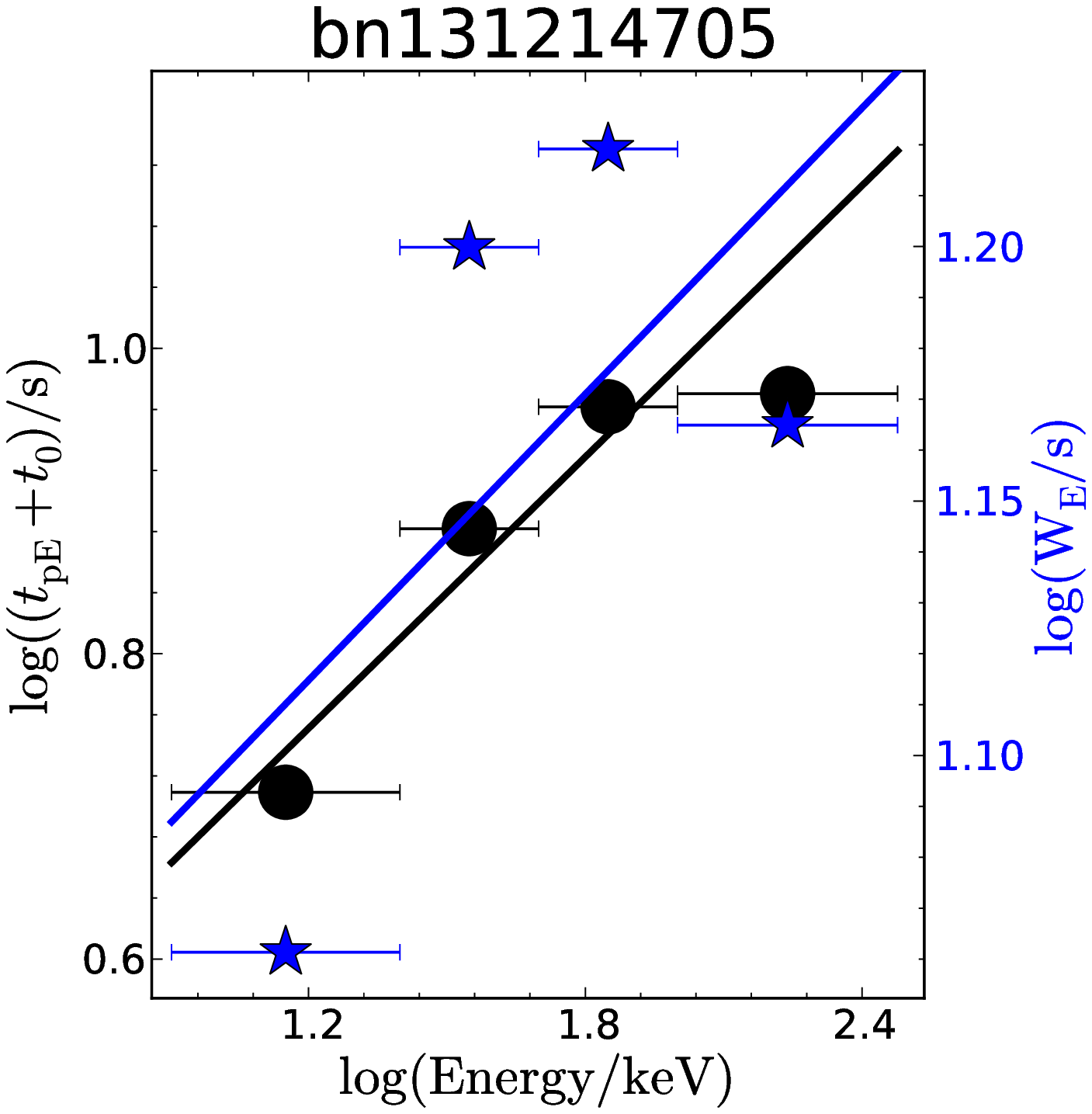} \\
\end{tabular}
\hfill
\vspace{4em}
\center{Fig. \ref{myfigF}---Continued.}
\end{figure*}

%%%%%%%%%%%%%%%%%%%%%%%%%%%%%%%%%%%%%%%%%%%%%%%%%%%%%%%%%%%%%%%%%%%%%%%%%%%%%%
%%%%%%%%%%%%%%%%%%%%%%%%%%%%%%%%%%%%%%%%%%%%%%%%%0%%%%%%%%%%%%%%%%%%%%%%%%%%%%%
%%%  Figure 7
\clearpage
\begin{figure*}\label{Twoslopes}
\centering
\begin{tabular}{cc}
\includegraphics[origin=c,angle=0,scale=1,width=0.48\textwidth,height=0.30\textheight]{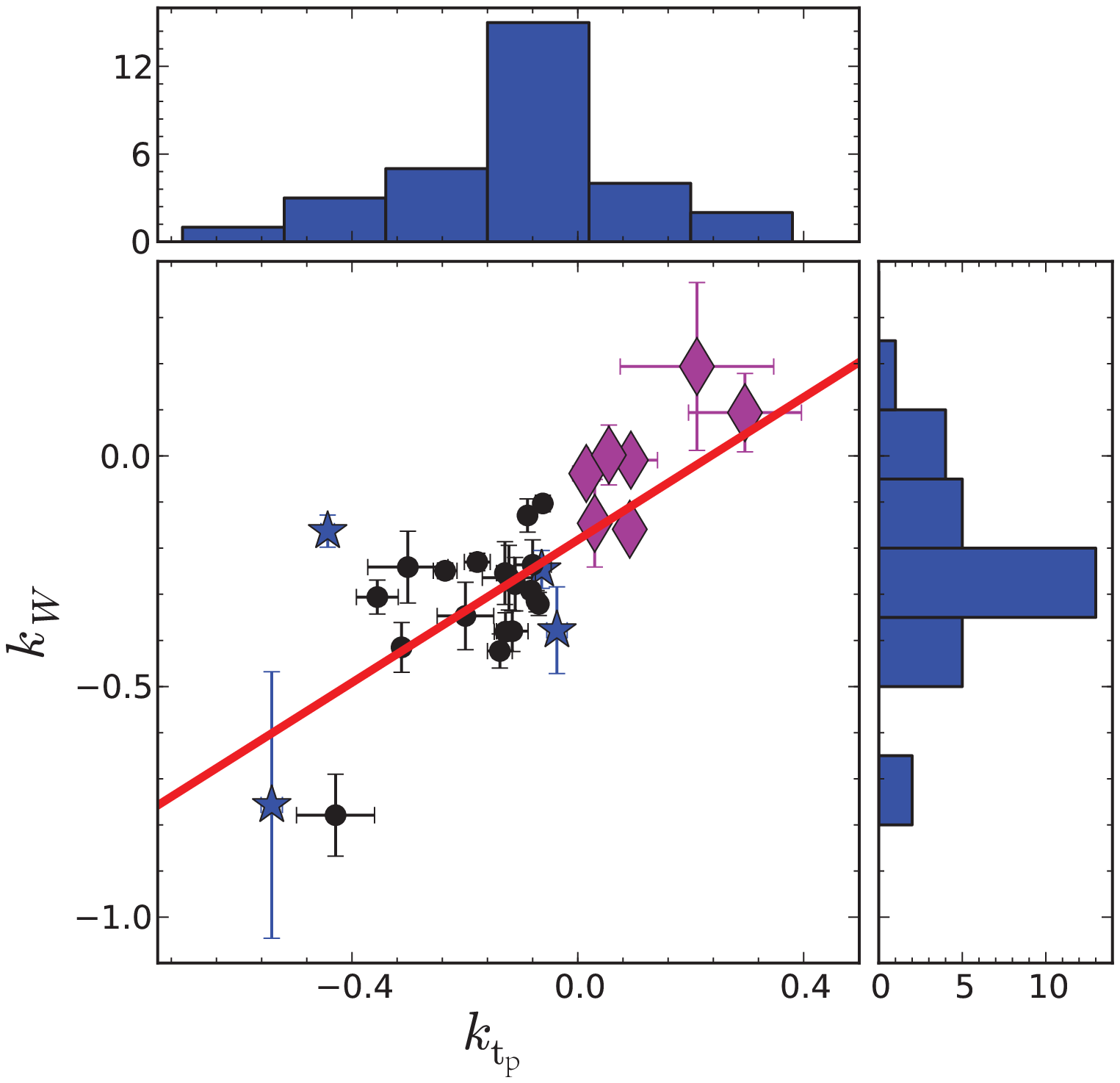} &
\includegraphics[origin=c,angle=0,scale=1,width=0.48\textwidth,height=0.30\textheight]{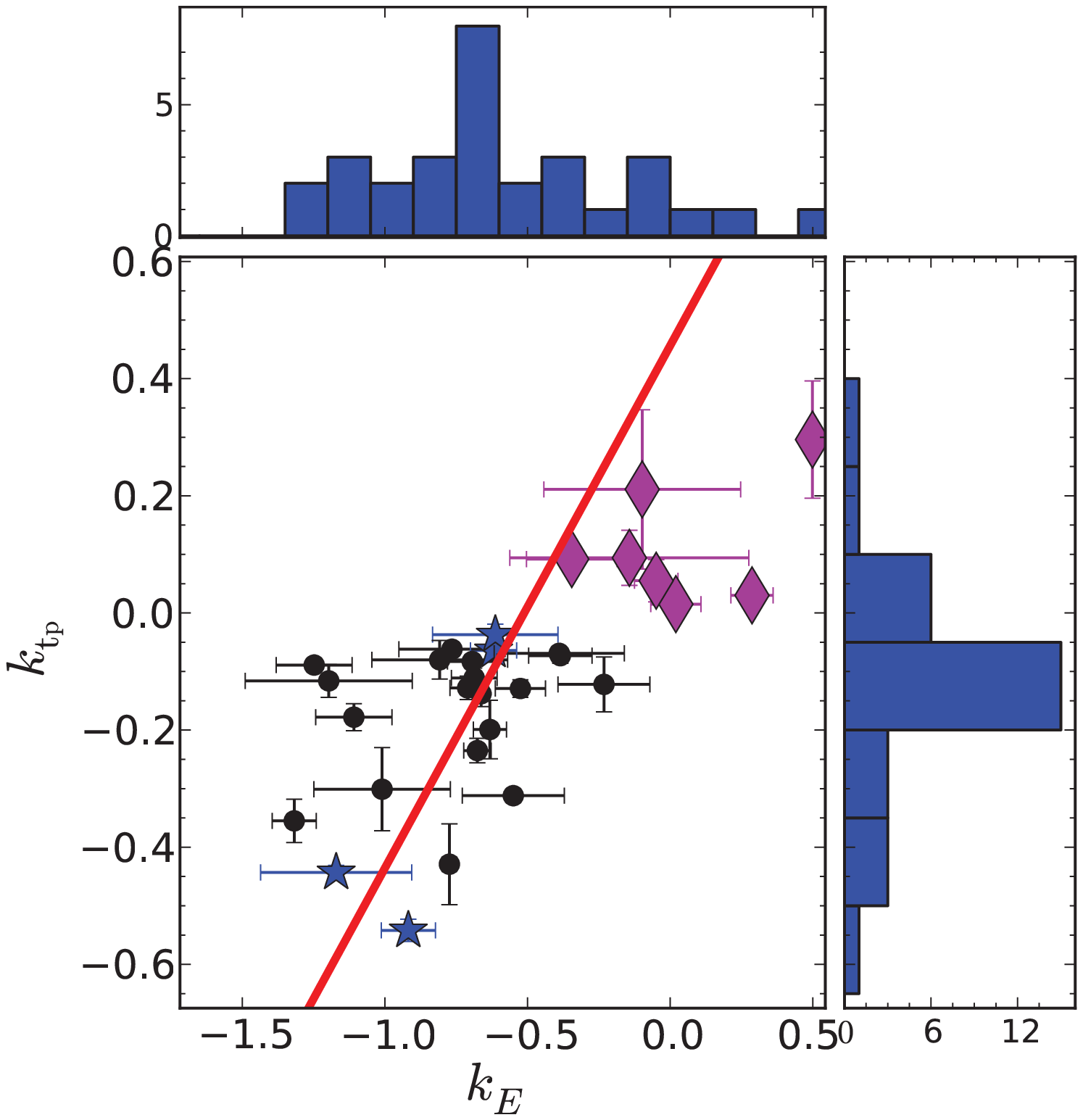} \\
\end{tabular}
\caption{Relations of $k_{W}$-$k_{t_p}$ and $k_{t_p}$-$k_{E}$
together with $k_{W}$, $k_{t_p}$, and $k_{E}$ distributions,
where the circles, stars and diamonds stand for the H2S pulses, H2S-dominated-tracking pulses, and the tracking pulses, respectively.
The red solid lines are the tentative correlations for H2S, H2S-dominated-tracking pulses and tracking pulses.}\label{myfigG}
\end{figure*}

%%%%%%%%%%%%%%%%%%%%%%%%%%%%%%%%%%%%%%%%%%%%%%%%%%%%%%%%%%%%%%%%%%%%%%%%%%%%%%
%%%%%%%%%%%%%%%%%%%%%%%%%%%%%%%%%%%%%%%%%%%%%%%%%%%%%%%%%%%%%%%%%%%%%%%%%%%%%%
%%%  Figure 8
\clearpage
\begin{figure*}\label{figH2S}
\vspace{-4em}\hspace{-4em}
\centering
\begin{tabular}{ccc}
\includegraphics[origin=c,angle=0,scale=1,width=0.30\textwidth,height=0.30\textheight]{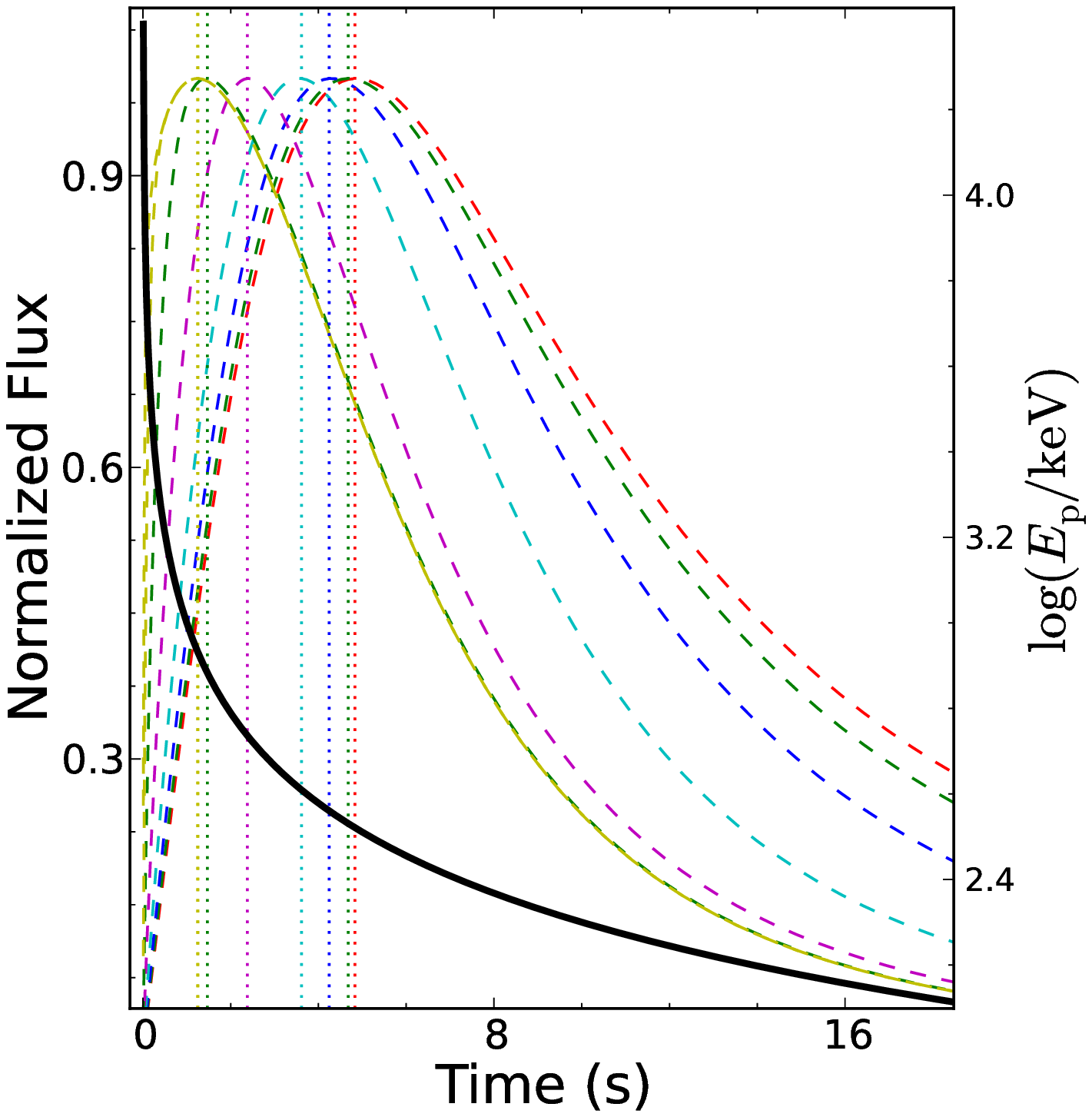} &
\includegraphics[origin=c,angle=0,scale=1,width=0.30\textwidth,height=0.30\textheight]{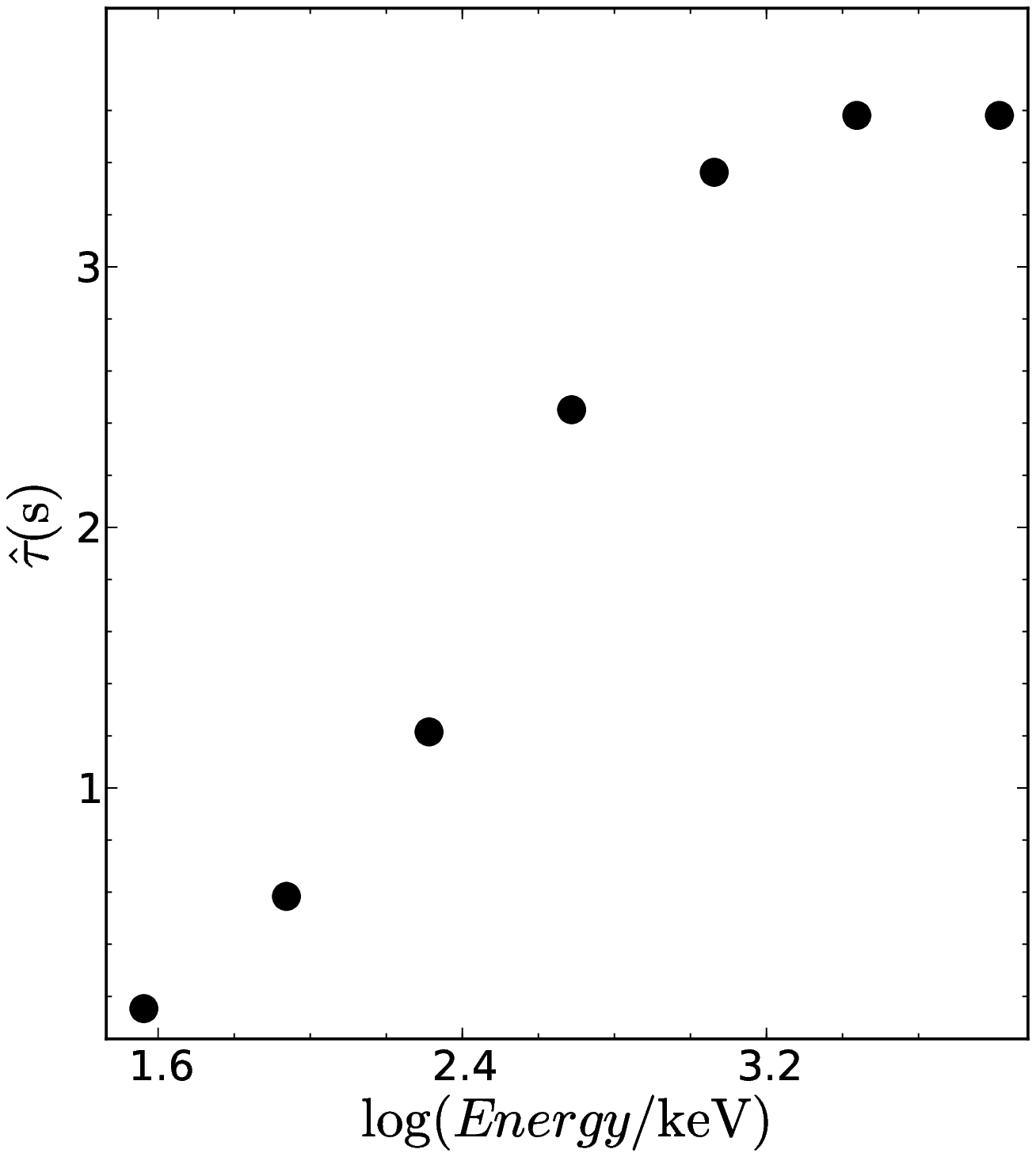} &
\includegraphics[origin=c,angle=0,scale=1,width=0.30\textwidth,height=0.30\textheight]{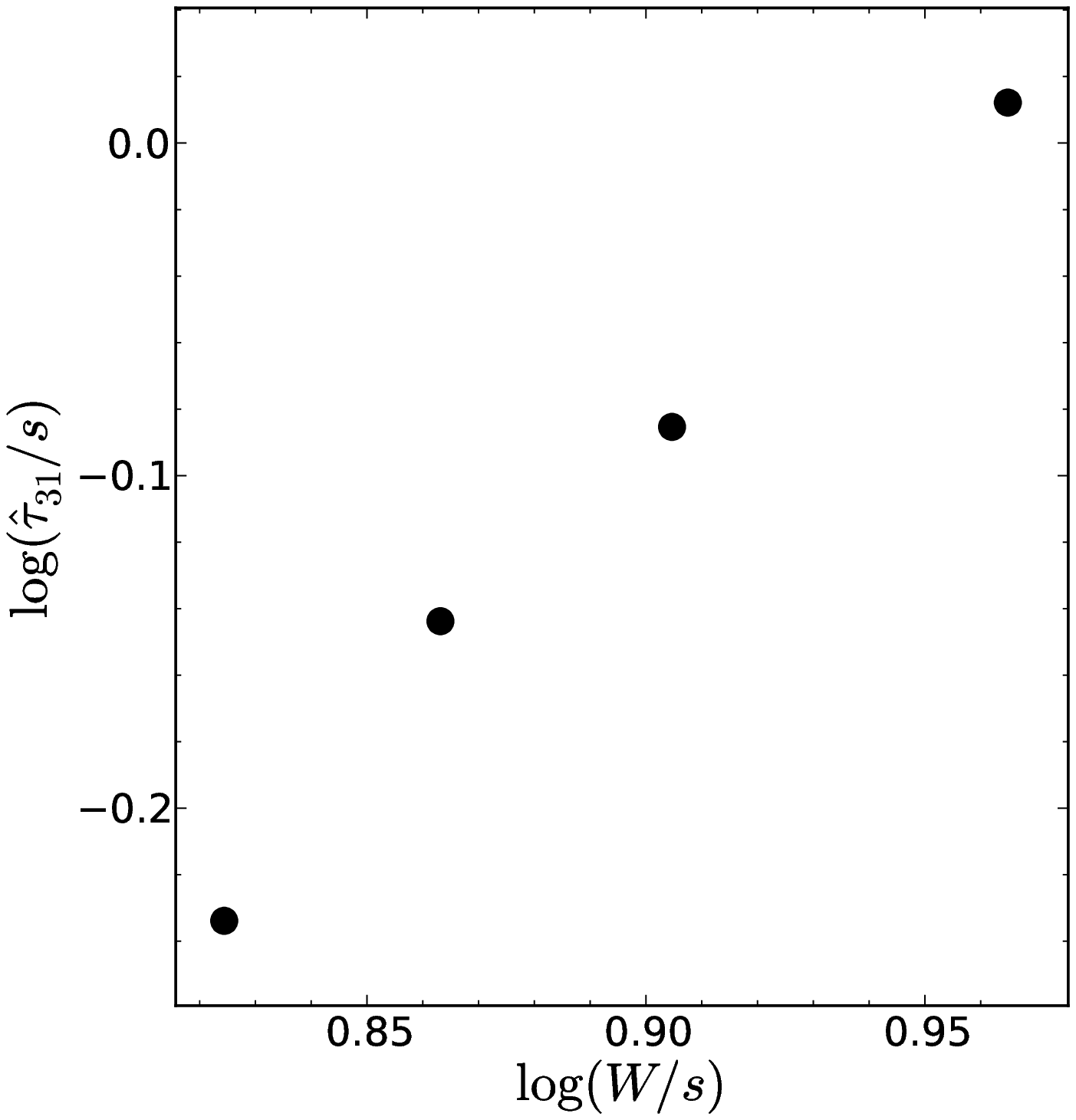} \\
\end{tabular}
\hfill
\vspace{4em}
\caption{Left panel: Simulated light curves of the pulse
(dashed lines, left $y$-axis) in different energy bands
along with H2S's $E_{p}$ evolution (thick solid line, right y-axis),
where the vertical dotted lines mark the peak times of each pulses.
Middle panel: the relation between the spectral lag (${\hat{\tau}}$) and photon energy.
Right panel: the relation of the spectral lag (${\hat{\tau}}_{31}$) and the pulse width $W$.
Simulations with Equation (\ref{H2S}) and the parameters of $F_{\rm m}=1$,
$t_{\rm m}=5$, $r=1$, $d=2$, $t_{\rm 0}=0$, $\alpha=-1$, $\beta=-2.3$, $k=-1$ are adopted.}\label{myfigH}
\end{figure*}

%%%%%%%%%%%%%%%%%%%%%%%%%%%%%%%%%%%%%%%%%%%%%%%%%%%%%%%%%%%%%%%%%%%%%%%%%%%%%%%
%%%%%%%%%%%%%%%%%%%%%%%%%%%%%%%%%%%%%%%%%%%%%%%%%%%%%%%%%%%%%%%%%%%%%%%%%%%%%%%
%%%  Figure 9
\clearpage
\begin{figure*}
\vspace{-4em}\hspace{-4em}
\centering
\begin{tabular}{cc}
\includegraphics[origin=c,angle=0,scale=1,width=0.45\textwidth,height=0.30\textheight]{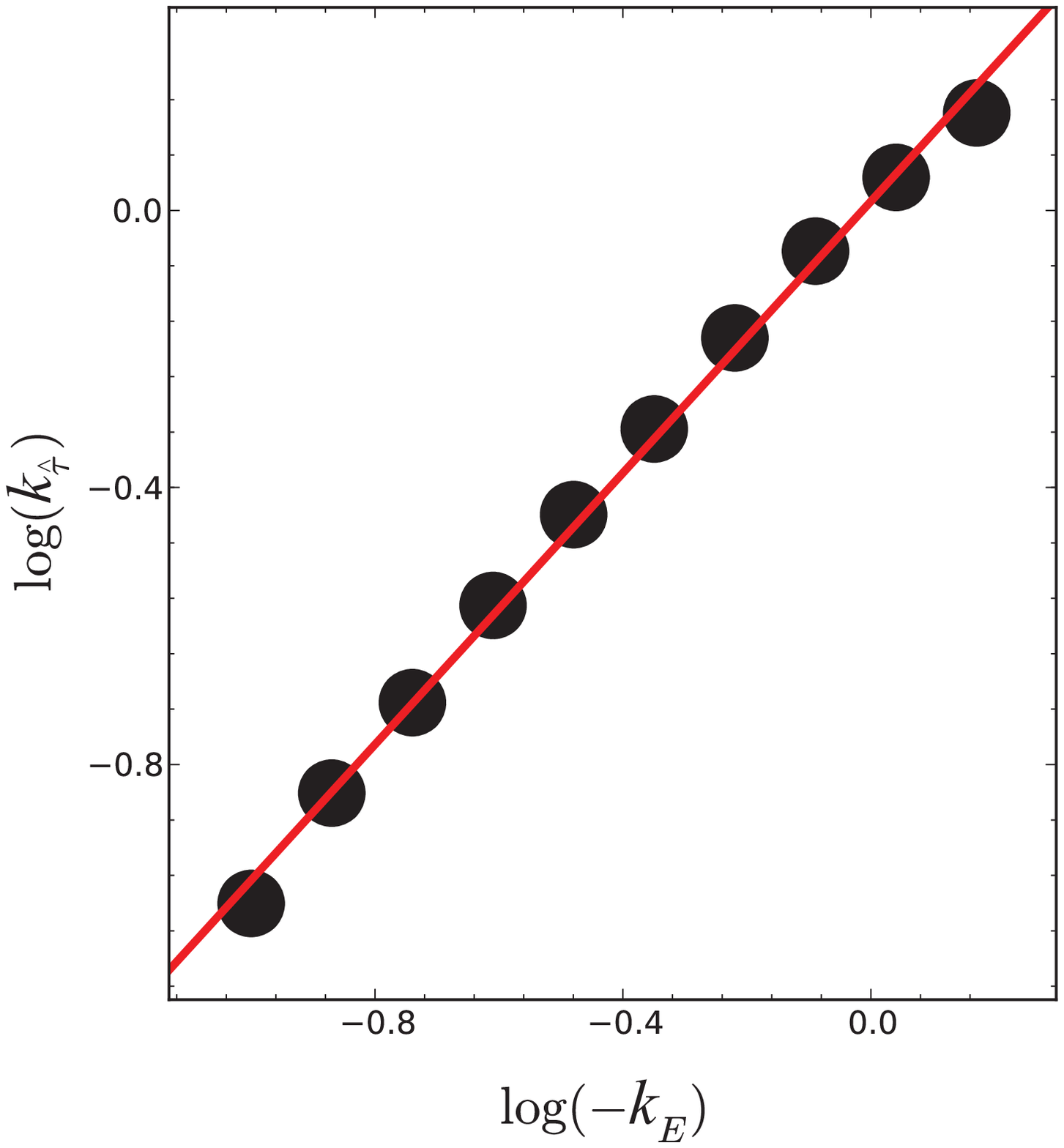} &
\includegraphics[origin=c,angle=0,scale=1,width=0.45\textwidth,height=0.30\textheight]{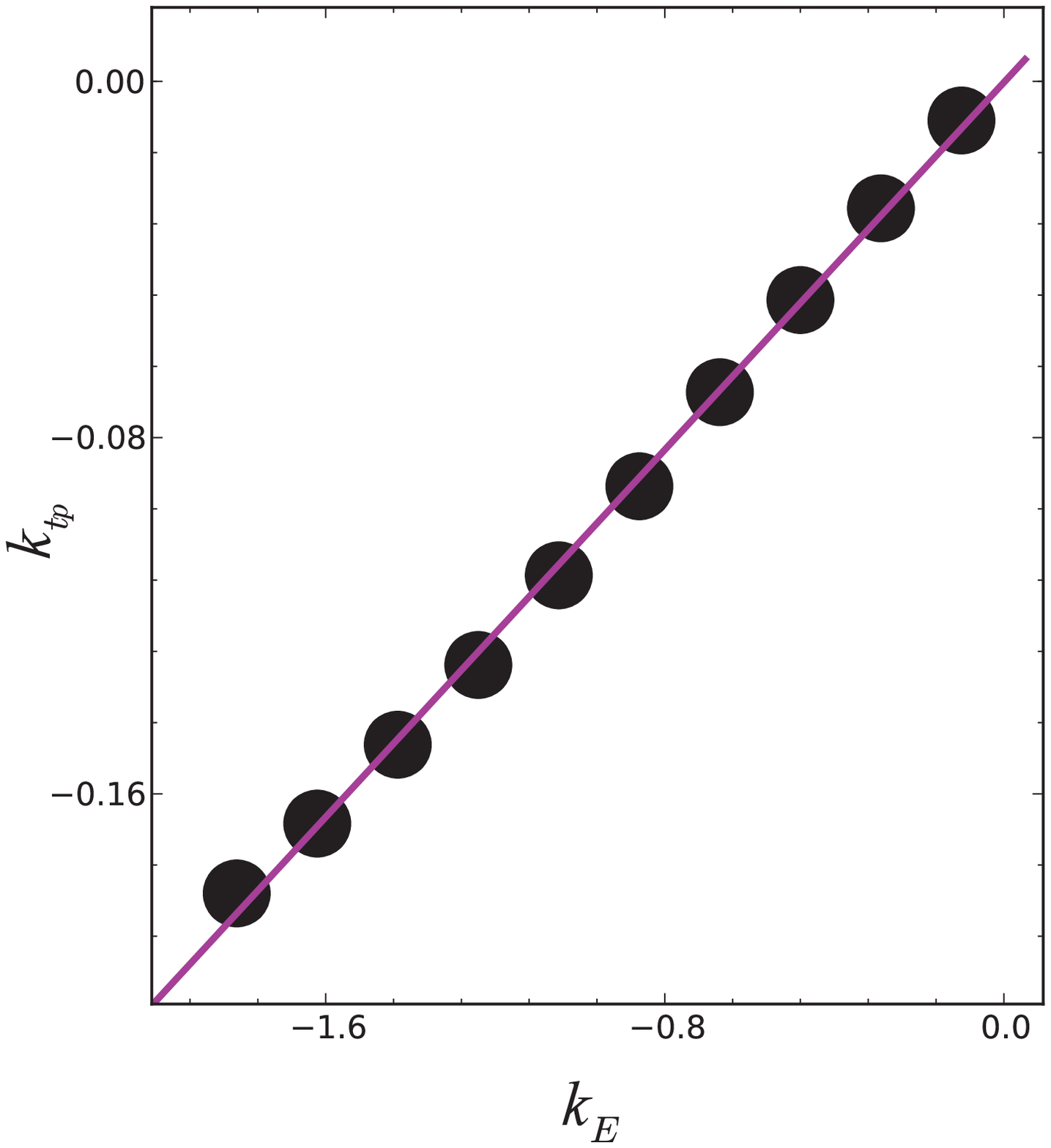} \\
\end{tabular}
\hfill
\vspace{4em}
\caption{Relations of $k_{\hat{\tau}}$-$k_E$ and $k_{t_p}$-$k_E$ from our numerical simulations with Case (I).}\label{myfigI}
\end{figure*}

%%%%%%%%%%%%%%%%%%%%%%%%%%%%%%%%%%%%%%%%%%%%%%%%%%%%%%%%%%%%%%%%%%%%%%%%%%%%%%
%%%%%%%%%%%%%%%%%%%%%%%%%%%%%%%%%%%%%%%%%%%%%%%%%%%%%%%%%%%%%%%%%%%%%%%%%%%%%%
%%  Figure 10

\clearpage
\begin{figure*}
\vspace{-4em}\hspace{-4em}
\centering
\begin{tabular}{c}
\includegraphics[origin=c,angle=0,scale=1,width=0.90\textwidth,height=0.30\textheight]{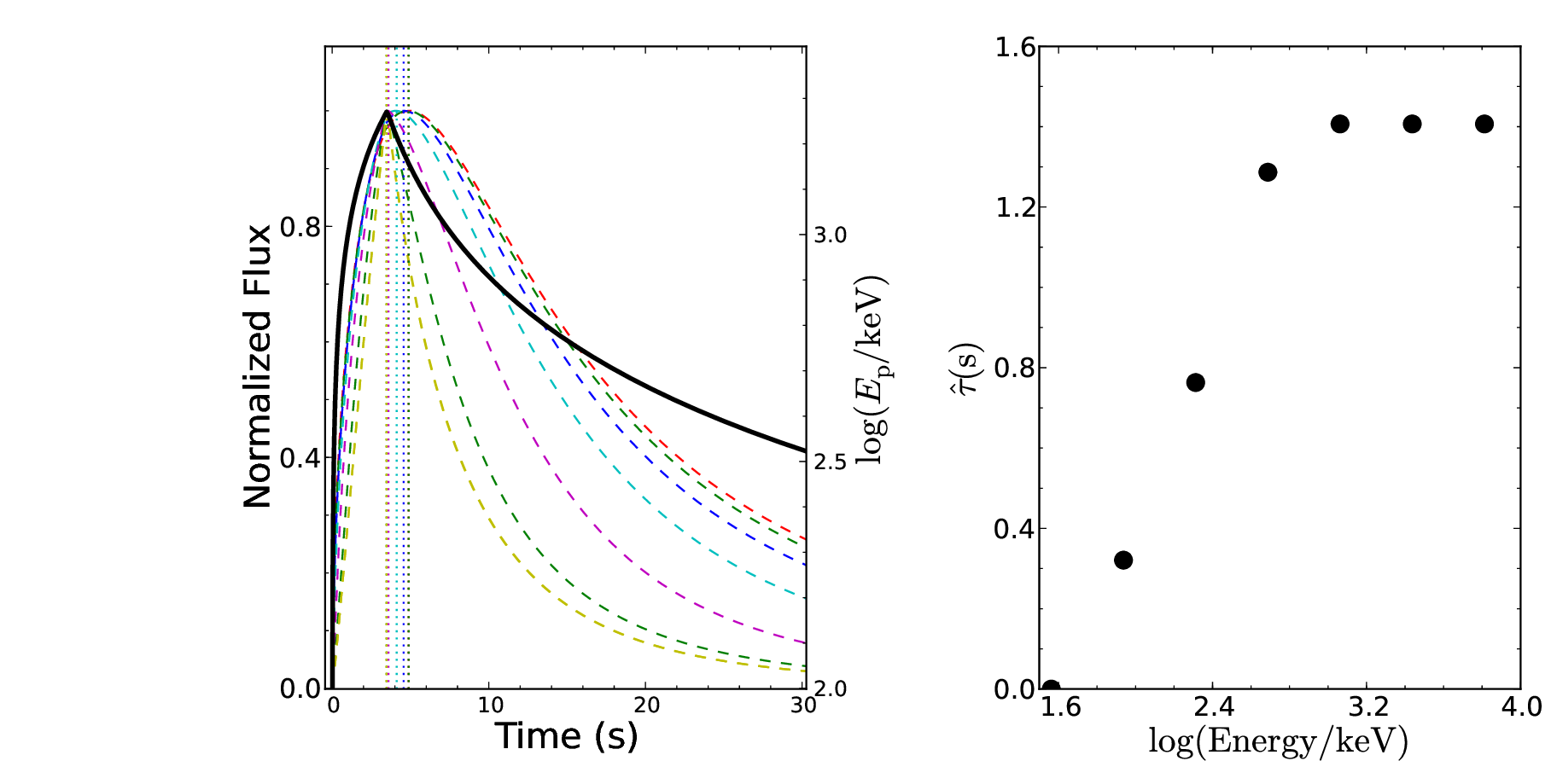} \\
\includegraphics[origin=c,angle=0,scale=1,width=0.90\textwidth,height=0.30\textheight]{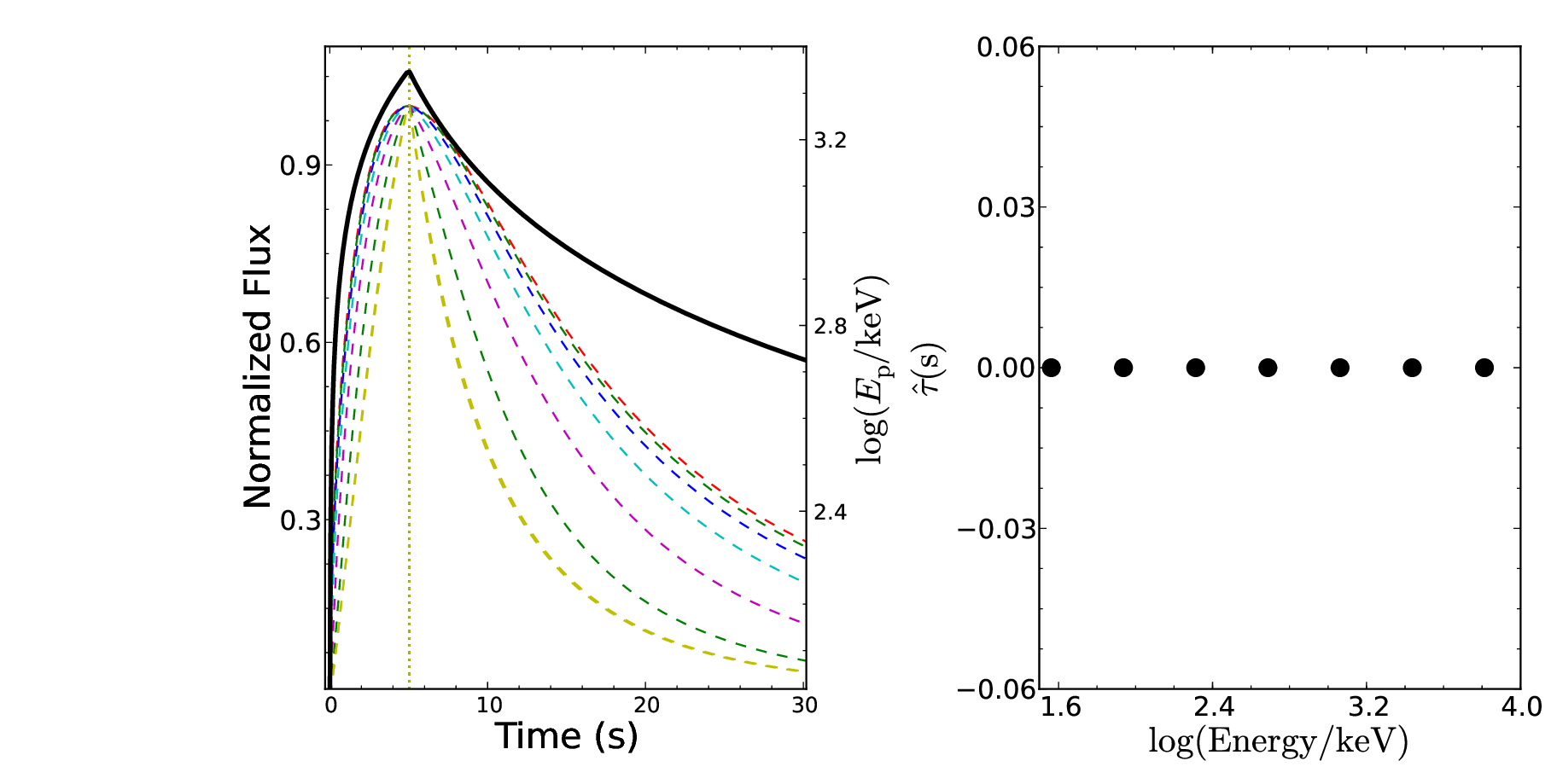} \\
\includegraphics[origin=c,angle=0,scale=1,width=0.90\textwidth,height=0.30\textheight]{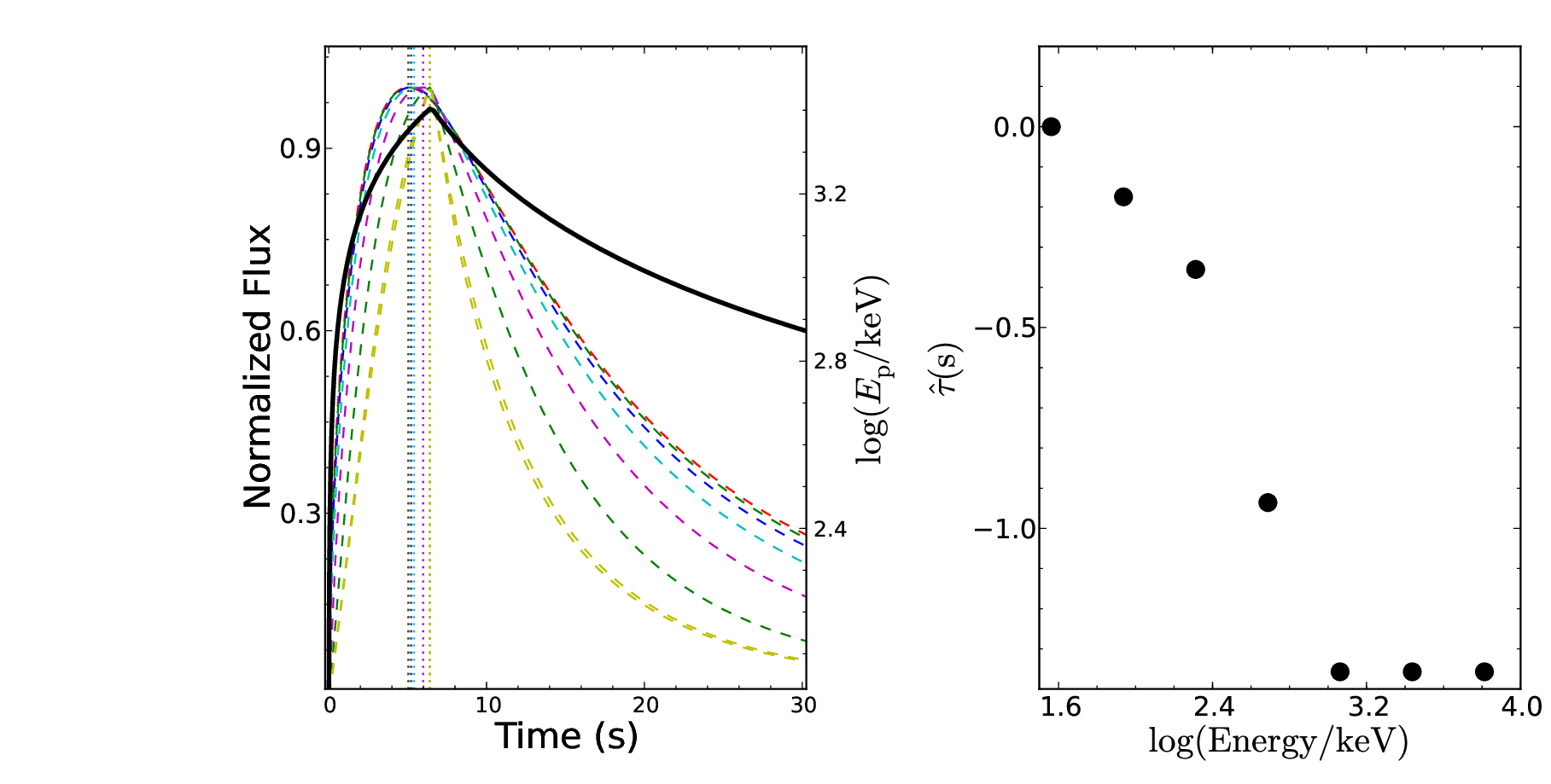} \\
\end{tabular}
\hfill
\vspace{4em}
\caption{Simulated light curves (left panels) and the obtained ${\hat{\tau}}$-$E$ relations (right panels) for Case (II),
where the upper, middle, and lower sub-figures are the results from the simulations with
$\xi$=0, -1.5, and 1.5, respectively.}\label{myfigJ}
\end{figure*}

%%%%%%%%%%%%%%%%%%%%%%%%%%%%%%%%%%%%%%%%%%%%%%%%%%%%%%%%%%%%%%%%%%%%%%%%%%%
%%%%%%%%%%%%%%%%%%%%%%%%%%%%%%%%%%%%%%%%%%%%%%%%%%%%%%%%%%%%%%%%%%%%%%%%%%
\begin{deluxetable}{ccccccc}
\tabletypesize{\tiny}
\tablecaption{Results of light curve fitting with Equation~(\ref{KRL}) for 92 pulses. }
\tablewidth{0pt}
\tabcolsep=2.5pt
\tablehead{ \colhead{GRB (Energy bin)}& \colhead{$I_{\rm m}$} & \colhead{$t_{\rm m}$}& \colhead{$r$}& \colhead{$d$} & \colhead{$t_{\rm 0}$} & \colhead{$\chi^2_{\rm r}$} }
\startdata
\hline
bn080906212  &  &   &   &   &   &  \\
NaI$^{a}$  &  156.04 $\pm$ 4.25 &  0.93 $\pm$ 0.03 &  1.12  $\pm$ 0.08 &  9.29  $\pm$ 3.20 &  0.23 &  1.17 \\
bn0810091402  &  &   &   &   &   &  \\
NaI  &  157.84 $\pm$ 2.31 &  42.23 $\pm$ 0.05 &  3.09  $\pm$ 0.10 &  6.69  $\pm$ 0.53 &  -35.06 &  1.34 \\
N1  &  107.40 $\pm$ 1.70 &  42.43 $\pm$ 0.06 &  2.32  $\pm$ 0.08 &  8.01  $\pm$ 0.86 &  -35.97 &  1.27 \\
N2  &  46.13 $\pm$ 1.35 &  41.85 $\pm$ 0.08 &  3.44  $\pm$ 0.22 &  7.33  $\pm$ 1.14 &  -35.56 &  1.00 \\
N3  &  8.66 $\pm$ 0.88 &  41.52 $\pm$ 0.25 &  4.11  $\pm$ 1.14 &  3.93  $\pm$ 1.44 &  -36.59 &  0.84 \\
bn081125496  &  &   &   &   &   &  \\
\enddata

\label{mytabA}
\tablenotetext{a}{\bf NaI: 8-1000 keV; N1: 8-25 keV; N2: 25-50 keV($\textcircled{1}$); N3: 50-100 keV ($\textcircled{2}$); N4: 100-300 keV ($\textcircled{3}$); N5: 300-1000 keV ($\textcircled{4}$); B1: 300-1000 keV ($\textcircled{5}$); B2: 1000-5000 keV ($\textcircled{6}$).}

\end{deluxetable}

%%%%%%%%%%%%%%%%%%%%%%%%%%%%%%%%%%%%%%%%%%%%%%%%%%%%%%%%%%%%%%%%%%%%%%%%%%%%%%%%%%
%%%%%%%%%%%%%%%%%%%%%%%%%%%%%%%%%%%%%%%%%%%%%%%%%%%%%%%%%%%%%%%%%%%%%%%%%%%%%%%%%%
\begin{deluxetable}{ccccccc}
\tabletypesize{\tiny}
\tablecaption{Results of the time-resolved spectral analysis for 92  pulses. }
\tablewidth{0pt}
\tabcolsep=2.5pt
\tablehead{ \colhead{Timebin (s)} & \colhead{Amplitude} & \colhead{$E_{p}$ (keV)} & \colhead{$\alpha$} & \colhead{$\beta$} & \colhead{Flux ($\rm erg/s.cm^2$)} & \colhead{$\chi^2_{\rm r}$}}
\startdata
\hline
GRB 080906B (bn080906212) & & & & & \\
-0.26  -0.59  & 0.06  $\pm$ 0.01  & 267.40  $\pm$ 46.36  & -0.53  $\pm$ 0.13  & -2.05  $\pm$ 0.18  & 2.50E-06 $\pm$ 1.08E-07 & 1.07  \\
0.59  -0.94  & 0.21  $\pm$ 0.06  & 160.72  $\pm$ 25.85  & -0.43  $\pm$ 0.16  & -1.98  $\pm$ 0.11  & 4.46E-06 $\pm$ 1.80E-07 & 1.06  \\
0.94  -1.37  & 0.16  $\pm$ 0.05  & 158.20  $\pm$ 26.18  & -0.47  $\pm$ 0.17  & -2.05  $\pm$ 0.13  & 3.44E-06 $\pm$ 1.58E-07 & 0.96  \\
1.37  -1.81  & 0.15  $\pm$ 0.04  & 171.11  $\pm$ 24.73  & -0.46  $\pm$ 0.15  & -2.27  $\pm$ 0.22  & 3.21E-06 $\pm$ 1.62E-07 & 0.87  \\
1.81  -5.82  & 0.04  $\pm$ 0.03  & 82.30  $\pm$ 28.90  & -0.65  $\pm$ 0.39  & -1.87  $\pm$ 0.11  & 4.86E-07 $\pm$ 3.59E-08 & 0.92  \\
GRB 081009A (1st pulse bn0810091401) & & & & & \\
-0.64  -1.31  & 1.37  $\pm$ 1.18  & 24.47  $\pm$ 1.42  & -0.39  $\pm$ 0.32  & -2.48  $\pm$ 0.05  & 1.18E-06 $\pm$ 3.45E-08 & 0.93  \\
\enddata
\label{mytabB}
\end{deluxetable}

%%%%%%%%%%%%%%%%%%%%%%%%%%%%%%%%%%%%%%%%%%%%%%%%%%%%%%%%%%%%%%%%%%%%%%%%%%%%%%%%%%
%%%%%%%%%%%%%%%%%%%%%%%%%%%%%%%%%%%%%%%%%%%%%%%%%%%%%%%%%%%%%%%%%%%%%%%%%%%%%%%%%%
\begin{deluxetable}{cccccc}
\tabletypesize{\tiny}
\tablecaption{Results of the spectral lag analysis for 92 pulses. }
\tablewidth{0pt}
\tabcolsep=2.5pt
\tablehead{\colhead{GRB (Timebin/s)} & \colhead{Energy band$^{a}$} & \colhead{${\hat{\tau}}$ (s)} & \colhead{GRB(Timebin/s)} & \colhead{Energy band} & \colhead{${\hat{\tau}}$ (s)}}
%\tablehead{\colhead {Timebin (s)} & & & \colhead{Timebin (s)} & &}
\startdata
\hline
bn080906212 & $\textcircled{1}$ & 0.20  $\pm$ 0.08  & bn0810091401 & $\textcircled{1}$ & 0.08  $\pm$ 0.02  \\
(0.0-6.0) & $\textcircled{2}$ & 0.35  $\pm$ 0.07  & (0.0-10.0) & $\textcircled{2}$ & 0.29  $\pm$ 0.03  \\
& $\textcircled{3}$ & 0.44  $\pm$ 0.07  & & $\textcircled{3}$ & 0.53  $\pm$ 0.04  \\
& $\textcircled{4}$ & 0.68  $\pm$ 0.14  & bn081125496 & $\textcircled{1}$ & 0.36  $\pm$ 0.17  \\
bn0810091402 & $\textcircled{1}$ & 0.47  $\pm$ 0.07  & (0.0-15.0) & $\textcircled{2}$ & 0.99  $\pm$ 0.16  \\
(36.0-52.0) & $\textcircled{2}$ & 0.83  $\pm$ 0.13  & & $\textcircled{3}$ & 1.70  $\pm$ 0.17  \\
& $\textcircled{3}$ & 0.58  $\pm$ 0.53  & & $\textcircled{4}$ & 2.33  $\pm$ 0.28  \\
bn081222204 & $\textcircled{1}$ & 0.04  $\pm$ 0.17  &bn081224887 & $\textcircled{1}$ & 0.36  $\pm$ 0.23  \\
\enddata
\label{mytabC}
\tablenotetext{a}{$\textcircled{1}$: 25-50 keV (NaI); $\textcircled{2}$: 50-100 keV (NaI); $\textcircled{3}$: 100-300 keV (NaI); $\textcircled{4}$: 300-1000 keV (NaI);
$\textcircled{5}$: 250-1000 keV (BGO); $\textcircled{6}$: 1000-5000 keV (BGO); $\textcircled{7}$: 25-40 keV (NaI); $\textcircled{8}$: 40-60 keV (NaI) ; $\textcircled{9}$: 60-100 keV (NaI); $\textcircled{10}$: 16-25 keV}
\end{deluxetable}

%%%%%%%%%%%%%%%%%%%%%%%%%%%%%%%%%%%%%%%%%%%%%%%%%%%%%%%%%%%%%%%%%%%%%%%%%%%%%%%%%%
%%%%%%%%%%%%%%%%%%%%%%%%%%%%%%%%%%%%%%%%%%%%%%%%%%%%%%%%%%%%%%%%%%%%%%%%%%%%%%%%%%
\begin{deluxetable}{ccccccc}
\tabletypesize{\tiny}
\tablecaption{Values of ${\hat{\tau}}_{31}$, $W$, $k_{{{\hat{\tau}}}}$ and $k_{E}$ for 92 pulses. }
\tablewidth{0pt}
\tabcolsep=2.5pt
\tablehead{ \colhead{GRB}& \colhead{Evolution$^{a}$}& \colhead{${{\hat{\tau}}}_{31}$}& \colhead{$W$}& \colhead{$k_{{{\hat{\tau}}}}$}& \colhead{$k_{E}$} }
\startdata
\hline
bn080906212 & H & 0.24  $\pm$ 0.05  & 1.86  & 0.37  $\pm$ 0.11  & -0.38  $\pm$ 0.14  & \\
bn0810091402 & H & 0.40  $\pm$ 13.99  & 5.97  & 1.36  $\pm$ 0.60  & -0.69  $\pm$ 0.09  & \\
bn081125496 & H & 1.37  $\pm$ 0.10  & 5.25  & 1.74  $\pm$ 0.25  & -0.71  $\pm$ 0.07  & \\
bn081222204 & H & 0.76  $\pm$ 0.16  & 6.84  & 0.98  $\pm$ 0.31  & -0.58  $\pm$ 0.26  & \\
bn081224887 & H & 1.47  $\pm$ 0.11  & 5.67  & 1.68  $\pm$ 0.22  & -0.68  $\pm$ 0.05  & \\
bn0901310901 & H & 0.18  $\pm$ 0.05  & 0.84  & 0.31  $\pm$ 0.10  & -0.83  $\pm$ 0.18  & \\
bn0901310903 & H & 0.12  $\pm$ 0.03  & 0.81  & 0.18  $\pm$ 0.07  & -0.28  $\pm$ 0.09  & \\
bn0905243461 & H & 0.47  $\pm$ 0.18  & 4.91  & 0.54  $\pm$ 0.54  & -0.24  $\pm$ 0.14  & \\
\enddata
\label{mytabD}
\tablenotetext{a}{H: H2S spectral evolution pattern; T: Tracking spectral evolution pattern.}
\end{deluxetable}

\clearpage
%%%%%%%%%%%%%%%%%%%%%%%%%%%%%%%%%%%%%%%%%%%%%%%%%%%%%%%%%%%%%%%%%%%%%%%%%%


\begin{thebibliography}{}
\bibitem[Ackermann et al.(2010)]{Ackermann_M-2010-Asano_K}Ackermann, M., Asano, K., Atwood, W.~B., et al.\ 2010, \apj, 716, 1178
     % Fermi Observations of GRB 090510: A Short-Hard Gamma-ray Burst with an Additional, Hard Power-law Component from 10 keV TO GeV Energies
     % http://adsabs.harvard.edu/abs/2010ApJ...716.1178A
\bibitem[Arimoto et al.(2010)]{Arimoto_M-2010-Kawai_N} Arimoto, M., Kawai, N., Asano, K., et al.\ 2010, \pasj, 62, 487
\bibitem[Band \& Trzhaskovskaya(1993)]{Band_IM-1993-Trzhaskovskaya_MB}Band, I.~M., \& Trzhaskovskaya, M.~B.\ 1993, Atomic Data and Nuclear Data Tables, 55, 43
     % Internal Conversion Coefficients for Low-Energy Nuclear Transitions
     % http://adsabs.harvard.edu/abs/1993ADNDT..55...43B
\bibitem[Band(1997)]{Band_DL-1997}Band, D.~L.\ 1997, \apj, 486, 928
     % Gamma-Ray Burst Spectral Evolution through Cross-Correlations of Discriminator Light Curves
     % http://adsabs.harvard.edu/abs/1997ApJ...486..928B
\bibitem[Bernardini et al.(2015)]{Bernardini_MG-2015-Ghirlanda_G}Bernardini, M.~G., Ghirlanda, G., Campana, S., et al.\ 2015, \mnras, 446, 1129
     % Comparing the spectral lag of short and long gamma-ray bursts and its relation with the luminosity
     % http://adsabs.harvard.edu/abs/2015MNRAS.446.1129B
\bibitem[Cheng et al.(1995)]{Cheng_LX-1995-Ma_YQ}Cheng, L.~X., Ma, Y.~Q., Cheng, K.~S., Lu, T., \& Zhou, Y.~Y.\ 1995, \aap, 300, 746
     % The time delay of gamma-ray bursts in the soft energy band.
     % http://adsabs.harvard.edu/abs/1995A&A...300..746C
\bibitem[Colgate(1974)]{Colgate_SA-1974}Colgate, S.~A.\ 1974, \apj, 187, 333
     % Early Gamma Rays from Supernovae
     % http://adsabs.harvard.edu/abs/1974ApJ...187..333C
\bibitem[Deng \& Zhang(2014)]{deng14}Deng, W., \& Zhang, B.\ 2014, \apj, 785, 112
\bibitem[Deng et al.(2015)]{deng15}Deng, W., Li, H., Zhang, B., \& Li, S.\ 2015, \apj, 805, 163
\bibitem[Eichler et al.(1989)]{Eichler_D-1989-Livio_M}Eichler, D., Livio, M., Piran, T., \& Schramm, D.~N.\ 1989, \nat, 340, 126
     % Nucleosynthesis, neutrino bursts and gamma-rays from coalescing neutron stars
     % http://adsabs.harvard.edu/abs/1989Natur.340..126E
\bibitem[Fishman et al.(1994)]{Fishman_GJ-1994-Bhat_PN}Fishman, G.~J., Bhat, P.~N., Mallozzi, R., et al.\ 1994, Science, 264, 1313
     % Discovery of Intense Gamma-Ray Flashes of Atmospheric Origin
     % http://adsabs.harvard.edu/abs/1994Sci...264.1313F
\bibitem[Gehrels et al.(2006)]{gehrels06}Gehrels, N., Norris, J.~P., Barthelmy, S.~D., et al. \ 2006, \nat, 444, 1044
\bibitem[Golenetskii et al.(1983)]{Golenetskii_SV-1983-Mazets_EP}Golenetskii, S.~V., Mazets, E.~P., Aptekar, R.~L., \& Ilinskii, V.~N.\ 1983, \nat, 306, 451
     % Correlation between luminosity and temperature in gamma-ray burst sources
     % http://adsabs.harvard.edu/abs/1983Natur.306..451G
\bibitem[Hakkila et al.(2008)]{Hakkila_J-2008-Giblin_TW}Hakkila, J., Giblin, T.~W., Norris, J.~P., Fragile, P.~C., \& Bonnell, J.~T.\ 2008, \apjl, 677, L81
     % Correlations between Lag, Luminosity, and Duration in Gamma-Ray Burst Pulses
% ---------------------------------------------------------------------------------
\bibitem[Hakkila \& Nemiroff(2009)]{Hakkila2009} Hakkila, J., \& Nemiroff, R.~J.\ 2009, \apj, 705, 372

\bibitem[Hakkila \& Preece(2011)]{Hakkila2011} Hakkila, J., \& Preece, R.~D.\ 2011, \apj, 740, 104

\bibitem[Hakkila \& Preece(2013)]{Hakkila2013} Hakkila, J.~E., \& Preece, R.~D.\ 2013, AAS/High Energy Astrophysics Division \#13, 13, 124.01

\bibitem[Hakkila \& Preece(2014)]{Hakkila2014} Hakkila, J., \& Preece, R.~D.\ 2014, \apj, 783, 88

\bibitem[Hakkila et al.(2015)]{Hakkila2015} Hakkila, J., Lien, A., Sakamoto, T., et al.\ 2015, \apj, 815, 134

\bibitem[Hakkila et al.(2018)]{Hakkila2018} Hakkila, J., Horv{\'a}th, I., Hofesmann, E., \& Lesage, S.\ 2018, \apj, 855, 101
% ----------------------------------------------------------------------------------
\bibitem[Hu et al.(2014)]{Hu_YD-2014-Liang_EW}Hu, Y.-D., Liang, E.-W., Xi, S.-Q., et al.\ 2014, \apj, 789, 145
     % Internal Energy Dissipation of Gamma-Ray Bursts Observed with Swift: Precursors, Prompt Gamma-Rays, Extended Emission, and Late X-Ray Flares
     % http://adsabs.harvard.edu/abs/2014ApJ...789..145H
\bibitem[Kocevski et al.(2003)]{Kocevski_D-2003-Ryde_F}Kocevski, D., Ryde, F., \& Liang, E.\ 2003, \apj, 596, 389
     % Search for Relativistic Curvature Effects in Gamma-Ray Burst Pulses
     % http://adsabs.harvard.edu/abs/2003ApJ...596..389K
\bibitem[Kumar \& Zhang(2015)]{Kumar_P-2015-Zhang_B}Kumar, P., \& Zhang, B.\ 2015, \physrep, 561, 1
     % The physics of gamma-ray bursts {\amp} relativistic jets
     % http://adsabs.harvard.edu/abs/2015PhR...561....1K
\bibitem[Lazarian et al.(2018)]{lazarian18}Lazarian, A., Zhang, B., \& Xu, S.\ 2018, \apj, submitted (arXiv:1801.04061)
\bibitem[Liang \& Kargatis(1996)]{Liang_E-1996-Kargatis_V}Liang, E., \& Kargatis, V.\ 1996, \nat, 381, 49
     % Dependence of the spectral evolution of {$\gamma$}-ray bursts on their photon fluence
     % http://adsabs.harvard.edu/abs/1996Natur.381...49L
\bibitem[Liang et al.(2006)]{Liang_EW-2006-Zhang_BB}Liang, E.-W., Zhang, B.-B., Stamatikos, M., et al.\ 2006, \apjl, 653, L81
     % Temporal Profiles and Spectral Lags of XRF 060218
     % http://adsabs.harvard.edu/abs/2006ApJ...653L..81L
\bibitem[Lu et al.(2010)]{Lu_RJ-2010-Hou_SJ}Lu, R.-J., Hou, S.-J., \& Liang, E.-W.\ 2010, \apj, 720, 1146
     % The E $_{p}$-flux Correlation in the Rising and Decaying Phases of gamma-ray Burst Pulses: Evidence for Viewing Angle Effect?
     % http://adsabs.harvard.edu/abs/2010ApJ...720.1146L
\bibitem[Lu et al.(2012)]{Lu_RJ-2012-Wei_JJ}Lu, R.-J., Wei, J.-J., Liang, E.-W., et al.\ 2012, \apj, 756, 112
     % A Comprehensive Analysis of Fermi Gamma-Ray Burst Data. II. E $_{p}$ Evolution Patterns and Implications for the Observed Spectrum-Luminosity Relations
     % http://adsabs.harvard.edu/abs/2012ApJ...756..112L
\bibitem[Margutti et al.(2010)]{Margutti_R-2010-Guidorzi_C}Margutti, R., Guidorzi, C., Chincarini, G., et al.\ 2010, \mnras, 406, 2149
     % Lag-luminosity relation in {$\gamma$}-ray burst X-ray flares: a direct link to the prompt emission
     % http://adsabs.harvard.edu/abs/2010MNRAS.406.2149M
\bibitem[McBreen et al.(2008)]{McBreen_S-2008-Foley_S}McBreen, S., Foley, S., Watson, D., et al.\ 2008, \apjl, 677, L85
     % The Spectral Lag of GRB 060505: A Likely Member of the Long-Duration Class
     % http://adsabs.harvard.edu/abs/2008ApJ...677L..85M
\bibitem[Narayan et al.(1992)]{Narayan_R-1992-Paczynski_B}Narayan, R., Paczynski, B., \& Piran, T.\ 1992, \apjl, 395, L83
     % Gamma-ray bursts as the death throes of massive binary stars
     % http://adsabs.harvard.edu/abs/1992ApJ...395L..83N
\bibitem[Norris \& Bonnell(2006)]{Norris_JP-2006-Bonnell_JT}Norris, J.~P., \& Bonnell, J.~T.\ 2006, \apj, 643, 266
     % Short Gamma-Ray Bursts with Extended Emission
     % http://adsabs.harvard.edu/abs/2006ApJ...643..266N
\bibitem[Norris et al.(1986)]{Norris_JP-1986-Share_GH}Norris, J.~P., Share, G.~H., Messina, D.~C., et al.\ 1986, \apj, 301, 213
     % Spectral evolution of pulse structures in gamma-ray bursts
     % http://adsabs.harvard.edu/abs/1986ApJ...301..213N
\bibitem[Norris et al.(1996)]{Norris_JP-1996-Nemiroff_RJ}Norris, J.~P., Nemiroff, R.~J., Bonnell, J.~T., et al.\ 1996, \apj, 459, 393
     % Attributes of Pulses in Long Bright Gamma-Ray Bursts
     % http://adsabs.harvard.edu/abs/1996ApJ...459..393N
\bibitem[Norris et al.(2000)]{Norris_JP-2000-Marani_GF}Norris, J.~P., Marani, G.~F., \& Bonnell, J.~T.\ 2000, \apj, 534, 248
     % Connection between Energy-dependent Lags and Peak Luminosity in Gamma-Ray Bursts
     % http://adsabs.harvard.edu/abs/2000ApJ...534..248N
\bibitem[Norris et al.(2005)]{Norris_JP-2005-Bonnell_JT}Norris, J.~P., Bonnell, J.~T., Kazanas, D., et al.\ 2005, \apj, 627, 324
     % Long-Lag, Wide-Pulse Gamma-Ray Bursts
     % http://adsabs.harvard.edu/abs/2005ApJ...627..324N
\bibitem[Paczynski(1986)]{Paczynski_B-1986}Paczynski, B.\ 1986, \apjl, 308, L43
     % Gamma-ray bursters at cosmological distances
     % http://adsabs.harvard.edu/abs/1986ApJ...308L..43P
\bibitem[Page et al.(2007)]{Page_KL-2007-Willingale_R}Page, K.~L., Willingale, R., Osborne, J.~P., et al.\ 2007, \apj, 663, 1125
     % GRB 061121: Broadband Spectral Evolution through the Prompt and Afterglow Phases of a Bright Burst
     % http://adsabs.harvard.edu/abs/2007ApJ...663.1125P
\bibitem[Preece et al.(2000)]{Preece_RD-2000-Briggs_MS}Preece, R.~D., Briggs, M.~S., Mallozzi, R.~S., et al.\ 2000, \apjs, 126, 19
     % The BATSE Gamma-Ray Burst Spectral Catalog. I. High Time Resolution Spectroscopy of Bright Bursts Using High Energy Resolution Data
     % http://adsabs.harvard.edu/abs/2000ApJS..126...19P
\bibitem[Preece et al.(2016)]{Preece2016} Preece, R., Goldstein, A., Bhat, N., et al.\ 2016, \apj, 821, 12
\bibitem[Qin et al.(2013)]{Qin_Y-2013-Liang_EW}Qin, Y., Liang, E.-W., Liang, Y.-F., et al.\ 2013, \apj, 763, 15
     % A Comprehensive Analysis of Fermi Gamma-Ray Burst Data. III. Energy-dependent T $_{90}$ Distributions of GBM GRBs and Instrumental Selection Effect on Duration Classification
     % http://adsabs.harvard.edu/abs/2013ApJ...763...15Q
\bibitem[R{\'a}cz et al.(2018)]{Racz_II-2018-Balazs_LG}R{\'a}cz, I.~I., Bal{\'a}zs, L.~G., Horvath, I., T{\'o}th, L.~V., \& Bagoly, Z.\ 2018, \mnras, 475, 306
     % Statistical properties of Fermi GBM GRBs' spectra
     % http://adsabs.harvard.edu/abs/2018MNRAS.475..306R
\bibitem[Schaefer(2007)]{Schaefer_BE-2007}Schaefer, B.~E.\ 2007, \apj, 660, 16
     % The Hubble Diagram to Redshift $\gt$6 from 69 Gamma-Ray Bursts
     % http://adsabs.harvard.edu/abs/2007ApJ...660...16S
\bibitem[Sonbas et al.(2013)]{Sonbas_E-2013-MacLachlan_GA}Sonbas, E., MacLachlan, G.~A., Shenoy, A., Dhuga, K.~S., \& Parke, W.~C.\ 2013, \apjl, 767, L28
     % A New Correlation between GRB X-Ray Flares and the Prompt Emission
     % http://adsabs.harvard.edu/abs/2013ApJ...767L..28S
\bibitem[Sultana et al.(2012)]{Sultana_J-2012-Kazanas_D}Sultana, J., Kazanas, D., \& Fukumura, K.\ 2012, \apj, 758, 32
     % Luminosity Correlations for Gamma-Ray Bursts and Implications for Their Prompt and Afterglow Emission Mechanisms
     % http://adsabs.harvard.edu/abs/2012ApJ...758...32S
\bibitem[Uhm \& Zhang(2016)]{uhm16}Uhm, Z.~L., \& Zhang, B.\ 2016, \apj, 825, 97
\bibitem[Uhm et al.(2018)]{uhm18}Uhm, Z.~L., Zhang, B., \& Racusin, J. L.\ 2018, \apj, submitted (arXiv:1801.09183)
\bibitem[Ukwatta et al.(2010)]{Ukwatta_TN-2010-Stamatikos_M}Ukwatta, T.~N., Stamatikos, M., Dhuga, K.~S., et al.\ 2010, \apj, 711, 1073
     % Spectral Lags and the Lag-Luminosity Relation: An Investigation with Swift BAT Gamma-ray Bursts
     % http://adsabs.harvard.edu/abs/2010ApJ...711.1073U
\bibitem[Ukwatta et al.(2012)]{Ukwatta_TN-2012-Dhuga_KS}Ukwatta, T.~N., Dhuga, K.~S., Stamatikos, M., et al.\ 2012, \mnras, 419, 614
     % The lag-luminosity relation in the GRB source frame: an investigation with Swift BAT bursts
     % http://adsabs.harvard.edu/abs/2012MNRAS.419..614U
\bibitem[Wheaton et al.(1973)]{Wheaton_WA-1973-Ulmer_MP}Wheaton, W.~A., Ulmer, M.~P., Baity, W.~A., et al.\ 1973, \apjl, 185, L57
     % The Direction and Spectral Variability of a Cosmic Gamma-Ray Burst
     % http://adsabs.harvard.edu/abs/1973ApJ...185L..57W
\bibitem[Woosley \& Bloom(2006)]{Woosley_SE-2006-Bloom_JS}Woosley, S.~E., \& Bloom, J.~S.\ 2006, \araa, 44, 507
     % The Supernova Gamma-Ray Burst Connection
     % http://adsabs.harvard.edu/abs/2006ARA&A..44..507W
\bibitem[Woosley(1993)]{Woosley_SE-1993}Woosley, S.~E.\ 1993, \apj, 405, 273
     % Gamma-ray bursts from stellar mass accretion disks around black holes
     % http://adsabs.harvard.edu/abs/1993ApJ...405..273W
\bibitem[Yi et al.(2006)]{Yi_T-2006-Liang_E}Yi, T., Liang, E., Qin, Y., \& Lu, R.\ 2006, \mnras, 367, 1751
     % On the spectral lags of the short gamma-ray bursts
     % http://adsabs.harvard.edu/abs/2006MNRAS.367.1751Y
\bibitem[Zhang \& Yan(2011)]{zhangyan11}Zhang, B., \& Yan, H.\ 2011, \apj, 726, 90
\bibitem[Zhang \& Zhang(2014)]{zhangzhang14}Zhang, B., \& Zhang, B.\ 2014, \apj, 782, 92
\bibitem[Zhang et al.(2009)]{zhangbb09}Zhang, B., Zhang, B.-B., Virgili, F. J., et al.\ 2009, \apj, 703, 1696
\bibitem[Zhang et al.(2011)]{zhangbb11}Zhang, B.-B., Zhang, B., Liang, E.-W., et al.\ 2011, \apj, 730, 141
\end{thebibliography}
\end{document}